\documentclass[12pt,a4,sans]{article}

\pdfoutput=1

\usepackage[latin1]{inputenc}
\usepackage{amssymb}
\usepackage{amsmath}
\usepackage{amsthm}
\usepackage{amsfonts}
\usepackage[pdftex]{graphicx}
\usepackage{geometry}
\usepackage{braket}
\usepackage{enumerate}
\usepackage[usenames,dvipsnames]{xcolor}
\usepackage{setspace}
\usepackage[backgroundcolor=white,linecolor=black]{todonotes}
\usepackage[in]{fullpage}
\usepackage{hyperref}
\usepackage{array}
\usepackage{wrapfig}
\usepackage[font=small,labelfont=bf]{caption}
\usepackage{anysize}
\usepackage{float}
\usepackage{graphicx}
\usepackage{xcolor}
\usepackage{mathtools}
\usepackage{youngtab}
\usepackage{setspace}
\usepackage{cite}

\numberwithin{equation}{section}

\marginsize{2.1cm}{2.1cm}{2cm}{2cm}
\hypersetup{
colorlinks=true,
linktoc=page,
citecolor=PineGreen,
linkcolor=BlueViolet,
urlcolor=MidnightBlue} 

\newcommand{\comments}[1]{}

\newcommand{\N}{\mathcal{N}}
\newcommand{\Q}{\mathcal{Q}}

\newcommand{\cO}{\mathcal{O}}
\renewcommand{\Q}{\mathcal{Q}}
\newcommand{\Tr}{\text{Tr}}

\newcommand{\nn}{\nonumber\\[3mm]}

\newcommand{\lara}[1]{\left\langle #1 \right\rangle}

\newcommand{\bea}{\begin{eqnarray}} 
\newcommand{\eea}{\end{eqnarray}}

\begin{document}

\makeatletter
\@addtoreset{equation}{section}
\makeatother
\renewcommand{\theequation}{\thesection.\arabic{equation}}

\rightline{QMUL-PH-16-03}
\vspace{1.8truecm}

\vspace{15pt}

%%%%%%%%%%%%%%%%%

{\LARGE{ 
\begin{center}
\bf Gauge Invariants and Correlators in \\ Flavoured Quiver Gauge Theories 
\end{center}
}} 

\vskip.5cm 

\thispagestyle{empty} \centerline{
 {\large \bf Paolo Mattioli
 \footnote{ {\tt p.mattioli@qmul.ac.uk}} }
 {\large \bf and Sanjaye Ramgoolam
 \footnote{ {\tt s.ramgoolam@qmul.ac.uk}} }}
 
\setcounter{footnote}{0} 
 
\vspace{.4cm}
\centerline{{\it Centre for Research in String Theory, School of Physics and Astronomy},}
\centerline{{ \it Queen Mary University of London},} \centerline{{\it
 Mile End Road, London E1 4NS, UK}}

\vspace{1truecm}

%%%%%%%%%%%%%%%%%
\thispagestyle{empty}

\centerline{\bf ABSTRACT}

\vskip.2cm 
In this paper we study the construction of holomorphic gauge invariant operators for general quiver gauge theories with flavour symmetries. 
Using a characterisation of the gauge invariants in terms of equivalence classes generated by permutation actions, along with representation theory results in symmetric groups and unitary groups, we give a diagonal basis for the 2-point functions of holomorphic and anti-holomorphic operators. 
This involves a generalisation of the previously constructed Quiver Restricted Schur operators to the flavoured case. The 3-point functions are derived and shown to be given in terms of networks of symmetric group branching coefficients. 
The networks are constructed through cutting and gluing operations on the quivers.

\newpage
%\pagenumbering{roman}
\tableofcontents

\pagebreak

\section{Introduction}

Finite $N$ aspects of AdS/CFT \cite{Maldacena:1997re,Witten:1998qj,Gubser:1998bc}, such as giant gravitons \cite{MST00}, the stringy exclusion principle \cite{MS98} and LLM geometries \cite{LLM}, have motivated the study of multi-matrix sectors of $\N=4$ SYM, associated with different BPS sectors of the theory. These multi-matrix systems are also of interest purely from the point of view of supersymmetric gauge theory and their moduli spaces (\emph{e.g.} \cite{MRW}).

In this paper we study correlation functions of holomorphic and anti-holomorphic gauge invariant operators in quiver gauge theories with flavour symmetries, in the zero coupling limit. 
This builds on the results in our previous paper focused on enumeration of gauge invariant operators \cite{quivwords} and proceeds to explicit construction of the operators and consideration of free field two and three point functions. 
These have non-trivial dependences on the structure of the operators and on the ranks of the gauge and flavour symmetries. 
Our results are exact in the ranks, and their expansions contain information about the planar limit as well as all order expansions. The techniques we use build on earlier work exploiting representation theory techniques in the context of $ \N =4$ SYM \cite{CJR01,CR02,BBFH04,KSS1,KSS2,BKS,KR1,BHR1,KR2,BHR2,Kimura:2012hp}. The zero coupling results contain information about a singular limit from the point of view of the dual AdS. For special BPS sectors, where non-renormalization theorems are available, the representation theory methods have made contact with branes and geometries in the semiclassical AdS background. 
These representation theoretic studies were extended beyond $\N=4$ SYM to ABJM \cite{ABJM} and conifolds \cite{Dey2011,DMMP1202,Mo2013,CapMoh2012}. The case of general quivers was studied in \cite{quivcalc} and 
related work on quivers has since appeared in \cite{Ber2015,dKS2014,dKN2014,LMP2013}.

In the context of AdS/CFT, adding matter to $ \N=4$ SYM introduces flavour symmetries \cite{KarKat02,EF1012,AFR1312, Ouyang:2003df, Levi:2005hh}. Typically, the added matter transforms in fundamental and anti-fundamental representations of these flavour symmetries.
Matrix invariants in flavoured gauge theories do not need to be invariant under the flavour group: on the contrary, they have free indices living in the representation carried by their constituent fields. In this paper, we consider a general class of flavoured free gauge theories parametrised by a quiver. A quiver is a directed graph comprising of round nodes (gauge groups) and square nodes (flavour groups). The directed edges which join the round nodes corresponds to fields transforming in the bi-fundamental representation of the gauge group, as illustrated in subsection \ref{Sec: definitions}.
Edges stretching between a round and a square node correspond to fields carrying a fundamental or antifundamental representation of the flavour group, depending on their orientation. We will call them simply quarks and antiquarks.

It was shown in \cite{quivcalc} that the quiver, besides being a compact way to encode all the gauge groups and the matter content of the theory, is a powerful computational tool for correlators of gauge invariants.
In that paper a generalisation of permutation group characters, called { \it quiver characters}, was introduced, involving branching coefficients of permutation groups in a non-trivial way. Obtaining the quiver character from the quiver diagram involves splitting each gauge node into \emph{two} nodes, called positive and negative nodes. The first one collects all the fields coming into the original node, while the second one collects all the fields outgoing from the original node. A new line is added to join the positive and the negative node of the split-node diagram. Each edge in this modified quiver is decorated with appropriate representation theory data, as will be explained in the following sections. 
The properties of these characters, which have natural pictorial representations, allowed the derivation of counting formulae for the gauge invariants 
and expressions for the correlation functions.

In this paper, we will be concerned with the construction of a basis for the Hilbert space of holomorphic matrix invariants for the class of quiver gauge theories with $\prod_a U(N_a)$ gauge group and $\prod_a SU(F_a)\times SU(\bar F_a)\times U(1)$ flavour group. This basis is obtained in terms of \emph{Quiver Restricted Schur Polynomials } $\cO_\Q(\pmb L)$, that we define in Section \ref{Sec: The Restricted Schur Basis}. 
These are a generalisation of the restricted Schur operators introduced in \cite{KSS1,KSS2,BKS,BCK08,BKS08}. In \cite{quivcalc}, the non-flavoured versions of these objects were called Generalised Restricted Schur operators, constructed in terms of quiver characters $\chi_{\Q} ( \pmb{L} )$
where $\pmb L$ is a collection of representation theory labels. In this flavoured case, we will find generalisations of these quiver characters, where 
the representation labels will include flavour states organised according to irreducible representations of the flavour groups. 
The advantages of using this approach is twofold. On the one hand, the Quiver Restricted Schur polynomials are orthogonal in the free field metric, as we will show, even for flavoured gauge theories. This leads to the simple expression for the two point function in eq. \eqref{ortho L}:
\begin{align}\label{orthogonality intro}
\left\langle
\mathcal O_{\mathcal Q}(\pmb L)\,
\mathcal O_{\mathcal Q}^\dagger(\pmb {L'})
\right\rangle
=\delta_{\pmb L,\pmb{L'}}\,\, c_{\vec n}\,\prod_a f_{N_a}(R_a)
\end{align}
In this equation $f_{N_a}(R_a)$ represents the product of weights of the $U(N_a)$ representation $R_a$, where $a$ runs over the gauge nodes of the quiver. $c_{\vec n}$ is a constant depending on the matter content of the matrix invariant $\mathcal O_{\mathcal Q}(\pmb L)$, 
given in (\ref{norm const c_n}). 
On the other hand, the Quiver Restricted Schur polynomial formalism offers a simple way to capture the finite $N$ constraints of matrix invariants. This can be seen directly from \eqref{orthogonality intro}: each $f_{N_a}(R_a)$ vanishes if the length of the first column of the $R_a$ Young diagram exceeds $N_a$.

In subsection \ref{sec: Holomorphic Gauge Invariant Operator Ring Structure Constants} we give an $N$-exact expression for the three point function of matrix invariants in the free limit. This computation is performed using the Quiver Restricted Schur polynomial basis. Specifically, we will derive the $G_{\pmb L^{(1)},\,\pmb L^{(2)},\,\pmb L^{(3)}}$ coefficients in
\begin{align}
\left\langle
\mathcal O_{\mathcal Q}(\pmb L^{(1)})\,
\mathcal O_{\mathcal Q}(\pmb {L}^{(2)})\,
\mathcal O_{\mathcal Q}^\dagger(\pmb L^{(3)})
\right\rangle
=c_{\vec n^{(3)}}\,\, G_{\pmb L^{(1)},\,\pmb L^{(2)},\,\pmb L^{(3)}}\,\prod_a f_{N_a}\left(R_a^{(3)}\right)
\end{align}
The analytical expression for $G_{\pmb L^{(1)},\,\pmb L^{(2)},\,\pmb L^{(3)}}$ looks rather complicated, but it can be easily understood in terms of diagrams. Although the identities we need appear somewhat complex, they all have a simple diagrammatic interpretation. Diagrammatics therefore play a central role in this paper: all the quantities we define and the calculational steps we perform can be visualised in terms of networks involving \emph{symmetric group branching coefficients} and \emph{Clebsch-Gordan coefficients}. Both these quantities are defined in Section \ref{Sec: The Restricted Schur Basis}. The quantity  $G_{\pmb L^{(1)},\,\pmb L^{(2)},\,\pmb L^{(3)}}$ is actually found to be a product 
over the gauge groups: for each gauge group there is a network of symmetric group branching coefficients and a 
single Clebsch-Gordan coefficient.

%%%%It is important to stress that these results are valid even at finite $N$.

The organisation of this paper is as follows. In the next section we establish the notation we will use throughout the paper, and we specify the class of quiver theories we will focus on.

In Section \ref{GIO's Quivers} we describe a permutation based approach to label matrix invariants of the flavoured gauge theories under study. A matrix invariant will be constructed using a set of permutations (schematically $ \sigma $) associated with gauge nodes of the quiver, and by a collection of fundamental and antifundamental states (schematically $ s , \bar s $ ) of the flavour group, associated with external flavour nodes. In this section we highlight how the simplicity of apparently complex formulae can be understood via diagrammatic techniques. We describe equivalence relations, generated by the action of 
permutations associated with edges of the quiver (schematically $ \eta$), acting on the gauge node permutations and flavour states. 
Equivalent data label the same matrix invariant. The equivalence is explained further and illustrated in Appendix \ref{app:Op Inv}.
The equivalences $\eta $ can be viewed as ``permutation gauge symmetries'', while the $( \sigma, s , \bar s ) $ can be viewed as ``matter fields'' for these permutation gauge symmetries. 

In Section \ref{Sec: The Restricted Schur Basis} we give a basis of the matrix invariants using representation theory data, $\pmb L$. 
This can be viewed as a {\it dual basis} where representation theory is used to perform a Fourier transformation on the equivalence classes of the permutation description. We refer to these gauge invariants, polynomial in the bi-fundamental and fundamental matter fields, as Quiver Restricted Schur polynomials. In this section we introduce the two main mathematical ingredients needed in this formalism. These are the symmetric group branching coefficients and the Clebsch-Gordan coefficients. Their definition will be accompanied by a corresponding diagram.
%They play a central role in the formulae that we will use.

In Section \ref{Sec: Two and Three Point Functions} we derive the results for the free field two and three point function of gauge invariants. 
In subsection \ref{sec: 2pt function} we show that the two point function which couples holomorphic and anti-holomorphic matrix invariants is diagonal in the basis of Quiver Restricted Schur polynomials. 
In subsection \ref{sec: Holomorphic Gauge Invariant Operator Ring Structure Constants} we give a diagrammatic description of the structure constants of the ring of Holomorphic Gauge Invariant Operators (GIOs). 
In particular, we present a step by step procedure to obtain such a diagram for the example of an $\N=2$ SQCD, starting from its split-node diagram.
Using these formulae, we identify four selection rules, all expressed in terms of symmetric group representation theory data. The analytical calculations are reported in Appendices \ref{ortho L derivation} and \ref{app:Holomorphic Ring SC}, and rely on the Quiver Restricted Schur polynomial technology introduced in the previous section. 

Finally, in Section \ref{Sec: Schur examples}, we give some examples of the matrix invariants we can build using our method, for the case of an $\N=2$ SQCD.

\subsection{Definitions and framework}\label{Sec: definitions}
In this paper we consider free quiver gauge theories with gauge group \(\prod_{a=1}^nU(N_a)\) and flavour symmetry of the general schematic form \(\prod_{a=1}^n U ( F_{a} ) \times U( \bar F_a) \). 
Specifically, to work in the most general configuration, we choose to focus our attention to the subgroup $ \prod_{a=1}^n[\times_{ \beta } U ( F_{a , \beta } ) \allowbreak \times_{ \gamma }  U( \bar F_{ a , \gamma}) ]$ of the flavour symmetry where $ F_a = \sum_{ \beta } F_{ a , \beta } $ and $ \bar F_a = \sum_{ \gamma } \bar F_{ a , \gamma }$. 
This more general flavour symmetry, where the $ U ( F_a ) \times U ( \bar F_a ) $ is broken to a product of unitary groups for the quarks and anti-quarks, is likely to be useful when interactions are turned on. 
Our calculations work without any significant modification for this case of product global symmetry, hence we will work in this generality. 

To recover the results for the global symmetry $ U( F_a ) \times U ( \bar F_a ) $ it is enough to drop the $ \beta , \gamma $ labels from all the equations that we are going to write. The constraint $F_a = \bar F_a $ solves chiral gauge anomaly conditions. 
As a last remark, notice that strictly speaking the global symmetry of the free theory contains only the determinant one part $ S ( U(F_{a, 1} ) \times U( F_{a,2} ) \times \cdots U (F_{a,M_a } ) \times U ( \bar F_{a,1} ) \times \cdots \times U ( \bar F_{a,\bar M_a} ) )$. This means that, although for simplicity we write $ \prod_{a=1}^n\left[\times_{ \beta } U ( F_{a , \beta } ) \times_{ \gamma } U( \bar F_{ a , \gamma}) \right]$ as the global symmetry, all the states we will write are neutral under the $ U(1) $ which acts with a phase on all of the chiral fields and with the opposite phase on all of the anti-chiral fields. This $U(1)$ is part of the $U(N_a)$ gauge symmetry. 

\begin{figure}[H]
\begin{center}\includegraphics[scale=0.9]{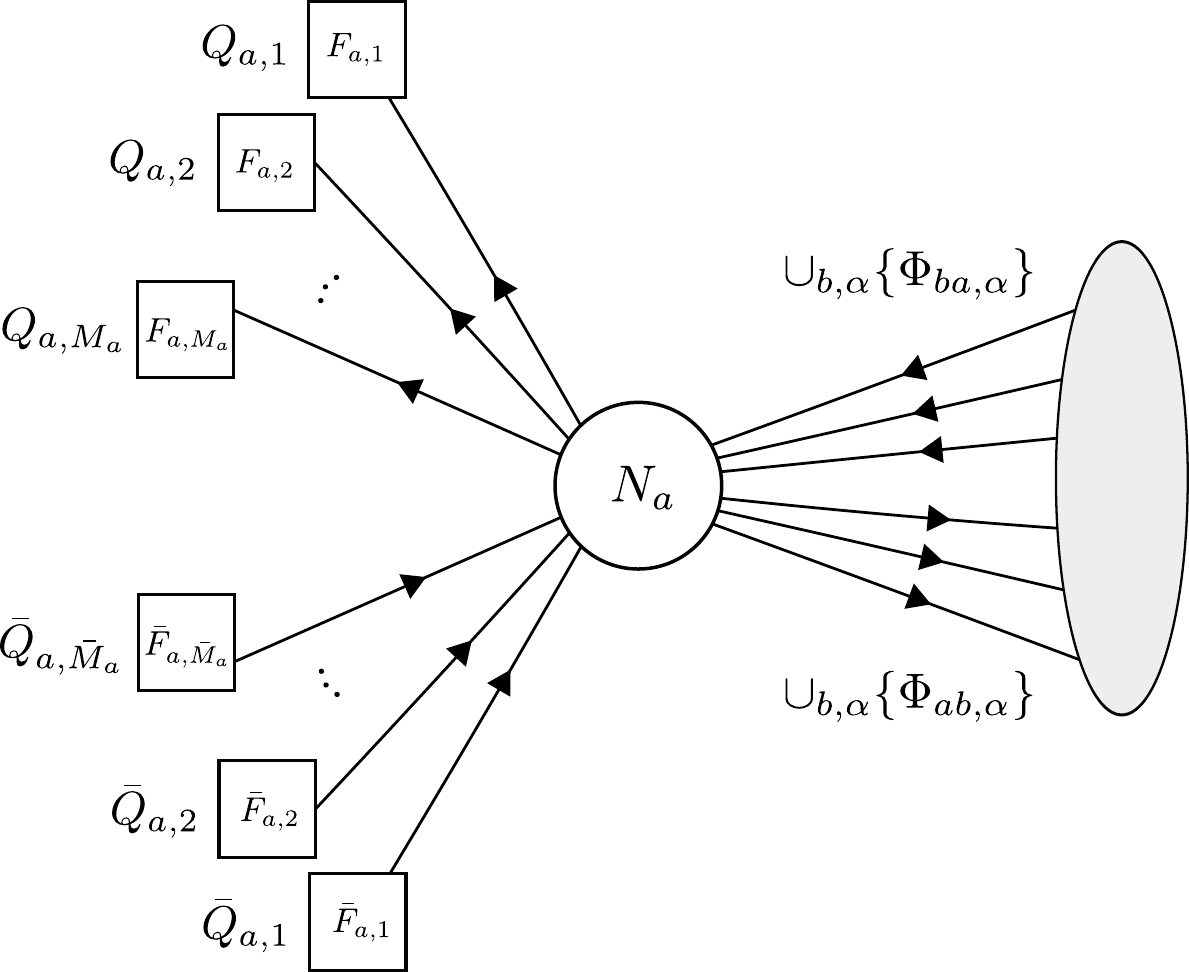}\\[1mm]
\caption{Pictorial representation of the fundamental fields (oriented edges), flavour group (square nodes) for a single gauge node labelled $a$.}\label{fig: quiver node}
\end{center}
\end{figure}

We now introduce the diagrammatic notation for the quivers. We follow the usual convention according to which round nodes in the quiver correspond to gauge groups, whereas square nodes correspond to global symmetries.
Fields leaving gauge node \(a\) and arriving at gauge node \(b\) are be denoted by \(\Phi_{ab,\alpha}\), and transform in the antifundamental representation $\bar V_{N_a}$ of \(U(N_a)\) and the fundamental representation $V_{N_b}$ of \(U(N_b)\). The third label $\alpha $ takes values in $ \{ 1,...,M_{ab}\}$, and is used to distinguishes between \(M_{ab}\) different fields that transform in the same way under the gauge group.
We can think of each \(\Phi_{ab,\alpha}\) as a map
\begin{align}
\Phi_{ab,\alpha}:\,\, V_{N_a}\rightarrow V_{N_b}
\end{align}
At every gauge node \(a\) we allow \(M_a\) different families of quarks $\{ Q_{a,\beta} , \beta=1,...,M_a \} $ transforming in the antifundamental of \(U(N_a)\) and \(\bar M_a\) different families of antiquarks $\{ \bar Q_{a,\gamma} , \gamma=1,...,\bar M_a \} $, transforming in the fundamental of \(U(N_a)\). As for the field $\Phi$, the greek letters \(\beta\) and \(\gamma\) distinguish the multiplicities of the quarks and antiquarks respectively. 
\(U(F_{a,\beta})\) and \(U(F_{a,\gamma})\) represent the flavour group of the quark \(Q_{a,\beta}\) and of the antiquark \(\bar Q_{a,\gamma}\) respectively. 
Figure \ref{fig: quiver node} explicitly show this field configuration for one node $a$ of the quiver. Table \ref{table: group charges} summarises instead all the gauge and flavour group representations carried by every field in the quiver.

\begin{center}
\begin{table}[H]
\centering
\begin{tabular}{|l|c|c||c|c|} 
\hline
 & $U(N_a)$ & $U(N_b)$& $U(F_{a,\beta})$ & $U( \bar F_{a,\gamma})$ \\[1mm] \hline
$\Phi_{ab,\alpha}$ & $\tiny\overline{\yng(1)}$ & $ \tiny{\yng(1)}$ & $\textbf{1}$ & $\textbf{1}$ \\[1mm] \hline
$\Phi_{aa,\alpha}$ & $\text{Adj}$ & $\textbf{1}$ & $\textbf{1}$ & $\textbf{1}$ \\[1mm] \hline
$ Q_{a,\beta}$ & $\tiny\overline{\yng(1)}$ &$\textbf{1}$& $\tiny{\yng(1)}$ & $\textbf{1}$ \\[1mm]\hline 
$\bar Q_{a,\gamma}$& $ \tiny{\yng(1)}$ & $\textbf{1}$ & $\textbf{1}$ & $\tiny\overline{\yng(1)}$ \\[1mm] \hline
\end{tabular}
\caption{Gauge and flavour group representations carried by \(\Phi_{ab,\alpha}\), \(Q_{a,\beta}\) and \(\bar Q_{a,\gamma}\). $\Box$, $\bar\Box$ and $\textbf{1}$ are respectively the fundamental, antifundamental and trivial representations of the corresponding group.}
\label{table: group charges}
\end{table}
\end{center}

\section{Gauge Invariant Operators and Permutations}\label{GIO's Quivers}
In this section we will present a systematic approach to list and label every holomorphic matrix invariant in quiver gauge theories of the type discussed above. We also allow for a flavour symmetry of the type discussed in Section \ref{Sec: definitions}.
The operators we consider are polynomial in the 
%$\{\Phi_{ab,\alpha}\}\cup\{Q_{a,\beta}\}\cup\{\bar Q_{a,\gamma}\}$ 
$\Phi$, $Q$ and $\bar Q$ type fields that are invariant under gauge transformations. Therefore, all colour indices are contracted to produce traces and products of traces of these fields. For example
\begin{equation}
\begin{array}{ll}
\cdot \left(\Phi_{ab}\Phi_{bc}\cdots\Phi_{ta}\right)&
\cdot \left(\Phi_{ab}\Phi_{bc}\right)\left(\Phi_{tt}\right)\\[2mm]
\cdot \left(\bar Q_a^k Q_{la}\right)&
\cdot (\bar Q_a^k \Phi_{ab}\Phi_{bc}\cdots\Phi_{qt}Q_{lq})
\end{array}
\end{equation}
and products thereof are suitable matrix invariants. In these examples round brackets denote contraction of gauge indices (\emph{i.e.} traces), while $k,l$ are flavour indices. The last two examples belong to the class of GIOs that in the literature has been called `generalised mesons' (see \emph{e.g.} \cite{Intriligator:2003mi}).
In order to label these matrix polynomials, the first ingredient we need to specify is the number of fundamental fields that they contain.
Let $n_{ab,\alpha}$ be the number of copies of $\Phi_{ab,\alpha}$ fields that are used to build the GIO. Similarly, let $n_{a,\beta}$ ($\bar n_{a,\gamma}$) be the number of copies of $Q_{a,\beta}$ quarks ($\bar Q_{a,\gamma}$ antiquarks) used in the GIO. 
In other words, the polynomial is characterised by degrees $ \vec {n}$ given by 
\begin{align}\label{vec n def}
\vec{n}=\cup_{a}\left\{\cup_{b,\alpha}\,n_{ab,\alpha};\cup_\beta\, n_{a,\beta};\cup_\gamma\, \bar n_{a,\gamma}\right\}
\end{align}
For fixed degrees there is a large number of gauge invariant polynomials, differing in how the gauge indices are contracted. 
To guarantee gauge invariance we have to impose that the GIO does not have any free gauge indices. This condition implies the constraint on \(\vec n\)
\begin{equation}\label{n_a def}
\left\{
\begin{array}{l}
n_a=\sum_{b,\alpha}n_{ab,\alpha}+\sum_\beta n_{a,\beta}=\sum_{b,\alpha}n_{ba,\alpha}+\sum_\gamma \bar n_{ a,\gamma}\qquad \forall a\\[2mm]
n_\alpha=\sum_a \sum_\beta n_{a,\beta}=\sum_a \sum_\gamma \bar n_{ a,\gamma}
\end{array}
\right.
\end{equation}

We now introduce a second vector-like quantity, $\vec s$. It will store the information about the states of the quarks and antiquarks in the matrix invariant.
To do so, let us first define the states
\begin{align}\label{bf s bar s}
|\pmb s_{a,\beta}\rangle\in V_{F_{a,\beta}}^{\otimes n_{a,\beta}}\,,\qquad
\langle\pmb {\bar s_{a,\gamma}}|\in \bar V_{\bar F_{a,\gamma}}^{\otimes \bar n_{a,\gamma}}
\end{align}
Here $V_{F_{a,\beta}}$ is the fundamental representation of $U(F_{a,\beta})$ and $\bar V_{\bar F_{a,\gamma}}$ is the antifundamental representation of $U(\bar F_{a,\gamma})$.
Therefore, $|\pmb s_{a,\beta}\rangle$ is the tensor product of all the $U(F_{a,\beta})$ fundamental representation states of the $n_{a,\beta}$ quarks $Q_{a,\beta}$. Similarly, $\langle\pmb s_{a,\gamma}|$ is the tensor product of all the $U(\bar F_{a,\gamma})$ antifundamental representation states of the $\bar n_{a,\gamma}$ quarks $\bar Q_{a,\gamma}$.
We define the vector $\vec s$ as the collection of these state labels:
\begin{align}
\vec{s}=\cup_a\left\{\cup_\beta\, \pmb s_{a,\beta};\cup_\gamma\, \pmb{\bar s}_{a,\gamma}\right\}
\end{align}
%For example, for the operator in \eqref{es vecn}, we would have $\pmb s_{1,1}=\{l,q\}$ and $\pmb{\bar s}_{1,1}=\{k,p\}$.

\begin{figure}[H]
\begin{center}\includegraphics[scale=0.76]{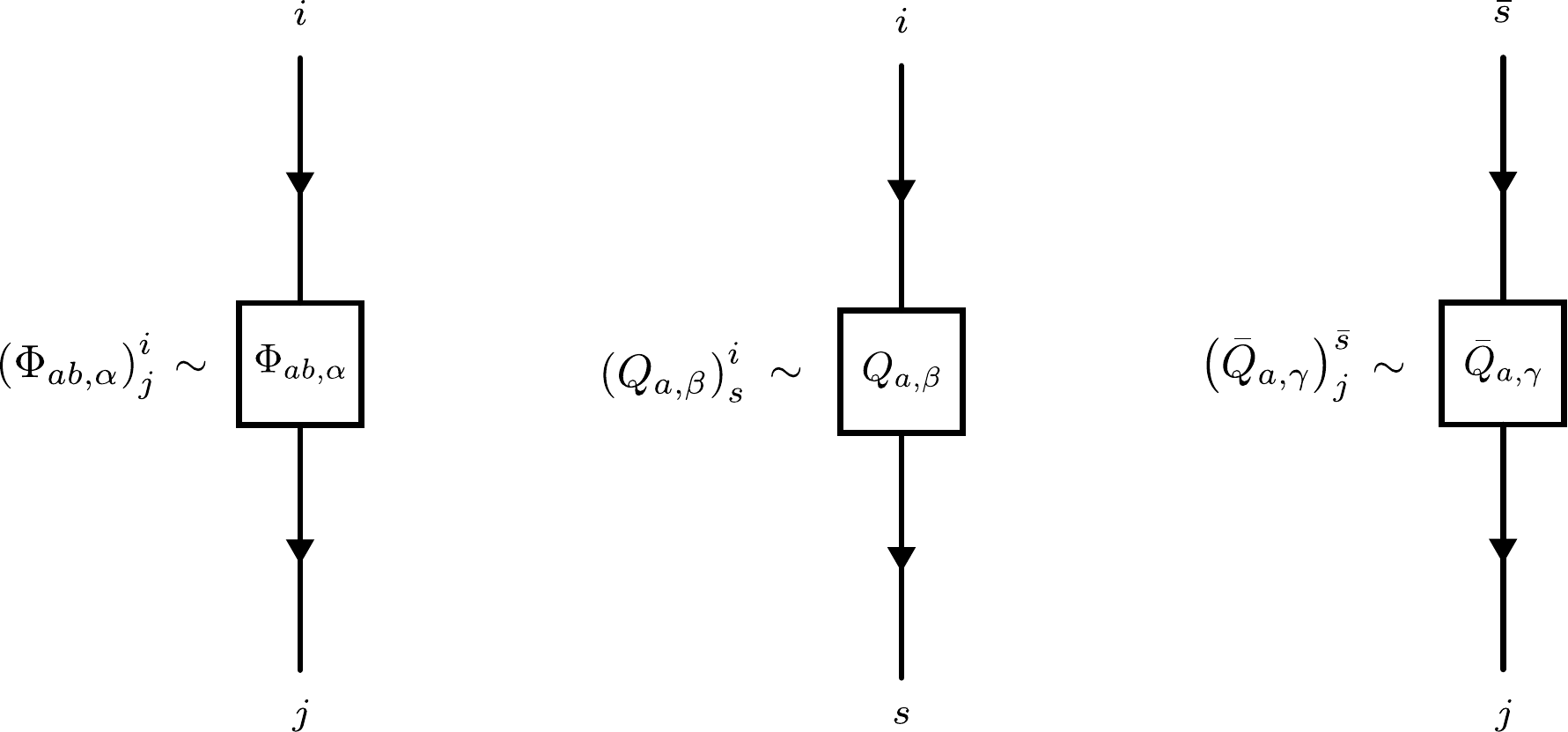}
\end{center}
\caption{Diagrammatic description of the matrix elements of the fundamental fields $\Phi$, $Q$ and $\bar Q$.}\label{fig: FundFieldDiagram}
\end{figure}

\begin{figure}[H]
\begin{center}\includegraphics[scale=0.76]{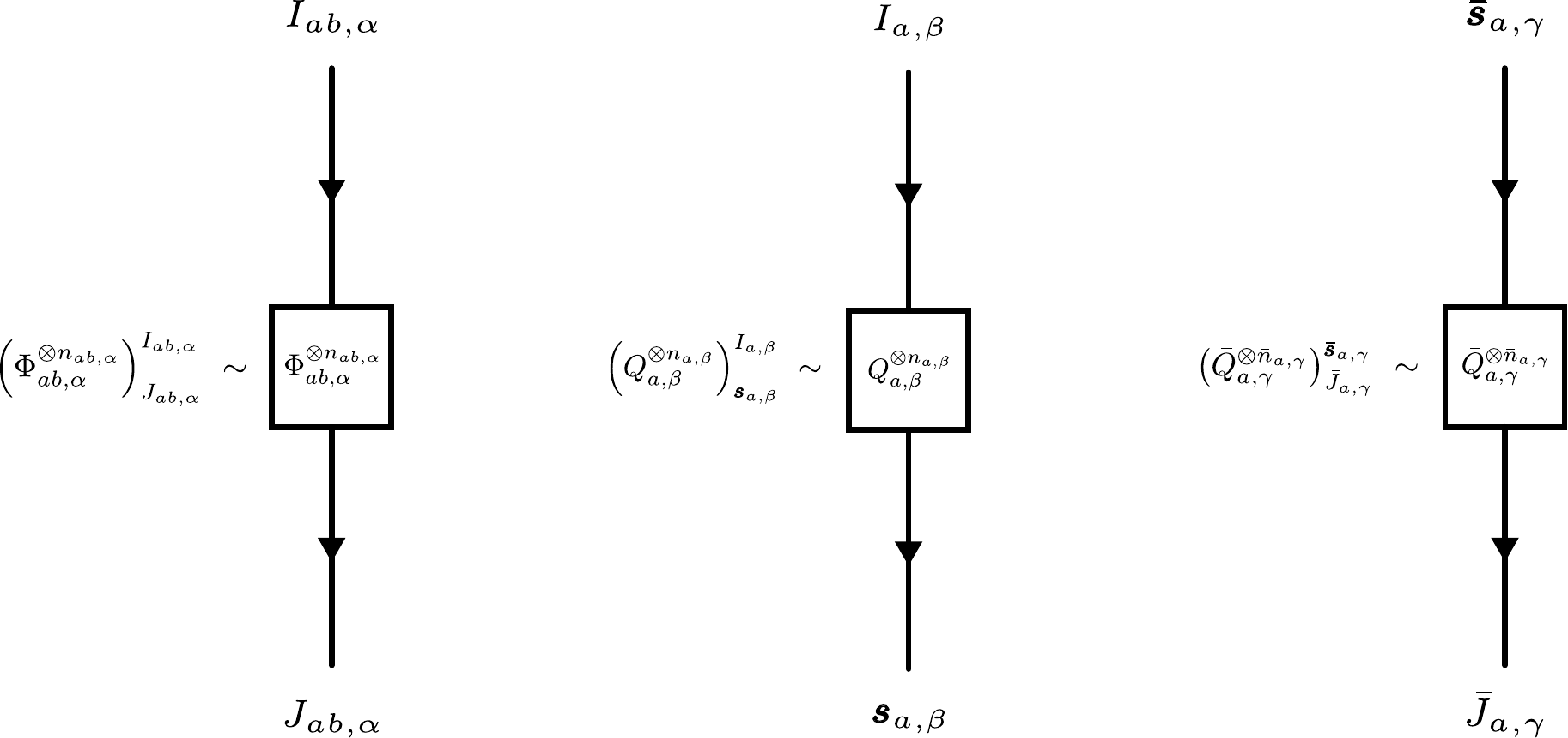}
\end{center}
\caption{Diagrammatic description of the matrix elements of the tensor products of the fundamental fields $\Phi$, $Q$ and $\bar Q$.}\label{fig: FundFieldDiagramTensor}
\end{figure}

In the framework that we are going to introduce in this section, the building blocks of any matrix invariant are the tensor products of the fundamental fields $\Phi_{ab,\alpha}^{\otimes n_{ab,\alpha}}$, $Q_{a,\beta}^{\otimes n_{a,\beta}}$ and $\bar Q_{a,\gamma}^{\otimes\bar n_{ {a,\gamma}}}$. Let us then introduce the states
% $I_{ab,\alpha}=\{i_1,...,i_{n_{ab,\alpha}}\}$, $I_{a,\beta}=\{i_1,...,i_{n_{a,\beta}}\}$ and $J_{ab,\alpha}=\{j_1,...,j_{n_{ab,\alpha}}\}$, $\bar J_{a,\gamma}=\{\bar j_1,...,\bar j_{\bar n_{a,\gamma}}\}$ are collections of antifundamental and fundamental $U(N_a)$ indices respectively. For each of these collections of labels, it is useful to define the associate states
\begin{equation}\label{fund states}
\begin{array}{lll}
&|I_{ab,\alpha}\rangle = |i_1,...,i_{n_{ab,\alpha}}\rangle \in V_{N_a}^{\otimes n_{ab,\alpha}}\,,\qquad\qquad
&|I_{a,\beta}\rangle = |i_1,...,i_{n_{a,\beta}}\rangle \in V_{N_a}^{\otimes n_{a,\beta}}\nn
&|J_{ab,\alpha}\rangle = |j_1,...,j_{n_{ab,\alpha}}\rangle \in V_{N_a}^{\otimes n_{ab,\alpha}}\,,\qquad
&|\bar J_{a,\gamma}\rangle = |\bar j_1,...,\bar j_{\bar n_{a,\gamma}}\rangle \in V_{N_a}^{\otimes \bar n_{a,\gamma}}
\end{array}
\end{equation}
Using these definitions, together with eq. \eqref{bf s bar s}, we can write the matrix elements of every $\Phi_{ab,\alpha}^{\otimes n_{ab,\alpha}}$ tensor product as
\begin{align}\label{phi matrix elements}
\left(\Phi_{ab,\alpha}^{\otimes n_{ab,\alpha}}\right)_{J_{ab,\alpha}}^{I_{ab,\alpha}}=
\lara{I_{ab,\alpha}\left|\Phi_{ab,\alpha}^{\otimes n_{ab,\alpha}}\right|J_{ab,\alpha}}
\end{align}
and similarly for $Q_{a,\beta}^{\otimes n_{a,\beta}}$ and $\bar Q_{a,\gamma}^{\otimes\bar n_{ {a,\gamma}}}$:
\begin{align}\label{Q bar Q matrix elements}
\left(Q_{a,\beta}^{\otimes n_{a,\beta}}\right)^{I_{a,\beta}}_{\pmb s_{a,\beta}}=
\lara{I_{a,\beta}\left|Q_{a,\beta}^{\otimes n_{a,\beta}}\right|\pmb s_{a,\beta}}\,,\qquad
\left(
\vphantom{Q_{a,\beta}^{\otimes n_{a,\beta}}}
\bar Q_{a,\gamma}^{\otimes\bar n_{ {a,\gamma}}}\right)_{\bar J_{ { a,\gamma}}}^{ {\pmb{\bar s}_{a,\gamma}}}=
\lara{
\pmb{\bar s}_{a,\gamma}\left|
\vphantom{Q_{a,\beta}^{\otimes n_{a,\beta}}}
\bar Q_{a,\gamma}^{\otimes\bar n_{ {a,\gamma}}}\right|\bar J_{ { a,\gamma}}}
\end{align}
We will now present the first of the many diagrammatic techniques that we will use throughout this paper. We draw the matrix components of fundamental fields $\left(\Phi_{ab,\alpha}\right)^i_j$, $\left(Q_{a,\beta}\right)^i_s$ and $\left(\bar Q_{a,\gamma}\right)^{\bar s}_j$ as in Fig. \ref{fig: FundFieldDiagram}.
This diagrammatic notation is then naturally extended to the tensor products $\left(\Phi_{ab,\alpha}^{\otimes n_{ab,\alpha}}\right)_{J_{ab,\alpha}}^{I_{ab,\alpha}}$, $\left(Q_{a,\beta}^{\otimes n_{a,\beta}}\right)^{I_{a,\beta}}_{\pmb s_{a,\beta}}$ and $\left(\bar Q_{a,\gamma}^{\otimes\bar n_{ {a,\gamma}}}\right)_{\bar J_{ { a,\gamma}}}^{ {\pmb{\bar s}_{a,\gamma}}}$, defined in eqs. \ref{phi matrix elements} and \ref{Q bar Q matrix elements}, as in Fig. \ref{fig: FundFieldDiagramTensor}. 

Permutations act on a tensor product of states by rearranging the order in which the states are tensored together. For example, given a permutation $\sigma\in S_k$ and a tensor product of $k$ states $|i_a\rangle$ ($1\leq a \leq k$) belonging to some vector space $V$, we have
\begin{align}\label{perm action on tens prod}
\sigma|i_1,i_2,...,i_k\rangle=
|i_{\sigma(1)},i_{\sigma(2)},...,i_{\sigma(k)}\rangle
\end{align}
Therefore, there is a natural permutation action on the states \eqref{bf s bar s} and \eqref{fund states}.

\begin{figure}[H]
\begin{center}\includegraphics[scale=0.8]{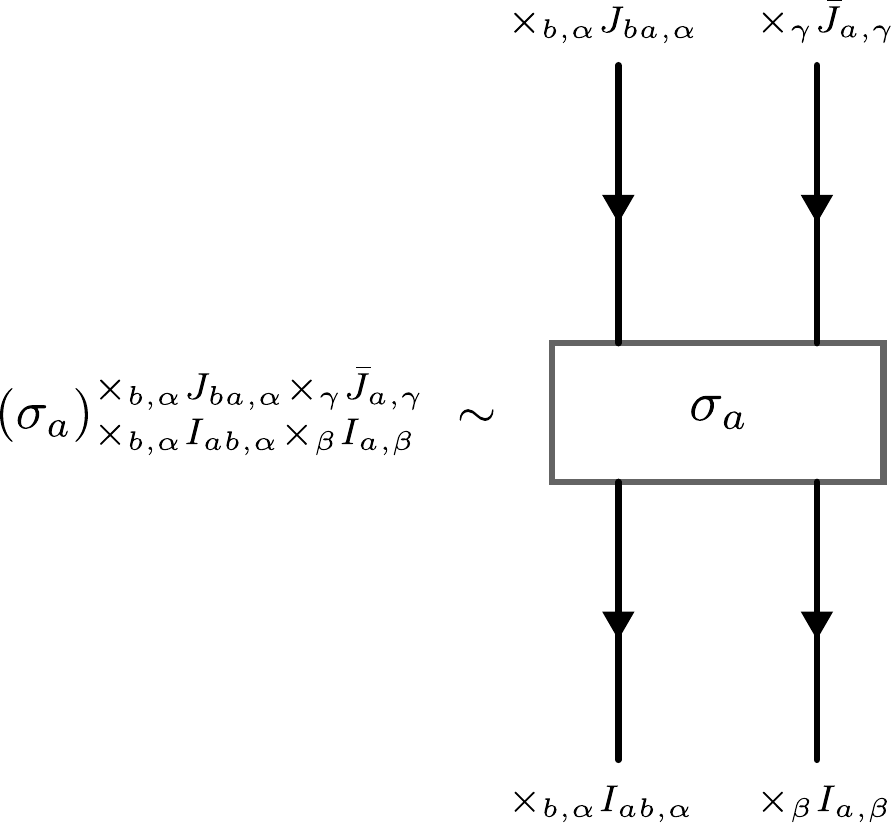}
\end{center}
\caption{Diagrammatic description of the matrix elements of the permutation $\sigma$.}\label{fig: permDiagram}
\end{figure}

The gauge invariant polynomial is constructed by contracting the upper $n_a$ indices 
of all the fields incident at the node $a$ with their lower $n_a$ indices. We describe these gauge invariants as follows. First we choose an ordering for all the fields with an upper $U(N_a)$ index. Then we fix a set of labelled upper indices: this means that we have picked an embedding of subsets into the set $ [ n_a] \equiv \{ 1, \cdots , n_a \}$, \emph{i.e.} 
\begin{align}\label{upper indices embedding}
 [ n_{a 1 , \alpha = 1 } ] \sqcup [ n_{a1, \alpha = 2} ] \sqcup\cdots\sqcup [n_{a 2, \alpha = 1 }] \sqcup [ n_{a2, \alpha = 2} ]\sqcup \cdots \sqcup[ n_{a , \beta =1 } ] \sqcup [ n_{a , \beta =2 } ]\sqcup \cdots \rightarrow [ n_a ]
\end{align}
which gives a set-partition of $[n_a]$.
Similarly, there is an embedding into $[n_a]$ corresponding to the ordering of the lower $U(N_a)$ indices, namely
\begin{align}\label{lower indices embedding}
 [ n_{1a , \alpha = 1 } ] \sqcup [ n_{1a, \alpha = 2} ] \sqcup\cdots\sqcup [n_{2 a, \alpha = 1 }] \sqcup [ n_{2 a, \alpha = 2} ]\sqcup \cdots \sqcup[ \bar n_{a , \gamma =1 } ] \sqcup [ \bar n_{a , \gamma =2 } ]\sqcup \cdots \rightarrow [ n_a ]
\end{align}
Now we contract the upper indices of these fields with their lower indices, \emph{after} a permutation $\sigma_a\in S_{n_a}$ of their labels.
We will therefore be considering permutations $\sigma_a\in S_{n_a}$, where $n_a=\sum_{b,\alpha}n_{ab,\alpha}+\sum_\beta n_{a,\beta} = \sum_{b,\alpha}n_{ba,\alpha}+\sum_\gamma n_{a,\gamma} $. Along the lines of eqs. \eqref{phi matrix elements} and \eqref{Q bar Q matrix elements} we can define the matrix elements of $\sigma_a$ as
\begin{align}
\left(\sigma_a\right)^{\times_{b,\alpha}J_{ba,\alpha}\times_\gamma\bar J_{ a,\gamma}}
_{\times_{b,\alpha}I_{ab,\alpha}\times_\beta I_{ {a,\beta}}}=
\left(\left.\otimes_{b,\alpha}\langle
J_{ba,\alpha}
\right|
\left.\otimes_{\gamma}\langle\bar J_{ { a,\gamma}}\right|\right)
\sigma_a
\left(\left.\otimes_{b,\alpha}|I_{ab,\alpha}\right\rangle\left.\otimes_{\beta}| I_{ { a,\beta}}
\right\rangle\vphantom{\bar J_a}\right)
\end{align}
where the product symbols appearing in the upper and lower indices of $ \sigma_a$ 
are ordered as in \eqref{upper indices embedding} and \eqref{lower indices embedding}. 
We depict these matrix elements as in Fig. \ref{fig: permDiagram}. 
%Notice that the order in which the collections of indices $I_{ab,\alpha}$, $I_{ {a,\beta}}$, $J_{ba,\alpha}$ and $\bar J_{ { a,\gamma}}$ appear in ${\times_{b,\alpha}J_{ba,\alpha}\times_\gamma\bar J_{ a,\gamma}}$, and similarly in ${\times_{b,\alpha}I_{ab,\alpha}\allowbreak\times_\beta I_{ {a,\beta}}}$, is important. We will refer to this string of states labels as an ordered set of labels.

\begin{figure}[H]
\begin{center}\includegraphics[scale=1.85]{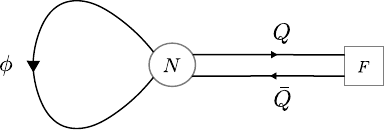}
\end{center}
\caption{The $\N=1$ quiver for an $\N=2$ SQCD model.}\label{fig: N=2 sqcd}
\end{figure}

\begin{figure}[H]
\begin{center}\includegraphics[scale=3]{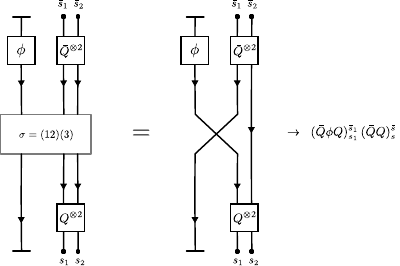}
\end{center}
\caption{Diagrammatic description of the GIO $(\bar Q\phi Q)^{\bar s_1}_{s_1}\,(\bar Q Q)^{\bar s_2}_{s_2}$ in an $\N=2$ SQCD. The horizontal bars are to be identified.}\label{fig: N=2 sqcd path-operator example}
\end{figure}

Following the approach of \cite{quivcalc}, we can write any GIO $\mathcal{O}_\mathcal{Q}$ of a quiver gauge theory $\Q$ with flavour symmetry as
\begin{align}\label{Q def}
&\mathcal{O}_\mathcal{Q}(\vec n;\, \vec{s};\,\vec\sigma)=
\prod_{a}\left[\prod_{b,\alpha}\left(\Phi_{ab,\alpha}^{\otimes n_{ab,\alpha}}\right)_{J_{ab,\alpha}}^{I_{ab,\alpha}}\right]
%quark
\otimes
\left[\prod_\beta \left(Q_{a,\beta}^{\otimes n_{a,\beta}}\right)^{I_{a,\beta}}_{\pmb s_{a,\beta}}\right]
\otimes
%antiquark
\left[\prod_\gamma\left(\bar Q_{a,\gamma}^{\otimes\bar n_{ {a,\gamma}}}\right)_{\bar J_{ { a,\gamma}}}^{ {\pmb{\bar s}_{a,\gamma}}}\right]\nonumber\\[5mm]
%permutation
&\qquad\qquad\qquad\qquad\qquad\qquad\qquad\qquad\times
\prod_a\left(\sigma_a\right)^{\times_{b,\alpha}J_{ba,\alpha}\times_\gamma\bar J_{ a,\gamma}}
_{\times_{b,\alpha}I_{ab,\alpha}\times_\beta I_{ {a,\beta}}}
\end{align}
Here $\vec\sigma=\cup_a\{\sigma_a\}$ is a collection of permutations $\sigma_a\in S_{n_a}$, where $n_a=\sum_{b,\alpha}n_{ab,\alpha}+\sum_\beta n_{a,\beta}$. 
The purpose of \(\vec \sigma\) is to contract all the gauge indices of the $\Phi$, $Q$ and $\bar Q$ fields to make a proper GIO. 
This formula looks rather complicated. However, it can be nicely interpreted in a diagrammatic way. We will now give an example of such a diagrammatic approach. Consider an $\N=2$ SCQD theory. The $\N=1$ quiver for this model is illustrated in Fig. \ref{fig: N=2 sqcd}.
We labelled the fields of this quiver by $\phi,\,Q$ and $\bar Q$, simplifying the notation of given in table \ref{table: group charges}. Consider now the GIO $(\bar Q\phi Q)^{\bar s_1}_{s_1}\,(\bar Q Q)^{\bar s_2}_{s_2}$. Here \(s_1,\,s_2\) and \(\bar s_1,\,\bar s_2\) are states of the fundamental and antifundamental representation of \(SU(F)\) respectively, and the round brackets denotes \(U(N)\) indices contraction. Figure \ref{fig: N=2 sqcd path-operator example} shows the diagrammatic interpretation of this GIO.

For fixed $\vec n$, the data $\vec\sigma , \vec s $ determines a gauge invariant. 
However changing $ \vec \sigma , \vec s $ can produce the same invariant. 
This fact can be described in terms of an equivalence relation generated by the action of permutations, associated with edges
of the quiver, on the data $ \vec \sigma , \vec s $. This has been discussed for the case without flavour symmetry in \cite{quivcalc}
and we will extend the discussion to flavours here. Continuing the example of the $\N=2$ SQCD introduced above, let us consider a matrix invariant built with $n$ adjoint fields $\phi$ and $n_q$ quarks and antiquarks $Q$ and $\bar Q$. 
We label the tensor product of all the $n_q$ quark states $|s_i\rangle \in V_{SU(F)}$ with the shorthand notation $|\pmb s\rangle=\otimes_{i=1}^{n_q} |s_i\rangle $. Here \(V_{SU(F)}\) is the fundamental representation of \(SU(F)\). Similarly, $\langle\pmb {\bar s}|=\otimes_{i=1}^{n_q} \langle \bar s_i| $ will be the tensor product of all the antiquarks states $\langle\bar s_i|\in \bar V_{SU(F)} $, where \(\bar V_{SU(F)}\) is the antifundamental representation of \(SU(F)\). 
In this model, a matrix invariant can be labelled by the triplet $(\sigma,\pmb s ,\pmb{\bar s})$. The redundancy discussed above is captured by the identification
\begin{align}\label{SQCDequivs}
(\sigma,\,\,\pmb s ,\,\,\pmb{\bar s}) 
\,\,\sim\,\,
\left((\eta\times \bar \rho)\sigma(\eta^{-1}\times \rho^{-1}),\,\,\rho(\pmb s),\,\,\bar\rho(\pmb{\bar s})\right) 
\end{align}
where $\eta\in S_n,\,\rho,\bar\rho\in S_{n_q}$ and $\rho(\pmb s)=(s_{\rho(1)},s_{\rho(2)},...,s_{\rho(n_q)})$, $\bar \rho(\pmb {\bar s})=(\bar s_{\bar \rho(1)},\bar s_{\bar \rho(2)},...,\bar s_{\bar \rho(n_q)})$. 
The last two equations are to be interpreted as the action of $\rho$ and $\bar\rho^-1$ on the states $|\pmb s\rangle $ and $\langle \pmb{\bar s}| $:
\begin{align}
\rho|\pmb s\rangle=|s_{\rho(1)},s_{\rho(2)},...,s_{\rho(n_q)}\rangle\,,\qquad\quad
\langle \pmb{\bar s}|\bar\rho^{-1} = \langle \bar s_{\bar \rho(1)},\bar s_{\bar \rho(2)},...,\bar s_{\bar \rho(n_q)}|
\end{align}
We refer to Appendix \ref{app:Op Inv} for a diagrammatic interpretation of this equivalence.

For the general case of a gauge theory with flavour symmetry, the degeneracy is described by the identity
\begin{align}\label{constr}
&\mathcal{O}_\mathcal{Q}(\vec n;\, \vec{s};\,\vec\sigma)
=
\mathcal{O}_\mathcal{Q}(\vec{n};\, \vec{\rho}\,(\vec{s}\,);\,\text{Adj}_{\vec\eta\times\vec\rho}(\vec\sigma))
\end{align}
Here we introduced the permutations
\begin{subequations}\label{perms eta and rho}
\begin{align}
&\vec\eta=\cup_{a,b, \alpha}\{\eta_{ab,\alpha}\}\,\,\,,\quad \eta_{ab,\alpha}\in S_{n_{ab,\alpha}}\\[4mm]
&\vec\rho=\cup_a\{\cup_\beta\, \rho_{a,\beta};\,\cup_\gamma\,\bar \rho_{a,\gamma}\}\,\,\,,\quad \rho_{a,\beta}\in S_{n_{a,\beta}}\,\,\,\,,\,\,\,\,\,\,\,\bar\rho_{a,\gamma}\in S_{\bar n_{a,\gamma}}
\end{align}
\end{subequations}
and we defined
\begin{align}
&\text{Adj}_{\vec\eta\times\vec\rho}(\vec\sigma)=
\cup_a\{
(\times_{b,\alpha}\eta_{ba,\alpha}\times_\gamma\bar\rho_{a,\gamma})
\sigma_a
(\times_{b,\alpha}\eta_{ab,\alpha}^{-1}\times_\beta \rho_{ a,\beta}^{-1})
\}\,\,\,,\label{adj action}\\[4mm]
&\vec\rho\,(\vec s)=\cup_a\{\cup_\beta\,\rho_{a,\beta}(\pmb s_{a,\beta});\,\cup_{\gamma}\, \bar \rho_{a,\gamma}(\pmb{\bar s}_{a,\gamma})\}
\end{align}
In Appendix \ref{app:Op Inv} we will derive the constraint \eqref{constr}. This is essentially a set of equivalences of the type 
\eqref{SQCDequivs}, iterated over all the nodes and edges of the quiver. The permutations $ \eta_{ ab, \alpha }, \rho_{a, \beta } , \bar \rho_{ a, \gamma} $ can be viewed as ``permutation gauge symmetries'', associated with the edges of the quiver. The permutations $ \vec \sigma$ and state labels $ \vec s $ can be viewed as ``matter fields'' for the permutation gauge symmetries, associated with the nodes of the quiver. 
It is very intriguing that, in terms of the original Lie group gauge symmetry, the round nodes were associated with gauge groups $U(N_a)$, while the edges were matter. In this world of permutations, these roles are reversed, with the edges being associated with gauge symmetries and the nodes with matter. 

%In the same appendix we also give a diagrammatic interpretation of this identity.\\
%Again, this identity is best understood in diagrammatic terms. Using the $\N=2$ SQCD model as an example, consider

\begin{figure}[H]
\begin{center}\includegraphics[scale=1]{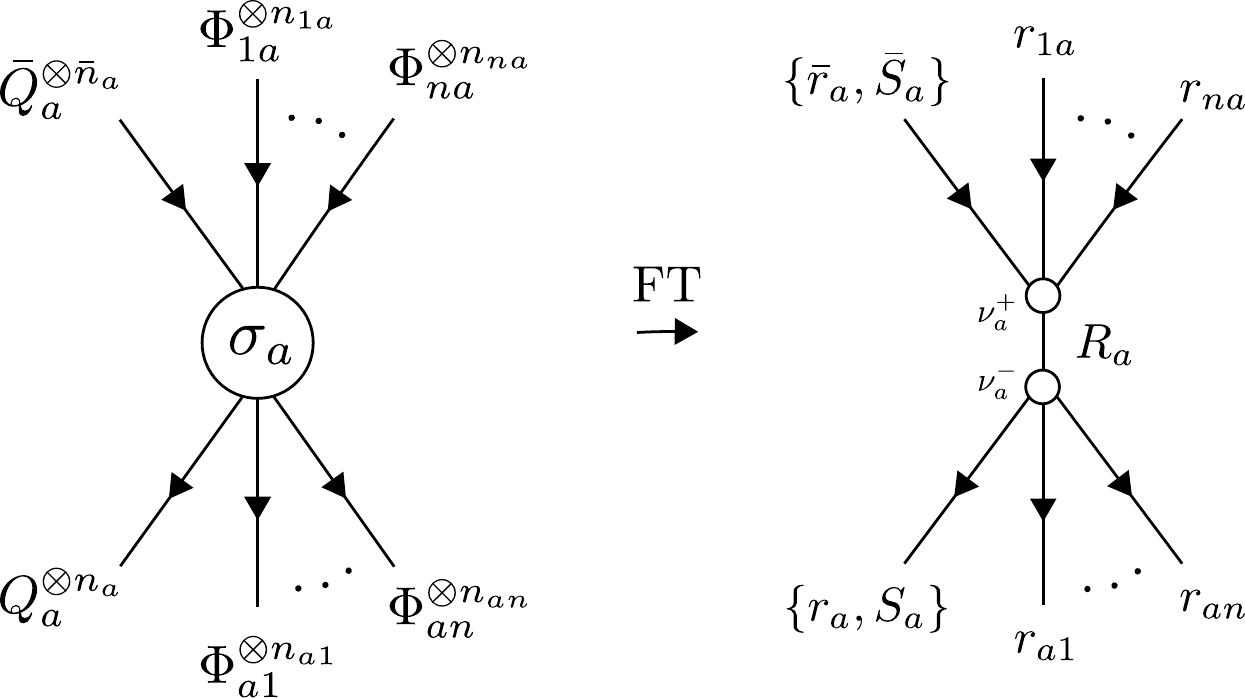}
\end{center}
\caption{Pictorial representation of the Fourier transform discussed in the text. The multiplicity labels of the fields are not displayed.}\label{fig: fourier}
\end{figure}

So far we have used a permutation basis approach to characterise the quiver matrix invariants. This has offered a nice diagrammatic interpretation, but on the other hand it is subject to the complicated constraint in eq. \eqref{constr}. 
In the following section we are going to introduce a Fourier Transformation (FT) from this permutation description to its dual space, which is described in terms of representation theory quantities. In other words, we are going to change the way we label the matrix invariants: instead of using permutation data, we are going to use representation theory data. The upshot of doing so is twofold. On one hand the new basis will not be subject to any equivalence relation such as the one in \eqref{constr}. On the other hand, as a consequence of the Schur-Weyl duality (see \emph{e.g.} \cite{FulHar}), it offers a simple way to capture the finite $N$ constraints of the GIOs.
Schematically, using this FT we trade the set of labels $\{\vec n;\, \vec{s};\,\vec\sigma\}$ of any GIO for the new set $\{R_a, r_{ab,\alpha},r_{a,\beta},S_{a,\beta},\bar r_{ a,\gamma},\bar S_{a,\gamma}, \nu_a^+,\nu_a^-\}$, that we denote with the shorthand notation $\pmb L$:
\begin{align}\label{pmb L ft def}
\text{FT}:\{\vec n;\, \vec{s};\,\vec\sigma\}\,\,\,\rightarrow\,\,\,\pmb L=\{R_a, r_{ab,\alpha},r_{a,\beta},S_{a,\beta},\bar r_{ a,\gamma},\bar S_{a,\gamma}, \nu_a^+,\nu_a^-\}
\end{align}
Each $R_a$ is a representation of the symmetric group $S_{n_a}$, where $n_a$ has been defined in \eqref{n_a def}. 
$ r_{ab,\alpha},r_{a,\beta},\bar r_{ a,\gamma}$ are partitions of $ n_{ab,\alpha},n_{a,\beta},\bar n_{ a,\gamma}$ respectively. $S_{a,\beta}$ and $\bar S_{ a,\gamma}$ are $U(F_{a,\beta})$ and $U(\bar F_{a,\gamma})$ states in the representation specified by the partitions $r_{a,\beta}$ and $\bar r_{a,\gamma}$ respectively. The integers $\nu_a^{\pm}$ are symmetric group multiplicity labels, a pair for each node in the quiver. Their meaning will be explained in the next section.
Graphically, at each node $a$ of the quiver we change the description of any matrix invariant as in Fig. \ref{fig: fourier}. The diagram on the right in this figure is also called a \emph{split-node} \cite{quivcalc}.
We call the Fourier transformed operators \emph{Quiver Restricted Schur polynomials}, or quiver Schurs for short. These are a generalisation of the Restricted Schur polynomials that first appeared in the literature in \cite{KSS1,KSS2,BKS,BCK08,BKS08}.
In section \ref{sec: 2pt function} we will show how the quiver Schurs form a basis for the Hilbert space of holomorphic operators.

\section{The Quiver Restricted Schur Polynomials}\label{Sec: The Restricted Schur Basis}
In this section we describe the FT introduced above. In other words, we will explicitly construct the map
\begin{align}\label{FT def}
\text{FT}:\,\,\cO_\Q(\vec n;\vec s;\vec\sigma) \,\,\rightarrow\,\,\cO_\Q(\pmb L)
\end{align}
In order to do so, we need to introduce two main mathematical ingredients. These are the symmetric group branching coefficients and the Clebsch-Gordan coefficients. For each of these quantities we give both an analytic and a diagrammatic description: the latter will aid to make notationally heavy formulae easier to understand.

We begin by focusing on the symmetric group branching coefficients. Consider the symmetric group restriction
\begin{align}\label{Sn restriction def}
\times_{i=1}^k \,S_{n_i}\rightarrow S_n\,,\qquad\sum_{i=1}^k n_i=n
\end{align}
For each representation $V_R^{S_n}$ of $S_n$, this restriction induces the representation branching 
\begin{align}\label{dec branch}
&\qquad V_{R}^{S_n} \simeq \bigoplus_{{{r_1\vdash n_1\atop r_2\vdash n_2}\atop \cdots}\atop r_k\vdash n_k }\left(\bigotimes_{i=1}^k\, V_{r_i}^{S_{n_i}}\right)\otimes V_R^{\vec r}\,,\qquad \vec{r}=(r_1,r_2,...,r_k)
\end{align}
$V_R^{\vec r}$ is the \emph{multiplicity} vector space, in case the representation $\otimes_i V_{r_i}$ appears more than once in the decomposition \eqref{dec branch}. The dimension of this space is $\text{dim}(V_R^{\vec r})=g\left(\cup_{i=1}^k r_i;R\right)$, where $g\left(\cup_{i=1}^k r_i;R\right)=g\left( r_1,r_2,...,r_k;R\right)$ are Littlewood-Richardson coefficients \cite{FulHar}.

\begin{figure}[H]
\begin{center}\includegraphics[scale=0.8]{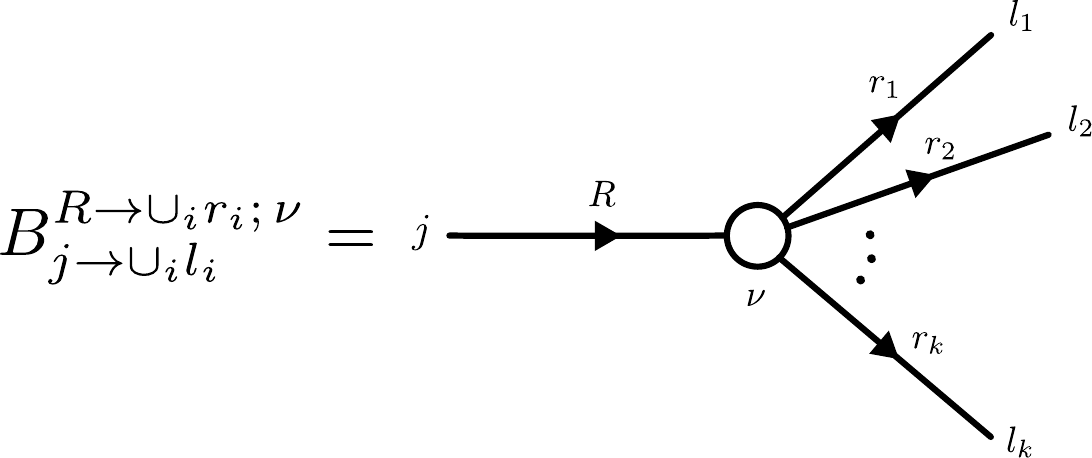}
\end{center}
\caption{Pictorial description of the symmetric group branching coefficients.}\label{fig: branching coeff}
\end{figure}

In the following, the vectors belonging to any vector space $V$ will be denoted using a bra-ket notation. The symbol $\langle\cdot|\cdot\rangle$ will indicate the inner product in $V$.
%We also equip every vector space $V$ with an inner product
%\begin{equation}\label{inner product}
%\begin{array}{cll}
%\langle\cdot|\cdot\rangle: &V\times V& \longrightarrow \mathbb{C}
%\end{array}
%\end{equation}
Let then the set of vectors $\{\otimes_{i=1}^k|r_i,l_i,\nu\rangle\}$ be an orthonormal basis for $\bigoplus_{\vec r}\left(\bigotimes_{i=1}^k\, V_{r_i}^{S_{n_i}}\right)\otimes V_R^{\vec r}$. Here $ l_i$ is a state in $V_{r_i}^{S_{n_i}}$ and $\nu=1,...,g\left(\cup_{i=1}^k r_i;R\right)$ is a multiplicity label. We adopt the convention that $\otimes_{i=1}^k|r_i,l_i,\nu\rangle \equiv |\cup_i r_i,\cup_i l_i,\nu\rangle$. Similarly, let the set of vectors $\{|R,j\rangle\,,\,j=1,...,\text{dim}(V_R^{S_n})\}$ be an orthonormal basis for $V_R^{S_n}$. 
The branching coefficients $B^{R\rightarrow \cup_i r_i;\,\nu}_{j\rightarrow\cup_i l_i}$ are the matrix entries of the linear invertible operator $B$, mapping
\begin{equation}\label{B map}
\begin{array}{cll}
B: &V_R^{S_n}\longrightarrow \bigoplus_{\vec r}\left(\bigotimes_{i=1}^k\, V_{r_i}^{S_{n_i}}\right)\otimes V_R^{\vec r}
\end{array}
\end{equation}
so that 
\begin{align}\label{B map action}
B^{R\rightarrow \cup_k r_k;\,\nu}_{j\rightarrow\cup_k l_k}\,\,|R,j\rangle
=
|\cup_i r_i,\cup_i l_i,\nu\rangle
\end{align}
The sum over repeated indices is understood. By acting with $\langle S,i|$ on the left of both sides of \eqref{B map action} we then have
\begin{align}\label{B map elements}
B^{S\rightarrow \cup_k r_k;\,\nu}_{i\rightarrow\cup_k l_k}=\langle S,i|\cup_i r_i,\cup_i l_i,\nu\rangle
\end{align}
Since $B$ is an automorphism that maps an orthonormal basis to an orthonormal basis, 
%\begin{align}
%\langle S,i|R,j\rangle=\delta^S_R\,\delta^i_j\,,\qquad\quad
%
%\langle \cup_i s_i,\cup_i q_i,\mu |\cup_i r_i,\cup_i l_i,\nu\rangle %=\left(\prod_i\delta^{s_i}_{r_i}\,\delta^{q_i}_{l_i}\right)\delta^\mu_\nu
%\end{align}
it follows that $B$ is an unitary operator, $B^\dagger=B^{-1}$. We can then write
\begin{align}\label{B orth id}
\sum_{j}B^{R\rightarrow \cup_i r_i;\,\nu}_{j\rightarrow\cup_i l_i}\,\,(B^\dagger)^{\cup_i s_i;\,\mu \rightarrow R}_{\cup_i q_i\rightarrow j}=\left(\prod_i\delta^{{s_i},{r_i}}\,\delta_{{q_i},{l_i}}\right)\delta^{\mu,\nu}
\end{align}
However, since all the irreducible representations of any symmetric group can be chosen to be real \cite{Hamermesh}, there exists a convention in which the branching coefficients \eqref{B map elements} are also real. Therefore $B^\dagger=B^{T}$, where $B^T$ is the transpose of the map \eqref{B map action}. Using this last fact we can write the chain of equalities
\begin{align}\label{Branching reversing}
\langle S,i|\cup_i r_i,\cup_i l_i,\nu\rangle=B^{S\rightarrow \cup_k r_k;\,\nu}_{i\rightarrow\cup_k l_k}=
(B^{T})^{\cup_k r_k;\,\nu \rightarrow S}_{\cup_k l_k\rightarrow i}=
(B^{-1})^{\cup_k r_k;\,\nu \rightarrow S}_{\cup_k l_k\rightarrow i}=
\langle \cup_i r_i,\cup_i l_i,\nu|S,i\rangle
\end{align}
We draw the branching coefficients \eqref{B map elements} as in Fig. \ref{fig: branching coeff}. The orientation of the arrows can be reversed because of the identities in \eqref{Branching reversing}.

\begin{figure}[H]
\begin{center}\includegraphics[scale=0.8]{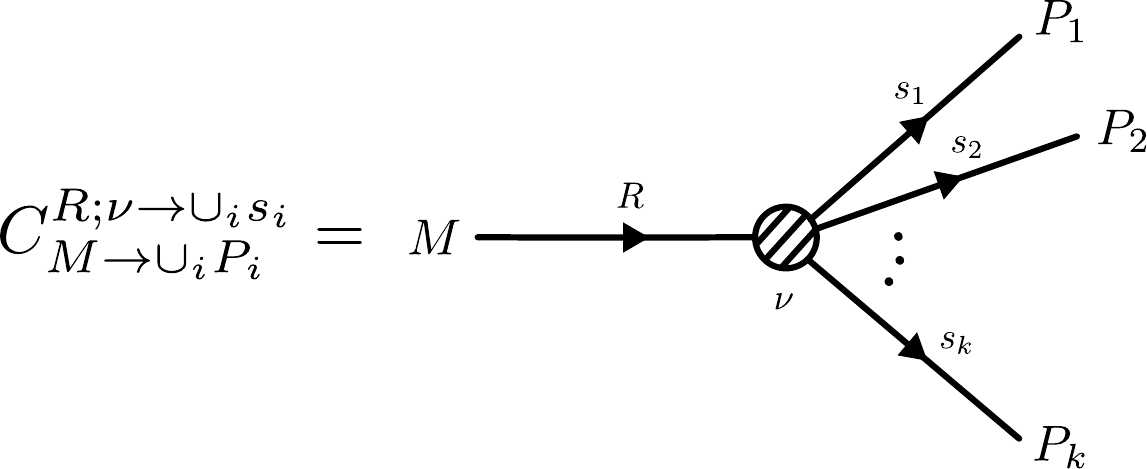}
\end{center}
\caption{Pictorial representation of the $U(N)$ Clebsch-Gordan coefficient in eq. \eqref{C map elements}.}\label{fig: U(N) CG}
\end{figure}

Consider now taking $k$ irreducible representations $V_{r_i}^{U(N)}$ of the unitary group $U(N)$, $i=1,2,...,k$. For each $V_{r_i}^{U(N)}$, $r_i$ is a partition of some integer $n_i$. This partition is associated with a Young diagram which is used to label the representation. If we tensor together all the $V_{r_i}^{U(N)}$'s, we generally end up with a reducible representation, and we have the isomorphism (see \emph{e.g.} \cite{Hamermesh})
\begin{align}\label{U(N) dec}
\bigotimes_{i=1}^k\,V_{r_i}^{U(N)} \simeq \bigoplus_{R\vdash n\atop c_1(R)\leq N}\,V_R^{U(N)}\otimes V_R^{\vec r}\,,\qquad\quad n=\sum_{i=1}^k n_i\,,
\end{align}
Here $R$ is a partition of $n=\sum_{i} n_i$. The direct sum on the RHS above is restricted to the Young diagrams $R$ whose first column length $c_1(R)$ does not exceed the rank $N$ of the gauge group.
$V_R^{\vec r}$, with $\vec{r}=(r_1,r_2,...,r_k)$, is the multiplicity vector space, satisfying $\text{dim}(V_R^{\vec r})=g\left(\cup_{i=1}^k r_i;R\right)$. The $g\left(\cup_{i=1}^k r_i;R\right)$ coefficients that appear in this formula are the same Littlewood-Richardson coefficients that we used in the above description of the symmetric group branching coefficients. Now let the set of vectors $\{|r_i,K_j\rangle\}$ be an orthonormal basis for $V_{r_i}^{U(N)}$, for $i=1,2,...,k$. Here $K_j$ is a state in $V_{r_i}^{U(N)}$. 
Also let $\{\left|R,M;\nu\right\rangle\}$ be an orthonormal basis for $\bigoplus_{R\vdash n}\,V_R^{U(N)}\otimes V_R^{\vec r}$. Here $M$ is a state in the $U(N)$ representation $V_{R}^{U(N)}$ and $\nu$ is a multiplicity index.
The Clebsch-Gordan coefficients $C^{R;\nu\rightarrow \cup_i r_i}_{M\rightarrow \cup_i K_i}$ are the matrix entries of the linear invertible operator $C$, mapping
\begin{equation}\label{C map}
\begin{array}{cll}
C: &\bigotimes_{i=1}^k\,V_{r_i}^{U(N)}\longrightarrow \bigoplus_{R\vdash n}\,V_R^{U(N)}\otimes V_R^{\vec r}
\end{array}
\end{equation}
so that 
\begin{align}\label{C map action}
C^{R;\nu\rightarrow \cup_i r_i}_{M\rightarrow \cup_i K_i}\,\,\left|
\cup_i r_i,\cup_i K_i
\right\rangle
=
\left|R,M;\nu\right\rangle
\end{align}
The sum over repeated indices is understood. By acting on the left of both sides of \eqref{C map action} with $\langle \cup_i s_i,\cup_i P_i|$, where $P_i$ are states of the $U(N)$ representations $V_{s_i}^{U(N)}$, we get
\begin{align}\label{C map elements}
C^{R;\nu\rightarrow \cup_i s_i}_{M\rightarrow \cup_i P_i}=\left\langle \cup_i s_i,\cup_i P_i |R,M;\nu\right\rangle
\end{align}
From \eqref{C map action}, we see that the automorphism $C$ maps an orthonormal basis to an orthonormal basis.
%Since we chose our bases to be orthonormal,
%\begin{align}
%\langle S,P,\mu|R,M;\nu\rangle=\delta^S_R\,\delta^P_M\,\delta^\mu_\nu\,,\qquad\quad
%
%\langle \cup_i s_i,\cup_i P_i |\cup_i r_i,\cup_i K_i\rangle =\prod_i\delta^{s_i}_{r_i}\, %\delta^{P_i}_{K_i}
%\end{align}
This makes $C$ an unitary operator, $C^\dagger=C^{-1}$, and we can therefore write
\begin{align}\label{unit C}
\sum_{\vec r,\,\vec K}\,C^{R;\nu\rightarrow \cup_i r_i}_{M\rightarrow \cup_i K_i}\,(C^\dagger)^{\cup_i r_i\rightarrow S;\mu}_{\cup_i K_i \rightarrow P}=
\delta^{S,R}\,\delta_{P,M}\,\delta^{\mu,\nu}
\end{align}
As with the branching coefficients, it is always possible to choose a consistent convention in which all the $U(N)$ Clebsch-Gordan coefficients \eqref{C map elements} are real. If we choose to work with such a convention, $C$ becomes an orthogonal operator: $C^T=C^{-1}$. We then have, in the same fashion of \eqref{Branching reversing}
\begin{align}\label{CG reversing}
\left\langle \cup_i s_i,\cup_i P_i |R,M;\nu\right\rangle=C^{R;\nu\rightarrow \cup_i s_i}_{M\rightarrow \cup_i P_i}=
(C^T)^{\cup_i s_i\rightarrow R;\nu}_{\cup_i P_i \rightarrow M}=
(C^{-1})^{\cup_i s_i\rightarrow R;\nu}_{\cup_i P_i \rightarrow M}=
\left\langle R,M;\nu | \cup_i s_i,\cup_i P_i\right\rangle
\end{align}
We draw the Clebsch-Gordan coefficients as in Fig. \ref{fig: U(N) CG}. Again, the orientation of the arrows can be reversed, due to \eqref{CG reversing}.

\begin{figure}[H]
\begin{center}\includegraphics[scale=0.6]{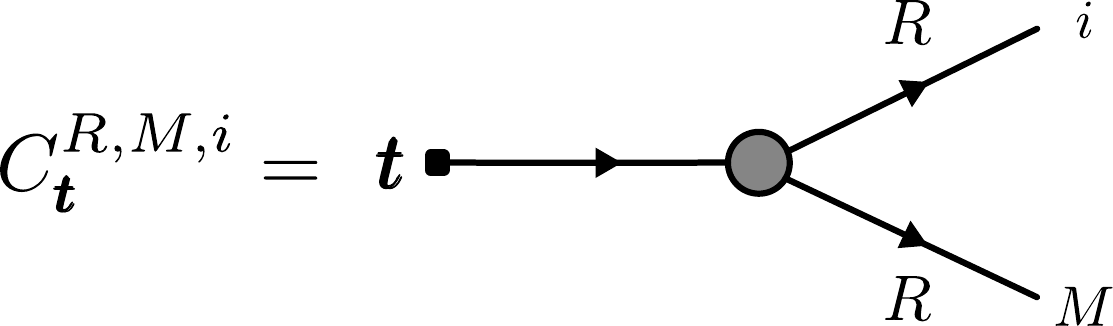}
\end{center}
\caption{Pictorial representation of the $U(N)$ Clebsch-Gordan coefficients \eqref{cg def} for the Schur-Weyl duality \eqref{SW d}.}\label{fig: Clebsch-Gordan coeff}
\end{figure}

Consider now the particular case of \eqref{U(N) dec} in which every representation $V^{U(N)}_{r_i}$ tensored on the LHS coincides with the $U(N)$ fundamental\footnote{We can get similar results by replacing the fundamental with the antifundamental representation of $U(N)$. The quantities we define here get modified accordingly.} representation, that for simplicity we just call $V$ for the remainder of this section.
This configuration allows us to use the Schur-Weyl duality to write
\begin{align}\label{SW d}
\overbrace{V\otimes \cdots\otimes V}^{k \text{ times}}=V^{\otimes k}& \simeq \bigoplus\limits_{R\vdash k\atop c_1(R)\leq N}\,\ V_R^{U(N)}\otimes V_R^{S_k}
\end{align}
where $ V_R^{U(N)}$ and $ V_R^{S_k}$ are irreducible representations of $U(N)$ and $S_k$ respectively. They correspond to the Young diagrams specified by the partition $R$ of $k$. By comparing \eqref{SW d} with \eqref{U(N) dec}, we see that the representation $V_R^{S_k}$ has now taken the place of the generic multiplicity vector space $V_R^{\vec r}$.
Since the Schur-Weyl decomposition will play a major role in this paper, we are now going to introduce a more compact notation for its Clebsch-Gordan coefficients. Let us consider the states 
\begin{align}
|\pmb s\rangle=\otimes_{j=1}^k|s_j\rangle\in V^{\otimes k}\,,\,\,|s_j\rangle\in V\,,
\qquad\,\,
|R;M,i\rangle=|R,M\rangle\otimes|R,i\rangle\in V_R^{U(N)}\otimes V_R^{S_k}
\end{align}
where $\{|R,M\rangle\,,\,M=1,...,\text{dim}(V_R^{U(N)})\}$ and $\{|R,i\rangle\,,\,i=1,...,\text{dim}(V_R^{S_k})\}$ are orthonormal bases of $V_R^{U(N)}$ and $V_R^{S_k}$ respectively. %In the notation we want to use for the Clebsch-Gordan coefficients for the Schur-Weyl duality, the equations corresponding to \eqref{C map action} and \eqref{C map elements} will read
The equations \eqref{C map action} and \eqref{CG reversing} imply
\begin{align}\label{C map action SW}
C^{R,M,i}_{\pmb s}\,\left|
\pmb s
\right\rangle
=
\left|R,M,i\right\rangle
\end{align}
and
\begin{align}\label{cg def}
C^{R,M,i}_{\pmb t}=\left\langle\pmb t|R,M,i\right\rangle=\left\langle R,M,i | \pmb t \right\rangle=C_{R,M,i}^{\pmb t}
\end{align}
respectively. 
%The second equality in \eqref{cg def} follows from \eqref{CG reversing}.
We draw these quantities as in Fig. \ref{fig: Clebsch-Gordan coeff}.
%\noindent The Shur-Weyl duality also hold if instead of the fundamental representation of a gauge group we consider its antifundamental representation. The Clebsch-Gordan coefficients get modified accordingly.

\subsection{The quiver characters}

We now have all the tools necessary to introduce a key quantity, the \emph{quiver characters} $\chi_\Q(\pmb L, \vec s,\vec \sigma)$. Here $\pmb L$ is the set of representation theory labels defined in \eqref{pmb L ft def}. The quiver characters are the expansion coefficients of the FT \eqref{FT def}:
\begin{align}\label{fourier2}
\cO_{\Q}(\pmb L)=\sum_{\vec s}\sum_{\vec\sigma}
\chi_\Q(\pmb L, \vec s,\vec \sigma)
\,
\mathcal{O}_\mathcal{Q}(\vec n,\vec s ,\vec \sigma)
\end{align}
We define them as
\begin{align}\label{quiv char}
\chi_\Q(\pmb L, \vec s,\vec \sigma)=&c_{\pmb L}
\sum_{\{ l_{ab,\alpha}\}\atop \{l_{a,\beta}\},\,\{\bar l_{a,\gamma}\}}
\prod_a\,
\sum_{i_a,j_a}\,
D^{R_a}_{i_a,j_a}(\sigma_a)\,\,
B
^{R_a\rightarrow\cup_{b,\alpha}r_{ab,\alpha}\cup_\beta r_{a,\beta};\nu_a^-}
_{j_a\rightarrow \cup_{b,\alpha}l_{ab,\alpha}\cup_\beta l_{a,\beta}}
\prod_\beta C_{ \pmb s_{a,\beta}}^{r_{a,\beta},S_{a,\beta},l_{a,\beta}}\nn
&\qquad\qquad\times
B
^{R_a\rightarrow\cup_{b,\alpha}r_{ba,\alpha}\cup_\gamma\bar r_{a,\gamma};\nu_a^+}
_{i_a\rightarrow \cup_{b,\alpha}l_{ba,\alpha}\cup_\gamma\bar l_{a,\gamma}}
\prod_\gamma C^{\pmb{\bar s}_{a,\gamma}}_{\bar r_{a,\gamma},\bar S_{a,\gamma},\bar l_{a,\gamma}}
\end{align}
where the coefficient $c_{\pmb L}$ is the normalisation constant
\begin{align}\label{norm const c_l copy}
c_{\pmb L}=
\prod_a\,\left(\frac{d(R_a)}{n_a!}\right)^{\frac{1}{2}}
\left(\prod_{b,\alpha}\frac{1}{d(r_{ab,\alpha})} \right)^{\frac{1}{2}}
\left(\prod_{\beta} \frac{1}{d(r_{a,\beta})} \right)^{\frac{1}{2}}
\left(\prod_{\gamma} \frac{1}{d(\bar r_{a,\gamma})}\right)^{\frac{1}{2}}
\end{align}
Since we chose to work in the convention in which all symmetric group representations and Clebsch-Gordan coefficients are real, then the quiver characters are real quantities as well:
\begin{align}
\chi_\Q(\pmb L, \vec s,\vec \sigma)=\chi_\Q^*(\pmb L, \vec s,\vec \sigma)
\end{align}
This convention will be convenient when we compute the 2-point functions of holomorphic and anti-holomorphic matrix invariants in section \ref{sec: 2pt function}. 

\begin{figure}[H]
\begin{center}\includegraphics[scale=0.7]{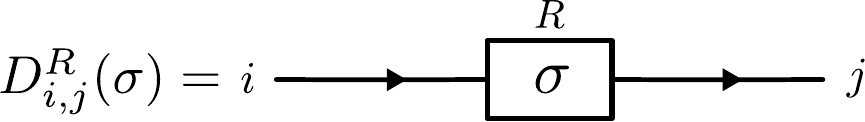}
\end{center}
\caption{Pictorial description of the matrix element \(D_{i,j}^R(\sigma)\) of the \(S_n\) symmetric group representation \(R\).}\label{fig: matrix element}
\end{figure}

\begin{figure}[H]
\begin{center}\includegraphics[scale=1.2]{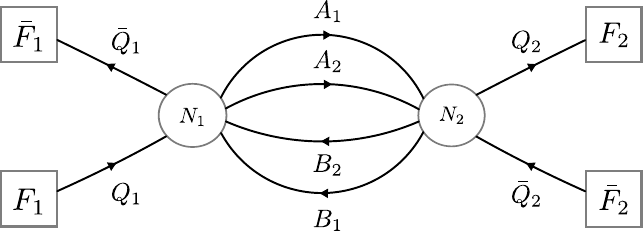}
\end{center}
\caption{$\mathcal N=1$ quiver for the flavoured conifold gauge theory.}\label{fig: flavoured conifold}
\end{figure}

\begin{figure}[H]
\begin{center}\includegraphics[scale=2.3]{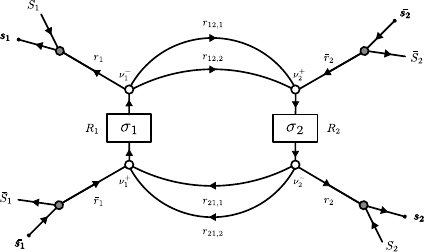}
\end{center}
\caption{The quiver character diagram for the flavoured conifold gauge theory.}\label{fig: flavoured conifold quiver character}
\end{figure}

These quantities have a pictorial interpretation. We have already introduced a diagrammatic notation for the branching and Clebsch-Gordan coefficients $B$ and $C$ in Fig. \ref{fig: branching coeff} and in Fig. \ref{fig: Clebsch-Gordan coeff} respectively. The pictorial notation for the $i,j$ matrix element of the permutation $\sigma$ in the irreducible representation $R$, \(D^R_{i,j}(\sigma)\), is displayed in Fig. \ref{fig: matrix element}. All the edges of these diagrams are to be contracted together as per instructions of formula \eqref{quiv char}.
Let us give an example of the diagrammatic of the quiver character of a well-known flavoured gauge theory. Consider the $\N=1$ quiver for the flavoured conifold \cite{Ouyang:2003df, Levi:2005hh, Benini:2006hh, Bigazzi:2008zt} in Fig. \ref{fig: flavoured conifold}. 
The quiver character for this model is depicted in Fig. \ref{fig: flavoured conifold quiver character}. This figure explicitly shows how all the symmetric group matrix elements, the branching coefficients and the Clebsch-Gordan coefficients are contracted together.

\begin{figure}[H]
\begin{center}\includegraphics[scale=2.3]{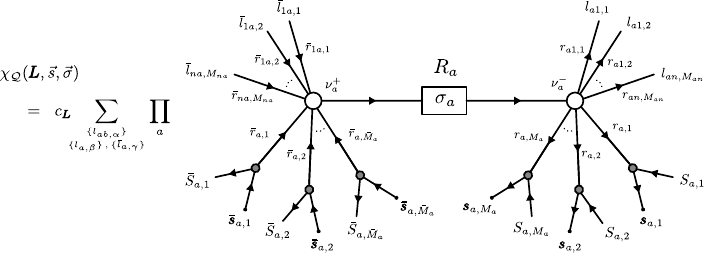}
\end{center}
\caption{Pictorial description of the quiver characters $\chi_\Q(\pmb L,\vec s,\vec\sigma)$.}\label{fig: quiv char}
\end{figure}

For completeness we also give a diagram for the the most generic quiver character $\chi_\Q(\pmb L, \vec s,\vec \sigma)$. This is done in Fig. \ref{fig: quiv char}. In this picture, we factored the quiver character into a product over the gauge nodes $a$ of the quiver. All the internal edges (that is, the ones that are not connected to a Clebsch-Gordan coefficient) are contracted following the prescription of \eqref{quiv char}.

The quiver characters \eqref{quiv char} satisfy the invariance relation
\begin{align}\label{invariance chi}
\chi_\Q(\pmb L, \vec s,\vec\sigma)=
\chi_\Q(\pmb L, \vec\rho\,(\vec s\,),\text{Adj}_{\vec\eta\times\vec\rho}(\vec\sigma))
\end{align}
where $\text{Adj}_{\vec\eta\times\vec\rho}(\vec\sigma)$ has been defined in \eqref{adj action}:
\begin{align}
&\text{Adj}_{\vec\eta\times\vec\rho}(\vec\sigma)=
\cup_a\{
(\times_{b,\alpha}\eta_{ba,\alpha}\times_\gamma\bar\rho_{a,\gamma})
\sigma_a
(\times_{b,\alpha}\eta_{ab,\alpha}^{-1}\times_\beta \rho_{ a,\beta}^{-1})
\}
\end{align}
They also satisfy the two orthogonality relations
\begin{align}\label{orthogonality chi}
\sum_{\vec s}\sum_{\vec\sigma}&\chi_\Q(\pmb L,\vec{s},\vec{\sigma})\,\chi_\Q(\pmb{\tilde L},\vec{s},\vec{\sigma})=\delta_{\pmb L,\tilde{\pmb L}}
\end{align}
and
\begin{align}\label{orthogonality chi 2}
\sum_{\pmb L}\chi_\Q(\pmb L,&\vec s,\vec\sigma)\,\chi_\Q(\pmb L,\vec t,\vec\tau)
=\frac{1}{c_{\vec{n}}}\,\sum_{\vec\eta\times\vec\rho} \delta\left(\text{Adj}_{\vec\eta\times\vec\rho}(\vec \sigma)\,\vec \tau^{\,-1}\right)
\delta_{\vec \rho(\vec{ s}),\vec{ t} } 
\end{align}
where we introduced the normalisation constant
\begin{align}\label{norm const c_n}
c_{\vec n}=\prod_a \left(\prod_{b,\alpha} n_{ab,\alpha}! \right)
\left(\prod_\beta n_{a,\beta}! \right)
\left(\prod_\gamma n_{a,\gamma}! \right)
\end{align}
It is worthwhile to note that this quantity can be interpreted as the order of the permutation gauge symmetry group.
All of these equations are derived in Appendix \ref{Quiver Characters}.

The set of operators \eqref{fourier2} form the Quiver Restricted Schur polynomial basis. Using \eqref{invariance chi} we can immediately check that such operators are invariant under the constraint \eqref{constr}. We have
\begin{align}
\cO_{\Q}(\pmb L)&=\sum_{\vec s}\sum_{\vec\sigma}\,
\chi_\Q(\pmb L, \vec s,\vec \sigma)
\,
\mathcal{O}_\mathcal{Q}(\vec n,\vec s ,\vec \sigma)\nn
&=
\sum_{\vec s}\sum_{\vec\sigma}\,
\chi_\Q(\pmb L, \vec s,\vec \sigma)
\,
\mathcal{O}_\mathcal{Q}(\vec{n},\, \vec{\rho}\,(\vec{s}\,),\,\text{Adj}_{\vec\eta\times\vec\rho}(\vec\sigma))\nn
&=
\sum_{\vec s}\sum_{\vec\sigma}\,
\chi_\Q(\pmb L, \vec\rho\,(\vec s\,),\text{Adj}_{\vec\eta\times\vec\rho}(\vec\sigma))
\,
\mathcal{O}_\mathcal{Q}(\vec{n},\, \vec{\rho}\,(\vec{s}\,),\,\text{Adj}_{\vec\eta\times\vec\rho}(\vec\sigma))\nn
&=
\sum_{\vec s}\sum_{\vec\sigma}\,
\chi_\Q(\pmb L, \vec s,\vec\sigma)
\,
\mathcal{O}_\mathcal{Q}(\vec{n},\, \vec{s},\,\vec\sigma)=\cO_{\Q}(\pmb L)
\end{align}
were in the second line we used the constraint \eqref{constr}, in the third one the invariance of the quiver characters \eqref{invariance chi} and in the fourth one we relabelled the dummy variables of the double sum.

Finally, the FT \eqref{fourier2} can be easily inverted. Starting from
\begin{align}\label{fourier to invert}
\cO_{\Q}(\pmb L)=\sum_{\vec t}\sum_{\vec\tau}
\chi_\Q(\pmb L, \vec t,\vec \tau)
\,
\mathcal{O}_\mathcal{Q}(\vec n,\vec t ,\vec \tau)
\end{align}
we multiply both sides by \(\chi_\Q(\pmb L,\vec s,\vec \sigma)\) and we take the sum over the set of labels in \(\pmb L\) to get
\begin{align}
\sum_{\pmb L}\,\chi_\Q(\pmb L,\vec s,\vec \sigma)\,\cO_{\Q}(\pmb L)=\sum_{\vec t}\sum_{\vec\tau}\left(\sum_{\pmb L}\,
\chi_\Q(\pmb L,\vec s,\vec \sigma)\,\chi_\Q(\pmb L, \vec t,\vec \tau)
\right)
\mathcal{O}_\mathcal{Q}(\vec n,\vec t ,\vec \tau)
\end{align}
Using the orthogonality relation \eqref{orthogonality chi 2}, the above equation becomes
\begin{align}
\sum_{\pmb L}\,\chi_\Q(\pmb L,\vec s,\vec \sigma)\,\cO_{\Q}(\pmb L)&=\sum_{\vec t}\sum_{\vec\tau}
\left(
\frac{1}{c_{\vec{n}}}\,\sum_{\vec\eta\times\vec\rho} \delta\left(\text{Adj}_{\vec\eta\times\vec\rho\,}(\vec \sigma)\,\vec \tau^{\,-1}\right)
\delta_{\vec \rho(\vec{ s}),\vec{ t} }
\right)
\mathcal{O}_\mathcal{Q}(\vec n,\vec t ,\vec \tau)\nn
&=
\frac{1}{c_{\vec{n}}}\,\sum_{\vec\eta\times\vec\rho}
\mathcal{O}_\mathcal{Q}\left(\vec n,\vec \rho(\vec s\,) ,\text{Adj}_{\vec\eta\times\vec\rho\,}(\vec \sigma)\right)
=
\frac{1}{c_{\vec{n}}}\,\sum_{\vec\eta\times\vec\rho}
\mathcal{O}_\mathcal{Q}(\vec n,\vec s ,\vec \sigma)
\end{align}
where in the last line we used the constraint \eqref{constr}. Now the sum over the permutations \(\vec\eta,\,\vec\rho\) is trivial, and it just gives a factor of \(c_{\vec n}\). We then have that the inverse of the map \eqref{fourier2} is simply
\begin{align}\label{inverse fourier}
\mathcal{O}_\mathcal{Q}(\vec n,\vec s ,\vec \sigma)=
\sum_{\pmb L}\chi_\Q(\pmb L,\vec s,\vec \sigma)\,\cO_{\Q}(\pmb L)
\end{align}

\section{Two and Three Point Functions}\label{Sec: Two and Three Point Functions}

In this section we will derive an expression for the two and three point function of matrix invariants, using the free field metric. All the computations are done using the Quiver Restricted Schur polynomials. The result for the two point function is rather compact, and offers a nice way to describe the Hilbert space of holomorphic GIOs. On the other hand, the expression for the three point function is still quite involved. We give a diagrammatic description of the answer in section \ref{sec: Holomorphic Gauge Invariant Operator Ring Structure Constants}, leaving the analytical expression and its derivation in Appendix \ref{app:Holomorphic Ring SC}.

\subsection{Hilbert space of holomorphic gauge invariant operators}\label{sec: 2pt function}
%As we have already anticipated, 
In the free field metric, the Quiver Restricted Schur polynomials \eqref{fourier2} form an orthogonal basis for the 2-point functions of holomorphic and anti-holomorphic matrix invariants. Explicitly, in Appendix \ref{ortho L derivation} we derive the equation
\begin{align}\label{ortho L} 
\left\langle
\mathcal O_{\mathcal Q}(\pmb L)\,
\mathcal O_{\mathcal Q}^\dagger(\pmb {L'})
\right\rangle
=\delta_{\pmb L,\pmb{L'}}\,\, c_{\vec n}\,\prod_a f_{N_a}(R_a)
\end{align}
where $ c_{\vec n} $ is given in \eqref{norm const c_n}. The quantity $f_{N_a}(R_a)$ is the product of weights of the $U(N_a)$ representation $R_a$, and it is defined as 
\begin{align}
f_{N_a}(R_a) = \prod_{i , j } ( N_a - i + j ) 
\end{align}
Here $i$ and $j$ label the row and column of the Young diagram $R_a$.
At finite $N_a$, this quantity vanishes if the length of the first column of its Young diagram exceeds $N_a$, that is if $c_1(R_a)> N_a$. This means that for a generic quiver $\mathcal Q$ the Hilbert space $\mathcal H_{\mathcal Q}$ of holomorphic GIOs can be described by
\begin{align}\label{Hilbert space}
\mathcal H_{\mathcal Q}=\text{Span}\left\{
\mathcal O_{\mathcal Q}(\pmb L)|\,\,\pmb L \,\, s.t. \,\,c_1(R_a)\leq N_a ,\,\forall a
\right\}
\end{align}
We can see how the finite $N_a$ constraints of any matrix invariant are captured by the simple rule $c_1(R_a)\leq N_a$.
We leave the formal proof of \eqref{ortho L} in Appendix \ref{ortho L derivation}, and we present here only the main steps.
In the free field metric, the only non-zero correlators are the ones that couple fields of the same kind (\emph{e.g.} $\Phi_{ab,\alpha}$ with $\Phi_{ab,\alpha}^\dagger$):
\begin{align}\label{free field metric}
\left\langle
\left(\Phi_{ab,\alpha}\right)^i_j
(\Phi_{ab,\alpha}^\dagger)^k_l
\right\rangle
=\delta^i_l\delta^k_j\,,\quad
\left\langle
\left(Q_{a,\beta}\right)^i_s
(Q_{a,\beta}^\dagger)^p_l
\right\rangle
=\delta^i_l\delta^p_s\,,\quad
\left\langle
\left(\bar Q_{a,\gamma}\right)^{\bar s}_j
(\bar Q_{a,\gamma}^\dagger)^k_{\bar p}
\right\rangle
=\delta^k_j\delta_{\bar p}^{\bar s}
\end{align}
Consequently, we can use Wick contractions to find the 2-point function of matrix invariants in the permutation basis of eq. \eqref{Q def}:
\begin{align}\label{corr K basis}
\left\langle
\mathcal O_{\mathcal Q}(\vec n,\vec s,\vec \sigma)
\mathcal O_{\mathcal Q}^\dagger(\vec n,\vec{ s}\,',\vec{ \sigma}\,')
\right\rangle=
\sum_{\vec\eta,\,\vec\rho}
\,\delta_{\vec s\,',\vec\rho\,(\vec s\,)}\,
\prod_a\,\Tr_{V_{N_a}^{\otimes n_a}}\left[\vphantom{\sum}\text{Adj}_{\vec\eta\times\vec\rho\,}(\sigma_a)\,
(\sigma_a')^{-1}
\right]
%
%(\prod_\beta\delta_{\pmb{s'}_{a,\beta}}^{\rho_{a,\beta}(\pmb{s}_{a,\beta})})
%(\prod_\gamma\delta_{\pmb{\bar s'}_{a,\gamma}}^{\bar \rho_{a,\gamma}(\pmb{\bar s}_{a,\gamma})})
\end{align}
The details of the computations are shown in Appendix \ref{ortho L derivation}.
Here the trace is taken over the product space $V_{N_a}^{\otimes n_a}$, $V_{N_a}$ being the fundamental representation of $U(N_a)$ and $n_a =\sum_{b,\alpha}n_{ab,\alpha}+\sum_\beta n_{a,\beta}$. The trace in the product is also equal to $N_a$ raised to the power of the number of cycles in the permutation appearing as an argument. 
Using the definition \eqref{fourier2} we get
\begin{align}
\left\langle
\mathcal O_{\mathcal Q}(\pmb L)
\mathcal O_{\mathcal Q}^\dagger(\pmb {L'})
\right\rangle&=
\sum_{\vec s,\vec{s}\,'}\sum_{\vec \sigma,\vec{\sigma}\,'}
\chi_{\Q}(\pmb L,\vec s,\vec \sigma)
\chi_\Q^*(\pmb L',\vec s\,',\vec\sigma\,')
\left\langle
\mathcal O_{\mathcal Q}(\vec n,\vec s,\vec \sigma)
\mathcal O_{\mathcal Q}^\dagger(\vec n,\vec{ s}\,',\vec{ \sigma}\,')
\right\rangle\\[3mm]
&=
\sum_{\vec s}\sum_{\vec \sigma,\vec{\sigma}\,'}\sum_{\vec\eta\,,\vec\rho}
\chi_{\Q}(\pmb L,\vec s,\vec \sigma)
\chi_\Q(\pmb L',\vec\rho(\vec s),\vec\sigma\,')\prod_a\Tr_{V_{N_a}^{\otimes n_a}}\left[\vphantom{\sum}\text{Adj}_{\vec\eta\times\vec\rho\,}(\sigma_a)\,
(\sigma_a')^{-1}
\right]\nonumber
\end{align}
where we also exploited the reality of the quiver characters.
Using the invariance and orthogonality properties \eqref{invariance chi} and \eqref{orthogonality chi} we get
\begin{align}\label{2pf text last}
\left\langle
\mathcal O_{\mathcal Q}(\pmb L)
\mathcal O_{\mathcal Q}^\dagger(\pmb {L'})
\right\rangle&=
\sum_{\vec s}\sum_{\vec \sigma,\vec{\sigma}\,'}\sum_{\vec\eta,\,\vec\rho}
\chi_\Q(\pmb L, \vec s,\vec\sigma)
\chi_\Q(\pmb L',\vec s,\vec\sigma\,')\prod_a\,\Tr_{V_{N_a}^{\otimes n_a}}\left[\vphantom{\sum}\sigma_a\,
(\sigma_a')^{-1}
\right]
\end{align}
Now the sum over the permutations $\vec\eta$ and $\vec\rho$ is trivial, and can be computed to give the factor $c_{\vec n}$ that appears in \eqref{ortho L}. Using the substitution $\sigma_a\rightarrow\tau_a\cdot\sigma_a'$, the identity \eqref{almost ortho}, and computing explicitly the trace in \eqref{2pf text last} allows us to obtain the result \eqref{ortho L}.

\subsection{Chiral ring structure constants and three point functions}\label{sec: Holomorphic Gauge Invariant Operator Ring Structure Constants}
In Appendix \ref{app:Holomorphic Ring SC} we derive an equation for the holomorphic GIO ring structure constants \(G_{\pmb L^{(1)},\,\pmb L^{(2)},\,\pmb L^{(3)}}\), defined as the coefficients of the operator product expansion
\begin{align}
\cO_\Q(\pmb L^{(1)})\,\cO_\Q(\pmb L^{(2)})=
\sum_{\pmb L^{(3)}}
G_{\pmb L^{(1)},\,\pmb L^{(2)},\,\pmb L^{(3)}}\,
\cO_\Q(\pmb L^{(3)})
\end{align}
Because of the orthogonality of the two point function \eqref{ortho L}, we also obtain an equation for the three point function:
\begin{align}\label{3pf}
\left\langle
\mathcal O_{\mathcal Q}(\pmb L^{(1)})\,
\mathcal O_{\mathcal Q}(\pmb {L}^{(2)})\,
\mathcal O_{\mathcal Q}^\dagger(\pmb L^{(3)})
\right\rangle
=c_{\vec n^{(3)}}\,\, G_{\pmb L^{(1)},\,\pmb L^{(2)},\,\pmb L^{(3)}}\,\prod_a f_{N_a}\left(R_a^{(3)}\right)
\end{align}
We only give here a pictorial interpretation of the equation we derived for \(G_{\pmb L^{(1)},\,\pmb L^{(2)},\,\pmb L^{(3)}}\), leaving the technicalities in Appendix \ref{app:Holomorphic Ring SC}. In particular, eq. \eqref{eq for G} gives the analytical formula for the \(G_{\pmb L^{(1)},\,\pmb L^{(2)},\,\pmb L^{(3)}}\) coefficients.

Let us begin by considering an example. We will show how to draw the diagram for the chiral ring structure constants for an $\N=2$ SCQD, through a step-by-step procedure.
The quiver for this theory is shown in Fig. \ref{fig: N=2 sqcd}.
As we discussed in the previous section, for any given model, a basis of GIOs is labelled by $\pmb L=\{R_a, r_{ab,\alpha},r_{a,\beta},S_{a,\beta},\bar r_{ a,\gamma},\bar S_{a,\gamma}, \nu_a^+,\nu_a^-\}$.
However, for an $\N=2$ SQCD theory, many of these $a,b,\alpha,\beta,\gamma$ indices are redundant: for this reason we can simplify \(\pmb L\) as
\begin{align}\label{N=2 L set}
\pmb L=\{R, r,\,r_q,S,\,\bar r_q,\bar S,\, \nu^+,\nu^-\}
\end{align}
Here \(r\) is the representation associated with the adjoint field \(\phi\); \(S\) denotes a state in the \(SU(F)\) representation \(r_q\) and \(\bar S\) denotes a state in the \(SU(F)\) representation \(\bar r_q\). \(R\) is the representation associated with the gauge group, \(U(N)\).
We therefore want to compute the three point function \eqref{3pf}, where all the $\pmb L^{(i)}$, $i=1,2,3$, are of the form given in \eqref {N=2 L set}. We split this process into five steps, that we now describe.

\begin{figure}[H]
\begin{center}\includegraphics[scale=1.82]{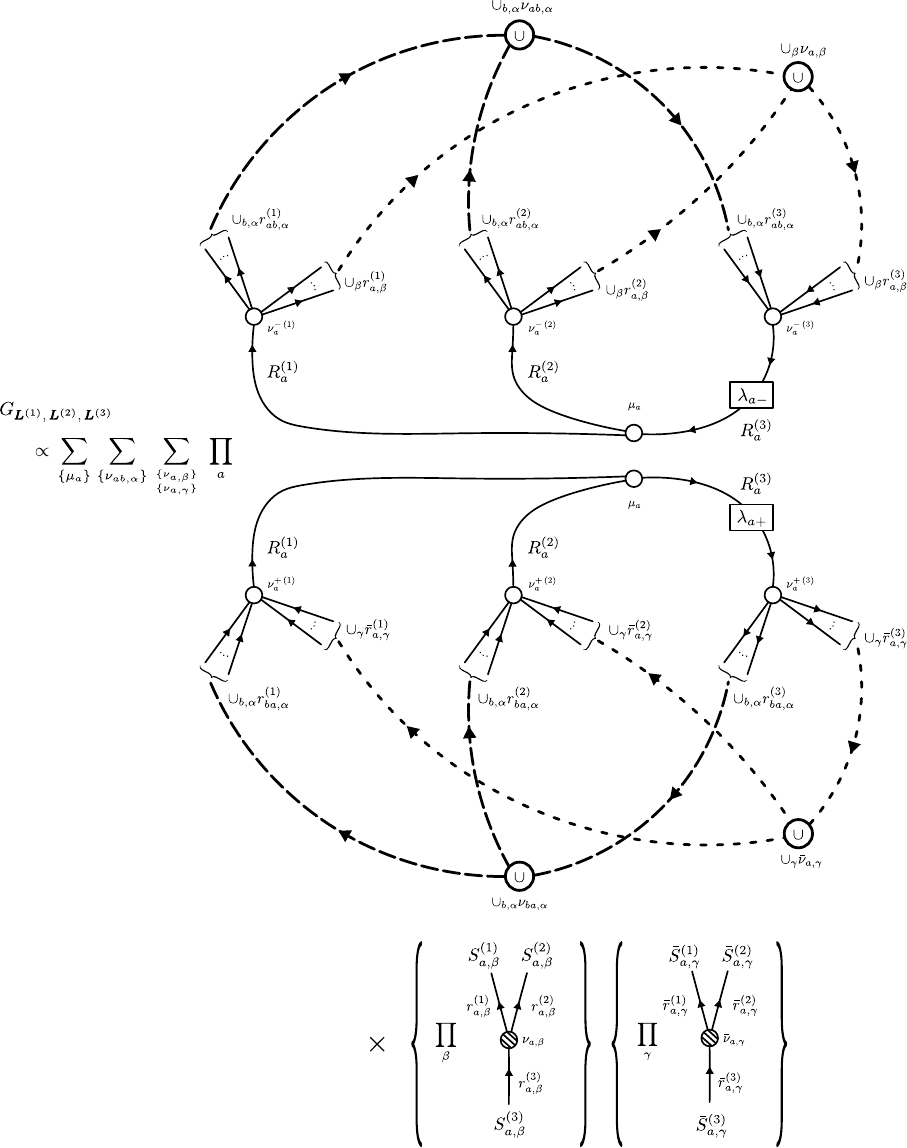}
\end{center}
\caption{Pictorial description of the expression for the holomorphic GIO ring structure constants \(G_{\pmb L^{(1)},\,\pmb L^{(2)},\,\pmb L^{(3)}}\), corresponding to eq. \eqref{eq for G}.}\label{fig: CRSC picture}
\end{figure}

\begin{itemize}
\item[i)]{\textbf{Create the split node quiver diagram.}}
The first step is to create the split-node quiver diagram from the $\N=2$ SCQD quiver of Fig. \ref{fig: N=2 sqcd}. This involves separating the gauge node into two components, one that collects all the incoming edges and one from which all the edges exit. The former is called a positive node of the split-node quiver, the latter is called a negative node. These two are then joined by an edge, called a gauge edge, directed from the 
positive to the negative node. We then decorate all the edges in the split-node quiver with symmetric group representation labels. The positive and negative nodes in the split-node diagram are points where the edges meet. Since the edges now carry a symmetric group representation, we interpret them as representation branching points, to which we associate a branching coefficient \eqref{B map}. To the positive node we associate the branching multiplicity $\nu^+$, to the negative node we associate the branching multiplicity $\nu^-$.
%The diagram we will use for these objects is the same as the one in Fig. \ref{fig: branching coeff}. 
Finally, we label the open endpoints of the quark and antiquark edges with $U(F)$ fundamental and antifundamental representation state labels, $S$ and $\bar S$.
The resulting diagram is shown on the left of Fig. \ref{fig: CRSC split node step1}. Notice that such a diagram contains all the labels in $\pmb L=\{R, r,\,r_q,S,\,\bar r_q,\bar S,\, \nu^+,\nu^-\}$.

\item[ii)]{\textbf{Cut the edges in the split-node quiver.}}
In this step we will cut all the edges in the split-node diagram, as shown in the middle picture of Fig. \ref{fig: CRSC split node step1}.
After all the cuts have been performed, we are left with two trivalent vertices and two edges corresponding to the quark and the antiquark fields. As previously stated, the trivalent vertices will be interpreted as branching coefficients (see Fig. \ref{fig: branching coeff}).
We group these four object into two pairs, depending whether their edges are connected to the positive or negative node of the split-node diagram. This is shown in the rightmost picture of Fig. \ref{fig: CRSC split node step1}.
\begin{figure}[H]
\begin{center}\includegraphics[scale=1.44]{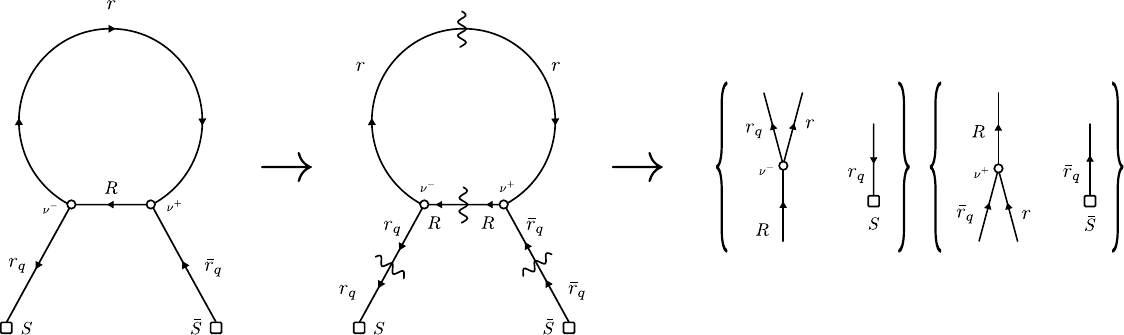}
\end{center}
\caption{From left to right: the split-node quiver for the $\N=2$ SQCD, the same diagram with the cut edges, and the two components of the negative and positive node of the split-node quiver.}\label{fig: CRSC split node step1}
\end{figure}

\item[iii)]{\textbf{Merge the edges connected to the negative node.}} We consider the set of edges connected to the negative node of the split-node quiver.
In order to compute the three point function \eqref{3pf}, we need three copies of these sets, one for each field $\mathcal O_{\mathcal Q}(\pmb L^{(1)})\,,\mathcal O_{\mathcal Q}(\pmb {L}^{(2)})$, $\mathcal O_{\mathcal Q}^\dagger(\pmb L^{(3)})$. These sets are shown in Fig. \ref{fig: CRSC split node step2}. The orientation of the edges in the last pair is reversed: this is because the third field on the LHS of \eqref{3pf} is hermitian conjugate.
\begin{figure}[H]
\begin{center}\includegraphics[scale=1.89]{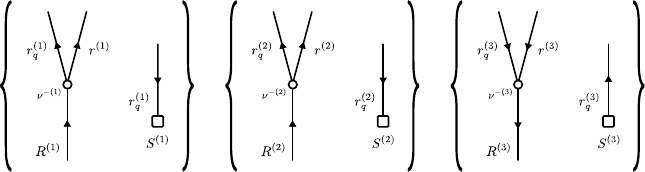}
\end{center}
\caption{The three sets of trivalent vertices and edges needed to construct part of the $\N=2$ SQCD three point function diagram.}\label{fig: CRSC split node step2}
\end{figure}
We will now suitably merge the three trivalent vertices (branching coefficients) in Fig. \ref{fig: CRSC split node step2}, and join the three edges corresponding to the quark fields. 
The outcome of this fusing process is shown in Fig. \ref{fig: CRSC split node step3}.
We introduced three new trivalent vertices, which as usual we interpret as branching coefficients: the labels $\mu$, $\nu_r $ and $\nu_q $ denote their multiplicity.
The fusing of the three quark edges has been achieved by introducing a Clebsch-Gordan coefficient, see Fig. \ref{fig: U(N) CG}.
We further impose that the label for the multiplicity of the representation branching $r_q^{(1)}\otimes r_q^{(2)}\rightarrow r_q^{(3)}$ is the same in both the Clebsch-Gordan coefficient and the branching coefficient that appear in Fig. \ref{fig: CRSC split node step2}.
In the figure we also inserted a permutation $\lambda_-$ in the edge carrying the representation $R^{(3)}$. The purpose of this permutation is to rearrange tensor factors given the two different factorisation of $R^{(3)}$, that is from $(r^{(1)}\otimes r^{(1)})\otimes(r_q^{(2)}\otimes r_q^{(2)})\rightarrow r^{(3)}\otimes r_q^{(3)}\rightarrow R^{(3)}$ to $R^{(3)}\rightarrow R^{(1)}\otimes R^{(2)}\rightarrow (r^{(1)}\otimes r_q^{(1)})\otimes(r^{(2)}\otimes r_q^{(2)})$. 
%The outcome of this merging process is shown in Fig. \ref{fig: CRSC split node step3}.
\begin{figure}[H]
\begin{center}\includegraphics[scale=1.6]{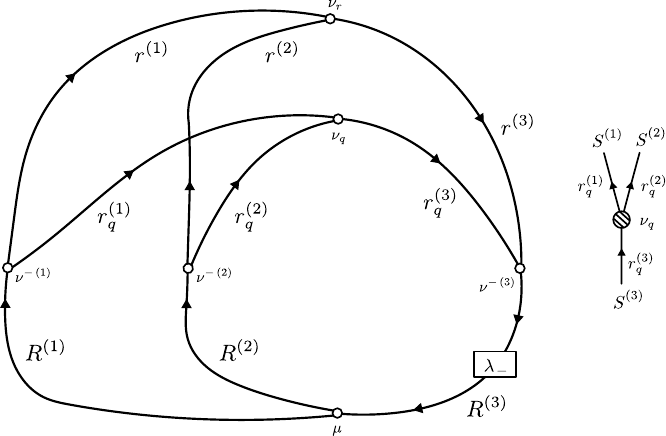}
\end{center}
\caption{Merging of branching coefficients and quarks labels for the three sets in Fig. \ref{fig: CRSC split node step2}.}\label{fig: CRSC split node step3}
\end{figure}

We thus obtained a closed network of branching coefficients, together with a single $SU(F)$ Clebsch-Gordan coefficient. All the edges involved into this process were the ones connected to the negative node of the split-node diagram they belonged to.

\item[iv)]{\textbf{Merge the edges connected to the positive node. }}
By repeating the fusing process presented in point iii) for all the edges connected to the positive node of the split-node quiver, we obtain a diagram very similar to the one in Fig. \ref{fig: CRSC split node step3}.
The only rule that we impose is that the multiplicity labels for representation branchings which appear in both these diagrams have to be the same.
In our example, the branching of $R^{(3)}$ into $R^{(1)}$ and $R^{(2)}$ will appear in both diagrams. This is because the edge carrying the representation label $R$ is connected to both the positive and negative node of the split-node quiver, as it can be seen from Fig. \ref{fig: CRSC split node step1}. Therefore these two branching coefficients will share the same multiplicity label, $\mu$. Similarly, the branching of $r^{(1)}$ and $r^{(2)}$ into $r^{(3)}$ will be present in both diagrams too. Following the same rule, these two branching coefficients will then have the same multiplicity label, $\nu_r$.

\item[v)]{\textbf{Combine the diagrams and sum over multiplicities.}}
To obtain the final expression for the three point function, we just need put together the two diagram we obtained in the steps iv) and v) and sum over the multiplicities $\mu$, $\nu_r $, $\nu_q $ and $\bar \nu_q $.
This final diagram is shown in Fig. \ref{fig: CRSC split node step4}.
\begin{figure}[H]
\begin{center}\includegraphics[scale=1.6]{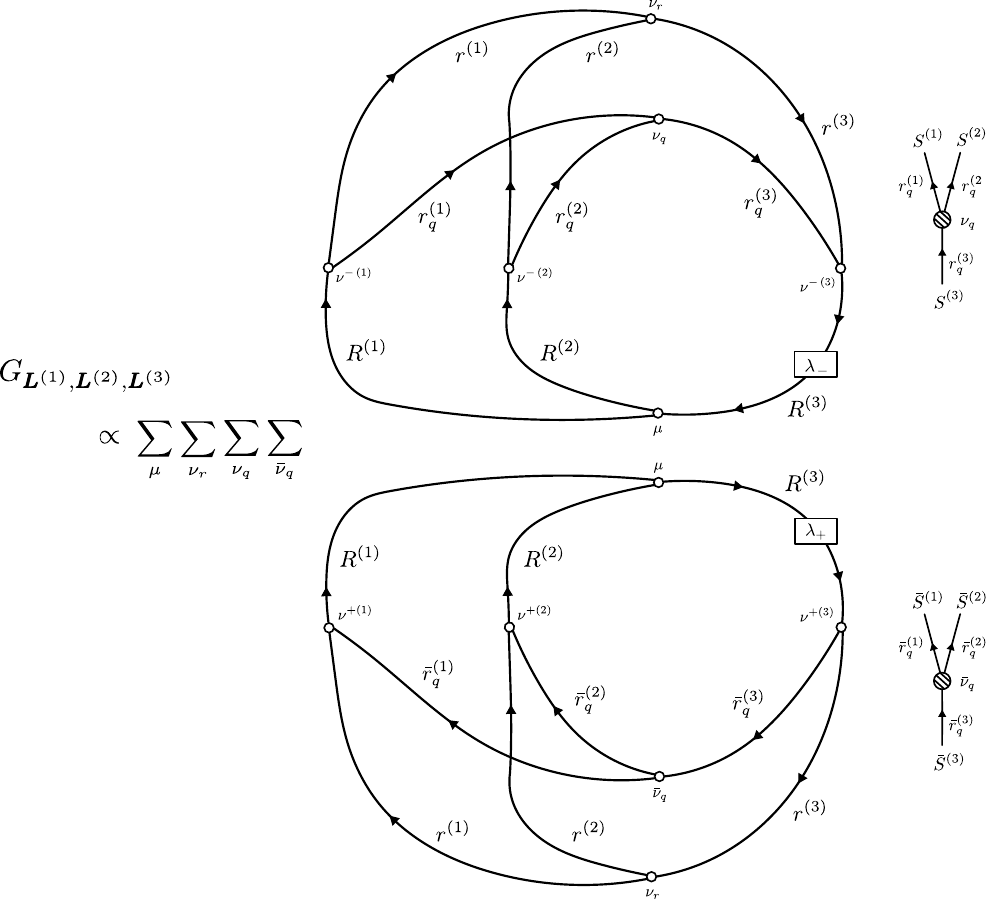}
\end{center}
\caption{The diagram of the three point function \eqref{3pf} for the $\N=2$ SQCD.}\label{fig: CRSC split node step4}
\end{figure}

\end{itemize} % end of the big itemize
In Appendix \ref{App_sub: CRSC derivation for N=2} we give a purely diagrammatic derivation of this result.
We can see how the answer for the three point function factorises into two components: the former features only edges connected to the negative node of the split-node diagram, the latter only involves edges connected to its positive node.
The same behaviour can be observed in the answer for the three point function of matrix invariants of generic quivers. We are now going to present this general result. The diagram for the three point function \eqref{3pf} is shown in Fig. \ref{fig: CRSC picture}.
\newpage

\begin{figure}[H]
\begin{center}\includegraphics[scale=2]{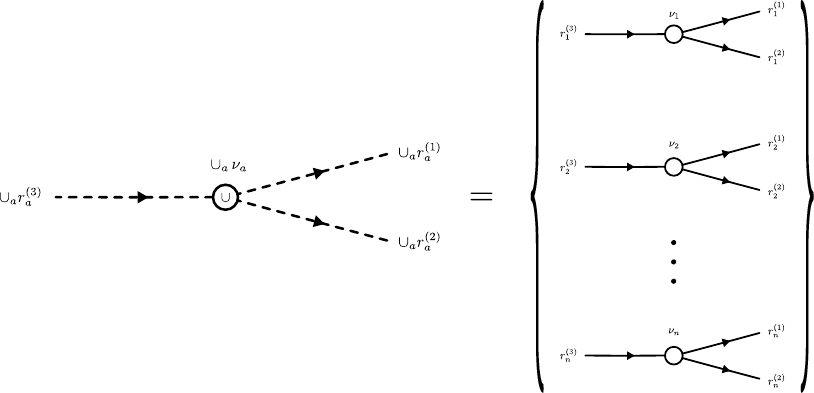}
\end{center}
\caption{A shorthand notation for a collection of branching coefficients.}\label{fig: shorthand branching}
\end{figure}

In drawing this picture we used the diagrammatic shorthand notation displayed in Fig. \ref{fig: shorthand branching}.

The $\lambda_{a-}$ and $\lambda_{a+}$ in Fig. \ref{fig: CRSC picture} are permutations of $n_{a}^{(3)}$ elements, defined by the equations \eqref{lamb_{a-}} and \eqref{lamb_{a+}}.
Figure \ref{fig: CRSC picture} shows that the holomorphic GIO ring structure constants factorise into a product over all the gauge nodes $a$ of the quiver. Each one of these terms, whose diagrammatic interpretation is drawn in the figure, further factorises into a product of two components. They correspond to the positive and negative nodes of the split node $a$, with $a=1,2,...,n$ (see also Fig. \ref{fig: fourier}). Notice that the multiplicity labels $\mu_a$, $\nu_{ab,\alpha}$, $\nu_{a,\beta}$ and $\bar \nu_{a,\gamma}$ always appear in pairs. For example, $\mu_a$ appears both in the upper and lower (disconnected) parts of the split-node $a$ diagram. In the same diagram, $\nu_{a,\beta}$ appears in both a symmetric group branching coefficient and in a Clebsch-Gordan coefficient.

By inspecting Fig. \ref{fig: CRSC picture} we can write four selection rules for the holomorphic GIO ring structure constants:
\begin{itemize}
\item[i)] upon the restriction $\left.S_{n_a^{(3)}}\right|_{H_a}$, where $H_a={S_{n_a^{(1)}}\times S_{n_a^{(2)}}}$, the $S_{n_a^{(3)}}$ representation $R_a^{(3)}$ becomes reducible. This reduction must contain the tensor product representation $R_a^{(1)}\otimes R_a^{(2)}$, $\forall\, a$. This implies the constraint $g(R_a^{(1)},R_a^{(2)};R_a^{(3)})\neq 0$, $\forall\, a$.
\item[ii)] upon the restriction $\left.S_{n_{ab,\alpha}^{(3)}}\right|_{H_{ab,\alpha}}$, where $H_{ab,\alpha}={S_{n_{ab,\alpha}^{(1)}}\times S_{n_{ab,\alpha}^{(2)}}}$, the $S_{n_{ab,\alpha}^{(3)}}$ representation $r_{ab,\alpha}^{(3)}$ becomes reducible. This reduction must contain the tensor product representation $r_{ab,\alpha}^{(1)}\otimes r_{ab,\alpha}^{(2)}$, $\forall\, a,\,b,\,\alpha$. This implies the constraint $g(r_{ab,\alpha}^{(1)},r_{ab,\alpha}^{(2)};r_{ab,\alpha}^{(3)})\neq 0$, $\forall\, a,\,b,\,\alpha$.
\item[iii)] upon the restriction $\left.S_{n_{a,\beta}^{(3)}}\right|_{H_{a,\beta}}$, where $H_{a,\beta}={S_{n_{a,\beta}^{(1)}}\times S_{n_{a,\beta}^{(2)}}}$, the $S_{n_{a,\beta}^{(3)}}$ representation $r_{a,\beta}^{(3)}$ becomes reducible. This reduction must contain the tensor product representation $r_{a,\beta}^{(1)}\otimes r_{a,\beta}^{(2)}$, $\forall\, a,\,\beta$. This implies the constraint $g(r_{a,\beta}^{(1)},r_{a,\beta}^{(2)};r_{a,\beta}^{(3)})\neq 0$, $\forall\, a,\,\beta$.
\item[iv)]upon the restriction $\left.S_{\bar n_{a,\gamma}^{(3)}}\right|_{H_{a,\gamma}}$, where $H_{a,\gamma}={S_{\bar n_{a,\gamma}^{(1)}}\times S_{\bar n_{a,\gamma}^{(2)}}}$, the $S_{\bar n_{a,\gamma}^{(3)}}$ representation $\bar r_{a,\gamma}^{(3)}$ becomes reducible. This reduction must contain the tensor product representation $\bar r_{a,\gamma}^{(1)}\otimes \bar r_{a,\gamma}^{(2)}$, $\forall\, a,\,\gamma$. This implies the constraint $g(\bar r_{a,\gamma}^{(1)},\bar r_{a,\gamma}^{(2)};\bar r_{a,\gamma}^{(3)})\neq 0$, $\forall\, a,\,\gamma$.
\end{itemize}
%With the notation $R^{(3)}\supseteq R^{(1)}\otimes R^{(2)}$ we mean that the representation \(R^{(3)}\) of $S_{n_1+n_2}$ contains the representation \(R^{(1)}\otimes R^{(2)}\) of $S_{n_1}\times S_{n_2}$, when we restrict $S_{n_1+n_2}\downarrow S_{n_1}\times S_{n_2}$.
All these rules are enforced by the branching coefficients networks in Fig. \ref{fig: CRSC picture}.
Given two matrix invariants labelled by $\pmb{L}^{(1)}$ and $\pmb{L}^{(2)}$ respectively, we conclude that $G_{\pmb L^{(1)},\,\pmb L^{(2)},\,\pmb L^{(3)}}\neq 0$ if and only if $\pmb L^{(3)}$ satisfies the selection rules i) - iv) above.

\section{An Example: Quiver Restricted Schur Polynomials for an $\mathcal N=2$ SQCD}\label{Sec: Schur examples}

We will now present some explicit examples of quiver Schurs for an $\mathcal N=2$ SQCD, whose $\N=1$ quiver is depicted in Fig. \ref{fig: N=2 sqcd}. We will begin by listing all the matrix invariants in the permutation basis \eqref{Q def} that it is possible to build using a fixed amount $\vec n$ of fundamental fields. We will then Fourier transform these operators to the quiver Schurs basis using \eqref{fourier2}. 
The set of representation theory labels needed to identify any matrix invariant in an $\mathcal N=2$ SQCD has been explicitly given in \eqref{N=2 L set}. In the following we will continue to use such a convention.

\begin{figure}[H]
\begin{center}\includegraphics[scale=1.9]{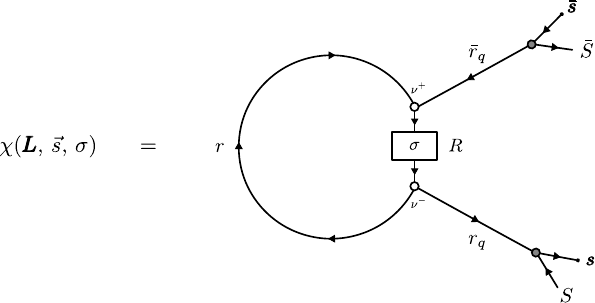}
\end{center}
\caption{Diagram for the $\N=2$ SQCD quiver character, corresponding to eq. \eqref{N=2 sqcd quiver character}.}\label{fig: N=2 SQCD quiv char}
\end{figure}

The permutation basis is generated by
\begin{align}
\cO(\vec n,\,\vec s,\,\sigma)= \left(\phi^{\otimes\, n}\right)^{I}_{J}\otimes\left(Q^{\otimes \,n_Q}\right)^{I_Q}_{\pmb s}\otimes\left(\bar Q^{\otimes\, \bar n_Q}\right)_{J_Q}^{\bar {\pmb s}}\,\,\left(\sigma \right)_{I\times I_Q}^{J\times J_Q}
\end{align}
where \(\vec n=\{n,\,n_Q,\,\bar n_Q\}\) specifies the field content of the operator $\cO$, and $\vec s=(\pmb s,\pmb{\bar s})$. 
As we previously stated, we construct the quiver Schurs $\cO(\pmb L)$ by using the Fourier transform \eqref{fourier2}:
\begin{align}\label{fourier for N=2 sqcd}
\cO(\pmb L)=\sum_{\sigma,\,\vec s}\,\chi(\pmb L,\,\vec s,\, \sigma)\,\cO(\vec n,\,\vec s,\,\sigma)
\end{align}
where \(\pmb L = \{R, r,\,r_q,S,\,\bar r_q,\bar S,\, \nu^+,\nu^-\}\) has been defined in eq. \eqref{N=2 L set}.
In this formula \(\chi(\pmb L,\,\vec s,\, \sigma)\) is the \(\N=2\) SQCD quiver character, which reads
\begin{align}\label{N=2 sqcd quiver character}
\chi(\pmb L,\,\vec s,\, \sigma)=c_{\pmb L}\,D_{i,j}^R(\sigma)\,\left\{B^{R\rightarrow\, r,\, r_q\,;\nu^-}_{j\rightarrow\, l ,\,p}\, C_{\pmb s}^{\,r_q,\,S,\,p}\right\}
\,\left\{B^{R\rightarrow\, r,\, \bar r_q\,;\nu^+}_{i\rightarrow\, l ,\,t}\, C_{\pmb{\bar s}}^{\,\bar r_q,\,\bar S,\,t}\right\}
\end{align}
Figure \ref{fig: N=2 SQCD quiv char} shows the diagram for this quantity.

We now focus on some fixed values of $\vec n$.

\subsection*{- $\vec n=(2,1,1)$ field content}

We start by listing the Fourier transformed holomorphic GIOs \eqref{fourier2} that we can build with the set of fields $\{\phi,\phi,Q,\bar Q\}$, that is with the choice \(\vec n=(2,1,1)\). In the permutation basis, these operators read
\begin{equation}
\begin{array}{ll}
\cO(\vec n\,,s,\bar s\,,(1))=(\phi)(\phi)(\bar Q Q)^{\bar s}_s\,,\qquad\quad
&\cO(\vec n\,,s,\bar s\,,(12))=(\phi\phi)(\bar Q Q)^{\bar s}_s\,,\\[3mm]
\cO(\vec n\,,s,\bar s\,,(13))=(\phi)(\bar Q \phi Q)^{\bar s}_s\,,\qquad\quad
&\cO(\vec n\,,s,\bar s\,,(23))=(\phi)(\bar Q \phi Q)^{\bar s}_s\,,\\[3mm]
\cO(\vec n\,,s,\bar s\,,(123))=(\bar Q\phi\phi Q)^{\bar s}_s\,,\qquad\quad
&\cO(\vec n\,,s,\bar s\,,(132))=(\bar Q\phi\phi Q)^{\bar s}_s
\end{array}
\end{equation}
where the round brackets denote $U(N)$ indices contraction. Notice that in this case $\vec s= (s,\bar s)$.
We will now construct the Fourier transformed operators. For this field content we do not have any branching multiplicity \(\nu^+,\,\nu^-\): we can drop them from the set of labels \(\pmb L\), which now reads $\pmb L=\{R, r,\,r_q,S,\,\bar r_q,\bar S\} $. We then look for the operators $\cO(\pmb L_{i})$, $i=1,2,3,4$, where
\begin{align}
&\Yvcentermath1
\pmb L_1=\{{\tiny\yng(3)}\,,{\tiny\yng(2)}\,,{\tiny\yng(1)}\,, S\,,{\tiny\overline{\yng(1)}}\,, \bar S\}\,,\qquad\quad
\pmb L_2=\left\{ \vphantom{\frac{1}{1}} {\tiny\yng(1,1,1)}\,,{\tiny\yng(1,1)}\,,{\tiny\yng(1)}\,, S\,,{\tiny\overline{\yng(1)}}\,, \bar S\right\}\,,\\[6mm]\nonumber
&\Yvcentermath1
\pmb L_3=\left\{ \vphantom{\sum} {\tiny\yng(2,1)}\,,{\tiny\yng(2)}\,,{\tiny\yng(1)}\,, S\,,{\tiny\overline{\yng(1)}}\,, \bar S\right\}\,,\qquad\quad
\pmb L_4=\left\{ \vphantom{\sum} {\tiny\yng(2,1)}\,,{\tiny\yng(1,1)}\,,{\tiny\yng(1)}\,, S\,,{\tiny\overline{\yng(1)}}\,, \bar S\right\}
\end{align}
We left the states $S,\,\bar S$ of the fundamental and antifundamental representation of $SU(F)$ implicit.

We first notice that, having one quark-antiquark pair only, the Clebsch-Gordan coefficients simplify as
\begin{align}
\Yvcentermath1 C_{\pmb s}^{\,r_q,\,S,\,p}=C_{ s}^{\,\scalebox{.4}{\yng(1)}\,,\,S,\,p}\equiv \delta^S_s\,,\qquad\quad
\Yvcentermath1 C_{\pmb {\bar s}}^{\,\bar r_q,\,\bar S,\,t}=C_{ \bar s}^{\, \bar{\scalebox{.4}{\yng(1)}}\,,\,\bar S,\,t}\equiv \delta^{\bar S}_{\bar s}
\end{align}
We can then easily compute \(\chi(\pmb L_1)\) and \(\chi(\pmb L_2)\). Both the symmetric group representation branching \(\Yvcentermath1 {\tiny\yng(3)}\rightarrow {\tiny\yng(2)}\otimes{\tiny\yng(1)}\) and \(\Yvcentermath1 {\tiny\yng(1,1,1)}\rightarrow {\tiny\yng(1,1)}\otimes{\tiny\yng(1)}\) describe the branching of a 1-dimensional space into itself: as such their associate branching coefficients equal 1 identically. On the other hand, \(D^{{\scalebox{.3}{\yng(3)}}}(\sigma)=1\) \(\forall\,\,\sigma\) and \( D^{\Yvcentermath1{\scalebox{.3}{\yng(1,1,1)}}}(\sigma)=\text{sign}(\sigma)\). We then have
\begin{align}\label{N=2 sqcd example, L1,L2}
\chi(\pmb L_1,\,s,\,\bar s,\,\sigma)=\frac{1}{\sqrt{3!}}\,\,\delta^{S}_s\,\delta^{\bar S}_{\bar s}\,,\qquad\quad
\chi(\pmb L_1,\,s,\,\bar s,\,\sigma)=\frac{1}{\sqrt{3!}}\,\,\text{sign}(\sigma)\,\delta^{S}_s\,\delta^{\bar S}_{\bar s}
\end{align}

The \(S_3\) irrep \( \tiny\yng(2,1)\) is two dimensional, and we work in an orthonormal basis \(\{e_1,\,e_2\}\) in which it reads\footnote{Note that this is \emph{not} the convention used in the SageMath software.}

\begin{align}\label{S3 standard}
\begin{array}{lll}
D^{\scalebox{.2}{\yng(2,1)}}((1))=\left(
\begin{array}{cc}
 1 & 0 \\
 0 & 1 \\
\end{array}
\right)\,,&
\,\,\,
D^{\scalebox{.2}{\yng(2,1)}}((12))=\left(
\begin{array}{cc}
 1 & 0 \\
 0 & -1 \\
\end{array}
\right)\,,&
\,\,\,
D^{\scalebox{.2}{\yng(2,1)}}((13))=\left(
\begin{array}{cc}
 -\frac{1}{2} & -\frac{\sqrt{3}}{2} \\
 -\frac{\sqrt{3}}{2} & \frac{1}{2} \\
\end{array}
\right)\\[12mm]
D^{\scalebox{.2}{\yng(2,1)}}((23))=\left(
\begin{array}{cc}
 -\frac{1}{2} & \frac{\sqrt{3}}{2} \\
 \frac{\sqrt{3}}{2} & \frac{1}{2} \\
\end{array}
\right)\,,&
\,\,\,
D^{\scalebox{.2}{\yng(2,1)}}((123))=\left(
\begin{array}{cc}
 -\frac{1}{2} & -\frac{\sqrt{3}}{2} \\
 \frac{\sqrt{3}}{2} & -\frac{1}{2} \\
\end{array}
\right)\,,&
\,\,\,
D^{\scalebox{.2}{\yng(2,1)}}((132))=\left(
\begin{array}{cc}
 -\frac{1}{2} & \frac{\sqrt{3}}{2} \\
 -\frac{\sqrt{3}}{2} & -\frac{1}{2} \\
\end{array}
\right)
\end{array}
\end{align}
If we restrict \(S_3\) to \(S_2\times S_1\), the ${\tiny\yng(2,1)}$ reduces as
\begin{align}
\Yvcentermath1 \left.{\small \yng(2,1)}\,\,\right|_{S_2\times S_1}= {\small\yng(2)}\otimes{\small\yng(1)}\,\,\,\,\oplus\,\,\,\, {\small\yng(1,1)}\otimes{\small\yng(1)}
\end{align}
The restricted group $\left.{S_3}\right|_{S_2\times S_1}$ only contains two elements: $\left.{S_3}\right|_{S_2\times S_1}=\{(1),\,(12)\}$. 
The branching coefficients for this restriction are the matrix elements of the orthogonal operator $B$ such that
\begin{align}
B^{-1}D^{\scalebox{.2}{\yng(2,1)}}((12))B=
D^{\scalebox{.2}{\yng(2)}}((12))\otimes 
D^{\scalebox{.2}{\yng(1)}}((1))\,\oplus \,
D^{\scalebox{.2}{\yng(1,1)}}((12))\otimes
D^{\scalebox{.2}{\yng(1)}}((1)) =\text{diag}(1,-1)
\end{align}
With our basis choice for ${\tiny\yng(2,1)}$ such a decomposition is already manifest, as it is clear from the matrix expression of the identity element and the $(12)$ transposition in \eqref{S3 standard}. Therefore, for this particular configuration, $B$ is just the two dimensional identity matrix: $B=1_2$. If we label \(f_1\) the only state in the ${\tiny\yng(2)}$ of \(S_2\) and \(f_2\) the only state in the ${\tiny\yng(1,1)}$ of \(S_2\), the branching coefficients read
\begin{align}
B_{j\rightarrow 1,1}^{\scalebox{.3}{\yng(2,1)}\,\rightarrow\,\scalebox{.3}{\yng(2)}\,,\,\scalebox{.3}{\yng(1)}}=(e_j,\,f_1)=\delta_{j,1}\,,\qquad\qquad\quad
B_{j\rightarrow 1,1}^{\scalebox{.3}{\yng(2,1)}\,\rightarrow\,\scalebox{.3}{\yng(1,1)}\,,\,\scalebox{.3}{\yng(1)}}=(e_j,\,f_2)=\delta_{j,2}
\end{align}
Inserting this result in \eqref{N=2 sqcd quiver character} we obtain an expression for \(\chi(\pmb L_3)\) and \(\chi(\pmb L_4)\):
\begin{align}\label{N=2 sqcd example, L3,L4}
&\chi(\pmb L_3,\,s,\,\bar s,\,\sigma)=\frac{1}{\sqrt{3}}\,\,\Tr\left[D^{\scalebox{.2}{\yng(2,1)}}(\sigma)\,P^{\scalebox{.2}{\yng(2,1)}\,\rightarrow\,\scalebox{.2}{\yng(2)}\,,\,\scalebox{.2}{\yng(1)}}\right]\,\,\delta^{S}_s\,\delta^{\bar S}_{\bar s}\,,\nn
&\chi(\pmb L_4,\,s,\,\bar s,\,\sigma)=\frac{1}{\sqrt{3}}\,\,\Tr\left[D^{\scalebox{.2}{\yng(2,1)}}(\sigma)\,P^{\scalebox{.2}{\yng(2,1)}\,\rightarrow\,\scalebox{.2}{\yng(1,1)}\,,\,\scalebox{.2}{\yng(1)}}\right]\,\,\delta^{S}_s\,\delta^{\bar S}_{\bar s}
\end{align}
Here \(P^{\scalebox{.2}{\yng(2,1)}\,\rightarrow\,\scalebox{.2}{\yng(2)}\,,\,\scalebox{.2}{\yng(1)}}\) and \(P^{\scalebox{.2}{\yng(2,1)}\,\rightarrow\,\scalebox{.2}{\yng(1,1)}\,,\,\scalebox{.2}{\yng(1)}}\) are the projection operators of the ${\tiny\yng(2,1)}$ of $S_3$ on the ${\tiny\yng(2)}\otimes {\tiny\yng(1)}$ of $S_2\times S_1$ and the ${\tiny\yng(2,1)}$ of $S_3$ on the ${\tiny\yng(1,1)}\otimes {\tiny\yng(1)}$ of $S_2\times S_1$:
\begin{align}
P^{\scalebox{.3}{\yng(2,1)}\,\rightarrow\,\scalebox{.3}{\yng(2)}\,,\,\scalebox{.3}{\yng(1)}}=
\left(
\begin{array}{cc}
1&0\\0&0
\end{array}
\right)\,,\qquad\qquad\quad
P^{\scalebox{.3}{\yng(2,1)}\,\rightarrow\,\scalebox{.3}{\yng(1,1)}\,,\,\scalebox{.3}{\yng(1)}}=
\left(
\begin{array}{cc}
0&0\\0&1
\end{array}
\right)
\end{align}
We are now ready to write down the Fourier transformed operators. Using the definition \eqref{fourier for N=2 sqcd} and the results \eqref{N=2 sqcd example, L1,L2} and \eqref{N=2 sqcd example, L3,L4}, we find that
\begin{align}\label{N=2 sqcd 211 example states}
&\cO(\pmb L_1)=\frac{1}{\sqrt{3!}}\left(
(\phi)(\phi)(\bar Q Q)^{\bar S}_S
+(\phi\phi)(\bar Q Q)^{\bar S}_S
+2(\phi)(\bar Q \phi Q)^{\bar S}_S
+2(\bar Q\phi\phi Q)^{\bar S}_S\,
\right)\,,\nn
&\cO(\pmb L_2)=\frac{1}{\sqrt{3!}}\left(
(\phi)(\phi)(\bar Q Q)^{\bar S}_S
-(\phi\phi)(\bar Q Q)^{\bar S}_S
-2(\phi)(\bar Q \phi Q)^{\bar S}_S
+2(\bar Q\phi\phi Q)^{\bar S}_S\,
\right)\,,\\[3mm]
&\cO(\pmb L_3)=\frac{1}{\sqrt{3}}\left(
(\phi)(\phi)(\bar Q Q)^{\bar S}_S
+(\phi\phi)(\bar Q Q)^{\bar S}_S
-(\phi)(\bar Q \phi Q)^{\bar S}_S
-(\bar Q\phi\phi Q)^{\bar S}_S\,
\right)\,,\nn
&\cO(\pmb L_4)=\frac{1}{\sqrt{3}}\left(
(\phi)(\phi)(\bar Q Q)^{\bar S}_S
-(\phi\phi)(\bar Q Q)^{\bar S}_S
+(\phi)(\bar Q \phi Q)^{\bar S}_S
-(\bar Q\phi\phi Q)^{\bar S}_S\,
\right)\nonumber
\end{align}

We can now perform some checks on this result. First of all, we expect to see the finite $N$ constraints to manifest themselves if the gauge group of the theory is either $N=1$ or $N=2$. In the former case, only $\cO(\pmb L_1)$ should remain, and it is in fact easy to see that for $N=1$ all the other operators are identically zero. For the latter case, we expect $\cO(\pmb L_2)$ to vanish, as \(l({\tiny\yng(1,1,1)})>2\), and as such it violates the finite $N$ constraints. Indeed, using the identity $\phi^2=(\phi)\phi-\det(\phi){1}_2$, which follows from the Cayley-Hamilton theorem, one can verify that $\cO(\pmb L_2)=0$ for a $U(2)$ gauge group.

We also expect these operators to be orthogonal in the free field metric. According to eq. \eqref{corr K basis}, the two point function in the permutation basis is simply
\begin{align}
\left\langle \cO(\vec n\,,s,\bar s\,,\sigma)\,\cO^{\dagger}(\vec n\,,t,\bar t\,,\tau ) \right\rangle=
\delta_{s,t}\,\delta_{\bar s, \bar t}\,\,\sum_{\eta\in S_2}\,N^{C\left[(\eta\times 1)\,\sigma\,(\eta\times 1)^{-1}\,\tau^{-1}\right]}\,,\qquad\, \vec n=(2,1,1)
\end{align}
were $C\left[\sigma\right]$ is the number of cycles in the permutation $\sigma$.
With this equation we can check that all the states in \eqref{N=2 sqcd 211 example states} are orthogonal, and that
\begin{equation}
\begin{array}{ll}
\left\langle \cO(\pmb L_1)\,\cO^{\dagger}(\pmb L_1 ) \right\rangle=2\,N(N+1)(N+2)\,,\qquad\qquad
&\left\langle \cO(\pmb L_2)\,\cO^{\dagger}(\pmb L_2 ) \right\rangle=2\,N(N-1)(N-2)\,,\\[3mm]
\left\langle \cO(\pmb L_3)\,\cO^{\dagger}(\pmb L_3 ) \right\rangle=2\,N(N^2-1)\,,
&\left\langle \cO(\pmb L_4)\,\cO^{\dagger}(\pmb L_4 ) \right\rangle=2\,N(N^2-1)
\end{array}
\end{equation}
in agreement with \eqref{ortho L}.

\subsection*{- $\vec n=(1,2,2)$ field content}

We now consider a different field content, that is $\{\phi,Q,Q,\bar Q,\bar Q\}$. This choice corresponds to \(\vec n=(1,2,2)\). In the permutation basis, the GIOs that we can form with these fields are
\begin{equation}\label{N=2 sqcd example 122 perm basis}
\begin{array}{ll}
\cO(\vec n\,,\vec s\,,(1))=(\phi)\,(\bar Q Q)^{\bar s_1}_{s_{1}}\,(\bar Q Q)^{\bar s_2}_{s_{2}}\,,\qquad\quad
&\cO(\vec n\,,\vec s\,,(12))=(\bar Q \phi Q)^{\bar s_1}_{s_1}\,(\bar Q Q)^{\bar s_2}_{s_{2}}\,,\\[3mm]
\cO(\vec n\,,\vec s\,,(13))=(\bar Q \phi Q)^{\bar s_2}_{s_2}\,(\bar Q Q)^{\bar s_1}_{s_{1}}\,,\qquad\quad
& \cO(\vec n\,,\vec s\,,(23))=(\phi)\,(\bar Q Q)^{\bar s_1}_{s_{2}}\,(\bar Q Q)^{\bar s_2}_{s_{1}}\,,\\[3mm]
\cO(\vec n\,,\vec s\,,(123))=(\bar Q \phi Q)^{\bar s_2}_{s_1}\,(\bar Q Q)^{\bar s_1}_{s_{2}}\,,\qquad\quad
&\cO(\vec n\,,\vec s\,,(132))=(\bar Q \phi Q)^{\bar s_1}_{s_2}\,(\bar Q Q)^{\bar s_2}_{s_{1}}
\end{array}
\end{equation}
Here $\vec s=(s_1,\,s_2\,,\bar s_1,\,\bar s_2)$, and the round brackets denote $U(N)$ indices contraction.

Let us now construct the Fourier transformed operators. As in the previous example, for this fields content we do not have any branching multiplicity \(\nu^+,\,\nu^-\), so that we will drop them from the set of labels in \(\pmb L\).
We will now write the expression for the six operators $\cO(\pmb L_{i})$, $i=1,2,...,6$, with 
\begin{equation}\label{N=2 sqcd 122 example}
\begin{array}{ll}
\pmb L_1=\{{\tiny\yng(3)}\,,{\tiny\yng(1)}\,,{\tiny\yng(2)}\,, S\,,{\tiny\overline{\yng(2)}}\,, \bar S\}\,,\qquad\quad
&\pmb L_2=\left\{ \vphantom{\frac{1}{1}} {\tiny\yng(1,1,1)}\,,{\tiny\yng(1)}\,,{\tiny\yng(1,1)}\,, S\,,{\tiny\overline{\yng(1,1)}}\,, \bar S\right\}\,,\\[6mm]
\pmb L_3=\left\{ \vphantom{\sum} {\tiny\yng(2,1)}\,,{\tiny\yng(1)}\,,{\tiny\yng(2)}\,, S\,,{\tiny\overline{\yng(2)}}\,, \bar S\right\}\,,\qquad\quad
&\pmb L_4=\left\{ \vphantom{\sum} {\tiny\yng(2,1)}\,,{\tiny\yng(1)}\,,{\tiny\yng(1,1)}\,, S\,,{\tiny\overline{\yng(1,1)}}\,, \bar S\right\}\,,\\[6mm]
\pmb L_5=\left\{ \vphantom{\sum} {\tiny\yng(2,1)}\,,{\tiny\yng(1)}\,,{\tiny\yng(2)}\,, S\,,{\tiny\overline{\yng(1,1)}}\,, \bar S\right\}\,,\qquad\quad
&\pmb L_6=\left\{ \vphantom{\sum} {\tiny\yng(2,1)}\,,{\tiny\yng(1)}\,,{\tiny\yng(1,1)}\,, S\,,{\tiny\overline{\yng(2)}}\,, \bar S\right\}
\end{array}
\end{equation}
As in the previous example, we leave the $SU(F)$ states $S,\,\bar S$ implicit.

The symmetric branching group coefficients are similar to the ones already introduced in the previous example. Both the branchings \({\tiny \yng(3)}\rightarrow{\tiny{\yng(1)}}\oplus{\tiny\yng(2)}\) and \({\tiny \yng(1,1,1)}\rightarrow{\tiny{\yng(1)}}\oplus{\tiny\yng(1,1)}\) are trivial, as they correspond to a branching of a 1-dimensional space into itself. These branching coefficients are therefore equal to 1 identically:
\begin{equation}
\begin{array}{ll}
B_{1\rightarrow 1,1}^{\scalebox{.3}{\yng(3)}\,\rightarrow\,\scalebox{.3}{\yng(1)}\,,\,\scalebox{.3}{\yng(2)}}\equiv 1\,,\qquad\qquad\quad
&B_{1\rightarrow 1,1}^{\scalebox{.3}{\yng(1,1,1)}\,\rightarrow\,\scalebox{.3}{\yng(1)}\,,\,\scalebox{.3}{\yng(1,1)}}\equiv 1
\end{array}
\end{equation}
We now turn to the reduction
\begin{align}\label{red 122}
\left.{\small \yng(2,1)}\,\,\right|_{S_1\times S_2}
=
{\small{\yng(1)}}\otimes{\small\yng(2)}
\,\,\,\,\oplus\,\,\,\,
{\small{\yng(1)}}\otimes{\small\yng(1,1)}
\end{align}
As in the previous example, the group $\left.{S_3}\right|_{S_1\times S_2}$ only contains two elements, but this time they are $\left.{S_3}\right|_{S_1\times S_2}=\{(1),\,(23)\}$. This is because the $(1)\times (12)\in S_{1}\times S_2$ has to be embedded into $S_3$, where it corresponds to the transposition $(23)$. The branching coefficients for the reduction in \eqref{red 122} will be the matrix elements of the orthogonal operator $B$ such that
\begin{align}
B^{-1}D^{\scalebox{.2}{\yng(2,1)}}((23))B=
D^{\scalebox{.2}{\yng(1)}}((1))\otimes 
D^{\scalebox{.2}{\yng(2)}}((12))\,\oplus \,
D^{\scalebox{.2}{\yng(1)}}((1))\otimes
D^{\scalebox{.2}{\yng(1,1)}}((12)) =\text{diag}(1,-1)
\end{align}
We equip the \({\tiny \yng(2,1)}\) of $S_3$ with a basis $\{e_1,\,e_2\}$, in which the representation takes the explicit form \eqref{S3 standard}. We then choose $f_1$ and $f_2$ to be the basis vectors of the \({\tiny \yng(2)}\) and the \({\tiny \yng(1,1)}\) of $S_2$ respectively. In this basis the orthogonal matrix $B$ must then take the form
\begin{equation}
B=\left(
\begin{array}{cc}
\frac{1}{2}&-\frac{\sqrt{3}}{2}\\
\frac{\sqrt{3}}{2}&\frac{1}{2}
\end{array}
\right)
\end{equation}
We then have, by construction, $Be_1=f_1$ and $Be_2=f_2$. The branching coefficients for the reduction \eqref{red 122} then read 
\begin{equation}
\begin{array}{llll}
B_{1\rightarrow 1,1}^{\scalebox{.3}{\yng(2,1)}\,\rightarrow\,\scalebox{.3}{\yng(1)}\,,\,\scalebox{.3}{\yng(2)}}&=(e_1,\,f_1)=\frac{1}{2}\,,\qquad\qquad\quad
&B_{1\rightarrow 1,1}^{\scalebox{.3}{\yng(2,1)}\,\rightarrow\,\scalebox{.3}{\yng(1)}\,,\,\scalebox{.3}{\yng(1,1)}}&=(e_1,\,f_2)=-\frac{\sqrt{3}}{2}\,,\\[3mm]
B_{2\rightarrow 1,1}^{\scalebox{.3}{\yng(2,1)}\,\rightarrow\,\scalebox{.3}{\yng(1)}\,,\,\scalebox{.3}{\yng(2)}}&=(e_2,\,f_1)=\frac{\sqrt{3}}{2}\,,\qquad\qquad\quad
&B_{2\rightarrow 1,1}^{\scalebox{.3}{\yng(2,1)}\,\rightarrow\,\scalebox{.3}{\yng(1)}\,,\,\scalebox{.3}{\yng(1,1)}}&=(e_2,\,f_2)=\frac{1}{2}
\end{array}
\end{equation}
It is useful to define the orthogonal projectors 
\begin{align}
P_{i,j}^{\scalebox{.3}{\yng(2,1)}\,\rightarrow\,\scalebox{.3}{\yng(1)}\,,\,\scalebox{.3}{\yng(2)}}=B_{i\rightarrow 1,1}^{\scalebox{.3}{\yng(2,1)}\,\rightarrow\,\scalebox{.3}{\yng(1)}\,,\,\scalebox{.3}{\yng(2)}}\,B_{j\rightarrow 1,1}^{\scalebox{.3}{\yng(2,1)}\,\rightarrow\,\scalebox{.3}{\yng(1)}\,,\,\scalebox{.3}{\yng(2)}}\,,\qquad
P_{i,j}^{\scalebox{.3}{\yng(2,1)}\,\rightarrow\,\scalebox{.3}{\yng(1)}\,,\,\scalebox{.3}{\yng(1,1)}}=B_{i\rightarrow 1,1}^{\scalebox{.3}{\yng(2,1)}\,\rightarrow\,\scalebox{.3}{\yng(1)}\,,\,\scalebox{.3}{\yng(1,1)}}\,B_{j\rightarrow 1,1}^{\scalebox{.3}{\yng(2,1)}\,\rightarrow\,\scalebox{.3}{\yng(1)}\,,\,\scalebox{.3}{\yng(1,1)}}
\end{align}
projecting the $\tiny \yng(2,1)$ of $S_3$ on the $\tiny \yng(1)\otimes\tiny\yng(2)$ and on the $\tiny \yng(1)\otimes\tiny\yng(1,1)$ of $S_1\times S_2$ respectively. We also define the linear operator $T$ through its matrix elements as
\begin{align}
T_{i,j}=
B_{i\rightarrow 1,1}^{\scalebox{.3}{\yng(2,1)}\,\rightarrow\,\scalebox{.3}{\yng(1)}\,,\,\scalebox{.3}{\yng(2)}}\,B_{j\rightarrow 1,1}^{\scalebox{.3}{\yng(2,1)}\,\rightarrow\,\scalebox{.3}{\yng(1)}\,,\,\scalebox{.3}{\yng(1,1)}}
\end{align}
Explicitly, these matrices read
\begin{equation}\label{projectors/op 122}
P^{\scalebox{.3}{\yng(2,1)}\,\rightarrow\,\scalebox{.3}{\yng(1)}\,,\,\scalebox{.3}{\yng(2)}}=
\frac{1}{4}\left(
\begin{array}{cc}
1&\sqrt 3\\\sqrt 3&3
\end{array}
\right)\,,\quad\,\,
P^{\scalebox{.3}{\yng(2,1)}\,\rightarrow\,\scalebox{.3}{\yng(1)}\,,\,\scalebox{.3}{\yng(1,1)}}=
\frac{1}{4}\left(
\begin{array}{cc}
3&-\sqrt{3}\\-\sqrt 3&1
\end{array}
\right)\,,\quad\,\,
T=
\frac{1}{4}\left(
\begin{array}{cc}
-\sqrt 3&1\\-3&\sqrt 3
\end{array}
\right)
\end{equation}
We will use these quantities to compactly write the quiver characters.

We now turn to the Clebsch-Gordan coefficients, $C_{s_1,\,s_2}^{r_q,\,S,\,p}$ and $C_{\bar s_1,\,\bar s_2}^{\bar r_q,\,\bar S ,\,t}$, where $r_q$ and $\bar r_q$ are both either $\tiny\yng(2)$ or $\tiny\yng(1,1)$.
First of all notice that we can drop the symmetric group state labels $p$ and $t$, because all the irreducible representation of $S_2$ are 1-dimensional. 
Let us call $V_F$ the the fundamental representation of $SU(F)$, and let us choose an orthonormal basis $e_{i}$, $i=1,2,...,F$. Consider now the $V_F\otimes V_F$ vector space, equipped with the induced basis $\{e_{i,j}=e_i\otimes e_j\}_{ij}$. The \({\tiny\yng(2)}\) of $SU(F)$ is spanned by every symmetric permutation of the $e_{i,j}=e_i\otimes e_j$ basis vectors of $V_F\otimes V_F$. We can label an orthonormal basis for this representation with the notation $\small \young(ij)$, where
\begin{subequations}\label{sqcd example 122}
\begin{align}
&\label{sqcd example 122 cg eq}\young(ii)= e_i\otimes e_i\,,\\[3mm]
&\label{sqcd example 122 cg neq}\young(ij)=\frac{1}{\sqrt 2}( e_i\otimes e_j+e_j\otimes e_i)\,,\qquad i\neq j
\end{align}
\end{subequations}
On the other hand, the \({\tiny\yng(1,1)}\) of $SU(F)$ is spanned by every antisymmetric permutation of the $e_{i,j}=e_i\otimes e_j$ basis vectors of $V_F\otimes V_F$. We can label an orthonormal basis for this representation with the notation $\small \young(i,j)$, where
\begin{align}\label{sqcd example 122 anti}
\young(i,j)=\frac{1}{\sqrt 2}(e_i\otimes e_j-e_j\otimes e_i)
\end{align}
We can therefore easily compute the Clebsch-Gordan coefficients \eqref{cg def}. 
To optimise the notation, we use the Young tableaux $\small \young(ij)$ and $\small \young(i,j)$ to label both the $SU(F)$ representations and their states. The Clebsch-Gordan coefficients then read
\begin{align}\label{sqcd ex cg1}
&C_{k,l}^{\scalebox{.6}{ \young(ii)}}=(e_{k,l},\,\scalebox{.8}{ \young(ii)}\,)=(e_k\otimes e_l,\,e_i\otimes e_i)=\delta_{k,i}\,\delta_{l,i}\,,\nn
&C_{k,l}^{\scalebox{.6}{ \young(ij)}}=(e_{k,l},\,\scalebox{.8}{ \young(ij)}\,)=\frac{1}{\sqrt 2}\,(e_k\otimes e_l,\,e_i\otimes e_j+e_j\otimes e_i)=\frac{1}{\sqrt{2}}\,\left(\delta_{k,i}\,\delta_{l,j}+\delta_{k,j}\,\delta_{l,i}\right)\,,\quad i\neq j\,,\nn
&C_{k,l}^{\scalebox{.6}{ \young(i,j)}}=(e_{k,l},\,\scalebox{.8}{ \young(i,j)}\,)=\frac{1}{\sqrt 2}\,(e_k\otimes e_l,\,e_i\otimes e_j-e_j\otimes e_i)=\frac{1}{\sqrt{2}}\,\left(\delta_{k,i}\,\delta_{l,j}-\delta_{k,j}\,\delta_{l,i}\right)
\end{align}
A similar approach can be used to derive the Clebsch-Gordan coefficients for the decomposition of the $\bar V_F\otimes \bar V_F$ representation of $SU(F)$, which gives similar results to the ones in \eqref{sqcd ex cg1}.

We can now write the quiver characters for the six states \eqref{N=2 sqcd 122 example}. 
%In the following we will denote the generic states $S_1$ in the $\tiny \yng(2)$ and $ S_2 $ in the $\tiny \yng(1,1)$ of $SU(F)$ by $\small \young(ij)$ and $\small \young(i,j)$ respectively. With this choice of notation, the labels in \eqref{N=2 sqcd 122 example} read now
Denoting the generic flavour state $|S\rangle\in V_{r_q}^{SU(F)}$ as in \eqref{sqcd example 122} for $r_q=\scriptsize{\yng(2)}$ and as in \eqref{sqcd example 122 anti} for $r_q=\scriptsize{\yng(1,1)}$ (and similarly for $|\bar S\rangle\in V_{\bar r_q}^{SU(F)}$), the labels in \eqref{N=2 sqcd 122 example} read now
\begin{equation}\label{N=2 sqcd 122 new notation}
\begin{array}{ll}
\pmb L_1=\{{\scriptsize\yng(3)}\,,{\scriptsize\yng(1)}\,,{\scriptsize\young(ij)}\,,{\scriptsize\overline{\young(pq)}}\,\}\,,\qquad\qquad
&\pmb L_2=\left\{ \vphantom{\frac{1}{1}} {\scriptsize\yng(1,1,1)}\,,{\scriptsize\yng(1)}\,,{\scriptsize\young(i,j)}\,\,,{\scriptsize\overline{\young(p,q)}}\,\right\}\,,\\[6mm]
\pmb L_3=\left\{ \vphantom{\sum} {\scriptsize\yng(2,1)}\,,{\scriptsize\yng(1)}\,,{\scriptsize\young(ij)}\,\,,{\scriptsize\overline{\young(pq)}}\,\right\}\,,\qquad\quad
&\pmb L_4=\left\{ \vphantom{\sum} {\scriptsize\yng(2,1)}\,,{\scriptsize\yng(1)}\,,{\scriptsize\young(i,j)}\,\,,{\scriptsize\overline{\young(p,q)}}\,\right\}\,,\\[6mm]
\pmb L_5=\left\{ \vphantom{\sum} {\scriptsize\yng(2,1)}\,,{\scriptsize\yng(1)}\,,{\scriptsize\young(ij)}\,\,,{\scriptsize\overline{\young(p,q)}}\,\right\}\,,\qquad\quad
&\pmb L_6=\left\{ \vphantom{\sum} {\scriptsize\yng(2,1)}\,,{\scriptsize\yng(1)}\,,{\scriptsize\young(i,j)}\,\,,{\scriptsize\overline{\young(pq)}}\,\right\}
\end{array}
\end{equation}
The quiver characters are
\begin{align}
&\chi(\pmb L_1,\vec s,\sigma)=\frac{1}{\sqrt{3!}}\,
C_{s_1,s_2}^{\scalebox{.6}{ \young(ij)}}\,\,C_{\bar s_1,\bar s_2}^{\overline{\scalebox{.6}{\young(pq)}}}\,,\nn
&\chi(\pmb L_2,\vec s,\sigma)=\frac{1}{\sqrt{3!}}\,\text{sign}(\sigma)\,
C_{s_1,s_2}^{\scalebox{.6}{ \young(i,j)}}\,\,C_{\bar s_1,\bar s_2}^{\overline{\scalebox{.6}{\young(p,q)}}}\,,\nn
&\chi(\pmb L_3,\vec s,\sigma)=\frac{1}{\sqrt{3}}\,\,\Tr\left[D^{\scalebox{.2}{\yng(2,1)}}(\sigma)\,P^{\scalebox{.2}{\yng(2,1)}\,\rightarrow\,\scalebox{.2}{\yng(1)}\,,\,\scalebox{.2}{\yng(2)}}\right]\,\,
C_{s_1,s_2}^{\scalebox{.6}{ \young(ij)}}\,\,C_{\bar s_1,\bar s_2}^{\overline{\scalebox{.6}{\young(pq)}}}\,,\nn
&\chi(\pmb L_4,\vec s,\sigma)=\frac{1}{\sqrt{3}}\,\,\Tr\left[D^{\scalebox{.2}{\yng(2,1)}}(\sigma)\,P^{\scalebox{.2}{\yng(2,1)}\,\rightarrow\,\scalebox{.2}{\yng(1)}\,,\,\scalebox{.2}{\yng(1,1)}}\,\right]\,\,
C_{s_1,s_2}^{\scalebox{.6}{ \young(i,j)}}\,\,C_{\bar s_1,\bar s_2}^{\overline{\scalebox{.6}{\young(p,q)}}}\,,\nn
&\chi(\pmb L_5,\vec s,\sigma)=\frac{1}{\sqrt{3}}\,\,\Tr\left[D^{\scalebox{.2}{\yng(2,1)}}(\sigma)\,T\right]\,\,
C_{s_1,s_2}^{\scalebox{.6}{ \young(ij)}}\,\,C_{\bar s_1,\bar s_2}^{\overline{\scalebox{.6}{\young(p,q)}}}\,,\nn
&\chi(\pmb L_6,\vec s,\sigma)=\frac{1}{\sqrt{3}}\,\,\Tr\left[D^{\scalebox{.2}{\yng(2,1)}}(\sigma)\,T^{\text{t}}\right]\,\,
C_{s_1,s_2}^{\scalebox{.6}{ \young(i,j)}}\,\,C_{\bar s_1,\bar s_2}^{\overline{\scalebox{.6}{\young(pq)}}}
\end{align}
where $T^{\text{t}}$ denotes the transpose of the matrix $T$, defined in \eqref{projectors/op 122}.

Defining the normalisation constants
\begin{equation}
\,f_{i,j}=\left\{
\begin{array}{ll}
1\qquad & \text{if}\quad i\neq j\\
\frac{1}{\sqrt 2} \qquad & \text{if}\quad i=j
\end{array}
\right.
\end{equation}
which keeps track of the different normalisation of the Clebsch-Gordan coefficients \eqref{sqcd example 122 cg eq} and \eqref{sqcd example 122 cg neq}, the Fourier transformed operators take the explicit form
\begin{align}\label{N=2 sqcd 122 example states}
&\cO(\pmb L_1)=\frac{f_{i,j}\,f_{\bar p,\bar q}}{\sqrt{3!}}\,
\left(\vphantom{\sum}
(\phi)\,(\bar Q Q)^{(\bar p}_{(i}\,(\bar Q Q)^{\bar q)}_{j)}+2(\bar Q \phi Q)^{(\bar p}_{(i}\,(\bar Q Q)^{\bar q)}_{j)}
\,\right)\,,\displaybreak[0]\nn
%\qquad (i,j)-\text{Sym}\,,\,\,\, (\bar p,\bar q)-\text{Sym}\,,\nn
%
%
&\cO(\pmb L_2)=\frac{1}{\sqrt{3!}}\,\left(\vphantom{\sum}
(\phi)\,(\bar Q Q)^{[\bar p}_{[i}\,(\bar Q Q)^{\bar q]}_{j]}-2(\bar Q \phi Q)^{[\bar p}_{[i}\,(\bar Q Q)^{\bar q]}_{j]}
\,\right)\,,\displaybreak[0]\nn%\qquad (i,j)-\text{Anti}\,,\,\,\, (\bar p,\bar q)-\text{Anti}\,,\nn
&\cO(\pmb L_3)=\frac{f_{i,j}\,f_{\bar p,\bar q}}{\sqrt{3}}\,
\left(\vphantom{\sum}
(\phi)\,(\bar Q Q)^{(\bar p}_{(i}\,(\bar Q Q)^{\bar q)}_{j)}-(\bar Q \phi Q)^{(\bar p}_{(i}\,(\bar Q Q)^{\bar q)}_{j)}\,\right)\,,\\[3mm]
%\qquad (i,j)-\text{Sym}\,,\,\,\, (\bar p,\bar q)-\text{Sym}\,,\nn
%
%
&\cO(\pmb L_4)=\frac{1}{\sqrt{3}}\,\left(\vphantom{\sum}
(\phi)\,(\bar Q Q)^{[\bar p}_{[i}\,(\bar Q Q)^{\bar q]}_{j]}+(\bar Q \phi Q)^{[\bar p}_{[i}\,(\bar Q Q)^{\bar q]}_{j]}
\,\right)\,,\displaybreak[0]\nn%\qquad (i,j)-\text{Anti}\,,\,\,\, (\bar p,\bar q)-\text{Anti}\,,\nn
&\cO(\pmb L_5)=-f_{i,j}\,
(\bar Q \phi Q)^{[\bar p}_{(i}\,(\bar Q Q)^{\bar q]}_{j)}
\,,\nn%\qquad (i,j)-\text{Sym}\,,\,\,\, (\bar p,\bar q)-\text{Anti}\,,\nn
&\cO(\pmb L_6)=-f_{\bar p,\bar q}\,
(\bar Q \phi Q)^{(\bar p}_{[i}\,(\bar Q Q)^{\bar q)}_{j]}\nonumber%\qquad (i,j)-\text{Anti}\,,\,\,\, (\bar p,\bar q)-\text{Sym}\,,\nn
\end{align}
Round brackets around the flavour indices denotes their symmetrisation, square brackets around them denotes their antisymmetrisation.

As in the previous case, we now run some tests on this result. 
It is easily seen that if the rank of the gauge group is $N=1$, then among these six operators only $\cO(\pmb L_1)$ is non-zero, in agreement with our finite $N$ constraints \eqref{Hilbert space}. Moreover, when $N=2$, by explicitly writing all the components of $\cO(\pmb L_2)$ it is possible to check that $\cO(\pmb L_2)= 0$. This is a nontrivial result, once again predicted by the finite $N$ constraints.
Let us now check the orthogonality of these operators, in the free field metric.
For this field content the two point function in the permutation basis, eq. \eqref{corr K basis}, reads 
\begin{align}
\left\langle \cO(\vec n\,,\vec s\,,\sigma)\,\cO^{\dagger}(\vec n\,,\vec{t}\,,\tau ) \right\rangle=
\sum_{\rho_1,\,\rho_2\in S_2}\,\delta_{\rho_1(\pmb s),\pmb t}\,\delta_{\rho_2(\pmb{\bar s}), \pmb{\bar t}}\,\,N^{ C\left[(1\times \rho_2)\,\sigma\,(1\times \rho_1)^{-1}\,\tau^{-1}\right]}\,,\qquad\, \vec n=(1,2,2)
\end{align}
As in the previous example, $C\left[\sigma\right]$ is the number of cycles in the permutation $\sigma$. Using this equation we can verify that the states in \eqref{N=2 sqcd 122 example states} are indeed orthogonal. Similarly, their squared norm are
\begin{equation}
\begin{array}{ll}
\left\langle \cO(\pmb L_1)\,\cO^{\dagger}(\pmb L_1 ) \right\rangle=4\,N(N+1)(N+2)\,,\qquad\qquad
&\left\langle \cO(\pmb L_2)\,\cO^{\dagger}(\pmb L_2 ) \right\rangle=4\,N(N-1)(N-2)\,,\\[3mm]
\left\langle \cO(\pmb L_3)\,\cO^{\dagger}(\pmb L_3 ) \right\rangle=4\,N(N^2-1)\,,
&\left\langle \cO(\pmb L_4)\,\cO^{\dagger}(\pmb L_4 ) \right\rangle=4\,N(N^2-1)\,,\\[3mm]
\left\langle \cO(\pmb L_5)\,\cO^{\dagger}(\pmb L_5 ) \right\rangle=4\,N(N^2-1)\,,\qquad\qquad
&\left\langle \cO(\pmb L_6)\,\cO^{\dagger}(\pmb L_6 ) \right\rangle=4\,N(N^2-1)
\end{array}
\end{equation}
in agreement with our prediction \eqref{ortho L}.

\subsection*{- $\vec n=(2,2,2)$ field content}

Consider now the field content $\{\phi,\phi,Q,Q,\bar Q,\bar Q\}$, that is \(\vec n=(2,2,2)\). Using the same notation of the previous examples, the quiver Schurs for this subspace can be labelled by the fourteen sets
\begin{equation}\label{sqcd ex 222 14 labels}
\begin{array}{ll}
\pmb L_1=\left\{\scriptsize{\yng(4)}\,,\scriptsize{{\yng(2)}}\,,\scriptsize{\young(ij)}\,,\overline{\scriptsize{{\young(pq)}}}\,\right\}\,,\qquad\quad
&\pmb L_2=\left\{ \vphantom{\frac{1}{1}} \scriptsize{\yng(1,1,1,1)}\,, \scriptsize{\yng(1,1)}\,, \scriptsize{\young(i,j)}\,, \overline{\scriptsize{{\young(p,q)}}}\,\right\}\,,\\[6mm]
\pmb L_3=\left\{{\scriptsize\yng(3,1)}\,,{\scriptsize\yng(2)}\,,{\scriptsize\young(ij)} \,,\overline{\scriptsize {\young(pq)}}\,\right\}\,,\qquad\quad
&\pmb L_4=\left\{ \vphantom{\frac{1}{1}} {\scriptsize\yng(3,1)}\,,{\scriptsize\yng(1,1)}\,,{\scriptsize\young(ij)} \,,\overline{\scriptsize {\young(pq)}}\,\right\}\,,\\[6mm]
\pmb L_5=\left\{{\scriptsize\yng(3,1)}\,,{\scriptsize\yng(2)}\,,{\scriptsize\young(i,j)} \,,\overline{\scriptsize {\young(p,q)}}\,\right\}\,,\qquad\quad
&\pmb L_6=\left\{ \vphantom{\frac{1}{1}} {\scriptsize\yng(3,1)}\,,{\scriptsize\yng(2)}\,,{\scriptsize\young(ij)} \,,\overline{\scriptsize {\young(p,q)}}\,\right\}\,,\\[6mm]
\pmb L_7=\left\{{\scriptsize\yng(3,1)}\,,{\scriptsize\yng(2)}\,,{\scriptsize\young(i,j)} \,,\overline{\scriptsize {\young(pq)}}\,\right\}\,,\qquad\quad
&\pmb L_8=\left\{ \vphantom{\frac{1}{1}} {\scriptsize\yng(2,1,1)}\,,{\scriptsize\yng(1,1)}\,,{\scriptsize\young(i,j)} \,,\overline{\scriptsize {\young(p,q)}}\,\right\}\,,\displaybreak[0]\\[6mm]
\pmb L_9=\left\{{\scriptsize\yng(2,1,1)}\,,{\scriptsize\yng(1,1)}\,,{\scriptsize\young(ij)} \,,\overline{\scriptsize {\young(pq)}}\,\right\}\,,\qquad\quad
&\pmb L_{10}=\left\{ \vphantom{\frac{1}{1}} {\scriptsize\yng(2,1,1)}\,,{\scriptsize\yng(2)}\,,{\scriptsize\young(i,j)} \,,\overline{\scriptsize {\young(p,q)}}\,\right\}\,,\\[6mm]
\pmb L_{11}=\left\{{\scriptsize\yng(2,1,1)}\,,{\scriptsize\yng(1,1)}\,,{\scriptsize\young(ij)} \,,\overline{\scriptsize {\young(p,q)}}\,\right\}\,,\qquad\quad
&\pmb L_{12}=\left\{ \vphantom{\frac{1}{1}} {\scriptsize\yng(2,1,1)}\,,{\scriptsize\yng(1,1)}\,,{\scriptsize\young(i,j)} \,,\overline{\scriptsize {\young(pq)}}\,\right\}\,,\\[6mm]
\pmb L_{13}=\left\{{\scriptsize\yng(2,2)}\,,{\scriptsize\yng(2)}\,,{\scriptsize\young(ij)} \,,\overline{\scriptsize {\young(pq)}}\,\right\}\,,\qquad\quad
&\pmb L_{14}=\left\{ \vphantom{\frac{1}{1}} {\scriptsize\yng(2,2)}\,,{\scriptsize\yng(1,1)}\,,{\scriptsize\young(i,j)} \,,\overline{\scriptsize {\young(p,q)}}\,\right\}
\end{array}
\end{equation}
As usual, we left the states $\scriptsize\young(ij)$ and $\scriptsize\young(i,j)$ (with $i,j=1,2,...,F$) of the symmetric and antisymmetric representation of $SU(F)$ unspecified.

The quiver Schurs explicitly read
\begin{align}\label{SQCD ex: 222 ops}
&\cO(\pmb L_1)=\frac{f_{i,j}\,f_{\bar p,\bar q}}{\sqrt{3!}}\,
\left(\vphantom{\sum}
2(\bar Q Q)^{(\bar p}_{(i}\,(\bar Q\phi\phi Q)^{\bar q)}_{j)}+(\bar Q\phi Q)^{(\bar p}_{(i}\,(\bar Q\phi Q)^{\bar q)}_{j)} + \frac{1}{2}(\bar Q Q)^{(\bar p}_{(i}\,(\bar Q Q)^{\bar q)}_{j)}\,(\phi)^2+\right.\nn
&\qquad\qquad\qquad\qquad\qquad\qquad\left.
+\frac{1}{2}(\bar Q Q)^{(\bar p}_{(i}\,(\bar Q Q)^{\bar q)}_{j)}\,(\phi\phi)+2(\bar Q Q)^{(\bar p}_{(i}\,(\bar Q\phi Q)^{\bar q)}_{j)}\,(\phi)
\,\right),\displaybreak[0]\nn
&\cO(\pmb L_2)=\frac{1}{\sqrt{3!}}\,
\left(\vphantom{\sum}
2(\bar Q Q)^{[\bar p}_{[i}\,(\bar Q\phi\phi Q)^{\bar q]}_{j]}+(\bar Q\phi Q)^{[\bar p}_{[i}\,(\bar Q\phi Q)^{\bar q]}_{j]} + \frac{1}{2}(\bar Q Q)^{[\bar p}_{[i}\,(\bar Q Q)^{\bar q]}_{j]}\,(\phi)^2+\right.\nn
&\qquad\qquad\qquad\qquad\qquad\qquad\left.
-\frac{1}{2}(\bar Q Q)^{[\bar p}_{[i}\,(\bar Q Q)^{\bar q]}_{j]}\,(\phi\phi)-2(\bar Q Q)^{[\bar p}_{[i}\,(\bar Q\phi Q)^{\bar q]}_{j]}\,(\phi)
\,\right),\displaybreak[0]\nn
&\cO(\pmb L_3)=\frac{f_{i,j}\,f_{\bar p,\bar q}}{2\sqrt{2}}\,
\left(\vphantom{\sum}
 -2(\bar Q\phi Q)^{(\bar p}_{(i}\,(\bar Q\phi Q)^{\bar q)}_{j)} + (\bar Q Q)^{(\bar p}_{(i}\,(\bar Q Q)^{\bar q)}_{j)}\,(\phi)^2 + (\bar Q Q)^{(\bar p}_{(i}\,(\bar Q Q)^{\bar q)}_{j)}\,(\phi\phi)
\,\right),\displaybreak[0]\nn
&\cO(\pmb L_4)=\frac{f_{i,j}\,f_{\bar p,\bar q}}{2\sqrt{2}}\,
\left(\vphantom{\sum}
-2(\bar Q Q)^{(\bar p}_{(i}\,(\bar Q\phi\phi Q)^{\bar q)}_{j)} + (\bar Q Q)^{(\bar p}_{(i}\,(\bar Q Q)^{\bar q)}_{j)}\,(\phi)^2+\right.\nn
&\qquad\qquad\qquad\qquad\qquad\qquad\left.
-(\bar Q Q)^{(\bar p}_{(i}\,(\bar Q Q)^{\bar q)}_{j)}\,(\phi\phi)+2(\bar Q Q)^{(\bar p}_{(i}\,(\bar Q\phi Q)^{\bar q)}_{j)}\,(\phi)
\,\right),\displaybreak[0]\nn
&\cO(\pmb L_5)=\frac{1}{2\sqrt{2}}\,
\left(\vphantom{\sum}
2(\bar Q Q)^{[\bar p}_{[i}\,(\bar Q\phi\phi Q)^{\bar q]}_{j]} + (\bar Q Q)^{[\bar p}_{[i}\,(\bar Q Q)^{\bar q]}_{j]}\,(\phi)^2+\right.\nn
&\qquad\qquad\qquad\qquad\qquad\qquad\left.
+(\bar Q Q)^{[\bar p}_{[i}\,(\bar Q Q)^{\bar q]}_{j]}\,(\phi\phi)+2(\bar Q Q)^{[\bar p}_{[i}\,(\bar Q\phi Q)^{\bar q]}_{j]}\,(\phi)
\,\right),\displaybreak[0]\\[3mm]
&\cO(\pmb L_6)=-f_{i,j}\,
\left(\vphantom{\sum}
(\bar Q Q)^{[\bar p}_{(i}\,(\bar Q\phi\phi Q)^{\bar q]}_{j)}+(\bar Q Q)^{[\bar p}_{(i}\,(\bar Q\phi Q)^{\bar q]}_{j)}\,(\phi)
\,\right),\displaybreak[0]\nn
&\cO(\pmb L_7)=-f_{\bar p,\bar q}\,
\left(\vphantom{\sum}
(\bar Q Q)^{(\bar p}_{[i}\,(\bar Q\phi\phi Q)^{\bar q)}_{j]}+(\bar Q Q)^{(\bar p}_{[i}\,(\bar Q\phi Q)^{\bar q)}_{j]}\,(\phi)
\,\right),\displaybreak[0]\nn
&\cO(\pmb L_8)=\frac{1}{2\sqrt{2}}\,
\left(\vphantom{\sum}
-2(\bar Q\phi Q)^{[\bar p}_{[i}\,(\bar Q\phi Q)^{\bar q]}_{j]} + (\bar Q Q)^{[\bar p}_{[i}\,(\bar Q Q)^{\bar q]}_{j]}\,(\phi)^2 - (\bar Q Q)^{[\bar p}_{[i}\,(\bar Q Q)^{\bar q]}_{j]}\,(\phi\phi)
\,\right),\displaybreak[0]\nn
&\cO(\pmb L_9)=\frac{f_{i,j}\,f_{\bar p,\bar q}}{2\sqrt{2}}\,
\left(\vphantom{\sum}
2(\bar Q Q)^{(\bar p}_{(i}\,(\bar Q\phi\phi Q)^{\bar q)}_{j)} + (\bar Q Q)^{(\bar p}_{(i}\,(\bar Q Q)^{\bar q)}_{j)}\,(\phi)^2+\right.\displaybreak[0]\nn
&\qquad\qquad\qquad\qquad\qquad\qquad\left.
-(\bar Q Q)^{(\bar p}_{(i}\,(\bar Q Q)^{\bar q)}_{j)}\,(\phi\phi)-2(\bar Q Q)^{(\bar p}_{(i}\,(\bar Q\phi Q)^{\bar q)}_{j)}\,(\phi)
\,\right),\displaybreak[0]\nn
&\cO(\pmb L_{10})=\frac{1}{2\sqrt{2}}\,
\left(\vphantom{\sum}
-2(\bar Q Q)^{[\bar p}_{[i}\,(\bar Q\phi\phi Q)^{\bar q]}_{j]} + (\bar Q Q)^{[\bar p}_{[i}\,(\bar Q Q)^{\bar q]}_{j]}\,(\phi)^2+\right.\displaybreak[0]\nn
&\qquad\qquad\qquad\qquad\qquad\qquad\left.
+(\bar Q Q)^{[\bar p}_{[i}\,(\bar Q Q)^{\bar q]}_{j]}\,(\phi\phi)-2(\bar Q Q)^{[\bar p}_{[i}\,(\bar Q\phi Q)^{\bar q]}_{j]}\,(\phi)
\,\right),\displaybreak[0]\nn
&\cO(\pmb L_{11})=-f_{i,j}\,
\left(\vphantom{\sum}
(\bar Q Q)^{[\bar p}_{(i}\,(\bar Q\phi\phi Q)^{\bar q]}_{j)}-(\bar Q Q)^{[\bar p}_{(i}\,(\bar Q\phi Q)^{\bar q]}_{j)}\,(\phi)
\,\right),\displaybreak[0]\nn
&\cO(\pmb L_{12})=-f_{\bar p,\bar q}\,
\left(\vphantom{\sum}
(\bar Q Q)^{(\bar p}_{[i}\,(\bar Q\phi\phi Q)^{\bar q)}_{j]}-(\bar Q Q)^{(\bar p}_{[i}\,(\bar Q\phi Q)^{\bar q)}_{j]}\,(\phi)
\,\right),\nn
&\cO(\pmb L_{13})=\frac{f_{i,j}\,f_{\bar p,\bar q}}{\sqrt{3}}\,
\left(\vphantom{\sum}
-(\bar Q Q)^{(\bar p}_{(i}\,(\bar Q\phi\phi Q)^{\bar q)}_{j)}+(\bar Q\phi Q)^{(\bar p}_{(i}\,(\bar Q\phi Q)^{\bar q)}_{j)} + \frac{1}{2}(\bar Q Q)^{(\bar p}_{(i}\,(\bar Q Q)^{\bar q)}_{j)}\,(\phi)^2+\right.\nn
&\qquad\qquad\qquad\qquad\qquad\qquad\left.
+\frac{1}{2}(\bar Q Q)^{(\bar p}_{(i}\,(\bar Q Q)^{\bar q)}_{j)}\,(\phi\phi) - (\bar Q Q)^{(\bar p}_{(i}\,(\bar Q\phi Q)^{\bar q)}_{j)}\,(\phi)
\,\right),\displaybreak[0]\nn
&\cO(\pmb L_{14})=\frac{1}{\sqrt{3}}\,
\left(\vphantom{\sum}
-(\bar Q Q)^{[\bar p}_{[i}\,(\bar Q\phi\phi Q)^{\bar q]}_{j]}+(\bar Q\phi Q)^{[\bar p}_{[i}\,(\bar Q\phi Q)^{\bar q]}_{j]} + \frac{1}{2}(\bar Q Q)^{[\bar p}_{[i}\,(\bar Q Q)^{\bar q]}_{j]}\,(\phi)^2+\right.\displaybreak[0]\nn
&\qquad\qquad\qquad\qquad\qquad\qquad\left.
-\frac{1}{2}(\bar Q Q)^{[\bar p}_{[i}\,(\bar Q Q)^{\bar q]}_{j]}\,(\phi\phi) + (\bar Q Q)^{[\bar p}_{[i}\,(\bar Q\phi Q)^{\bar q]}_{j]}\,(\phi)
\,\right)\nonumber
\end{align}
The convention for round and square brackets around flavour indices is the same as the one used in the previous example.
The computation that leads to this result is summarised in Appendix \ref{app: SQCD example 222}. Using Mathematica, we checked that all these operators are orthogonal in the free field metric, that their norm satisfy \eqref{ortho L}, and that they obey the finite $N$ constraints \eqref{Hilbert space}.

\section{Conclusions and Outlook}

In this paper we considered free quiver gauge theories with gauge group \(\prod_{a=1}^nU(N_a)\) and flavour group \(\prod_{a=1}^n U ( F_{a} ) \times U( \bar F_a) \). We found that the basis of Quiver Restricted Schur polynomials \eqref{fourier2} diagonalises the two point function \eqref{ortho L}. Relying on diagrammatic methods, we also provided an analytical finite $N$ expression for the three point function of holomorphic matrix invariants. The relevant diagram is shown in Fig. \ref{fig: CRSC picture}.

For quiver gauge theories with bi-fundamental matter (no fundamental matter), the counting and correlators of gauge invariant operators can be expressed in terms of defect observables in two dimensional topological field theories (TFT2). These theories are based on lattice gauge theory where permutation groups play the role of gauge groups \cite{quivcalc}. The relevant two dimensional surfaces were obtained by a process of thickening the quiver. This leads us to expect that the counting and correlators for the present case can be expressed in terms of defect observables in TFT2 on Riemann surfaces with boundary. It will be very interesting to elaborate on this in the future. Another interesting future direction is the relation of gauge invariant correlators to the counting of branched covers. This has been discussed for the case of a single gauge group and one or more adjoint fields \cite{RS-Galois,Brown:2010pb,Gopakumar:2011ev,Gopakumar:2012ny,Koch:2014hsa,Freidel:2014aqa}. The equation \eqref{corr K basis} giving the formula for the 2-point function in the permutation basis would be a good starting point. By tracing the flavour indices, we expect to see that powers of the flavour rank are related to the counting of covering surfaces with boundaries (see for example \cite{Gadde:2009dj}). 

For the case of a single gauge group but multi-matrices (quiver with one node and multiple edges), a complete set of charges measuring the group theoretic labels of orthogonal bases for gauge invariant operators were given in \cite{KR2}. They were constructed 
from Noether charges for enhanced symmetries in the zero coupling limit. A minimal set of charges can be characterised by using properties of Permutation Centralizer Algebras (PCAs) \cite{CentralizerAlgebras}. 
We expect similar applications of PCAs to gauge invariant operators in general quiver theories (without fundamental matter) to proceed in a fairly similar manner. For the case of quivers with fundamental matter, we may expect that appropriate PCAs along with modules over these algebras will play a role. There are in fact two ways one might associate a PCA to quiver with fundamentals. One is to excise the flavour legs of the quiver to be left with a quiver with bi-fundamentals only. Putting back the legs might correspond to going from algebra to a broader construction involving modules over the algebra. The other way is to tie all the incoming and outgoing legs to a single new node, preserving their orientation.
This latter procedure was useful in consideration of the counting of gauge invariant operators \cite{quivwords}.

Another interesting line of research would be to study the action of the one-loop dilatation operator on the basis of matrix invariants \eqref{fourier2} for flavoured theories, possibly in some simple subsector. The action of the one-loop dilatation operator on the Schur basis for $\N=4$ SYM has already been studied \cite{DeComarmond:2010ie,KDGM11}. For example, in the giant graviton sector of $\N=4$ SYM, the explicit action of the one-loop dilatation operator corresponds to moving a single box in the Young diagram that parametrises the giant graviton. It is an open problem to find analogous results in flavoured theories: an interesting starting point would be $\N=2$ SQCD with gauge group $SU(N)$ and flavour symmetry $SU(2N)$, which is a conformal theory. An explicit basis for its matrix invariants is given in \eqref{fourier for N=2 sqcd}.

The broad summary of the results of the present paper and of a number of future directions is that the quiver, combined with associated permutation algebras and topological field theories, can be a powerful device in constructing correlators of gauge invariant observables and exposing hidden geometrical structures associated with these.

\vskip2cm

\begin{centerline}
{\bf Acknowledgements} 
\end{centerline} 

\vskip.4cm 

SR is supported by STFC consolidated grant ST/L000415/1 ``String Theory, Gauge Theory \& Duality." 
PM is supported by a Queen Mary University of London studentship.

\vskip2cm

%\pagebreak

\addtocontents{toc}{\vspace{2cm}}
\addcontentsline{toc}{section}{Appendices}

\appendix

\section{Operator Invariance}\label{app:Op Inv}
In this appendix we will derive the identity \eqref{constr}. Let us consider a matrix $\Phi$ in the bi-fundamental $(\Box,\bar\Box )$ representation of $U(N_a)\times U(N_b)$, and a permutation $\eta\in S_n$. Eq. \eqref{constr} arises from the equivalence
\begin{align}\label{equivPhi}
\eta^{-1}\left(\Phi^{\otimes n}\right)\eta=\Phi^{\otimes n}
\quad \Rightarrow\quad \left[\Phi^{\otimes n},\eta\right]=0
%\,,\qquad\qquad\eta\in S_n
\end{align}
which follows from the identities
\begin{align}
\langle e^{i_1},e^{i_2},\cdots, e^{i_n}|&\Phi^{\otimes n}|e_{j_1},e_{j_2},\cdots,e_{j_n}\rangle=
(\Phi^{\otimes n})^{i_1,i_2,...i_n}_{j_1,j_2,...,j_n}=
\Phi^{i_1}_{j_1}\Phi^{i_2}_{j_2}\cdots \Phi^{i_n}_{j_n}=
\Phi^{i_{\eta(1)}}_{j_{\eta(1)}}\Phi^{i_{\eta(2)}}_{j_{\eta(2)}}\cdots \Phi^{i_{\eta(n)}}_{j_{\eta(n)}}\nn
&=
(\Phi^{\otimes n})^{i_{\eta(1)},i_{\eta(2)},...i_{\eta(n)}}_{j_{\eta(1)},j_{\eta(2)},...,j_{\eta(n)}}=
\langle e^{i_{\eta(1)}},e^{i_{\eta(2)}},\cdots, e^{i_{\eta(n)}}|\Phi^{\otimes n}|e_{j_{\eta(1)}},e_{j_{\eta(2)}},\cdots,e_{j_{\eta(n)}}\rangle\nn
&=
\langle e^{i_1},e^{i_2},\cdots, e^{i_n}|\eta^{-1}\Phi^{\otimes n}\eta|e_{j_1},e_{j_2},\cdots,e_{j_n}\rangle\,,\qquad\quad \eta\in S_n
\end{align}
Here $|e_{j_1},e_{j_2},\cdots,e_{j_n}\rangle\in V^{\otimes n}_{N_a}$ and $\langle e^{j_1},e^{j_2},\cdots,e^{j_n}|\in \bar V^{\otimes n}_{N_b}$, $V_{N_a}$ and $\bar V_{N_b}$ being the fundamental and antifundamental representations of $U(N_a)$ and $U(N_b)$ respectively. In the following, we will need the two identities
\begin{align}\label{Qrho}
\left(Q^{\otimes n}\rho\right)^{I}_{\pmb s}&=
\langle e^{i_1},e^{i_2},\cdots, e^{i_n}|Q^{\otimes n}\rho|e_{s_1},e_{s_2},\cdots,e_{s_n}\rangle\nn
&=
\langle e^{i_1},e^{i_2},\cdots, e^{i_n}|Q^{\otimes n}|e_{s_{\rho(1)}},e_{s_{\rho(2)}},\cdots,e_{s_{\rho(n)}}\rangle
=\left(Q^{\otimes n}\right)^{I}_{\rho(\pmb s)}
\end{align}
and 
\begin{align}\label{barQrho}
\left(\bar \rho^{-1}\bar Q^{\otimes n}\right)_{J}^{\pmb {\bar s}}&=
\langle e^{\bar s_1},e^{\bar s_2},\cdots,e^{\bar s_n}|\bar \rho^{-1}\bar Q^{\otimes n}| e_{j_1},e_{j_2},\cdots, e_{j_n}\rangle\nn
&=
\langle e^{\bar s_{\bar\rho(1)}},e^{\bar s_{\bar\rho(2)}},\cdots,e^{\bar s_{\bar\rho(n)}}|\bar Q^{\otimes n}| e_{j_1},e_{j_2,}\cdots, e_{j_n}\rangle
=\left(\bar Q^{\otimes n}\right)_{J}^{\bar \rho(\pmb {\bar s})}
\end{align}

Now let us consider a generic GIO $\cO_\Q(\vec n;\, \vec{s};\,\vec\sigma)$, built with $n_{ab,\alpha}$ type $\Phi_{ab,\alpha}$ fields, $n_{a,\beta}$ type $Q_{a,\beta}$ fields and $\bar n_{a,\gamma}$ type $\bar Q_{a,\gamma}$ fields. We also introduce the permutations 
\begin{subequations}
\begin{align}
&\vec\eta=\cup_{a,b, \alpha}\{\eta_{ab,\alpha}\}\,\,\,,\quad \eta_{ab,\alpha}\in S_{n_{ab,\alpha}}\\[4mm]
&\vec\rho=\cup_a\{\cup_\beta\, \rho_{a,\beta};\,\cup_\gamma\,\bar \rho_{a,\gamma}\}\,\,\,,\quad \rho_{a,\beta}\in S_{n_{a,\beta}}\,\,\,\,,\,\,\,\,\,\,\,\bar\rho_{a,\gamma}\in S_{\bar n_{a,\gamma}}
\end{align}
\end{subequations}
From \eqref{equivPhi}, we then have the equivalences
\begin{align}
\eta_{ab,\alpha}^{-1}\left(\Phi_{ab,\alpha}^{\otimes n_{ab,\alpha}}\right)\eta_{ab,\alpha}=\Phi_{ab,\alpha}^{\otimes n_{ab,\alpha}}\,,\quad
\rho_{a,\beta}^{-1}\left(Q_{a,\beta}^{\otimes n_{a,\beta}}\right)\rho_{a,\beta}=Q_{a,\beta}^{\otimes n_{a,\beta}}\,,\quad
\bar\rho_{a,\gamma}^{-1}\left(\bar Q_{a,\gamma}^{\otimes \bar n_{a,\gamma}}\right)\bar \rho_{a,\gamma}=\bar Q_{a,\gamma}^{\otimes \bar n_{a,\gamma}}
\end{align}
for every $a,\,b,\,\alpha,\,\beta,\,\gamma$. Inserting these identities in \eqref{Q def} gives
\begin{align}\label{Q w id}
&\mathcal{O}_\mathcal{Q}(\vec n;\, \vec{s};\,\vec\sigma)=
\prod_{a}\left[\prod_{b,\alpha}\left(\eta_{ab,\alpha}^{-1}\left(\Phi_{ab,\alpha}^{\otimes n_{ab,\alpha}}\right)\eta_{ab,\alpha}\right)_{J_{ab,\alpha}}^{I_{ab,\alpha}}\right]
%quark
\otimes
\left[\prod_\beta\left( \rho_{a,\beta}^{-1}\left(Q_{a,\beta}^{\otimes n_{a,\beta}}\right)\rho_{a,\beta}\right)^{I_{a,\beta}}_{\pmb s_{a,\beta}}\right]\nn
&\qquad\qquad\qquad\qquad\otimes
%antiquark
\left[\prod_\gamma \left(\bar\rho_{a,\gamma}^{-1}\left( \bar Q_{a,\gamma}^{\otimes\bar n_{ {a,\gamma}}}\right)\bar \rho_{a,\gamma}\right)_{\bar J_{ { a,\gamma}}}^{ {\pmb{\bar s}_{a,\gamma}}}\right]\,
\left(\sigma_a\right)^{\times_{b,\alpha}J_{ba,\alpha}\times_\gamma\bar J_{ a,\gamma}}
_{\times_{b,\alpha}I_{ab,\alpha}\times_\beta I_{ {a,\beta}}}\nn
%
%
%
%		% Commented this line because it is a superfluous step!
%&=
%\prod_{a}\left(\prod_{b,\alpha}\eta_{ab,\alpha}^{-1}\right)^{I_{ab,\alpha}}_{K_{ab,\alpha}}\left(\prod_{b,\alpha}\Phi_{ab,\alpha}^{\otimes n_{ab,\alpha}}\right)^{K_{ab,\alpha}}_{L_{ab,\alpha}}\left(\prod_{b,\alpha}\eta_{ab,\alpha}\right)_{J_{ab,\alpha}}^{L_{ab,\alpha}}\nn
%
%
%
%%quark
%&\qquad\qquad\otimes
%\left(\prod_\beta\rho_{a,\beta}^{-1}\right)^{I_{a,\beta}}_{K_{a,\beta}}\left(\prod_\beta Q_{a,\beta}^{\otimes n_{a,\beta}}\rho_{a,\beta}\right)^{K_{a,\beta}}_{\pmb s_{a,\beta}}
%%antiquark
%\otimes\left(\prod_\gamma \bar\rho_{a,\gamma}^{-1} \bar Q_{a,\gamma}^{\otimes\bar n_{ {a,\gamma}}}\right)^{ {\pmb{\bar s}_{a,\gamma}}}_{\bar L_{a,\gamma}}\left(\prod_\gamma\bar \rho_{a,\gamma}\right)_{\bar J_{ { a,\gamma}}}^{\bar L_{a,\gamma}}\nn
%
%
%&\qquad\qquad\qquad\qquad\times
%\left(\sigma_a\right)^{\cup_{b,\alpha}J_{ba,\alpha}\cup_\gamma\bar J_{ a,\gamma}}
%_{\cup_{b,\alpha}I_{ab,\alpha}\cup_\beta I_{ {a,\beta}}}\nn
%
%
%
%
&=
\prod_{a}\left[\prod_{b,\alpha}\left(\Phi_{ab,\alpha}^{\otimes n_{ab,\alpha}}\right)^{K_{ab,\alpha}}_{L_{ab,\alpha}}\right]%
%quark
\otimes
\left[\prod_\beta\left( Q_{a,\beta}^{\otimes n_{a,\beta}}\rho_{a,\beta}\right)^{K_{a,\beta}}_{\pmb s_{a,\beta}}\right]
%antiquark
\otimes
\left[\prod_\gamma\left( \bar\rho_{a,\gamma}^{-1}\bar Q_{a,\gamma}^{\otimes\bar n_{ {a,\gamma}}}\right)^{ {\pmb{\bar s}_{a,\gamma}}}_{\bar L_{a,\gamma}}\right]\nn
&\quad\times
\left[\prod_{b,\alpha}\left(\eta_{ab,\alpha}\right)_{J_{ab,\alpha}}^{L_{ab,\alpha}}\right]
\left[\prod_\gamma\left(\bar \rho_{a,\gamma}\right)_{\bar J_{ { a,\gamma}}}^{\bar L_{a,\gamma}}\right]\left(\sigma_a\right)^{\times_{b,\alpha}J_{ba,\alpha}\times_\gamma\bar J_{ a,\gamma}}
_{\times_{b,\alpha}I_{ab,\alpha}\times_\beta I_{ {a,\beta}}}
\left[\prod_{b,\alpha}\left(\eta_{ab,\alpha}^{-1}\right)^{I_{ab,\alpha}}_{K_{ab,\alpha}}\right]
\left[\prod_\beta\left(\rho_{a,\beta}^{-1}\right)^{I_{a,\beta}}_{K_{a,\beta}}\right]%\nn\displaybreak
\end{align}
Now we use the equations \eqref{Qrho} and \eqref{barQrho} to obtain
\begin{align}
&\mathcal{O}_\mathcal{Q}(\vec n;\, \vec{s};\,\vec\sigma)=
\prod_{a}\left[\prod_{b,\alpha}\left(\Phi_{ab,\alpha}^{\otimes n_{ab,\alpha}}\right)^{K_{ab,\alpha}}_{L_{ab,\alpha}}\right]
%quark
\otimes
\left[\prod_\beta \left( Q_{a,\beta}^{\otimes n_{a,\beta}}\right)^{K_{a,\beta}}_{\rho_{a,\beta}(\pmb s_{a,\beta})}\right]
%antiquark
\otimes
\left[\prod_\gamma\left(\bar Q_{a,\gamma}^{\otimes\bar n_{ {a,\gamma}}}\right)^{ \bar\rho_{a,\gamma}({\pmb{\bar s}_{a,\gamma}})}_{\bar L_{a,\gamma}}\right]\nn
&\qquad\times
\left((\times_{b,\alpha}\eta_{ba,\alpha}\times_\gamma\bar \rho_{a,\gamma})\sigma_a(\times_{b,\alpha}\eta_{ab,\alpha}^{-1}\times_\beta\rho_{a,\beta}^{-1})
\right)^{\times_{b,\alpha}L_{ba,\alpha}\times_\gamma \bar L_{a,\gamma}}_{\times_{b,\alpha}K_{ab,\alpha}\times_\beta K_{a,\beta}}\nn
&\qquad\qquad\qquad=\mathcal{O}_\mathcal{Q}(\vec{n};\, \vec{\rho}\,(\vec{s}\,);\,\text{Adj}_{\vec\eta\times\vec\rho}(\vec\sigma))
\end{align}
where we also used the definition of $\text{Adj}_{\vec\eta\times\vec\rho}(\vec\sigma)$, eq. \eqref{adj action}:
\begin{align}
\text{Adj}_{\vec\eta\times\vec\rho}(\vec\sigma)=
\cup_a\{
(\times_{b,\alpha}\eta_{ba,\alpha}\times_\gamma\bar\rho_{a,\gamma})
\sigma_a
(\times_{b,\alpha}\eta_{ab,\alpha}^{-1}\times_\beta \rho_{ a,\beta}^{-1})
\}
\end{align}
We thus have explicitly shown the equivalence \eqref{constr}.

\begin{figure}[H]
\begin{center}\includegraphics[scale=0.90]{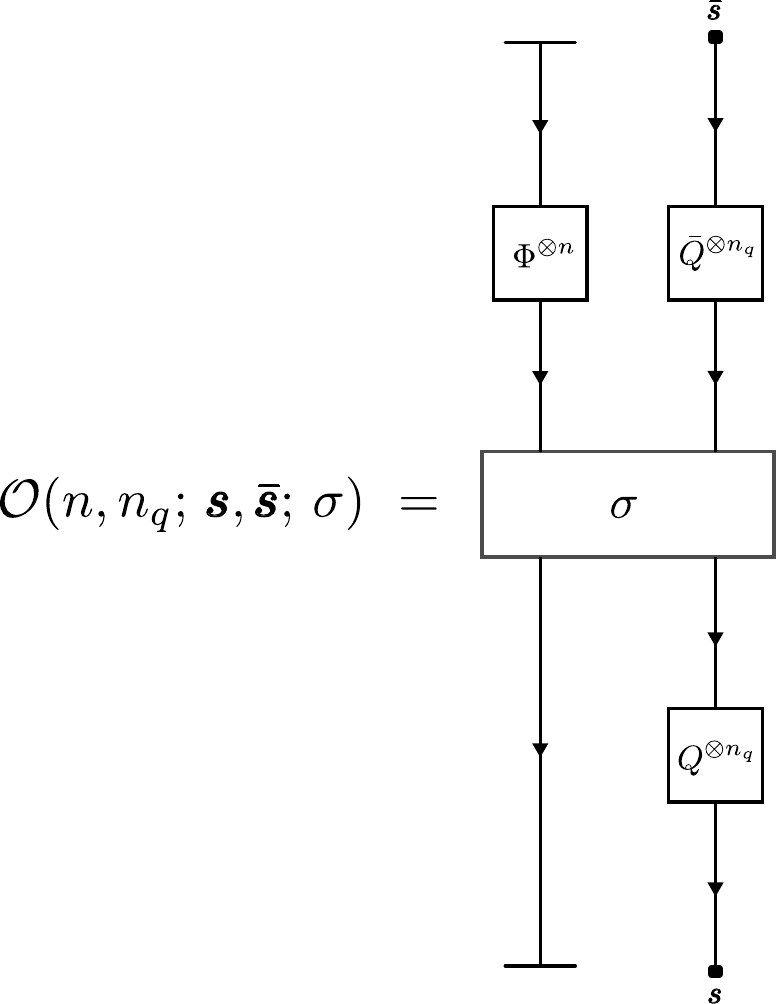}
\end{center}
\caption{Diagram corresponding to a generic $\N=2$ SQCD GIO.}\label{fig: App: op N=2 SQCD}
\end{figure}

\begin{figure}[H]
\begin{center}\includegraphics[scale=0.90]{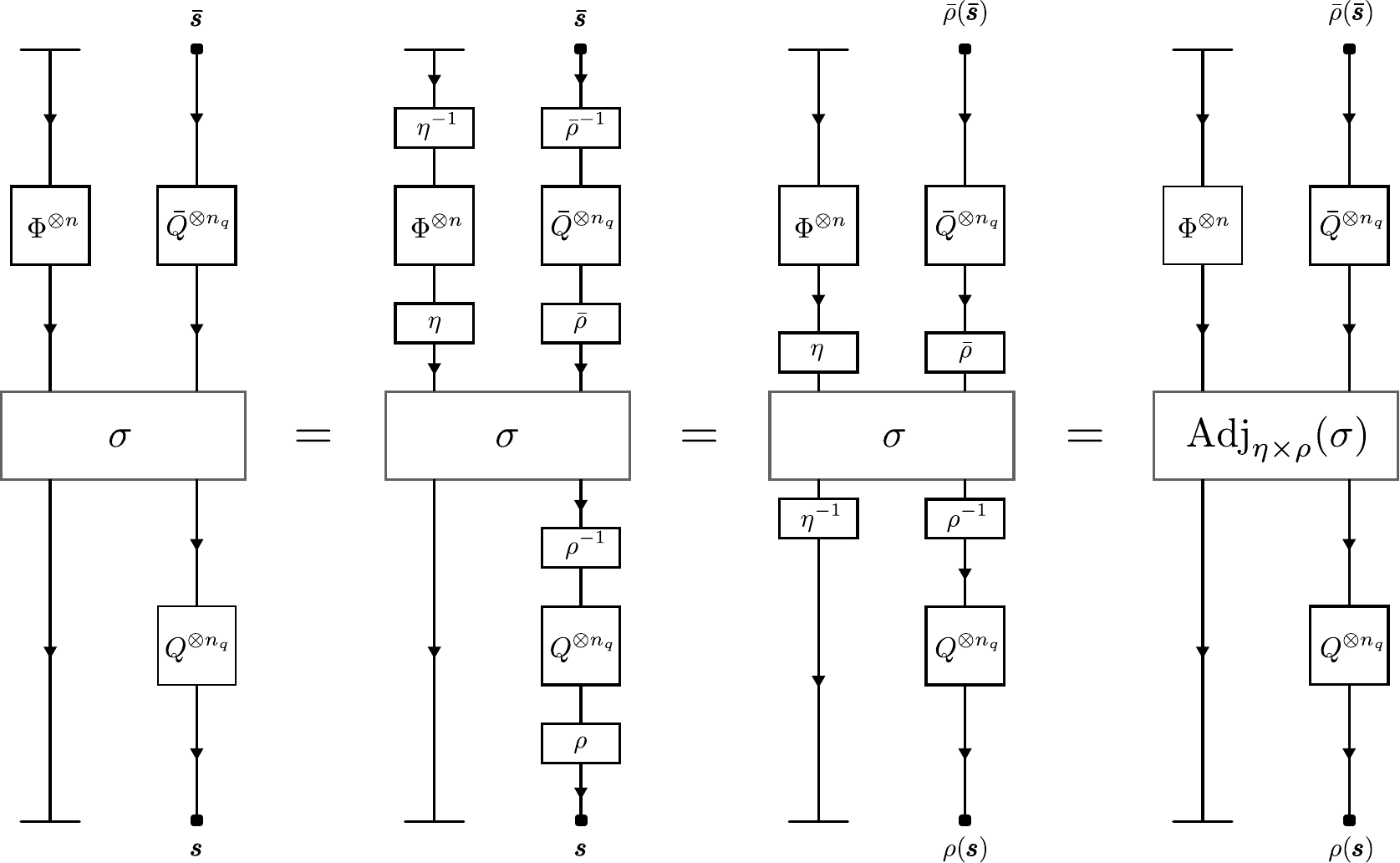}
\end{center}
\caption{Diagrammatic interpretation of the identity \eqref{constr app N=2 SQCD}.}\label{fig: App: op invariance N=2 SQCD}
\end{figure}

As it usually is the case when working in this framework, \eqref{constr} has a pictorial interpretation. We now give an example of this diagrammatic interpretation, for the simple case of an $\N=2$ SQCD. The $\N=1$ quiver for this model is the one depicted in Fig. \ref{fig: N=2 sqcd}. Let us then consider an $\N=2$ SQCD GIO built with $n$ adjoint fields $\phi$ and $n_q$ quarks $Q$ and antiquarks $\bar Q$. Each quark comes with a fixed state $s_i$ state belonging to the fundamental representation of the flavour group $SU(F)$. We label the collection of these $n_q$ states as $\pmb s=(s_1,s_2,...,s_{n_q})$. Similarly, $\pmb {\bar s}=(\bar s_1,\bar s_2,...,\bar s_{n_q})$ is the collection of the $SU(F)$ antifundamental states of the antiquarks $\bar Q$. The generic GIO $\mathcal{O}_\mathcal{Q}(n,n_q;\, \pmb {s},\pmb {\bar s};\,\sigma)$ can be drawn as in Fig. \ref{fig: App: op N=2 SQCD}.
The horizontal bars denotes the identification of the indices. Specialising eq. \eqref{constr} to this case, we have the identity
\begin{align}\label{constr app N=2 SQCD}
\mathcal{O}( n,n_q;\, \pmb s, \pmb {\bar s};\,\sigma) = 
\mathcal{O}( n,n_q;\, \rho(\pmb s), \bar\rho(\pmb {\bar s});\,\text{Adj}_{\eta\times\rho}(\sigma))
\end{align}
for $\sigma\in S_{n+n_q}$, $\eta \in S_n$ and $\rho\,\,\bar \rho\in S_{n_q}$. This equivalence is described in diagrammatic terms in Fig. \ref{fig: App: op invariance N=2 SQCD}.

\section{Quiver Character Identities}\label{Quiver Characters}

In this appendix we will derive equations \eqref{invariance chi}, \eqref{orthogonality chi} and \eqref{orthogonality chi 2}. Many of the symmetric group identities that we will use in this appendix were already introduced and discussed in Appendix A of \cite{quivcalc}.

\subsection{Invariance Relation}
In this section we will prove formula \eqref{invariance chi}:
\begin{align}\label{invariance chi copy}
\chi_\Q(\pmb L,\vec s,\vec\sigma)=
\chi_\Q(\pmb L, \vec\rho\,(\vec s\,),\text{Adj}_{\vec\rho\times\vec\eta}(\vec\sigma))
\end{align}
Using the definition of $\text{Adj}_{\vec\rho\times\vec\eta}(\vec\sigma)$ given in \eqref{adj action}
\begin{align}\label{adj action copy}
\text{Adj}_{\vec\rho\times\vec\eta}(\vec\sigma)=
\cup_a\{
(\times_{b,\alpha}\eta_{ba,\alpha}\times_\gamma\bar\rho_{a,\gamma})
\sigma_a
(\times_{b,\alpha}\eta_{ab,\alpha}^{-1}\times_\beta \rho_{ a,\beta}^{-1})
\}
\end{align}
we start by writing
\begin{align}\label{quiv char adv}
\chi_\Q(\pmb L, \vec s,&\text{Adj}_{\vec\rho\times\vec\eta}(\vec\sigma))=c_{\pmb L}\prod_a\,
\sum_{i_a,j_a}\,\sum_{ l_{ab,\alpha}\atop l_{a,\beta},\bar l_{a,\gamma}} D^{R_a}_{i_a,j_a}((\times_{b,\alpha}\eta_{ba,\alpha}\times_\gamma\bar\rho_{a,\gamma})
\sigma_a
(\times_{b,\alpha}\eta_{ab,\alpha}^{-1}\times_\beta \rho_{ a,\beta}^{-1}))\nn
&\times
B^{R_a\rightarrow\cup_{b,\alpha}r_{ab,\alpha}\cup_\beta r_{a,\beta};\nu_a^-}
_{j_a\rightarrow \cup_{b,\alpha}l_{ab,\alpha}\cup_\beta l_{a,\beta}}
\left(\prod_\beta C_{ \pmb s_{a,\beta}}^{r_{a,\beta},S_{a,\beta},l_{a,\beta}}\right)\,\,
B^{R_a\rightarrow\cup_{b,\alpha}r_{ba,\alpha}\cup_\gamma\bar r_{a,\gamma};\nu_a^+}
_{i_a\rightarrow \cup_{b,\alpha}l_{ba,\alpha}\cup_\gamma\bar l_{a,\gamma}}
\left(\prod_\gamma C^{\pmb{\bar s}_{a,\gamma}}_{\bar r_{a,\gamma},\bar S_{a,\gamma},\bar l_{a,\gamma}}\right)\displaybreak[0]\nn
=&c_{\pmb L}\prod_a\,
\sum_{i_a,j_a}\,\sum_{ l_{ab,\alpha}\atop l_{a,\beta},\bar l_{a,\gamma}} D^{R_a}_{i_a,i_a'}(\times_{b,\alpha}\eta_{ba,\alpha}\times_\gamma\bar\rho_{a,\gamma})
D^{R_a}_{i_a',j_a'}(\sigma_a)
D^{R_a}_{j_a',j_a}(\times_{b,\alpha}\eta_{ab,\alpha}^{-1}\times_\beta \rho_{ a,\beta}^{-1})\nn
&\times
B^{R_a\rightarrow\cup_{b,\alpha}r_{ab,\alpha}\cup_\beta r_{a,\beta};\nu_a^-}
_{j_a\rightarrow \cup_{b,\alpha}l_{ab,\alpha}\cup_\beta l_{a,\beta}}
\left(\prod_\beta C_{ \pmb s_{a,\beta}}^{r_{a,\beta},S_{a,\beta},l_{a,\beta}}\right)\,\,
B^{R_a\rightarrow\cup_{b,\alpha}r_{ba,\alpha}\cup_\gamma\bar r_{a,\gamma};\nu_a^+}
_{i_a\rightarrow \cup_{b,\alpha}l_{ba,\alpha}\cup_\gamma\bar l_{a,\gamma}}
\left(\prod_\gamma C^{\pmb{\bar s}_{a,\gamma}}_{\bar r_{a,\gamma},\bar S_{a,\gamma},\bar l_{a,\gamma}}\right)
\end{align}
To ease the notation, for the remainder of this section we will drop the summation symbol in our equations. The sum over repeated symmetric group state indices will therefore be implicit. Notice however that there is no summation over the repeated representation labels $r_{ab,\alpha}$, $r_{a,\beta}$, $\bar r_{a,\gamma}$. Using the equivariance property of the branching coefficients \cite{Hamermesh}
\begin{align}\label{B a D}
D^R_{k,j}(\times_a\gamma_a)\,B_{j\rightarrow\cup_a l_a}^{R\rightarrow \cup_a r_a;\nu_a}
=\left(\prod_a\,D_{l_a',l_a}^{r_a}(\gamma_a)\right)\,B_{k\rightarrow \cup_{a} l_a'}^{R\rightarrow\cup_a r_a;\nu_a}
\end{align}
we can write
\begin{align}
D^{R_a}_{j_a',j_a}(\times_{b,\alpha}\eta_{ab,\alpha}^{-1}\times_\beta \rho_{ a,\beta}^{-1})\,
&B
^{R_a\rightarrow\cup_{b,\alpha}r_{ab,\alpha}\cup_\beta r_{a,\beta};\nu_a^-}
_{j_a\rightarrow \cup_{b,\alpha}l_{ab,\alpha}\cup_\beta l_{a,\beta}}\nn
&=
\left(\prod_{b,\alpha}D_{l_{ab,\alpha}',l_{ab,\alpha}}^{r_{ab,\alpha}}(\eta_{ab,\alpha}^{-1})
\prod_\beta D_{l_{a,\beta}',l_{a,\beta}}^{r_{a,\beta}}(\rho_{a,\beta}^{-1})
\right)
B_{j_a'\rightarrow\cup_{b,\alpha}l_{ab,\alpha}'\cup_\beta l_{a,\beta}'}^{R_a\rightarrow \cup_{b,\alpha} r_{ab,\alpha}\cup_\beta r_{a,\beta},\nu_a^-}
\end{align}
and
\begin{align}
D^{R_a}_{i_a,i_a'}(\times_{b,\alpha}\eta_{ba,\alpha}\times_\gamma\bar\rho_{a,\gamma})\,
&B
^{R_a\rightarrow\cup_{b,\alpha}r_{ba,\alpha}\cup_\gamma\bar r_{a,\gamma};\nu_a^+}
_{i_a\rightarrow \cup_{b,\alpha}l_{ba,\alpha}\cup_\gamma\bar l_{a,\gamma}}\\[3mm]
&=
\left(\prod_{b,\alpha}D_{l_{ba,\alpha},l_{ba,\alpha}{''}}^{r_{ba,\alpha}}(\eta_{ba,\alpha})\prod_\gamma D_{\bar l_{a,\gamma},\bar l_{a,\gamma}{''}}^{\bar r_{a,\gamma}}(\bar \rho_{a,\gamma})%
\right)B_{i_a'\rightarrow \cup_{b,\alpha}l_{ba,\alpha}{''}\cup_\gamma l_{a,\gamma}{''}}^{R_a\rightarrow\cup_{b,\alpha}r_{ba,\alpha}\cup_\gamma \bar r_{a,\gamma};\nu_a^+}\nonumber
\end{align}
Inserting the last two equations in \eqref{quiv char adv} gives
\begin{align}\label{quiv char adv 2}
\chi_\Q(\pmb L, \vec s,\text{Adj}_{\vec\rho\times\vec\eta}(\vec\sigma))
=&c_{\pmb L}\prod_a\,%\sum_{i_a,j_a\atop l_{ab,\alpha},l_{a,\beta}.\bar l_{a,\gamma}} 
D^{R_a}_{i_a',j_a'}(\sigma_a)
\left\{
\prod_{b,\alpha}
D_{l_{ab,\alpha}',l_{ab,\alpha}}^{r_{ab,\alpha}}(\eta_{ab,\alpha}^{-1})
D_{l_{ba,\alpha},l_{ba,\alpha}{''}}^{r_{ba,\alpha}}(\eta_{ba,\alpha})
\right\}\nn
&\times
B_{j_a'\rightarrow\cup_{b,\alpha}l_{ab,\alpha}'\cup_\beta l_{a,\beta}'}^{R_a\rightarrow \cup_{b,\alpha} r_{ab,\alpha}\cup_\beta r_{a,\beta},\nu_a^-}\,
\left\{
\prod_\beta D_{l_{a,\beta},l_{a,\beta}'}^{r_{a,\beta}}(\rho_{a,\beta})\,
C_{ \pmb s_{a,\beta}}^{r_{a,\beta},S_{a,\beta},l_{a,\beta}}
\right\}\nn
&\times
B_{i_a'\rightarrow \cup_{b,\alpha}l_{ba,\alpha}{''}\cup_\gamma l_{a,\gamma}{''}}^{R_a\rightarrow\cup_{b,\alpha}r_{ba,\alpha}\cup_\gamma \bar r_{a,\gamma};\nu_a^+}\,
\left\{
\prod_\gamma D_{\bar l_{a,\gamma},\bar l_{a,\gamma}{''}}^{\bar r_{a,\gamma}}(\bar \rho_{a,\gamma})\,
C^{\pmb{\bar s}_{a,\gamma}}_{\bar r_{a,\gamma},\bar S_{a,\gamma},\bar l_{a,\gamma}}
\right\}
\end{align}
A first simplification comes from noticing that
\begin{align}
\prod_{a,b,\alpha}D_{l_{ab,\alpha}',l_{ab,\alpha}}^{r_{ab,\alpha}}(\eta_{ab,\alpha}^{-1})
D_{l_{ba,\alpha},l_{ba,\alpha}{''}}^{r_{ba,\alpha}}(\eta_{ba,\alpha})
=
%
%\prod_{a,b,\alpha}D_{l_{ab,\alpha}',l_{ab,\alpha}}^{r_{ab,\alpha}}(\eta_{ab,\alpha})^{-1}
%
%D_{l_{ab,\alpha},l_{ab,\alpha}{''}}^{r_{ab,\alpha}}(\eta_{ab,\alpha})
%=
%
\prod_{a,b,\alpha}\delta_{l_{ab,\alpha}',l_{ab,\alpha}{''}}
\end{align}
%and since $D_{l_{a,\beta},l_{a,\beta}'}^{r_{a,\beta}}(\rho_{a,\beta}^{-1})=D_{l_{a,\beta}',l_{a,\beta}}^{r_{a,\beta}}(\rho_{a,\beta})$, together with the definition of the Clebsch-Gordan coefficients given in \eqref{cg def}:
We now focus on the Clebsch-Gordan coefficients. Let us first consider the chain of equalities
\begin{align}
D^R_{i,i'}(\sigma)&C_{\pmb s}^{R,M,i}=D^R_{i,i'}(\sigma)\langle \pmb s|R,M,i\rangle=
\langle \pmb s|D(\sigma)|R,M,i'\rangle\nn
&=\langle D(\sigma)^{-1} \pmb s|R,M,i'\rangle=
\langle \sigma^{-1}(\pmb s)|R,M,i'\rangle=C_{\sigma^{-1}(\pmb s)}^{R,M,i'}
\end{align}
We can use this identity to write
\begin{align}\label{clbs D 1}
D_{l_{a,\beta},l_{a,\beta}'}^{r_{a,\beta}}(\rho_{a,\beta})\,C_{ \pmb s_{a,\beta}}^{r_{a,\beta},S_{a,\beta},l_{a,\beta}}=
C_{ \rho_{a,\beta}^{-1}(\pmb s_{a,\beta})}^{r_{a,\beta},S_{a,\beta},l_{a,\beta}'}
\end{align}
and
\begin{align}\label{clbs D 2}
D_{\bar l_{a,\gamma},\bar l_{a,\gamma}{''}}^{\bar r_{a,\gamma}}(\bar \rho_{a,\gamma})\,
C^{\pmb{\bar s}_{a,\gamma}}_{\bar r_{a,\gamma},\bar S_{a,\gamma},\bar l_{a,\gamma}}=
C^{\bar\rho_{a,\gamma}^{-1}(\pmb{\bar s}_{a,\gamma})}_{\bar r_{a,\gamma},\bar S_{a,\gamma},\bar l_{a,\gamma}{''}}
\end{align}
Using these results in \eqref{quiv char adv 2} we then get
\begin{align}\label{quiv char adv 3}
\chi_\Q(\pmb L, \vec s,\text{Adj}_{\vec\rho\times\vec\eta}(\vec\sigma))
=&c_{\pmb L}\prod_a\,%\sum_{i_a,j_a\atop l_{ab,\alpha},l_{a,\beta}.\bar l_{a,\gamma}} 
D^{R_a}_{i_a',j_a'}(\sigma_a)
\left(\prod_{b,\alpha}\delta_{l_{ab,\alpha}',l_{ab,\alpha}{''}}\right)\nn
&\times
B_{j_a'\rightarrow\cup_{b,\alpha}l_{ab,\alpha}'\cup_\beta l_{a,\beta}'}^{R_a\rightarrow \cup_{b,\alpha} r_{ab,\alpha}\cup_\beta r_{a,\beta},\nu_a^-}\,
\left\{
\prod_\beta C_{ \rho_{a,\beta}^{-1}(\pmb s_{a,\beta})}^{r_{a,\beta},S_{a,\beta},l_{a,\beta}'}
\right\}\nn
&\times
B_{i_a'\rightarrow \cup_{b,\alpha}l_{ba,\alpha}{''}\cup_\gamma l_{a,\gamma}{''}}^{R_a\rightarrow\cup_{b,\alpha}r_{ba,\alpha}\cup_\gamma \bar r_{a,\gamma};\nu_a^+}\,
\left\{
\prod_\gamma C^{\bar\rho_{a,\gamma}^{-1}(\pmb{\bar s}_{a,\gamma})}_{\bar r_{a,\gamma},\bar S_{a,\gamma},\bar l_{a,\gamma}{''}}
\right\}\displaybreak[0]\\[3mm]
=&c_{\pmb L}\prod_a\,
D^{R_a}_{i_a',j_a'}(\sigma_a)\,
B_{j_a'\rightarrow\cup_{b,\alpha}l_{ab,\alpha}'\cup_\beta l_{a,\beta}'}^{R_a\rightarrow \cup_{b,\alpha} r_{ab,\alpha}\cup_\beta r_{a,\beta},\nu_a^-}\,
\prod_\beta C_{ \rho_{a,\beta}^{-1}(\pmb s_{a,\beta})}^{r_{a,\beta},S_{a,\beta},l_{a,\beta}'}
\nn
&\times
B_{i_a'\rightarrow \cup_{b,\alpha}l_{ba,\alpha}'\cup_\gamma l_{a,\gamma}{''}}^{R_a\rightarrow\cup_{b,\alpha}r_{ba,\alpha}\cup_\gamma \bar r_{a,\gamma};\nu_a^+}\,
\prod_\gamma C^{\bar\rho_{a,\gamma}^{-1}(\pmb{\bar s}_{a,\gamma})}_{\bar r_{a,\gamma},\bar S_{a,\gamma},\bar l_{a,\gamma}{''}}
=\chi_\Q(\pmb L, \vec\rho^{\,\,-1}\,(\vec s),\vec\sigma)\nonumber
\end{align}
Substituting $\vec s\rightarrow \vec\rho\,(\vec s)$, we finally get
\begin{align}\label{invariance chi copy2}
\chi_\Q(\pmb L,\vec s,\vec\sigma)=
\chi_\Q(\pmb L, \vec\rho\,(\vec s\,),\text{Adj}_{\vec\rho\times\vec\eta}(\vec\sigma))
\end{align}
Our proposition is thus proven.

\subsection{Orthogonality Relations}
In this section we will prove the quiver character orthogonality equations \eqref{orthogonality chi} and \eqref{orthogonality chi 2}.
\subsubsection{Orthogonality in $\pmb L$}
Let us start with eq. \eqref{orthogonality chi}:
\begin{align}\label{orth in app}
\sum_{\vec s}\sum_{\vec\sigma}&\,\chi_\Q(\pmb L,\vec{s},\vec{\sigma})\,\chi_\Q(\pmb{\tilde L},\vec{s},\vec{\sigma})=\delta_{\pmb L,\tilde{\pmb L}}
\end{align}
This formula is actually a particular case of the more general identity
\begin{align}\label{almost ortho}
\sum_{\vec s}\sum_{\vec\sigma}& \chi_\Q(\pmb L,\vec{s},\vec{\sigma'}\cdot\vec{\sigma})\,\chi_\Q(\pmb{\tilde L},\vec{s},\vec{\sigma})\\[3mm]
&=c_{\pmb L}\,c_{\pmb {\tilde L}}\prod_a\,
\frac{n_a!}{d(R_a)}\,\Tr\left(D^{R_a}(\sigma_a') P_{R_a\rightarrow\cup_{b,\alpha} r_{ba,\alpha}\cup_\gamma {\bar r}_{a,\gamma}}^{\nu_a^+,\tilde\nu_a^+}\right)\,
\delta_{R_a,\tilde R_a}\nn
&\quad\times \left(\prod_{b,\alpha}\delta_{r_{ab,\alpha},\tilde r_{ab,\alpha}}\right)\left(\prod_\beta\,d(r_{a,\beta})\delta_{r_{a,\beta},\tilde r_{a,\beta}}\delta_{S_{a,\beta},\tilde S_{a,\beta}}\right)\left(\prod_\gamma\,\delta_{\bar r_{a,\gamma},\tilde{\bar r}_{a,\gamma}}\delta_{\bar S_{a,\gamma},\tilde{\bar S}_{a,\gamma}}\right)\delta_{\nu_a^-,\tilde\nu_a^-}\nonumber
\end{align}
Here $P_{R_a\rightarrow\cup_{b,\alpha} r_{ba,\alpha}\cup_\gamma {\bar r}_{a,\gamma}}^{\nu_a^+,\tilde\nu_a^+}$ is a linear operator whose matrix elements are
\begin{align}\label{mat el proj}
\left.P_{R_a\rightarrow\cup_{b,\alpha} r_{ba,\alpha}\cup_\gamma {\bar r}_{a,\gamma}}^{\nu_a^+,\tilde\nu_a^+}\right|_{\tilde i_a, i_a}=
B_{\tilde i_a\rightarrow \cup_{b,\alpha} l_{ba,\alpha}\cup_\gamma {\bar l}_{a,\gamma}}^{ R_a\rightarrow \cup_{b,\alpha} r_{ba,\alpha}\cup_\gamma {\bar r}_{a,\gamma};\tilde \nu_a^+}
B_{i_a\rightarrow \cup_{b,\alpha}l_{ba,\alpha}\cup_\gamma \bar l_{a,\gamma}}^{R_a\rightarrow \cup_{b,\alpha}r_{ba,\alpha}\cup_\gamma \bar r_{a,\gamma};\nu_a^+}
\end{align}
Let us prove eq. \eqref{almost ortho}. As a first step we expanding its LHS to get
\begin{align}\label{orth 1}
\sum_{\vec s}\sum_{\vec\sigma}& \chi_\Q(\pmb L,\vec{s},\vec{\sigma'}\cdot\vec{\sigma})\,\chi_\Q(\pmb{\tilde L},\vec{s},\vec{\sigma})\nn
&=c_{\pmb L}\,c_{\pmb {\tilde L}}\sum_{\vec s}\sum_{\vec\sigma}\,\prod_a\,
D_{i_a,j_a}^{R_a}(\sigma_a'\cdot\sigma_a)D_{\tilde i_a,\tilde j_a}^{\tilde R_a}(\sigma_a)\nn
&\quad\times
B_{j_a\rightarrow \cup_{b,\alpha}l_{ab,\alpha}\cup_\beta l_{a,\beta}}^{R_a\rightarrow \cup_{b,\alpha}r_{ab,\alpha}\cup_\beta r_{a,\beta};\nu_a^-}
B_{\tilde j_a\rightarrow \cup_{b,\alpha}\tilde l_{ab,\alpha}\cup_\beta \tilde l_{a,\beta}}^{\tilde R_a\rightarrow \cup_{b,\alpha}\tilde r_{ab,\alpha}\cup_\beta \tilde r_{a,\beta};\tilde \nu_a^-}
\prod_\beta
C_{\pmb s_{a,\beta}}^{r_{a,\beta},S_{a,\beta},l_{a,\beta}}
C_{\pmb s_{a,\beta}}^{\tilde r_{a,\beta},\tilde S_{a,\beta},\tilde l_{a,\beta}}\nn
&\quad\times
B_{i_a\rightarrow \cup_{b,\alpha}l_{ba,\alpha}\cup_\gamma \bar l_{a,\gamma}}^{R_a\rightarrow \cup_{b,\alpha}r_{ba,\alpha}\cup_\gamma \bar r_{a,\gamma};\nu_a^+}
B_{\tilde i_a\rightarrow \cup_{b,\alpha}\tilde l_{ba,\alpha}\cup_\gamma \tilde {\bar l}_{a,\gamma}}^{\tilde R_a\rightarrow \cup_{b,\alpha}\tilde r_{ba,\alpha}\cup_\gamma \tilde {\bar r}_{a,\gamma};\tilde \nu_a^+}
\prod_\gamma
C^{\pmb{\bar s}_{a,\gamma}}_{\bar r_{a,\gamma},\bar S_{a,\gamma},\bar l_{a,\gamma}}
C^{\pmb{\bar s}_{a,\gamma}}_{\tilde{\bar r}_{a,\gamma},\tilde {\bar S}_{a,\gamma},\tilde {\bar l}_{a,\gamma}}
\end{align}
The next step is to rewrite the known relation
\begin{align}
\sum_{\sigma\in S_n}D^R_{i,j}(\sigma)D^{R'}_{p,q}(\sigma)=\frac{n!}{d(R)}\delta_{R,R'}\delta_{i,p}\delta_{j,q}
\end{align}
into the form
\begin{align}
\sum_{\sigma_a}
D_{i_a,j_a}^{R_a}(\sigma_a'\cdot\sigma_a)D_{\tilde i_a,\tilde j_a}^{\tilde R_a}(\sigma_a)&=
\sum_{k_a}D_{i_a,k_a}^{R_a}(\sigma_a')\frac{n_a!}{d(R_a)}\delta_{R_a,\tilde R_a}\delta_{k_a,\tilde i_a}\delta_{j_a,\tilde j_a}\nn
&=
D_{i_a,\tilde i_a}^{R_a}(\sigma_a')\frac{n_a!}{d(R_a)}\delta_{R_a,\tilde R_a}\delta_{j_a,\tilde j_a}
\end{align}
This identity can be inserted into eq. \eqref{orth 1} to get
\begin{align}\label{orth 2}
\sum_{\vec s}&\sum_{\vec\sigma} \chi_\Q(\pmb L,\vec{s},\vec{\sigma'}\cdot\vec{\sigma})\,\chi_\Q(\pmb{\tilde L},\vec{s},\vec{\sigma})\nn
&=c_{\pmb L}\,c_{\pmb {\tilde L}}\sum_{\vec s}\,\prod_a\,
\frac{n_a!}{d(R_a)}\delta_{R_a,\tilde R_a}\,D_{i_a,\tilde i_a}^{R_a}(\sigma_a')\nn
&\quad\times
\delta_{j_a,\tilde j_a}B_{j_a\rightarrow \cup_{b,\alpha}l_{ab,\alpha}\cup_\beta l_{a,\beta}}^{R_a\rightarrow \cup_{b,\alpha}r_{ab,\alpha}\cup_\beta r_{a,\beta};\nu_a^-}
B_{\tilde j_a\rightarrow \cup_{b,\alpha}\tilde l_{ab,\alpha}\cup_\beta \tilde l_{a,\beta}}^{ R_a\rightarrow \cup_{b,\alpha}\tilde r_{ab,\alpha}\cup_\beta \tilde r_{a,\beta};\tilde \nu_a^-}
\prod_\beta
C_{\pmb s_{a,\beta}}^{r_{a,\beta},S_{a,\beta},l_{a,\beta}}
C_{\pmb s_{a,\beta}}^{\tilde r_{a,\beta},\tilde S_{a,\beta},\tilde l_{a,\beta}}\nn
&\quad\times
B_{i_a\rightarrow \cup_{b,\alpha}l_{ba,\alpha}\cup_\gamma \bar l_{a,\gamma}}^{R_a\rightarrow \cup_{b,\alpha}r_{ba,\alpha}\cup_\gamma \bar r_{a,\gamma};\nu_a^+}
B_{\tilde i_a\rightarrow \cup_{b,\alpha}\tilde l_{ba,\alpha}\cup_\gamma \tilde {\bar l}_{a,\gamma}}^{ R_a\rightarrow \cup_{b,\alpha}\tilde r_{ba,\alpha}\cup_\gamma \tilde {\bar r}_{a,\gamma};\tilde \nu_a^+}
\prod_\gamma
C^{\pmb{\bar s}_{a,\gamma}}_{\bar r_{a,\gamma},\bar S_{a,\gamma},\bar l_{a,\gamma}}
C^{\pmb{\bar s}_{a,\gamma}}_{\tilde{\bar r}_{a,\gamma},\tilde {\bar S}_{a,\gamma},\tilde {\bar l}_{a,\gamma}}
\end{align}
Now using the orthogonality relation \eqref{B orth id} in eq. \eqref{orth 2}, we further obtain
%
%	----------------------------------> This was id {B orth id}, simplified in notation
%\begin{align}\label{BB rel}
%\sum_j B_{j\rightarrow\cup_a i_a}^{R\rightarrow \cup_a r_a;\nu}
%B_{j\rightarrow\cup_a \tilde i_a}^{R\rightarrow \cup_a \tilde r_a;\tilde \nu}=\delta_{\nu,\tilde \nu}\prod_a \delta_{r_a,\tilde r_a}\delta_{i_a,\tilde i_a}
%\end{align}
%
\begin{align}\label{orth 3}
\sum_{\vec s}\sum_{\vec\sigma}& \chi_\Q(\pmb L,\vec{s},\vec{\sigma'}\cdot\vec{\sigma})\,\chi_\Q(\pmb{\tilde L},\vec{s},\vec{\sigma})\nn
&=c_{\pmb L}\,c_{\pmb {\tilde L}}\sum_{\vec s}\,\prod_a\,
\frac{n_a!}{d(R_a)}\delta_{R_a,\tilde R_a}\left(\prod_{b,\alpha}\delta_{r_{ab,\alpha},\tilde r_{ab,\alpha}}\delta_{l_{ab,\alpha},\tilde l_{ab,\alpha}}\right)\delta_{\nu_a^-,\tilde\nu_a^-}\,D_{i_a,\tilde i_a}^{R_a}(\sigma_a')\displaybreak[0]\nn
&\quad\times
\left(\prod_\beta\delta_{r_{a,\beta},\tilde r_{a,\beta}}\delta_{l_{a,\beta},\tilde l_{a,\beta}}
C_{\pmb s_{a,\beta}}^{r_{a,\beta},S_{a,\beta},l_{a,\beta}}
C_{\pmb s_{a,\beta}}^{\tilde r_{a,\beta},\tilde S_{a,\beta},\tilde l_{a,\beta}}\right)\nn
&\quad\times
B_{i_a\rightarrow \cup_{b,\alpha}l_{ba,\alpha}\cup_\gamma \bar l_{a,\gamma}}^{R_a\rightarrow \cup_{b,\alpha}r_{ba,\alpha}\cup_\gamma \bar r_{a,\gamma};\nu_a^+}
B_{\tilde i_a\rightarrow \cup_{b,\alpha}\tilde l_{ba,\alpha}\cup_\gamma \tilde {\bar l}_{a,\gamma}}^{ R_a\rightarrow \cup_{b,\alpha}\tilde r_{ba,\alpha}\cup_\gamma \tilde {\bar r}_{a,\gamma};\tilde \nu_a^+}
\prod_\gamma
C^{\pmb{\bar s}_{a,\gamma}}_{\bar r_{a,\gamma},\bar S_{a,\gamma},\bar l_{a,\gamma}}
C^{\pmb{\bar s}_{a,\gamma}}_{\tilde{\bar r}_{a,\gamma},\tilde {\bar S}_{a,\gamma},\tilde {\bar l}_{a,\gamma}}\displaybreak[0]\nn
%
%
% NEW LINE
%
&=c_{\pmb L}\,c_{\pmb {\tilde L}}\sum_{\vec s}\,\prod_a\,
\frac{n_a!}{d(R_a)}\delta_{R_a,\tilde R_a}\left(\prod_{b,\alpha}\delta_{r_{ab,\alpha},\tilde r_{ab,\alpha}}\right)\left(\prod_\beta\delta_{r_{a,\beta},\tilde r_{a,\beta}}\right)\delta_{\nu_a^-,\tilde\nu_a^-}\,D_{i_a,\tilde i_a}^{R_a}(\sigma_a')\nn
&\quad\times
\prod_\beta
C_{\pmb s_{a,\beta}}^{r_{a,\beta},S_{a,\beta},l_{a,\beta}}
C_{\pmb s_{a,\beta}}^{ r_{a,\beta},\tilde S_{a,\beta}, l_{a,\beta}}\nn
&\quad\times
B_{i_a\rightarrow \cup_{b,\alpha}l_{ba,\alpha}\cup_\gamma \bar l_{a,\gamma}}^{R_a\rightarrow \cup_{b,\alpha}r_{ba,\alpha}\cup_\gamma \bar r_{a,\gamma};\nu_a^+}
B_{\tilde i_a\rightarrow \cup_{b,\alpha} l_{ba,\alpha}\cup_\gamma \tilde {\bar l}_{a,\gamma}}^{ R_a\rightarrow \cup_{b,\alpha} r_{ba,\alpha}\cup_\gamma \tilde {\bar r}_{a,\gamma};\tilde \nu_a^+}
\prod_\gamma
C^{\pmb{\bar s}_{a,\gamma}}_{\bar r_{a,\gamma},\bar S_{a,\gamma},\bar l_{a,\gamma}}
C^{\pmb{\bar s}_{a,\gamma}}_{\tilde{\bar r}_{a,\gamma},\tilde {\bar S}_{a,\gamma},\tilde {\bar l}_{a,\gamma}}
\end{align}
Let us focus on the pair of Clebsch-Gordan coefficients $C_{\pmb s_{a,\beta}}^{r_{a,\beta},S_{a,\beta},l_{a,\beta}}
C_{\pmb s_{a,\beta}}^{ r_{a,\beta},\tilde S_{a,\beta}, l_{a,\beta}}$ in this formula. It is immediate to verify that, for a $U(F) $ Clebsch-Gordan coefficient $C_{\pmb s}^{r,S,i}$
\begin{align}\label{otho of CG app}
\sum_{\pmb s}C_{\pmb s}^{r,S,i}\,C_{\pmb s}^{r',S',i'}&=
\sum_{\pmb s}\left\langle r,S,i|\pmb s\right\rangle\,\left\langle r',S',i'|\pmb s\right\rangle=
\langle r,S,i|\left(\sum_{\pmb s}|\pmb s\rangle\,\langle \pmb s|\right)|r',S',i'\rangle\nn
&=\langle r,S,i|{1}|r',S',i'\rangle=\delta_{r,r'}\,\delta_{S,S'}\,\delta_{i,i'}
\end{align}
Therefore we can write
\begin{align}
\sum_{l_{a,\beta}}\,\sum_{\pmb s_{a,\beta}}\,C_{\pmb s_{a,\beta}}^{r_{a,\beta},S_{a,\beta},l_{a,\beta}}
C_{\pmb s_{a,\beta}}^{ r_{a,\beta},\tilde S_{a,\beta}, l_{a,\beta}}
&=\sum_{l_{a,\beta}}\,\delta_{S_{a,\beta},\tilde S_{a,\beta}}=d(r_{a,\beta})\,\delta_{S_{a,\beta},\tilde S_{a,\beta}}
\end{align}
Inserting this in \eqref{orth 3} we obtain
\begin{align}\label{orth 4}
\sum_{\vec s}&\sum_{\vec\sigma} \chi_\Q(\pmb L,\vec{s},\vec{\sigma'}\cdot\vec{\sigma})\,\chi_\Q(\pmb{\tilde L},\vec{s},\vec{\sigma})\nn
&=c_{\pmb L}\,c_{\pmb {\tilde L}}\prod_a\,
\frac{n_a!}{d(R_a)}\delta_{R_a,\tilde R_a}\left(\prod_{b,\alpha}\delta_{r_{ab,\alpha},\tilde r_{ab,\alpha}}\right)\left(\prod_\beta\,d(r_{a,\beta})\delta_{r_{a,\beta},\tilde r_{a,\beta}}\delta_{S_{a,\beta},\tilde S_{a,\beta}}\right)\delta_{\nu_a^-,\tilde\nu_a^-}\,D_{i_a,\tilde i_a}^{R_a}(\sigma_a')\nn
&\quad\times
B_{i_a\rightarrow \cup_{b,\alpha}l_{ba,\alpha}\cup_\gamma \bar l_{a,\gamma}}^{R_a\rightarrow \cup_{b,\alpha}r_{ba,\alpha}\cup_\gamma \bar r_{a,\gamma};\nu_a^+}
B_{\tilde i_a\rightarrow \cup_{b,\alpha} l_{ba,\alpha}\cup_\gamma \tilde {\bar l}_{a,\gamma}}^{ R_a\rightarrow \cup_{b,\alpha} r_{ba,\alpha}\cup_\gamma \tilde {\bar r}_{a,\gamma};\tilde \nu_a^+}
\prod_\gamma\,\sum_{\pmb {\bar s}_{a,\gamma}}\,
C^{\pmb{\bar s}_{a,\gamma}}_{\bar r_{a,\gamma},\bar S_{a,\gamma},\bar l_{a,\gamma}}
C^{\pmb{\bar s}_{a,\gamma}}_{\tilde{\bar r}_{a,\gamma},\tilde {\bar S}_{a,\gamma},\tilde {\bar l}_{a,\gamma}}\displaybreak[0]\nn
%
% NEW LINE
%
&=c_{\pmb L}\,c_{\pmb {\tilde L}}\prod_a\,
\frac{n_a!}{d(R_a)}\delta_{R_a,\tilde R_a}\left(\prod_{b,\alpha}\delta_{r_{ab,\alpha},\tilde r_{ab,\alpha}}\right)\left(\prod_\beta\,d(r_{a,\beta})\delta_{r_{a,\beta},\tilde r_{a,\beta}}\delta_{S_{a,\beta},\tilde S_{a,\beta}}\right)\delta_{\nu_a^-,\tilde\nu_a^-}\,D_{i_a,\tilde i_a}^{R_a}(\sigma_a')\nn
&\quad\times
B_{i_a\rightarrow \cup_{b,\alpha}l_{ba,\alpha}\cup_\gamma \bar l_{a,\gamma}}^{R_a\rightarrow \cup_{b,\alpha}r_{ba,\alpha}\cup_\gamma \bar r_{a,\gamma};\nu_a^+}
B_{\tilde i_a\rightarrow \cup_{b,\alpha} l_{ba,\alpha}\cup_\gamma \tilde {\bar l}_{a,\gamma}}^{ R_a\rightarrow \cup_{b,\alpha} r_{ba,\alpha}\cup_\gamma \tilde {\bar r}_{a,\gamma};\tilde \nu_a^+}
\left(\prod_\gamma\,\delta_{\bar r_{a,\gamma},\tilde{\bar r}_{a,\gamma}}\delta_{\bar S_{a,\gamma},\tilde{\bar S}_{a,\gamma}}\delta_{\bar l_{a,\gamma},\tilde{\bar l}_{a,\gamma}}\right)\displaybreak[0]\nn
%
% NEW LINE 2
%
&=c_{\pmb L}\,c_{\pmb {\tilde L}}\prod_a\,
\frac{n_a!}{d(R_a)}\delta_{R_a,\tilde R_a}\left(\prod_{b,\alpha}\delta_{r_{ab,\alpha},\tilde r_{ab,\alpha}}\right)\left(\prod_\beta\,d(r_{a,\beta})\delta_{r_{a,\beta},\tilde r_{a,\beta}}\delta_{S_{a,\beta},\tilde S_{a,\beta}}\right)\nn
&\quad\times \left(\prod_\gamma\,\delta_{\bar r_{a,\gamma},\tilde{\bar r}_{a,\gamma}}\delta_{\bar S_{a,\gamma},\tilde{\bar S}_{a,\gamma}}\right)\delta_{\nu_a^-,\tilde\nu_a^-}\,\,D_{i_a,\tilde i_a}^{R_a}(\sigma_a')
B_{i_a\rightarrow \cup_{b,\alpha}l_{ba,\alpha}\cup_\gamma \bar l_{a,\gamma}}^{R_a\rightarrow \cup_{b,\alpha}r_{ba,\alpha}\cup_\gamma \bar r_{a,\gamma};\nu_a^+}
B_{\tilde i_a\rightarrow \cup_{b,\alpha} l_{ba,\alpha}\cup_\gamma {\bar l}_{a,\gamma}}^{ R_a\rightarrow \cup_{b,\alpha} r_{ba,\alpha}\cup_\gamma {\bar r}_{a,\gamma};\tilde \nu_a^+}
\end{align}
In the second equality above we again used \eqref{otho of CG app}:
\begin{align}
\sum_{\pmb {\bar s}_{a,\gamma}}\,
C^{\pmb{\bar s}_{a,\gamma}}_{\bar r_{a,\gamma},\bar S_{a,\gamma},\bar l_{a,\gamma}}
C^{\pmb{\bar s}_{a,\gamma}}_{\tilde{\bar r}_{a,\gamma},\tilde {\bar S}_{a,\gamma},\tilde {\bar l}_{a,\gamma}}=
\delta_{\bar r_{a,\gamma},\tilde{\bar r}_{a,\gamma}}\delta_{\bar S_{a,\gamma},\tilde{\bar S}_{a,\gamma}}\delta_{\bar l_{a,\gamma},\tilde{\bar l}_{a,\gamma}}
\end{align}
We now define the projector-like operator $P_{R_a\rightarrow\cup_{b,\alpha} r_{ba,\alpha}\cup_\gamma {\bar r}_{a,\gamma}}^{\nu_a^+,\tilde\nu_a^+}$, whose matrix elements are% the ones defined in \eqref{mat el proj}: 
\begin{align}
\left.P_{R_a\rightarrow\cup_{b,\alpha} r_{ba,\alpha}\cup_\gamma {\bar r}_{a,\gamma}}^{\nu_a^+,\tilde\nu_a^+}\right|_{\tilde i_a, i_a}=
B_{\tilde i_a\rightarrow \cup_{b,\alpha} l_{ba,\alpha}\cup_\gamma {\bar l}_{a,\gamma}}^{ R_a\rightarrow \cup_{b,\alpha} r_{ba,\alpha}\cup_\gamma {\bar r}_{a,\gamma};\tilde \nu_a^+}
B_{i_a\rightarrow \cup_{b,\alpha}l_{ba,\alpha}\cup_\gamma \bar l_{a,\gamma}}^{R_a\rightarrow \cup_{b,\alpha}r_{ba,\alpha}\cup_\gamma \bar r_{a,\gamma};\nu_a^+}
\end{align}
For \(\nu_a^+=\tilde \nu_a^+\) the operator $P_{R_a\rightarrow\cup_{b,\alpha} r_{ba,\alpha}\cup_\gamma {\bar r}_{a,\gamma}}^{\nu_a^+,\tilde\nu_a^+}$ is the projector on the \((\cup_{b,\alpha} r_{ba,\alpha}\cup_\gamma {\bar r}_{a,\gamma},\nu_a^+)\) subspace of \(R_a\), but when \(\nu_a^+\neq\tilde \nu_a^+\) it is rather an intertwining operator mapping the copies \(\nu_a^+\) and \(\tilde\nu_a^+\) of the same subspace \(\cup_{b,\alpha} r_{ba,\alpha}\cup_\gamma {\bar r}_{a,\gamma}\subset R_a\) one to another. 
With this definition, we can finally write
\begin{align}\label{orth 5}
\sum_{\vec s}\sum_{\vec\sigma}& \chi_\Q(\pmb L,\vec{s},\vec{\sigma'}\cdot\vec{\sigma})\,\chi_\Q(\pmb{\tilde L},\vec{s},\vec{\sigma})\nn
&=c_{\pmb L}\,c_{\pmb {\tilde L}}\prod_a\,
\frac{n_a!}{d(R_a)}\,\Tr\left(D^{R_a}(\sigma_a') P_{R_a\rightarrow\cup_{b,\alpha} r_{ba,\alpha}\cup_\gamma {\bar r}_{a,\gamma}}^{\nu_a^+,\tilde\nu_a^+}\right)\,
\delta_{R_a,\tilde R_a}\nn
&\quad\times \left(\prod_{b,\alpha}\delta_{r_{ab,\alpha},\tilde r_{ab,\alpha}}\right)\left(\prod_\beta\,d(r_{a,\beta})\delta_{r_{a,\beta},\tilde r_{a,\beta}}\delta_{S_{a,\beta},\tilde S_{a,\beta}}\right)\left(\prod_\gamma\,\delta_{\bar r_{a,\gamma},\tilde{\bar r}_{a,\gamma}}\delta_{\bar S_{a,\gamma},\tilde{\bar S}_{a,\gamma}}\right)\delta_{\nu_a^-,\tilde\nu_a^-}
\end{align}
which is eq. \eqref{almost ortho}. Consider now the case in which $\vec\sigma'=\vec{{1}}$. Then
\begin{align}
&\Tr\left(D^{R_a}({1}) P_{R_a\rightarrow\cup_{b,\alpha} r_{ba,\alpha}\cup_\gamma {\bar r}_{a,\gamma}}^{\nu_a^+,\tilde\nu_a^+}\right)=
\Tr\left(P_{R_a\rightarrow\cup_{b,\alpha} r_{ba,\alpha}\cup_\gamma {\bar r}_{a,\gamma}}^{\nu_a^+,\tilde\nu_a^+}\right)\nn
&\qquad=
\sum_{l_{ba,\alpha}\atop\bar l_{a,\gamma}}
\left(\sum_{i_a}B_{ i_a\rightarrow \cup_{b,\alpha} l_{ba,\alpha}\cup_\gamma {\bar l}_{a,\gamma}}^{ R_a\rightarrow \cup_{b,\alpha} r_{ba,\alpha}\cup_\gamma {\bar r}_{a,\gamma};\tilde \nu_a^+}
B_{i_a\rightarrow \cup_{b,\alpha}l_{ba,\alpha}\cup_\gamma \bar l_{a,\gamma}}^{R_a\rightarrow \cup_{b,\alpha}r_{ba,\alpha}\cup_\gamma \bar r_{a,\gamma};\nu_a^+}\right)\\[3mm]
&\qquad=\delta_{\nu_a^+,\tilde \nu_a^+}
\sum_{l_{ba,\alpha}\atop\bar l_{a,\gamma}}
\left(
\prod_{b,\alpha}\delta_{l_{ba,\alpha},l_{ba,\alpha}}\right)\left(\prod_\gamma
\delta_{\bar l_{a,\gamma},\bar l_{a,\gamma}}\right)=
\delta_{\nu_a^+,\tilde \nu_a^+}\,\left(\prod_{b,\alpha}d(r_{ba,\alpha})\right)\left(\prod_\gamma
d(\bar r_{a,\gamma})\right)\nonumber
\end{align}
where the third equality follows from the orthogonality relation \eqref{B orth id}. Using this identity in \eqref{almost ortho} we get
\begin{align}\label{orth 6 almost}
\sum_{\vec s}\sum_{\vec\sigma}& \chi_\Q(\pmb L,\vec{s},\vec{\sigma})\,\chi_\Q(\pmb{\tilde L},\vec{s},\vec{\sigma})\nn
&=c_{\pmb L}\,c_{\pmb {\tilde L}}\prod_a\,
\frac{n_a!}{d(R_a)}\,
\delta_{R_a,\tilde R_a}\delta_{\nu_a^-,\tilde\nu_a^-}\delta_{\nu_a^+,\tilde \nu_a^+}\left(\prod_{b,\alpha}\,d(r_{ab,\alpha})\delta_{r_{ab,\alpha},\tilde r_{ab,\alpha}}\right)\nn
&\quad\times \left(\prod_\beta\,d(r_{a,\beta})\delta_{r_{a,\beta},\tilde r_{a,\beta}}\delta_{S_{a,\beta},\tilde S_{a,\beta}}\right)\left(\prod_\gamma\,d(\bar r_{a,\gamma})\delta_{\bar r_{a,\gamma},\tilde{\bar r}_{a,\gamma}}\delta_{\bar S_{a,\gamma},\tilde{\bar S}_{a,\gamma}}\right)
\end{align}
Recalling the definition of the set of labels $\pmb L=\{R_a, r_{ab,\alpha},r_{a,\beta},S_{a,\beta},\bar r_{ a,\gamma},\bar S_{a,\gamma}, \nu_a^+,\nu_a^-\}$, we can thus write
\begin{align}\label{orth 7}
\sum_{\vec s}\sum_{\vec\sigma}& \chi_\Q(\pmb L,\vec{s},\vec{\sigma})\,\chi_\Q(\pmb{\tilde L},\vec{s},\vec{\sigma})\nn
&=\delta_{\pmb L,\tilde{\pmb L}}\,c_{\pmb L}\,c_{\pmb {\tilde L}}\prod_a\,
\frac{n_a!}{d(R_a)}\,
\left(\prod_{b,\alpha}\,d(r_{ab,\alpha})\right) \left(\prod_\beta\,d(r_{a,\beta})\right)\left(\prod_\gamma\,d(\bar r_{a,\gamma})\right)
=
\delta_{\pmb L,\tilde{\pmb L}}
\end{align}
The identity \eqref{orth in app} is proven.
\subsubsection{Orthogonality in $\vec s,\,\vec \sigma$}
In this section we are going to prove \eqref{orthogonality chi 2}:
\begin{align}\label{orthogonality chi 2 copy}
\sum_{\pmb L} \chi_\Q(\pmb L,&\vec s,\vec\sigma)\,\chi_\Q(\pmb L,\vec t,\vec\tau)
=\frac{1}{c_{\vec{n}}}\,\sum_{\vec\eta\times\vec\rho} \delta\left(\text{Adj}_{\vec\eta\times\vec\rho}(\vec \sigma)\,\vec \tau^{\,-1}\right)
\delta_{\vec \rho(\vec{ s}),\vec{ t} } 
\end{align}
We start by writing two useful identities, which will allow us to connect state indices appearing in the 
first quiver character with state indices appearing in the second quiver character. Consider contracting both sides of the equation \cite{Hamermesh} 
\begin{align}\label{ham id}
\sum_{\sigma\in S_n}D_{i,j}^r(\sigma)\,D_{k,l}^{r'}(\sigma)=\frac{n!}{d(r)}\delta_{r,r'}\delta_{i,k}\delta_{j,l}
\end{align}
with \(B^{R\rightarrow r,\cdots;\nu^-}_{I\rightarrow i,\cdots}\,B^{R\rightarrow r,\cdots;\nu^+}_{J\rightarrow j,\cdots}\,B^{R\rightarrow r',\cdots;\nu^+}_{K\rightarrow k,\cdots}\,B^{R\rightarrow r',\cdots;\nu^-}_{L\rightarrow l,\cdots}\) and then summing over the representation \(r'\vdash n\). By doing so, we get the identity
\begin{align}\label{App:id1}
B^{R\rightarrow r,\cdots;\nu^-}_{I\rightarrow i,\cdots}\,& B^{R\rightarrow r,\cdots;\nu^+}_{K\rightarrow i,\cdots}\,B^{R\rightarrow r,\cdots;\nu^-}_{L\rightarrow l,\cdots}\,B^{R\rightarrow r,\cdots;\nu^+}_{J\rightarrow l,\cdots}\\[3mm]
&=
\frac{d(r)}{n!}\sum_{\sigma\in S_n}\sum_{r' \vdash n} \,
B^{R\rightarrow r,\cdots;\nu^-}_{I\rightarrow i,\cdots}\,B^{R\rightarrow r,\cdots;\nu^+}_{J\rightarrow j,\cdots}\,B^{R\rightarrow r',\cdots;\nu^+}_{K\rightarrow k,\cdots}\,B^{R\rightarrow r',\cdots;\nu^-}_{L\rightarrow l,\cdots}
D_{i,j}^r(\sigma)\,D_{k,l}^{r'}(\sigma)\nonumber
\end{align}
Alternatively, contracting both sides of \eqref{ham id} with \(C_{\pmb s}^{r',S,k}\,C_{\pmb t}^{r',S,l}\) and summing over the representations $r'\vdash n$, we obtain
\begin{align}\label{App:id2}
C_{\pmb s}^{r,S,i}\,C_{\pmb t}^{r,S,j}=
\frac{d(r)}{n!}\sum_{\sigma\in S_n}
\,D_{i,j}^r(\sigma)\,\left(\sum_{r'\vdash n}\,D_{k,l}^{r'}(\sigma)\, C_{\pmb s}^{r',S,k}\,C_{\pmb t}^{r',S,l}\right)
\end{align}
This is the second identity we are going to need.

Let us then consider the product
\begin{align}\label{orth L be}
\chi_\Q&(\pmb L,\vec s,\vec\sigma)\,\chi_\Q(\pmb L,\vec t,\vec\tau)=
c_{\pmb L}^2\prod_a D_{i_a,j_a}^{R_a}(\sigma_a) \,D_{i_a',j_a'}^{R_a}(\tau_a) \nn
&\times B_{j_a\rightarrow \cup_{b,\alpha} l_{ab,\alpha}\cup_{\beta } l_{a,\beta}}^{R_a\rightarrow \cup_{b,\alpha}r_{ab,\alpha}\cup_\beta r_{a,\beta};\nu_a^-}
\left(\prod_\beta C^{ r_{a,\beta}, S_{a,\beta}, l_{a,\beta}}_{\pmb{ s}_{a,\beta}}\right)
 B_{i_a\rightarrow \cup_{b,\alpha} l_{ba,\alpha}\cup_{\gamma } \bar l_{a,\gamma}}^{R_a\rightarrow \cup_{b,\alpha}r_{ba,\alpha}\cup_\gamma \bar r_{a,\gamma};\nu_a^+}
\left(\prod_\gamma C_{\bar r_{a,\gamma},\bar S_{a,\gamma},\bar l_{a,\gamma}}^{\pmb{\bar s}_{a,\gamma}}\right)\nn
&\times B_{j_a'\rightarrow \cup_{b,\alpha} l_{ab,\alpha}'\cup_{\beta } l_{a,\beta}'}^{R_a\rightarrow \cup_{b,\alpha}r_{ab,\alpha}\cup_\beta r_{a,\beta};\nu_a^-}
\left(\prod_\beta C^{ r_{a,\beta}, S_{a,\beta}, l_{a,\beta}'}_{\pmb{ t}_{a,\beta}}\right)
B_{i_a'\rightarrow \cup_{b,\alpha} l_{ba,\alpha}'\cup_{\gamma } \bar l_{a,\gamma}'}^{R_a\rightarrow \cup_{b,\alpha}r_{ba,\alpha}\cup_\gamma \bar r_{a,\gamma};\nu_a^+}
\left(\prod_\gamma C_{\bar r_{a,\gamma},\bar S_{a,\gamma},\bar l_{a,\gamma}'}^{\pmb{\bar t}_{a,\gamma}}\right)
\end{align}
Using \eqref{App:id1} and \eqref{App:id2} in \eqref{orth L be} we find %, together with the reality of the Clebsch-Gordan coefficients, we find
\begin{align}\label{orth disc}
\chi_\Q&(\pmb L,\vec s,\vec\sigma)\,\chi_\Q(\pmb L,\vec t,\vec\tau)=c_{\pmb L}^2\sum_{\vec\eta,\,\vec\rho}\,\sum_{\{r'_{ab,\alpha}\}}\,\sum_{\{r'_{a,\beta}\}}\,\sum_{\{\bar r'_{a,\gamma}\}} \nn
&\times 
\prod_a
\left(\prod_{b,\alpha}\frac{d(r_{ab,\alpha})}{n_{ab,\alpha}!}\right)
\left(\prod_{\beta}\frac{d(r_{a,\beta})}{n_{a,\beta}!}\right) 
\left(\prod_{\gamma}\frac{d(\bar r_{a,\gamma})}{\bar n_{a,\gamma}!}\right)\,
 D_{i_a,j_a}^{R_a}(\sigma_a) \,D_{i_a',j_a'}^{R_a}(\tau_a)\nn
&\times
\left[
\left(
\prod_{b,\alpha} 
D^{r_{ab,\alpha}}_{p_{ab,\alpha},p_{ab,\alpha}'}(\eta_{ab,\alpha})
\right)
\left(
\prod_{\beta}
D^{r_{a,\beta}}_{p_{a,\beta},p_{a,\beta}'}(\rho_{a,\beta})
\right)
B_{j_a\rightarrow \cup_{b,\alpha} p_{ab,\alpha}\cup_{\beta } p_{a,\beta}}^{R_a\rightarrow \cup_{b,\alpha}r_{ab,\alpha}\cup_\beta r_{a,\beta};\nu_a^-}
\right]\nn
&\times 
\left(\prod_\beta D^{r'_{a,\beta}}_{q_{a,\beta},q_{a,\beta}'}(\rho_{a,\beta})\,
C^{ r'_{a,\beta}, S_{a,\beta}, q_{a,\beta}}_{\pmb{ s}_{a,\beta}}\,C^{ r'_{a,\beta}, S_{a,\beta}, q_{a,\beta}'}_{\pmb{ t}_{a,\beta}} \right)\nn
 &\times
\left[
\left(
\prod_{b,\alpha} 
D^{r'_{ba,\alpha}}_{q_{ba,\alpha},q_{ba,\alpha}'}(\eta_{ba,\alpha})
\right)
\left(
\prod_{\gamma}
D^{\bar r_{a,\gamma}}_{\bar p_{a,\gamma},\bar p_{a,\gamma}'}(\bar \rho_{a,\gamma})
\right)
B_{i_a\rightarrow \cup_{b,\alpha} q_{ba,\alpha}\cup_{\gamma } \bar p_{a,\gamma}}^{R_a\rightarrow \cup_{b,\alpha}r'_{ba,\alpha}\cup_\gamma \bar r_{a,\gamma};\nu_a^+}
\right]
\nn
&\times
\left(\prod_\gamma D^{\bar r'_{a,\gamma}}_{\bar q_{a,\gamma},\bar q_{a,\gamma}'}(\bar \rho_{a,\gamma})\,C_{\bar r'_{a,\gamma},\bar S_{a,\gamma},\bar q_{a,\gamma}}^{\pmb{\bar s}_{a,\gamma}}\,C_{\bar r'_{a,\gamma},\bar S_{a,\gamma},\bar q_{a,\gamma}'}^{\pmb{\bar t}_{a,\gamma}}\right)\nn
&\times B_{j_a'\rightarrow \cup_{b,\alpha} p_{ab,\alpha}'\cup_{\beta } p_{a,\beta}'}^{R_a\rightarrow \cup_{b,\alpha}r_{ab,\alpha}\cup_\beta r_{a,\beta};\nu_a^-}
 B_{i_a'\rightarrow \cup_{b,\alpha} q_{ba,\alpha}'\cup_{\gamma } \bar p_{a,\gamma}'}^{R_a\rightarrow \cup_{b,\alpha}r'_{ba,\alpha}\cup_\gamma \bar r_{a,\gamma};\nu_a^+}
\end{align}
where \(\{r'_{ab,\alpha}\}\), \(\{r'_{a,\beta}\}\) and \(\{\bar r'_{a,\gamma}\}\) are shorthands for \(\cup_{a,b,\alpha}\{r'_{ab,\alpha}\}\), \(\cup_{a,\beta}\{r'_{a,\beta}\}\) and \(\cup_{a,\gamma}\{\bar r'_{a,\gamma}\}\) respectively.
We now use the equivariance property of the branching coefficients (eq. \eqref{B a D}) to rewrite the terms in the square brackets above as
\begin{align}
\left(
\prod_{b,\alpha} 
D^{r_{ab,\alpha}}_{p_{ab,\alpha},p_{ab,\alpha}'}(\eta_{ab,\alpha})
\right)&
\left(
\prod_{\beta}
D^{r_{a,\beta}}_{p_{a,\beta},p_{a,\beta}'}(\rho_{a,\beta})
\right)
B_{j_a\rightarrow \cup_{b,\alpha} p_{ab,\alpha}\cup_{\beta } p_{a,\beta}}^{R_a\rightarrow \cup_{b,\alpha}r_{ab,\alpha}\cup_\beta r_{a,\beta};\nu_a^-}\nn
&=D_{j_a,l_a}^{R_a}\left(\times_{b,\alpha} \eta_{ab,\alpha}\times_\beta \rho_{a,\beta}\right)B_{l_a\rightarrow \cup_{b,\alpha}p_{ab,\alpha}'\cup_\beta p_{a,\beta}'}^{R_a\rightarrow \cup_{b,\alpha}r_{ab,\alpha}\cup_\beta r_{a,\beta};\nu_a^-}
\end{align}
and
\begin{align}\label{equivDDB}
\left(
\prod_{b,\alpha} 
D^{r'_{ba,\alpha}}_{q_{ba,\alpha},q_{ba,\alpha}'}(\eta_{ba,\alpha})
\right)&
\left(
\prod_{\gamma}
D^{\bar r_{a,\gamma}}_{\bar p_{a,\gamma},\bar p_{a,\gamma}'}(\bar \rho_{a,\gamma})
\right)
B_{i_a\rightarrow \cup_{b,\alpha} q_{ba,\alpha}\cup_{\gamma } \bar p_{a,\gamma}}^{R_a\rightarrow \cup_{b,\alpha}r'_{ba,\alpha}\cup_\gamma \bar r_{a,\gamma};\nu_a^+}\nn
&=D_{i_a,l_a'}^{R_a}\left(\times_{b,\alpha} \eta_{ba,\alpha}\times_\gamma \rho_{a,\gamma}\right)B_{l_a'\rightarrow \cup_{b,\alpha}q_{ba,\alpha}'\cup_\gamma \bar p_{a,\gamma}'}^{R_a\rightarrow \cup_{b,\alpha}r'_{ba,\alpha}\cup_\gamma \bar r_{a,\gamma};\nu_a^+}
\end{align}
On the other hand, we can use eqs. \eqref{clbs D 1} and \eqref{clbs D 2} to write the Clebsch-Gordan coefficient terms as
\begin{align}
\prod_\beta D^{r'_{a,\beta}}_{q_{a,\beta},q_{a,\beta}'}(\rho_{a,\beta})\,
C^{ r'_{a,\beta}, S_{a,\beta}, q_{a,\beta}}_{\pmb{ s}_{a,\beta}}\,C^{ r'_{a,\beta}, S_{a,\beta}, q_{a,\beta}'}_{\pmb{ t}_{a,\beta}}
=
\prod_\beta
C^{ r'_{a,\beta}, S_{a,\beta}, q_{a,\beta}'}_{\rho_{a,\beta}^{-1}(\pmb{ s}_{a,\beta})}\,C^{ r'_{a,\beta}, S_{a,\beta}, q_{a,\beta}'}_{\pmb{ t}_{a,\beta}}
\end{align}
(there is no sum over the \(r_{a,\beta}\) and \(S_{a,\beta}\) labels) and
\begin{align}
\prod_\gamma D^{\bar r'_{a,\gamma}}_{\bar q_{a,\gamma},\bar q_{a,\gamma}'}(\bar \rho_{a,\gamma})\,C_{\bar r'_{a,\gamma},\bar S_{a,\gamma},\bar q_{a,\gamma}}^{\pmb{\bar s}_{a,\gamma}}\,C_{\bar r'_{a,\gamma},\bar S_{a,\gamma},\bar q_{a,\gamma}'}^{\pmb{\bar t}_{a,\gamma}}
=\prod_\gamma 
C_{\bar r'_{a,\gamma},\bar S_{a,\gamma},\bar q_{a,\gamma}'}^{\bar \rho_{a,\gamma}^{-1}(\pmb{\bar s}_{a,\gamma})}\,C_{\bar r'_{a,\gamma},\bar S_{a,\gamma},\bar q_{a,\gamma}'}^{\pmb{\bar t}_{a,\gamma}}
\end{align}
(again no sum over the \(\bar r_{a,\gamma}\) and \(\bar S_{a,\gamma}\) labels).

Inserting the last four equations in \eqref{orth disc}, taking the transpose of the matrix element on the RHS of 
\eqref{equivDDB} and relabelling the dummy permutation variables as $\vec\eta\rightarrow \vec\eta^{\,\,-1}, \vec\rho\rightarrow\vec\rho^{\,\,-1}$ gives
\begin{align}\label{o i L alm}
\chi_\Q&(\pmb L,\vec s,\vec\sigma)\,\chi_\Q(\pmb L,\vec t,\vec\tau)=\frac{1}{c_{\vec{n}}}\,\sum_{\vec\eta,\,\vec\rho}\,\sum_{\{s_{ab,\alpha}\}}\,\sum_{\{s_{a,\beta}\}}\,\sum_{\{\bar s_{a,\gamma}\}}\,\prod_a
\frac{d(R_a)}{n_a!}\nn
%\left(\prod_{b,\alpha}\frac{d(r_{ab,\alpha})}{n_{ab,\alpha}!}\right)
%\left(\prod_{\beta}\frac{d(r_{a,\beta})}{n_{a,\beta}!}\right) 
%\left(\prod_{\gamma}\frac{d(\bar r_{a,\gamma})}{\bar n_{a,\gamma}!}\right) \nn
%
%
%
&\times 
D_{l_a',i_a}^{R_a}\left(\times_{b,\alpha} \eta_{ba,\alpha}\times_\gamma \rho_{a,\gamma}\right) D_{i_a,j_a}^{R_a}(\sigma_a)D_{j_a,l_a}^{R_a}\left(\times_{b,\alpha} \eta_{ab,\alpha}^{-1}\times_\beta \rho_{a,\beta}^{-1}\right) \,D_{i_a',j_a'}^{R_a}(\tau_a)\nn
&\times
\left[B_{l_a\rightarrow \cup_{b,\alpha}p_{ab,\alpha}'\cup_\beta p_{a,\beta}'}^{R_a\rightarrow \cup_{b,\alpha}r_{ab,\alpha}\cup_\beta r_{a,\beta};\nu_a^-}\,
B_{j_a'\rightarrow \cup_{b,\alpha} p_{ab,\alpha}'\cup_{\beta } p_{a,\beta}'}^{R_a\rightarrow \cup_{b,\alpha}r_{ab,\alpha}\cup_\beta r_{a,\beta};\nu_a^-}\right]\nn
&\times
\left[B_{l_a'\rightarrow \cup_{b,\alpha}q_{ba,\alpha}'\cup_\gamma \bar p_{a,\gamma}'}^{R_a\rightarrow \cup_{b,\alpha}s_{ba,\alpha}\cup_\gamma \bar r_{a,\gamma};\nu_a^+}\,
B_{i_a'\rightarrow \cup_{b,\alpha} q_{ba,\alpha}'\cup_{\gamma } \bar p_{a,\gamma}'}^{R_a\rightarrow \cup_{b,\alpha}s_{ba,\alpha}\cup_\gamma \bar r_{a,\gamma};\nu_a^+} 
\right]\nn
&\times
\left( \prod_\beta
C^{ s_{a,\beta}, S_{a,\beta}, q_{a,\beta}}_{\rho_{a,\beta}(\pmb{ s}_{a,\beta})}\,C^{ s_{a,\beta}, S_{a,\beta}, q_{a,\beta}}_{\pmb{ t}_{a,\beta}} \right)\,
\left( \prod_\gamma 
C_{\bar s_{a,\gamma},\bar S_{a,\gamma},\bar q_{a,\gamma}}^{\bar \rho_{a,\gamma}(\pmb{\bar s}_{a,\gamma})}\,C_{\bar s_{a,\gamma},\bar S_{a,\gamma},\bar q_{a,\gamma}}^{\pmb{\bar t}_{a,\gamma}}\right)
\end{align}
where we also used the definitions of \(c_{\pmb L}\) and \(c_{\vec n}\) given in \eqref{norm const c_l copy} and \eqref{norm const c_n}. Now, from eq. \eqref{mat el proj} we have
\begin{align}
\left[B_{l_a\rightarrow \cup_{b,\alpha}p_{ab,\alpha}'\cup_\beta p_{a,\beta}'}^{R_a\rightarrow \cup_{b,\alpha}r_{ab,\alpha}\cup_\beta r_{a,\beta};\nu_a^-}\,
 B_{j_a'\rightarrow \cup_{b,\alpha} p_{ab,\alpha}'\cup_{\beta } p_{a,\beta}'}^{R_a\rightarrow \cup_{b,\alpha}r_{ab,\alpha}\cup_\beta r_{a,\beta};\nu_a^-}\right]
=\left. P_{R_a\rightarrow \cup_{b,\alpha}r_{ab,\alpha}\cup_\beta r_{a,\beta}}^{\nu_a^-,\nu_a^-}\right|_{l_a,j_a'}\\[3mm]
\left[B_{l_a'\rightarrow \cup_{b,\alpha}q_{ba,\alpha}'\cup_\gamma \bar p_{a,\gamma}'}^{R_a\rightarrow \cup_{b,\alpha}r'_{ba,\alpha}\cup_\gamma \bar r_{a,\gamma};\nu_a^+}\,
B_{i_a'\rightarrow \cup_{b,\alpha} q_{ba,\alpha}'\cup_{\gamma } \bar p_{a,\gamma}'}^{R_a\rightarrow \cup_{b,\alpha}r'_{ba,\alpha}\cup_\gamma \bar r_{a,\gamma};\nu_a^+} 
\right]
=\left. P_{R_a\rightarrow \cup_{b,\alpha}r'_{ba,\alpha}\cup_\gamma \bar r_{a,\gamma}}^{\nu_a^+,\nu_a^+}\right|_{l_a',i_a'}
\end{align}
so that we can write
\begin{align}\label{orthoL mid}
\chi_\Q&(\pmb L,\vec s,\vec\sigma)\,\chi_\Q(\pmb L,\vec t,\vec\tau)=\frac{1}{c_{\vec{n}}}\,\sum_{\vec\eta,\,\vec\rho}\,\sum_{\{r'_{ab,\alpha}\}}\,\prod_a
\frac{d(R_a)}{n_a!}\,
D_{l_a',l_a}^{R_a}\left( \text{Adj}_{\vec\eta\times\vec\rho}(\sigma_a) \right) \,D_{i_a',j_a'}^{R_a}(\tau_a)\nn
&\qquad\quad\times
\left. P_{R_a\rightarrow \cup_{b,\alpha}r_{ab,\alpha}\cup_\beta r_{a,\beta}}^{\nu_a^-,\nu_a^-}\right|_{l_a,j_a'}\,\left. P_{R_a\rightarrow \cup_{b,\alpha}r'_{ba,\alpha}\cup_\gamma \bar r_{a,\gamma}}^{\nu_a^+,\nu_a^+}\right|_{l_a',i_a'}\\[2mm]
&\qquad\quad\times
\left(\sum_{\{r'_{a,\beta}\}}\, \prod_\beta
C^{ r'_{a,\beta}, S_{a,\beta}, q_{a,\beta}}_{\rho_{a,\beta}(\pmb{ s}_{a,\beta})}\,C^{ r'_{a,\beta}, S_{a,\beta}, q_{a,\beta}}_{\pmb{ t}_{a,\beta}} \right)\,
\left(\sum_{\{\bar r'_{a,\gamma}\}}\, \prod_\gamma 
C_{\bar r'_{a,\gamma},\bar S_{a,\gamma},\bar q_{a,\gamma}}^{\bar \rho_{a,\gamma}(\pmb{\bar s}_{a,\gamma})}\,C_{\bar r'_{a,\gamma},\bar S_{a,\gamma},\bar q_{a,\gamma}}^{\pmb{\bar t}_{a,\gamma}}\right)\nonumber
\end{align}
where we defined
\begin{align}
\text{Adj}_{\vec\eta\times\vec\rho}(\sigma_a)=(\times_{b,\alpha} \eta_{ba,\alpha}\times_\gamma \rho_{a,\gamma})(\sigma_a)(\times_{b,\alpha} \eta_{ab,\alpha}^{-1}\times_\beta \rho_{a,\beta}^{-1})
\end{align}

Now we can proceed to sum over \(\pmb L=\{R_a, r_{ab,\alpha},r_{a,\beta},S_{a,\beta},\bar r_{ a,\gamma},\bar S_{a,\gamma}, \nu_a^+,\nu_a^-\}\). This introduces, among others, a summation over the flavour states \(S_{a,\beta}\) and \(\bar S_{a,\gamma}\). Consider then a pair of Clebsch-Gordan coefficients like the ones appearing in the last line of eq. \eqref{orthoL mid}. It is easy to write the relation
\begin{align}
\sum_{r,S,i}C_{\rho(\pmb s)}^{r,S,i}\,C_{\pmb t}^{r,S,i}=
\langle \rho(\pmb s)|\left(\sum_{r,S,i}|r,S,i\rangle\langle r,S,i|\right)|\pmb t\rangle=
\langle \rho(\pmb s)|{1}|\pmb t\rangle=\delta_{\rho(\pmb s),\pmb t}
\end{align}
We then have the identity
\begin{align}
\sum_{r'_{a,\beta},\,S_{a,\beta},\,q_{a,\beta}}\,C^{r'_{a,\beta}, S_{a,\beta}, q_{a,\beta}}_{\rho_{a,\beta}(\pmb{ s}_{a,\beta})}\, C^{ r'_{a,\beta}, S_{a,\beta}, q_{a,\beta}}_{\pmb{ t}_{a,\beta}}
&=
\delta_{\rho_{a,\beta}(\pmb{ s}_{a,\beta}),\pmb{ t}_{a,\beta} }
\end{align}
and similarly
\begin{align}
\sum_{\bar r'_{a,\gamma},\,\bar S_{a,\gamma},\,\bar q_{a,\gamma}} C_{\bar r'_{a,\gamma},\bar S_{a,\gamma},\bar q_{a,\gamma}}^{\bar \rho_{a,\gamma}(\pmb{\bar s}_{a,\gamma})}\,C_{\bar r'_{a,\gamma},\bar S_{a,\gamma},\bar q_{a,\gamma}}^{\pmb{\bar t}_{a,\gamma}}
=\delta_{\bar \rho_{a,\gamma}(\pmb{\bar s}_{a,\gamma}),\pmb{\bar t}_{a,\gamma}}
\end{align}
Inserting this result in eq. \eqref{orthoL mid} we obtain
\begin{align}
\sum_{\pmb L} &\chi_\Q(\pmb L,\vec s,\vec\sigma)\,\chi_\Q(\pmb L,\vec t,\vec\tau)\nn
=&\frac{1}{c_{\vec{n}}}\,\sum_{\vec\eta,\,\vec\rho}\, \prod_a \,\sum_{R_a}\,\frac{d(R_a)}{n_a!} 
D_{l_a',l_a}^{R_a}\left(\text{Adj}_{\vec\eta\times\vec\rho}(\sigma_a)\right) \,D_{i_a',j_a'}^{R_a}(\tau_a)\nn
&\times\sum_{\cup_{b,\alpha}\{r_{ab,\alpha}\}}\,\sum_{\cup_\beta \{r_{a,\beta}\}\atop \nu_a^-}
\,\left. P_{R_a\rightarrow \cup_{b,\alpha}r_{ab,\alpha}\cup_\beta r_{a,\beta}}^{\nu_a^-,\nu_a^-}\right|_{l_a,j_a'}
\sum_{\cup_{b,\alpha}\{r'_{ba,\alpha}\}}\,\sum_{\cup_\gamma\{\bar r_{a,\gamma}\}\atop \nu_a^+}\,\left. P_{R_a\rightarrow \cup_{b,\alpha}r'_{ba,\alpha}\cup_\gamma \bar r_{a,\gamma}}^{\nu_a^+,\nu_a^+}\right|_{l_a',i_a'}\nn
&\qquad\qquad\times
\left( \prod_\beta \delta_{\rho_{a,\beta}(\pmb{ s}_{a,\beta}),\pmb{ t}_{a,\beta} }
 \right)\,
\left( \prod_\gamma \delta_{\bar \rho_{a,\gamma}(\pmb{\bar s}_{a,\gamma}),\pmb{\bar t}_{a,\gamma}}
\right)
\end{align}
Now using the projector identity
\begin{align} 
 \sum_{\cup_i \{r_i\},\,\nu} \left.P_{R\rightarrow \cup_i r_i}^{\nu,\nu}\right|_{k,l}=\delta_{k,l}
\end{align}
we further get
\begin{align}\label{ortho L nl}
\sum_{\pmb L}\chi_\Q&(\pmb L,\vec s,\vec\sigma)\,\chi_\Q(\pmb L,\vec t,\vec\tau)=\frac{1}{c_{\vec{n}}}\,\sum_{\vec\eta,\,\vec\rho}\,\prod_a
\sum_{R_a}\,\frac{d(R_a)}{n_a!}
\chi_{R_a}\left(\text{Adj}_{\vec\eta\times\vec\rho}(\sigma_a) \,\tau_a^{-1}\right)\nn
&\qquad\qquad\qquad\qquad\qquad\qquad\times
\left( \prod_\beta \delta_{\rho_{a,\beta}(\pmb{ s}_{a,\beta}),\pmb{ t}_{a,\beta} }
 \right)\,
\left( \prod_\gamma \delta_{\bar \rho_{a,\gamma}(\pmb{\bar s}_{a,\gamma}),\pmb{\bar t}_{a,\gamma}}
\right)
\end{align}
Finally, through the identity
\begin{align}
\sum_{R\vdash n}\frac{d(R)}{n!}\chi_R(\sigma)=\delta(\sigma)
\end{align}
we can rewrite \eqref{ortho L nl} as
\begin{align}
\sum_{\pmb L}\chi_\Q&(\pmb L,\vec s,\vec\sigma)\,\chi_\Q(\pmb L,\vec t,\vec\tau)=\nn
&=
\frac{1}{c_{\vec{n}}}\,\sum_{\vec\eta,\,\vec\rho}\,
\prod_a\,\delta\left(\text{Adj}_{\vec\eta\times\vec\rho}(\sigma_a) \,\tau_a^{-1}\right)
\,
\left( \prod_\beta \delta_{\rho_{a,\beta}(\pmb{ s}_{a,\beta}),\pmb{ t}_{a,\beta} }
 \right)\,
\left( \prod_\gamma \delta_{\bar \rho_{a,\gamma}(\pmb{\bar s}_{a,\gamma}),\pmb{\bar t}_{a,\gamma}}
\right)\nn
&=
\frac{1}{c_{\vec{n}}}\,\sum_{\vec\eta,\,\vec\rho}\,
\delta\left(\text{Adj}_{\vec\eta\times\vec\rho\,}(\vec \sigma) \,\vec\tau^{\,-1}\right)
\,
\delta_{\vec \rho(\vec{ s}),\vec{ t} }
\end{align}
This last equation is exactly \eqref{orthogonality chi 2}.

\section{Two Point Function: Proof of Orthogonality}\label{ortho L derivation}
In this section we will prove the orthogonality formula \eqref{ortho L}:
\begin{align}\label{ortho L copy}
\left\langle
\mathcal O_{\mathcal Q}(\pmb L)\,
\mathcal O_{\mathcal Q}^\dagger(\pmb {L'})
\right\rangle
=\delta_{\pmb L,\pmb{L'}}\,\, c_{\vec n}\,\prod_a f_{N_a}(R_a)
\end{align}
The first ingredient we need is the Hermitean conjugated version of the operator defined in \eqref{Q def}, which is simply
\begin{align}\label{Q conj def}
\mathcal{O}^{\dagger}_\mathcal{Q}(\vec n;\, \vec{s};\,\vec\sigma)%&=
&=
\prod_{a}\left[\prod_{b,\alpha}\left(\Phi_{ab,\alpha}^{\dagger\,\,\otimes n_{ab,\alpha}}\right)^{J_{ab,\alpha}}_{I_{ab,\alpha}}\right]
%quark
\otimes
\left[\prod_\beta \left(Q_{a,\beta}^{\dagger\,\,\otimes n_{a,\beta}}\right)_{I_{a,\beta}}^{\pmb s_{a,\beta}}\right]
\otimes
%antiquark
\left[\prod_\gamma\left(\bar Q_{a,\gamma}^{\dagger\,\,\otimes\bar n_{ {a,\gamma}}}\right)^{\bar J_{ { a,\gamma}}}_{ {\pmb{\bar s}_{a,\gamma}}}\right]\nonumber\\[5mm]
%permutation
&\qquad\qquad\qquad\qquad\qquad\times
\prod_c\left(\sigma_c^{-1}\right)_{\cup_{b,\alpha}J_{bc,\alpha}\cup_\gamma\bar J_{ a,\gamma}}
^{\cup_{b,\alpha}I_{cb,\alpha}\cup_\beta I_{ {a,\beta}}}
\end{align}
Here we used $\left(\sigma\vphantom{^{-1}}\right)^j_i=\left(\sigma^{-1}\right)_j^i$. Using the free field metric
\begin{align}\label{free field metric copy}
\left\langle
\left(\Phi_{ab,\alpha}\right)^i_j
(\Phi_{ab,\alpha}^\dagger)^k_l
\right\rangle
=\delta^i_l\delta^k_j\,,\quad
\left\langle
\left(Q_{a,\beta}\right)^i_s
(Q_{a,\beta}^\dagger)^p_l
\right\rangle
=\delta^i_l\delta^p_s\,,\quad
\left\langle
\left(\bar Q_{a,\gamma}\right)^{\bar s}_j
(\bar Q_{a,\gamma}^\dagger)^k_{\bar p}
\right\rangle
=\delta^k_j\delta_{\bar p}^{\bar s}
\end{align}
(the remaining correlators are zero) we get
\begin{align}
\left\langle
\left(\Phi_{ab,\alpha}^{\otimes n_{ab,\alpha}}\right)^{I_{ab,\alpha}}_{J_{ab,\alpha}}
\left(\Phi_{ab,\alpha}^{\dagger\,\,\otimes n_{ab,\alpha}}\right)^{J_{ab,\alpha}'}_{I_{ab,\alpha}'}
\right\rangle
=
\sum_{\eta\in S_{n_{ab,\alpha}}}\delta^{\eta(I_{ab,\alpha})}_{I'_{ab,\alpha}}\delta^{J_{ab,\alpha}'}_{\eta(J_{ab,\alpha})}
\end{align}
In this formula the sum over permutations represents all possible Wick contractions of the labels $I_{ab,\alpha}=\{i_1,...,i_{n_{ab,\alpha}}\}$, $J_{ab,\alpha}=\{j_1,...,j_{n_{ab,\alpha}}\}$. Denoting the states belonging to the fundamental and the antifundamental representation of $U(N)$ by \(|e_{j}\rangle\) and \(\langle e^j|\)respectively, we have the identities
\begin{align}
\delta^{J_{ab,\alpha}'}_{\eta(J_{ab,\alpha})}=
\langle e^{j_1'},...,e^{j_{n_{ab,\alpha}}'}|
e_{j_{\eta(1)}},...,e_{j_{\eta( n_{ab,\alpha})}}\rangle=
\langle e^{j_1'},...,e^{j_{n_{ab,\alpha}}'}|
\eta |e_{j_{1}},...,e_{j_{ n_{ab,\alpha}}}\rangle=
\left(\eta\right)_{J_{ab,\alpha}}^{J_{ab,\alpha}'}
\end{align}
and
\begin{align}
\delta_{\eta(I_{ab,\alpha})}^{I_{ab,\alpha}'}=
\left(\eta\right)_{I_{ab,\alpha}}^{I_{ab,\alpha}'}= \left(\eta^{-1}\right)^{I_{ab,\alpha}}_{I_{ab,\alpha}'}=\delta_{\eta^{-1}(I_{ab,\alpha}')}^{I_{ab,\alpha}}
\end{align}
Performing similar steps on the correlators of quarks and antiquarks, we can then write
\begin{subequations}
\begin{align}
&\left\langle
\left(\Phi_{ab,\alpha}^{\otimes n_{ab,\alpha}}\right)^{I_{ab,\alpha}}_{J_{ab,\alpha}}
\left(\Phi_{ab,\alpha}^{\dagger\,\,\otimes n_{ab,\alpha}}\right)^{J_{ab,\alpha}'}_{I_{ab,\alpha}'}
\right\rangle
=\sum_{\eta\in S_{n_{ab,\alpha}}}
\left(\eta^{-1}\right)^{I_{ab,\alpha}}_{I_{ab,\alpha}'}\,\left(\eta\right)_{J_{ab,\alpha}}^{J_{ab,\alpha}'}\\[3mm]
%
%
%
%QUARK
%
%
&\left\langle
\left(Q_{a,\beta}^{\otimes n_{a,\beta}}\right)^{I_{a,\beta}}_{\pmb{s}_{a,\beta}}
\left(Q_{a,\beta}^{\dagger\,\,\otimes n_{a,\beta}}\right)^{\pmb{s'}_{a,\beta}}_{I_{a,\beta}'}
\right\rangle
=
\sum_{\rho\in S_{n_{a,\beta}}} \left(\rho^{-1}\right)^{I_{a,\beta}}_{I_{a,\beta}'}\,(\rho)^{\pmb{s'}_{a,\beta}}_{\pmb{s}_{a,\beta}}
=
\sum_{\rho\in S_{n_{a,\beta}}} \left(\rho^{-1}\right)^{I_{a,\beta}}_{I_{a,\beta}'}\,\delta^{\pmb{s'}_{a,\beta}}_{\rho(\pmb{s}_{a,\beta})}\displaybreak[0]\\[3mm]
%
%
%
%ANTIQUARK
%
%
&\left\langle
\left(\vphantom{\sum}\bar Q_{a,\gamma}^{\otimes \bar n_{a,\gamma}}\right)_{\bar J_{a,\gamma}}^{\pmb{\bar s}_{a,\gamma}}
\left(\vphantom{\sum}\bar Q_{a,\gamma}^{\dagger\,\,\otimes \bar n_{a,\gamma}}\right)_{\pmb{\bar s'}_{a,\gamma}}^{\bar J_{a,\gamma}'}
\right\rangle
=\sum_{\bar \rho\in S_{\bar n_{a,\gamma}}}
(\bar \rho^{-1})^{\pmb{\bar s}_{a,\gamma}}_{\pmb{\bar {s}'}_{a,\gamma}}
\left(\bar\rho\vphantom{^{-1}}\right)^{\bar J_{a,\gamma}'}_{\bar J_{a,\gamma}}\,
=\sum_{\bar \rho\in S_{\bar n_{a,\gamma}}}\left(\bar\rho\vphantom{^{-1}}\right)^{\bar J_{a,\gamma}'}_{\bar J_{a,\gamma}}\,
\delta^{\bar\rho(\pmb{\bar s}_{a,\gamma})}_{\pmb{\bar {s}'}_{a,\gamma}}
\end{align}
\end{subequations}
We therefore get
\begin{align}
&\left\langle
\mathcal O_{\mathcal Q}(\vec n,\vec s,\vec \sigma)
\mathcal O_{\mathcal Q}^\dagger(\vec n,\vec{ s}\,',\vec{ \sigma}\,')
\right\rangle=
\sum_{\vec\eta,\,\vec\rho}\,\prod_a
\left[\prod_{b,\alpha}
\left(\eta_{ab,\alpha}^{-1}\right)^{I_{ab,\alpha}}_{I_{ab,\alpha}'}\,\left(\eta_{ab,\alpha}\right)_{J_{ab,\alpha}}^{J_{ab,\alpha}'}\right]
\left[\prod_\beta\left(\rho_{a,\beta}^{-1}\right)^{I_{a,\beta}}_{I_{a,\beta}'}\,
\delta_{\pmb{s'}_{a,\beta}}^{\rho_{a,\beta}(\pmb{s}_{a,\beta})}
\right]\nn
&\qquad\qquad\qquad\times
\left[\prod_\gamma
\left(\bar\rho_{a,\gamma}\vphantom{^{-1}}\right)^{\bar J_{a,\gamma}'}_{\bar J_{a,\gamma}}\,
\delta^{\bar\rho_{a,\gamma}(\pmb{\bar s}_{a,\gamma})}_{\pmb{\bar {s}'}_{a,\gamma}}\right]
\left(\sigma_a\vphantom{^{-1}}\right)^{\cup_{b,\alpha}J_{ba,\alpha}\cup_\gamma\bar J_{ a,\gamma}}
_{\cup_{b,\alpha}I_{ab,\alpha}\cup_\beta I_{ {a,\beta}}}
\left((\sigma_a')^{-1}\right)_{\cup_{b,\alpha}J'_{ba,\alpha}\cup_\gamma\bar J'_{ a,\gamma}}
^{\cup_{b,\alpha}I'_{ab,\alpha}\cup_\beta I'_{ {a,\beta}}}\nn
&\qquad\qquad\qquad\qquad\qquad\qquad=
\sum_{\vec\eta,\,\vec\rho}\,\prod_a 
\Tr_{V_{N_a}^{\otimes n_a}}\left[\vphantom{\sum}
\left(\times_{b,\alpha}\eta_{ba,\alpha}\times_\gamma\bar\rho_{a,\gamma}\right)\sigma_a\left(\times_{b,\alpha}\eta_{ab,\alpha}^{-1}\times_\beta\rho_{a,\beta}^{-1}\right)(\sigma_a')^{-1}
\right]\nn
&\qquad\qquad\qquad\qquad\qquad\qquad\qquad\qquad\qquad\qquad
\times\,\left[\prod_\beta \delta_{\pmb{s'}_{a,\beta}}^{\rho_{a,\beta}(\pmb{s}_{a,\beta})}\right]
\left[\prod_\gamma\delta_{\pmb{\bar s'}_{a,\gamma}}^{\bar \rho_{a,\gamma}(\pmb{\bar s}_{a,\gamma})}\right]
\end{align}
where, as we defined in \eqref{perms eta and rho},
\begin{subequations}\label{perms eta and rho copy}
\begin{align}
&\vec\eta=\cup_{a,b, \alpha}\{\eta_{ab,\alpha}\}\,\,\,,\quad \eta_{ab,\alpha}\in S_{n_{ab,\alpha}}\\[4mm]
&\vec\rho=\cup_a\{\cup_\beta\, \rho_{a,\beta};\,\cup_\gamma\,\bar \rho_{a,\gamma}\}\,\,\,,\quad \rho_{a,\beta}\in S_{n_{a,\beta}}\,\,\,\,,\,\,\,\,\,\,\,\bar\rho_{a,\gamma}\in S_{\bar n_{a,\gamma}}
\end{align}
\end{subequations}
The trace is taken over the product space $V_{N_a}^{\otimes n_a}$, $V_{N_a}$ being the fundamental representation of $U(N_a)$ and $n_a =\sum_{b,\alpha}n_{ab,\alpha}+\sum_\beta n_{a,\beta}$. Recalling \eqref{adj action},
\begin{align}
\text{Adj}_{\vec\eta\times\vec\rho}(\vec\sigma)=
\cup_a\{
(\times_{b,\alpha}\eta_{ba,\alpha}\times_\gamma\bar\rho_{a,\gamma})
\sigma_a
(\times_{b,\alpha}\eta_{ab,\alpha}^{-1}\times_\beta \rho_{ a,\beta}^{-1})
\}
\end{align}
we can finally write
\begin{align}\label{corr K basis copy}
\left\langle
\mathcal O_{\mathcal Q}(\vec n,\vec s,\vec \sigma)
\mathcal O_{\mathcal Q}^\dagger(\vec n,\vec{ s}\,',\vec{ \sigma}\,')
\right\rangle=
\sum_{\vec\eta,\,\vec\rho}\prod_a\,&\Tr_{V_{N_a}^{\otimes n_a}}\left[\vphantom{\sum}\text{Adj}_{\vec\eta\times\vec\rho}(\sigma_a)\,
(\sigma_a')^{-1}
\right]\nn
&\times
\left[\prod_\beta \delta_{\pmb{s'}_{a,\beta}}^{\rho_{a,\beta}(\pmb{s}_{a,\beta})}\right]
\left[\prod_\gamma\delta_{\pmb{\bar s'}_{a,\gamma}}^{\bar \rho_{a,\gamma}(\pmb{\bar s}_{a,\gamma})}\right]
\end{align}
which is eq. \eqref{corr K basis}.

Using the definition of the Fourier transformed operator \eqref{fourier2} we then get
\begin{align}\label{otho mid}
\left\langle
\mathcal O_{\mathcal Q}(\pmb L)
\mathcal O_{\mathcal Q}^\dagger(\pmb {L'})
\right\rangle&=
\sum_{\vec s,\vec{s}\,'}\sum_{\vec \sigma,\vec{\sigma}\,'}
\chi_{\Q}(\pmb L,\vec s,\vec \sigma)
\chi_\Q^\dagger(\pmb L',\vec s\,',\vec\sigma\,')
\left\langle
\mathcal O_{\mathcal Q}(\vec n,\vec s,\vec \sigma)
\mathcal O_{\mathcal Q}^\dagger(\vec n,\vec{ s}\,',\vec{ \sigma}\,')
\right\rangle\\[3mm]
&=
\sum_{\vec s}\sum_{\vec \sigma,\vec{\sigma}\,'}\sum_{\vec\eta,\,\vec\rho}
\chi_{\Q}(\pmb L,\vec s,\vec \sigma)
\chi_\Q(\pmb L',\vec\rho(\vec s),\vec\sigma\,')
\,\prod_a\,\Tr_{V_{N_a}^{\otimes n_a}}\left[\vphantom{\sum}\text{Adj}_{\vec\eta\times\vec\rho}(\sigma_a)\,
(\sigma_a')^{-1}
\right]\nonumber
\end{align}
where to get the second equality we summed over $\vec s\,'$, used the Kronecker delta functions and used the reality of the quiver characters.
Redefining the dummy variable \(\vec s\rightarrow \vec \rho^{\,\,-1}(\vec s\,)\) in \eqref{otho mid} we obtain
\begin{align}\label{otho L mid}
&\left\langle
\mathcal O_{\mathcal Q}(\pmb L)
\mathcal O_{\mathcal Q}^\dagger(\pmb {L'})
\right\rangle\nn
&\qquad=\sum_{\vec s}\sum_{\vec \sigma,\vec{\sigma}\,'}\sum_{\vec\eta,\,\vec\rho}
\chi_{\Q}(\pmb L,\vec\rho^{\,\,-1}(\vec s\,),\vec \sigma)
\chi_\Q(\pmb L',\vec s,\vec\sigma\,') \prod_a\,\Tr_{V_{N_a}^{\otimes n_a}}\left[\vphantom{\sum}\text{Adj}_{\vec\eta\times\vec\rho}(\sigma_a)\,
(\sigma_a')^{-1}
\right]\nn
&\qquad=
\sum_{\vec s}\sum_{\vec \sigma,\vec{\sigma}\,'}\sum_{\vec\eta,\,\vec\rho}
\chi_\Q(\pmb L, \vec s,\text{Adj}_{\vec\eta\times\vec\rho}(\vec\sigma))
\chi_\Q(\pmb L',\vec s,\vec\sigma\,')\prod_a\,\Tr_{V_{N_a}^{\otimes n_a}}\left[\vphantom{\sum}\text{Adj}_{\vec\eta\times\vec\rho}(\sigma_a)\,
(\sigma_a')^{-1}
\right]\nn
&\qquad=
\sum_{\vec s}\sum_{\vec \sigma,\vec{\sigma}\,'}\sum_{\vec\eta,\,\vec\rho}
\chi_\Q(\pmb L, \vec s,\vec\sigma)
\chi_\Q(\pmb L',\vec s,\vec\sigma\,')\prod_a\,\Tr_{V_{N_a}^{\otimes n_a}}\left[\vphantom{\sum}\sigma_a\,
(\sigma_a')^{-1}
\right]
\end{align}
To get the second equality we used the invariance relation \eqref{invariance chi}, and in the third we relabelled the dummy variable $\vec\sigma\rightarrow\text{Adj}_{\vec\rho\times\vec\eta}(\vec\sigma)$. We then see that the dependence on the permutations $\vec\eta$ and $\vec\rho$ drops out, so that their sums can be trivially computed to obtain
\begin{align}
\left\langle
\mathcal O_{\mathcal Q}(\pmb L)
\mathcal O_{\mathcal Q}^\dagger(\pmb {L'})
\right\rangle&=
\sum_{\vec s}\sum_{\vec \sigma,\vec{\sigma}\,'}
\chi_\Q(\pmb L, \vec s,\vec\sigma)
\chi_\Q(\pmb L',\vec s,\vec\sigma\,')\nn
&\qquad\times
\prod_a
\left(\prod_{b,\alpha}n_{ab,\alpha}!\right)
\left(\prod_{\beta}n_{a,\beta}!\right)\left(\prod_{\gamma}\bar n_{a,\gamma}!\right)
\,\Tr_{V_{N_a}^{\otimes n_a}}\left[\vphantom{\sum}\sigma_a\,
(\sigma_a')^{-1}
\right]
\end{align}
Now let us further relabel $\sigma_a\rightarrow\tau_a\cdot\sigma_a'$ and use the definition of $c_{\vec n}$ given in \eqref{norm const c_n} to get
\begin{align}\label{ortho L mid almost}
\left\langle
\mathcal O_{\mathcal Q}(\pmb L)
\mathcal O_{\mathcal Q}^\dagger(\pmb {L'})
\right\rangle&= c_{\vec n}
\sum_{\vec s}\sum_{\vec \tau,\vec{\sigma}\,'}
\chi_\Q(\pmb L, \vec s,\vec\tau\cdot\vec\sigma\,')
\chi_\Q(\pmb L',\vec s,\vec\sigma\,')\,
\prod_a
\Tr_{V_{N_a}^{\otimes n_a}}\left[\vphantom{\sum}\tau_a
\right]
\end{align}
The only dependence on $\vec\sigma'$ and $\vec s$ is now inside the two quiver characters, so that we can use \eqref{almost ortho} and write
\begin{align}
\sum_{\vec s}\sum_{\vec\sigma\,'}& \chi_\Q(\pmb L,\vec{s},\vec{\tau}\cdot\vec{\sigma}\,')\,\chi_\Q(\pmb{L'},\vec{s},\vec{\sigma}\,')\nn
&=c_{\pmb L}c_{\pmb{L'}}\prod_a\,
\frac{n_a!}{d(R_a)}\,\Tr\left(D^{R_a}(\tau_a) P_{R_a\rightarrow\cup_{b,\alpha} r_{ba,\alpha}\cup_\gamma {\bar r}_{a,\gamma}}^{\nu_a^+,{\nu_a^+}'}\right)\,
\delta_{R_a, R'_a}\nn
&\quad\times \left(\prod_{b,\alpha}\delta_{r_{ab,\alpha}, r'_{ab,\alpha}}\right)\left(\prod_\beta\,d(r_{a,\beta})\delta_{r_{a,\beta}, r'_{a,\beta}}\delta_{S_{a,\beta},S'_{a,\beta}}\right)\left(\prod_\gamma\,\delta_{\bar r_{a,\gamma},{\bar r'}_{a,\gamma}}\delta_{\bar S_{a,\gamma},{\bar S'}_{a,\gamma}}\right)\delta_{\nu_a^-,{\nu_a^{-}}'}
\end{align}
Inserting this equation into \eqref{ortho L mid almost} and recalling the identity 
\begin{align}\label{N cicle ortho}
\Tr_{V_N^{\otimes n}}\left(\sigma \right)=
N^{C[\sigma]}
\end{align}
we get
\begin{align}\label{ortho L mid almost almost} 
\left\langle
\mathcal O_{\mathcal Q}(\pmb L)
\mathcal O_{\mathcal Q}^\dagger(\pmb {L'})
\right\rangle&=c_{\vec n}\,
\sum_{\vec \tau}
c_{\pmb L}c_{\pmb L'}
\prod_a\,
\frac{n_a!}{d(R_a)}\delta_{R_a, R'_a}\delta_{\nu_a^-,{\nu_a^{-}}'} \left(\prod_{b,\alpha}\delta_{r_{ab,\alpha}, r'_{ab,\alpha}}\right) \,\,
\nn
&\quad\times\left(\prod_\beta\,d(r_{a,\beta})\delta_{r_{a,\beta}, r'_{a,\beta}}\delta_{S_{a,\beta},S'_{a,\beta}}\right)
\left(\prod_\gamma\delta_{\bar r_{a,\gamma},{\bar r'}_{a,\gamma}}\delta_{\bar S_{a,\gamma},{\bar S'}_{a,\gamma}}\right)\,\nn
&\qquad\times
\Tr\left(D^{R_a}(\tau_a) P_{R_a\rightarrow\cup_{b,\alpha} r_{ba,\alpha}\cup_\gamma {\bar r}_{a,\gamma}}^{\nu_a^+,{\nu_a^+}'}\right)N_a^{c[\tau_a]}
\end{align}
The last piece we need to obtain eq. \eqref{ortho L copy} is the identity
\begin{align}
\sum_{\tau_a}\Tr\left(D^{R_a}(\tau_a) P_{R_a\rightarrow\cup_{b,\alpha} r_{ba,\alpha}\cup_\gamma {\bar r}_{a,\gamma}}^{\nu_a^+,{\nu_a^+}'}\right)N_a^{c[\tau_a]}=
\delta_{\nu_a^+,{\nu_a^+}'}\,\left(\prod_{b,\alpha}d(r_{ba,\alpha})\right)
\left(\prod_{\gamma}d(\bar r_{a,\gamma})\right)\,f_{N_a}(R_a)
\end{align}
a proof of which can be found in \emph{e.g.} \cite{Hamermesh}. Inserting it in \eqref{ortho L mid almost almost} we finally get
\begin{align}\label{ortho L mid almost last} 
\left\langle
\mathcal O_{\mathcal Q}(\pmb L)
\mathcal O_{\mathcal Q}^\dagger(\pmb {L'})
\right\rangle&=
c_{\vec n}\,c_{\pmb L}c_{\pmb L'}
\prod_a\,
\frac{n_a!}{d(R_a)}\delta_{R_a, R'_a}\delta_{\nu_a^-,{\nu_a^{-}}'}\delta_{\nu_a^+,{\nu_a^+}'} \left(\prod_{b,\alpha}d(r_{ab,\alpha})\delta_{r_{ab,\alpha}, r'_{ab,\alpha}}\right) \,\,
\nn
&\quad\times\left(\prod_\beta\,d(r_{a,\beta})\delta_{r_{a,\beta}, r'_{a,\beta}}\delta_{S_{a,\beta},S'_{a,\beta}}\right)
\left(\prod_\gamma\bar d(\bar r_{a,\gamma})\delta_{\bar r_{a,\gamma},{\bar r'}_{a,\gamma}}\delta_{\bar S_{a,\gamma},{\bar S'}_{a,\gamma}}\right)\, f_{N_a}(R_a)\displaybreak[0]\nn
%
%
% NEW LINE
%
%
&=\delta_{\pmb L,\pmb{L'}}\,c_{\vec n}\,c_{\pmb L}^2
\prod_a\,
\frac{n_a!}{d(R_a)} \left(\prod_{b,\alpha}d(r_{ab,\alpha})\right) \,
\left(\prod_\beta\,d(r_{a,\beta})\right)
\left(\prod_\gamma\bar d(\bar r_{a,\gamma})\right)\, f_{N_a}(R_a)
\end{align}
which, using the normalisation constant $c_{\pmb L}$ defined in \eqref{norm const c_l copy}, reduces to eq. \eqref{ortho L}:
%\begin{align}\label{ortho L last} 
%\left\langle
%\mathcal O_{\mathcal Q}(\pmb L)\,
%\mathcal O_{\mathcal Q}^\dagger(\pmb {L'})
%\right\rangle
%
%=\delta_{\pmb L,\pmb{L'}}\,\prod_a \left(\prod_{b,\alpha}n_{ab,\alpha}!\right)\left(\prod_\beta n_{a,\beta}!\right)\left(\prod_\gamma \bar n_{a,\gamma}!\right)f_{N_a}(R_a)
%\end{align}
\begin{align}\label{ortho L last} 
\left\langle
\mathcal O_{\mathcal Q}(\pmb L)\,
\mathcal O_{\mathcal Q}^\dagger(\pmb {L'})
\right\rangle
=\delta_{\pmb L,\pmb{L'}}\,\, c_{\vec n}\,\prod_a f_{N_a}(R_a)
\end{align}
The orthogonality of the Fourier transformed operators is thus proven.

\section{Deriving the Holomorphic Gauge Invariant Operator Ring Structure Constants}\label{app:Holomorphic Ring SC}
In this appendix we will derive the analytical expression for the holomorphic GIO ring structure constants \(G_{\pmb L^{(1)},\,\pmb L^{(2)},\,\pmb L^{(3)}}\), corresponding to the diagram given in Fig. \ref{fig: CRSC picture}. We will divide the computation into five main steps, for improved clarity.
In the following subsection \ref{App_sub: CRSC derivation for N=2} we will explicitly derive the chiral ring structure constants for an \(\N=2\) SQCD, by using diagrammatic techniques alone.

\subsubsection*{1) The permutation basis product}
In this first step we are going to rewrite the product of two operators in the permutation basis, \(\cO_\Q(\vec n_1,\vec s^{\,(1)},\vec \sigma^{(1)})\,\cO_\Q(\vec n_2,\vec s^{\,(2)},\vec \sigma^{(2)})\), as a single operator \(\cO_\Q(\vec n_3,\vec s^{\,(3)},\vec\sigma^{(3)})\), specified by appropriate labels $\vec n_3,\,\vec s^{\,(3)}$ and $\vec\sigma^{(3)}$. We use the defining equation \eqref{Q def} for \(\cO_\Q(\vec n,\vec s,\vec\sigma)\) to write this product as
\begin{align}\label{OP product in CR}
\cO_\Q&(\vec n_1,\vec s^{\,(1)},\vec \sigma^{(1)})\,\cO_\Q(\vec n_2,\vec s^{\,(2)},\vec \sigma^{(2)})\nn
&=
\prod_{a}\left[\prod_{b,\alpha}\left(\Phi_{ab,\alpha}^{\otimes n_{ab,\alpha}^{(1)}}\right)_{J_{ab,\alpha}^{(1)}}^{I_{ab,\alpha}^{(1)}}\right]
%quark
\otimes
\left[\prod_\beta \left(Q_{a,\beta}^{\otimes n_{a,\beta}^{(1)}}\right)^{I_{a,\beta}^{(1)}}_{\pmb s_{a,\beta}^{(1)}}\right]
\otimes
%antiquark
\left[\prod_\gamma\left(\bar Q_{a,\gamma}^{\otimes\bar n_{ {a,\gamma}}^{(1)}}\right)_{\bar J_{ { a,\gamma}}^{(1)}}^{ {\pmb{\bar s}_{a,\gamma}^{(1)}}}\right]%\nonumber\\[5mm]
%permutation
%&\qquad\qquad\qquad\times
\left(\sigma_a^{(1)}\right)^{\times_{b,\alpha}J_{ba,\alpha}^{(1)}\times_\gamma\bar J_{ a,\gamma}^{(1)}}
_{\times_{b,\alpha}I_{ab,\alpha}^{(1)}\times_\beta I_{ {a,\beta}}^{(1)}}\nn
%
%
%		Second operator
%
%
&\,\,\times\prod_{a}\left[\prod_{b,\alpha}\left(\Phi_{ab,\alpha}^{\otimes n_{ab,\alpha}^{(2)}}\right)_{J_{ab,\alpha}^{(2)}}^{I_{ab,\alpha}^{(2)}}\right]
%quark
\otimes
\left[\prod_\beta \left(Q_{a,\beta}^{\otimes n_{a,\beta}^{(2)}}\right)^{I_{a,\beta}^{(2)}}_{\pmb s_{a,\beta}^{(2)}}\right]
\otimes
%antiquark
\left[\prod_\gamma\left(\bar Q_{a,\gamma}^{\otimes\bar n_{ {a,\gamma}}^{(2)}}\right)_{\bar J_{ { a,\gamma}}^{(2)}}^{ {\pmb{\bar s}_{a,\gamma}^{(2)}}}\right]%\nonumber\\[5mm]
%permutation
%&\qquad\qquad\qquad\times
\left(\sigma_a^{(2)}\right)^{\times_{b,\alpha}J_{ba,\alpha}^{(2)}\times_\gamma\bar J_{ a,\gamma}^{(2)}}
_{\times_{b,\alpha}I_{ab,\alpha}^{(2)}\times_\beta I_{ {a,\beta}}^{(2)}}\nn
&=
\prod_{a}\left[\prod_{b,\alpha}\left(\Phi_{ab,\alpha}^{\otimes \left(n_{ab,\alpha}^{(1)}+n_{ab,\alpha}^{(2)}\right)}\right)_{J_{ab,\alpha}^{(1)}\times J_{ab,\alpha}^{(2)}}^{I_{ab,\alpha}^{(1)}\times I_{ab,\alpha}^{(2)}}\right]%\nn
%quark
%&\quad\otimes
\left[\prod_\beta \left(Q_{a,\beta}^{\otimes \left(n_{a,\beta}^{(1)}+n_{a,\beta}^{(2)}\right)}\right)^{I_{a,\beta}^{(1)}\times I_{a,\beta}^{(2)}}_{\pmb s_{a,\beta}^{(1)}\times \pmb s_{a,\beta}^{(2)}}\right]\nn
&\qquad\qquad\otimes
%antiquark
\left[\prod_\gamma\left(\bar Q_{a,\gamma}^{\otimes\left(\bar n_{ {a,\gamma}}^{(1)}+\bar n_{ {a,\gamma}}^{(1)}\right)}\right)_{\bar J_{ { a,\gamma}}^{(1)}\times \bar J_{ { a,\gamma}}^{(2)}}^{ {\pmb{\bar s}_{a,\gamma}^{(1)}}\times\pmb{\bar s}_{a,\gamma}^{(2)}}\right]%\nonumber\\[5mm]
%permutation
%&\qquad\qquad\qquad\times
\,\,\left(\sigma_a^{(1)}\times \sigma_a^{(2)}\right)^{\times_{b,\alpha}J_{ba,\alpha}^{(1)}\times_\gamma\bar J_{ a,\gamma}^{(1)}\times_{b,\alpha}J_{ba,\alpha}^{(2)}\times_\gamma\bar J_{ a,\gamma}^{(2)}}
_{\times_{b,\alpha}I_{ab,\alpha}^{(1)}\times_\beta I_{ {a,\beta}}^{(1)}\times_{b,\alpha}I_{ab,\alpha}^{(2)}\times_\beta I_{ {a,\beta}}^{(2)}}
\end{align}
In the following we will continue to use the shorthand notation
\begin{equation}
\begin{array}{ll}
|e_{i_1},e_{i_2},...,e_{i_n}\rangle =|I\rangle \,,\qquad\qquad &I=(i_1,i_2,...,i_n)\,,\nn
\langle e^{j_1},e^{j_2},...,e^{j_n}| =\langle J| \,,\qquad &J=(j_1,j_2,...,j_n)

\end{array}
\end{equation}
which was already introduced in the previous sections.
For each gauge node $a$, let us now define the \(\lambda_{a-}\) and \(\lambda_{a+}\) permutations such that
\begin{align}\label{lamb_{a-}}
\lambda_{a-}
\left|\times_{b,\alpha}I_{ab,\alpha}^{(1)}\times_\beta I_{ {a,\beta}}^{(1)}\times_{b,\alpha}I_{ab,\alpha}^{(2)}\times_\beta I_{ {a,\beta}}^{(2)}
\right\rangle
=
\left|
\times_{b,\alpha}\left(I_{ab,\alpha}^{(1)}\times I_{ab,\alpha}^{(2)}\right)\times_\beta \left(I_{ {a,\beta}}^{(1)} \times I_{ {a,\beta}}^{(2)}\right)
\right\rangle
\end{align}
and 
\begin{align}\label{lamb_{a+}}
\lambda_{a+}^{-1}
\left|\times_{b,\alpha}J_{ba,\alpha}^{(1)}\times_\gamma\bar J_{ a,\gamma}^{(1)}\times_{b,\alpha}J_{ba,\alpha}^{(2)}\times_\gamma\bar J_{ a,\gamma}^{(2)}
\right\rangle
=
\left|
\times_{b,\alpha}\left(J_{ba,\alpha}^{(1)}\times J_{ba,\alpha}^{(2)}\right)\times_\gamma\left(\bar J_{ a,\gamma}^{(1)}\times \bar J_{ a,\gamma}^{(2)}\right)
\right\rangle
\end{align}
These permutations have been chosen such that, when suitably acting on the $\sigma_a^{(1)}\times \sigma_a^{(2)}$ component of \eqref{OP product in CR}, the resulting term has the right index structure to match the index structure of the associated field component,
\begin{align}\label{app: field index structure}
\left[\prod_{b,\alpha}\Phi_{ab,\alpha}^{\otimes \left(n_{ab,\alpha}^{(1)}+n_{ab,\alpha}^{(2)}\right)}\right]%\nn
%quark
\left[\prod_\beta Q_{a,\beta}^{\otimes \left(n_{a,\beta}^{(1)}+n_{a,\beta}^{(2)}\right)}\right]
%antiquark
\left[\prod_\gamma\bar Q_{a,\gamma}^{\otimes\left(\bar n_{ {a,\gamma}}^{(1)}+\bar n_{ {a,\gamma}}^{(1)}\right)}\right]
_{\times_{b,\alpha}\left(J_{ba,\alpha}^{(1)}\times J_{ba,\alpha}^{(2)}\right)\times_\gamma\left(\bar J_{ a,\gamma}^{(1)}\times \bar J_{ a,\gamma}^{(2)}\right)}
^{\times_{b,\alpha}\left(I_{ab,\alpha}^{(1)}\times I_{ab,\alpha}^{(2)}\right)\times_\beta \left(I_{ {a,\beta}}^{(1)} \times I_{ {a,\beta}}^{(2)}\right)}
\end{align}
We have in fact
\begin{align}\label{lambdas}
&\left(\sigma_a^{(1)}\times \sigma_a^{(2)}\right)^{\times_{b,\alpha}J_{ba,\alpha}^{(1)}\times_\gamma\bar J_{ a,\gamma}^{(1)}\times_{b,\alpha}J_{ba,\alpha}^{(2)}\times_\gamma\bar J_{ a,\gamma}^{(2)}}
_{\times_{b,\alpha}I_{ab,\alpha}^{(1)}\times_\beta I_{ {a,\beta}}^{(1)}\times_{b,\alpha}I_{ab,\alpha}^{(2)}\times_\beta I_{ {a,\beta}}^{(2)}}
&=
\left(\lambda_{a+}^{-1}\left(\sigma_a^{(1)}\times \sigma_a^{(2)}\right)\lambda_{a-}^{-1}\right)^{\times_{b,\alpha}\left(J_{ba,\alpha}^{(1)}\times J_{ba,\alpha}^{(2)}\right)\times_\gamma\left(\bar J_{ a,\gamma}^{(1)}\times \bar J_{ a,\gamma}^{(2)}\right)}
_{\times_{b,\alpha}\left(I_{ab,\alpha}^{(1)}\times I_{ab,\alpha}^{(2)}\right)\times_\beta \left(I_{ {a,\beta}}^{(1)} \times I_{ {a,\beta}}^{(2)}\right)}
\end{align}
The purpose of \(\lambda_{a-}\) and \(\lambda_{a+}\) is therefore to change the embedding into $[n_a]$ corresponding to the ordering of the upper (lower) $U(N_a)$ indices of the fields coming into (departing from) node $a$, eq. \eqref{upper indices embedding} (eq. \eqref{lower indices embedding}).
It can be seen that the index structure of the RHS of \eqref{lambdas} now matches the one in \eqref{app: field index structure}.
Inserting \eqref{lambdas} into \eqref{OP product in CR}, we then obtain
\begin{align}\label{CRST perm basis prod}
\cO_\Q(\vec n_1,&\vec s^{\,(1)},\vec \sigma^{(1)})\,\cO_\Q(\vec n_2,\vec s^{\,(2)},\vec \sigma^{(2)})=
\cO_\Q(\vec n_{1+2},\vec s^{\,(1)}\cup\vec s^{\,(2)},\vec\lambda_+^{\,-1} \left(\vec \sigma^{(1)}\times\vec \sigma^{(2)}\right) \vec \lambda_-^{\,-1})
\end{align}
where
\begin{align}
&\vec n_{1+2}=\cup_{a}\left\{\cup_{b,\alpha}\,\{n_{ab,\alpha}^{(1)},n_{ab,\alpha}^{(2)}\};\cup_\beta\, \{n_{a,\beta}^{(1)},n_{a,\beta}^{(2)}\};\cup_\gamma\, \{\bar n_{a,\gamma}^{(1)},\bar n_{a,\gamma}^{(2)}\}\right\}\,,\qquad\nn
%
%&\vec s^{\,(1)}\cup\vec s^{\,(2)}=\cup_a\{\cup_\beta \pmb s^{(1)}_{a,\beta},\cup_\beta\pmb s^{(2)}_{a,\beta}\,;\,\cup_\gamma \pmb {\bar s}^{(1)}_{a,\gamma},\cup_\gamma\pmb {\bar s}^{(2)}_{a,\gamma}\}\nn <- OLD VERSION
&\vec s^{\,(1)}\cup\vec s^{\,(2)}=\cup_a\{\cup_\beta\{ \pmb s^{(1)}_{a,\beta},\,\pmb s^{(2)}_{a,\beta}\}\,;\,\cup_\gamma\{ \pmb {\bar s}^{(1)}_{a,\gamma},\,\pmb {\bar s}^{(2)}_{a,\gamma}\}\}\,,\displaybreak[0]\\[3mm]
&\vec\lambda_+^{\,-1} \left(\vec \sigma^{(1)}\times\vec \sigma^{(2)}\right) \vec \lambda_-^{\,-1}=\cup_a \{\lambda_{a+}^{-1}\left(\sigma_a^{(1)}\times \sigma_a^{(2)} \right)\lambda_{a-}^{\,-1}\}\nonumber
\end{align}

\subsubsection*{2) Using the inversion formula}
In this step we are going to use eq. \eqref{CRST perm basis prod} to write a first expression for the \(G_{\pmb L^{(1)},\,\pmb L^{(2)},\,\pmb L^{(3)}}\) coefficients.
Let us start form the product \(\cO_\Q(\pmb L^{(1)})\,\cO_\Q(\pmb L^{(2)})\), that we expand as
\begin{align}\label{CR prod ops}
\cO_\Q&(\pmb L^{(1)})\,\cO_\Q(\pmb L^{(2)})\nn
&=\sum_{\vec s^{\,(1)},\vec s^{\,(2)}}\sum_{\vec \sigma^{(1)},\vec \sigma^{(2)}}\,\chi_Q(\pmb L^{(1)},\vec s^{\,(1)},\vec \sigma^{(1)})\,\chi_Q(\pmb L^{(2)},\vec s^{\,(2)},\vec \sigma^{(2)})\, \cO_\Q(\vec n_1,\vec s^{\,(1)},\vec \sigma^{(1)})\,\cO_\Q(\vec n_2,\vec s^{\,(2)},\vec \sigma^{(2)})
\end{align}
Plugging eq. \eqref{CRST perm basis prod} into this equation we get
\begin{align}\label{CRSC prod perm basis}
\cO_\Q(\pmb L^{(1)})\,\cO_\Q(\pmb L^{(2)})
=\sum_{\vec s^{\,(1)},\vec s^{\,(2)}}\sum_{\vec \sigma^{(1)},\vec \sigma^{(2)}}\,\chi_Q&(\pmb L^{(1)},\vec s^{\,(1)},\vec \sigma^{(1)})\,\chi_Q(\pmb L^{(2)},\vec s^{\,(2)},\vec \sigma^{(2)})\,\nn
&\times\cO_\Q(\vec n_{1+2},\vec s^{\,(1)}\cup\vec s^{\,(2)},\vec\lambda_+^{\,-1} \left(\vec \sigma^{(1)}\times\vec \sigma^{(2)}\right) \vec \lambda_-^{\,-1})
\end{align}
We now use the inversion formula \eqref{inverse fourier} to get
\begin{align}
\cO_\Q(\pmb L^{(1)})&\,\cO_\Q(\pmb L^{(2)})
=\sum_{\pmb L^{(3)}}\left\{\sum_{\vec s^{\,(1)},\vec s^{\,(2)}}\sum_{\vec \sigma^{(1)},\vec \sigma^{(2)}}\,\chi_Q(\pmb L^{(1)},\vec s^{\,(1)},\vec \sigma^{(1)})\chi_Q(\pmb L^{(2)},\vec s^{\,(2)},\vec \sigma^{(2)})\right.\nn
&\qquad\qquad\qquad\qquad\qquad\qquad
\left.\vphantom{\sum_{\vec n}}\times \chi_\Q(\pmb L^{(3)},\vec s^{\,(1)}\cup\vec s^{\,(2)},\vec\lambda_+^{\,-1} \left(\vec \sigma^{(1)}\times\vec \sigma^{(2)}\right) \vec \lambda_-^{\,-1})\right\}\,\cO_\Q(\pmb L^{(3)})
\end{align}
from which we obtain an expression for \(G_{\pmb L^{(1)},\,\pmb L^{(2)},\,\pmb L^{(3)}}\):
\begin{align}\label{first stcs}
&G_{\pmb L^{(1)},\,\pmb L^{(2)},\,\pmb L^{(3)}}\nn
&=\sum_{\vec s^{\,(1)},\vec s^{\,(2)}}\sum_{\vec \sigma^{\,(1)},\vec \sigma^{\,(2)}}\,\chi_Q(\pmb L^{(1)},\vec s^{\,(1)},\vec \sigma^{\,(1)})\chi_Q(\pmb L^{(2)},\vec s^{\,(2)},\vec \sigma^{\,(2)})\chi_\Q(\pmb L^{(3)},\vec s^{\,(1)}\cup\vec s^{\,(2)},\vec\lambda_+^{\,-1} \left(\vec \sigma^{\,(1)}\times\vec \sigma^{\,(2)}\right) \vec \lambda_-^{\,-1})\nn
&=
c_{\pmb L^{(1)}}\,c_{\pmb L^{(2)}}\,c_{\pmb L^{(3)}}\,\sum_{\vec s^{\,(1)},\vec s^{\,(2)}}\sum_{\vec \sigma^{(1)},\vec \sigma^{(2)}}\,\prod_a\,
D^{R_a^{(1)}}_{i_a^{(1)},j_a^{(1)}}(\sigma_a^{(1)})
D^{R_a^{(2)}}_{i_a^{(2)},j_a^{(2)}}(\sigma_a^{(2)})
D^{R_a^{(3)}}_{i_a^{(3)},j_a^{(3)}}(\lambda_{a+}^{-1}\left(\sigma_a^{(1)}\times \sigma_a^{(2)} \right)\lambda_{a-}^{\,-1})\nn
&\quad\times\left(\prod_{p=1}^3 B
^{R_a^{(p)}\rightarrow\cup_{b,\alpha}r_{ab,\alpha}^{(p)}\cup_\beta r_{a,\beta}^{(p)};\nu_a^{-(p)}}
_{j_a^{(p)}\rightarrow \cup_{b,\alpha}l_{ab,\alpha}^{(p)}\cup_\beta l_{a,\beta}^{(p)}}
B
^{R_a^{(p)}\rightarrow\cup_{b,\alpha}r_{ba,\alpha}^{(p)}\cup_\gamma\bar r_{a,\gamma}^{(p)};\nu_a^{+(p)}}
_{i_a^{(p)}\rightarrow \cup_{b,\alpha}l_{ba,\alpha}^{(p)}\cup_\gamma\bar l_{a,\gamma}^{(p)}}
\right)\\[3mm]
&\quad\times\left(\prod_\beta C_{ \pmb s_{a,\beta}^{(1)}}^{r_{a,\beta}^{(1)},S_{a,\beta}^{(1)},l_{a,\beta}^{(1)}}\,C_{ \pmb s_{a,\beta}^{(2)}}^{r_{a,\beta}^{(2)},S_{a,\beta}^{(2)},l_{a,\beta}^{(2)}}\,C_{ \pmb s_{a,\beta}^{(1)}\cup\pmb s_{a,\beta}^{(2)}}^{r_{a,\beta}^{(3)},S_{a,\beta}^{(3)},l_{a,\beta}^{(3)}}\right)%\nn
%
%&\qquad\qquad\times
\left(\prod_\gamma C^{\pmb{\bar s}_{a,\gamma}^{(1)}}_{\bar r_{a,\gamma}^{(1)},\bar S_{a,\gamma}^{(1)},\bar l_{a,\gamma}^{(1)}}\,
C^{\pmb{\bar s}_{a,\gamma}^{(2)}}_{\bar r_{a,\gamma}^{(2)},\bar S_{a,\gamma}^{(2)},\bar l_{a,\gamma}^{(2)}}\,
C^{\pmb{\bar s}_{a,\gamma}^{(1)}\cup\pmb{\bar s}_{a,\gamma}^{(2)}}_{\bar r_{a,\gamma}^{(3)},\bar S_{a,\gamma}^{(3)},\bar l_{a,\gamma}^{(3)}}\right)\nonumber
\end{align}

\subsubsection*{3) Fusing of gauge edges}
\begin{figure}[H]
\begin{center}\includegraphics[scale=0.9]{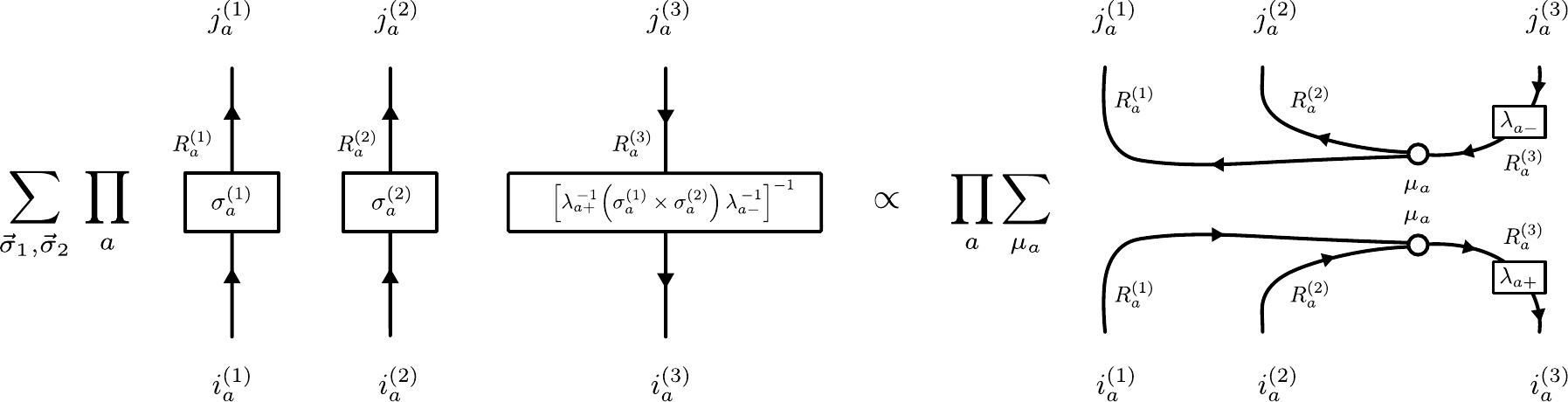}
\end{center}
\caption{Diagrammatic interpretation of eq. \eqref{ddd sigma}.}\label{fig: App_Chiral_Rings_recombining}
\end{figure}

At this stage the chiral ring structure constants are given as a product of three definite quantities, \emph{i.e.} three quiver characters. We now proceed to fuse together their gauge edges, by using standard representation theory identities.
Let us then focus on the permutation dependent piece of eq. \eqref{first stcs}, namely
\begin{align}
\sum_{\vec \sigma^{(1)},\vec \sigma^{(2)}}&\,\prod_a\,
D^{R_a^{(1)}}_{i_a^{(1)},j_a^{(1)}}(\sigma_a^{(1)})
D^{R_a^{(2)}}_{i_a^{(2)},j_a^{(2)}}(\sigma_a^{(2)})
D^{R_a^{(3)}}_{i_a^{(3)},j_a^{(3)}}(\lambda_{a+}^{-1}\left(\sigma_a^{(1)}\times \sigma_a^{(2)} \right)\lambda_{a-}^{\,-1})\nn
&=\sum_{\vec \sigma^{(1)},\vec \sigma^{(2)}}\,\prod_a\,
D^{R_a^{(1)}}_{i_a^{(1)},j_a^{(1)}}(\sigma_a^{(1)})
D^{R_a^{(2)}}_{i_a^{(2)},j_a^{(2)}}(\sigma_a^{(2)})
D^{R_a^{(3)}}_{i_a^{(3)},h_a^{(3)}}(\lambda_{a+}^{-1})
D^{R_a^{(3)}}_{h_a^{(3)},g_a^{(3)}}(\sigma_a^{(1)}\times \sigma_a^{(2)})
D^{R_a^{(3)}}_{g_a^{(3)},j_a^{(3)}}(\lambda_{a-}^{\,-1})
\end{align}
Using the identity
\begin{align}
D^R_{ij}(\sigma^{(1)}\times\sigma^{(2)})=\sum_{r_1,r_2,\,\mu}B_{i\rightarrow l_1,l_2}^{R\rightarrow r_1,r_2;\mu}\,
B^{R\rightarrow r_1,r_2;\mu}_{j\rightarrow k_1,k_2}
D^{r_1}_{l_1,k_1}(\sigma^{(1)})\,D^{r_2}_{l_2,k_2}(\sigma^{(2)})
\end{align}
we can write
\begin{align}\label{ddd sigma}
\sum_{\vec \sigma^{(1)},\vec \sigma^{(2)}}&\,\prod_a\,
D^{R_a^{(1)}}_{i_a^{(1)},j_a^{(1)}}(\sigma_a^{(1)})
D^{R_a^{(2)}}_{i_a^{(2)},j_a^{(2)}}(\sigma_a^{(2)})
D^{R_a^{(3)}}_{i_a^{(3)},h_a^{(3)}}(\lambda_{a+}^{-1})
D^{R_a^{(3)}}_{h_a^{(3)},g_a^{(3)}}(\sigma_a^{(1)}\times \sigma_a^{(2)})
D^{R_a^{(3)}}_{g_a^{(3)},j_a^{(3)}}(\lambda_{a-}^{\,-1})
\nn
&=
\sum_{\vec \sigma^{(1)},\vec \sigma^{(2)}}\,\prod_a\,
D^{R_a^{(1)}}_{i_a^{(1)},j_a^{(1)}}(\sigma_a^{(1)})
D^{R_a^{(2)}}_{i_a^{(2)},j_a^{(2)}}(\sigma_a^{(2)})\,
D^{R_a^{(3)}}_{i_a^{(3)},h_a^{(3)}}(\lambda_{a+}^{-1})
D^{R_a^{(3)}}_{g_a^{(3)},j_a^{(3)}}(\lambda_{a-}^{\,-1})
\nn
&\times 
\left(
\sum_{S_a^{(1)},S_a^{(2)},\,\mu_a}
B^{R_a^{(3)}\rightarrow S_a^{(1)},S_a^{(2)};\mu_a}_{h_a^{(3)}\rightarrow l_a^{(1)},l_{a}^{(2)}}\,
B^{R_a^{(3)}\rightarrow S_a^{(1)},S_a^{(2)};\mu_a}_{g_a^{(3)}\rightarrow k_a^{(1)},k_{a}^{(2)}}\,\,
D^{S_a^{(1)}}_{l_a^{(1)},k_a^{(1)}}(\sigma_a^{(1)})
D^{S_a^{(2)}}_{l_a^{(2)},k_a^{(2)}}(\sigma_a^{(2)})
\right)\nn
&=
\prod_a\,\frac{n_a^{(1)}!\,n_{a}^{(2)}!}{d(R_a^{(1)})\,d(R_a^{(2)})}\,
D^{R_a^{(3)}}_{i_a^{(3)},h_a^{(3)}}(\lambda_{a+}^{-1})
D^{R_a^{(3)}}_{g_a^{(3)},j_a^{(3)}}(\lambda_{a-}^{\,-1})\\[3mm]
&\qquad \times\sum_{S_a^{(1)},S_a^{(2)},\,\mu_a}
B^{R_a^{(3)}\rightarrow S_a^{(1)},S_a^{(2)};\mu_a}_{h_a^{(3)}\rightarrow l_a^{(1)},l_{a}^{(2)}}\,
B^{R_a^{(3)}\rightarrow S_a^{(1)},S_a^{(2)};\mu_a}_{g_a^{(3)}\rightarrow k_a^{(1)},k_{a}^{(2)}}
\prod_{q=1}^2\delta_{R_a^{(q)},S_{a}^{(q)}}%\,\delta_{R_a^{(2)},S_{a}^{(2)}}\,
\,\delta_{i_a^{(q)},l_{a}^{(q)}}%\,\delta_{i_a^{(2)},l_{a}^{(2)}}\,
\,\delta_{j_a^{(q)},k_{a}^{(q)}}%\,\delta_{j_a^{(2)},k_{a}^{(2)}}\,
\nn
&=
\prod_a\,\frac{n_a^{(1)}!\,n_{a}^{(2)}!}{d(R_a^{(1)})\,d(R_a^{(2)})}\nn
&\qquad\qquad\times
\sum_{\mu_a}
\left(D^{R_a^{(3)}}_{i_a^{(3)},h_a^{(3)}}(\lambda_{a+}^{-1})\,B^{R_a^{(3)}\rightarrow R_a^{(1)},R_a^{(2)};\mu_a}_{h_a^{(3)}\rightarrow i_a^{(1)},i_{a}^{(2)}}\right)
\left(D^{R_a^{(3)}}_{j_a^{(3)},g_a^{(3)}}(\lambda_{a-})B^{R_a^{(3)}\rightarrow R_a^{(1)},R_a^{(2)};\mu_a}_{g_a^{(3)}\rightarrow j_a^{(1)},j_{a}^{(2)}}\right)\nonumber
\end{align}
where in the second equality we used
\begin{align}
\sum_{\sigma\in S_n}D^R_{ij}(\sigma)D^S_{kl}(\sigma)=\frac{n!}{d(R)}\delta_{R,S}\,\delta_{i,k}\,\delta_{j,l}
\end{align}
It is important to stress that all the steps that we will be describing in this appendix can be also interpreted diagrammatically. For example, \eqref{ddd sigma} can be understood trough the diagram in Fig. \ref{fig: App_Chiral_Rings_recombining}.
Similar pictures can be drawn for all the following steps. In equation \eqref{ddd sigma} (or equivalently, in Fig. \ref{fig: App_Chiral_Rings_recombining}) we see the emergence of the first of the selection rules already anticipated in section \ref{sec: Holomorphic Gauge Invariant Operator Ring Structure Constants}. This selection rule is expressed by the terms
\begin{align}
B^{R_a^{(3)}\rightarrow R_a^{(1)},R_a^{(2)};\mu_a}_{h_a^{(3)}\rightarrow i_a^{(1)},i_{a}^{(2)}}\,,\qquad
B^{R_a^{(3)}\rightarrow R_a^{(1)},R_a^{(2)};\mu_a}_{g_a^{(3)}\rightarrow j_a^{(1)},j_{a}^{(2)}}
\end{align}
These coefficients are non-zero only if the restriction of the $S_{n_1+n_2}$ representation $R^{(3)}_a$ to $S_{n_1}\times S_{n_2}$ contains the representation $R_a^{(1)}\otimes R_a^{(2)}$, $\forall\, a$.

\subsubsection*{4) Fusing of the quark/antiquark edges}
In this step we will perform the fusing of the edges corresponding to the fundamental/anti\-fundamental matter fields. %quarks and antiquarks.
This involves summing over the quark/antiquark states $\pmb{s}_{a,\beta}^{(1,2)}$ and $\pmb{\bar s}_{a,\gamma}^{(1,2)}$.
Let us then turn to the Clebsch-Gordan parts of equation \eqref{first stcs}, that is
\begin{align}\label{cg stcs}
&\sum_{\pmb s_{a,\beta}^{(1)},\,\pmb s_{a,\beta}^{(2)}}\, C_{ \pmb s_{a,\beta}^{(1)}}^{r_{a,\beta}^{(1)},S_{a,\beta}^{(1)},l_{a,\beta}^{(1)}}\,C_{ \pmb s_{a,\beta}^{(2)}}^{r_{a,\beta}^{(2)},S_{a,\beta}^{(2)},l_{a,\beta}^{(2)}}\,C_{ \pmb s_{a,\beta}^{(1)}\cup\pmb s_{a,\beta}^{(2)}}^{r_{a,\beta}^{(3)},S_{a,\beta}^{(3)},l_{a,\beta}^{(3)}}
\end{align}
and 
\begin{align}\label{cg g stcs}
\sum_{\pmb{\bar s}_{a,\gamma}^{(1)},\,\pmb{\bar s}_{a,\gamma}^{(2)}}\,C^{\pmb{\bar s}_{a,\gamma}^{(1)}}_{\bar r_{a,\gamma}^{(1)},\bar S_{a,\gamma}^{(1)},\bar l_{a,\gamma}^{(1)}}\,
C^{\pmb{\bar s}_{a,\gamma}^{(2)}}_{\bar r_{a,\gamma}^{(2)},\bar S_{a,\gamma}^{(2)},\bar l_{a,\gamma}^{(2)}}\,
C^{\pmb{\bar s}_{a,\gamma}^{(1)}\cup\pmb{\bar s}_{a,\gamma}^{(2)}}_{\bar r_{a,\gamma}^{(3)},\bar S_{a,\gamma}^{(3)},\bar l_{a,\gamma}^{(3)}}
\end{align}
Consider for example the former. Aiming at simplifying notation, we rewrite it here dropping the \(a,\beta\) labels:
\begin{align}
&\sum_{\pmb s^{(1)},\,\pmb s ^{(2)}}\, C_{ \pmb s ^{(1)}}^{r ^{(1)},S ^{(1)},l^{(1)}}\,C_{ \pmb s ^{(2)}}^{r ^{(2)},S ^{(2)},l ^{(2)}}\,C_{ \pmb s ^{(1)}\cup\pmb s ^{(2)}}^{r ^{(3)},S^{(3)},l^{(3)}}
\end{align}
We can expand this quantity as
\begin{align}\label{3pf matter part}
\sum_{\pmb s ^{(1)},\,\pmb s ^{(2)}}&\, C_{ \pmb s ^{(1)}}^{r ^{(1)},S ^{(1)},l ^{(1)}}\,C_{ \pmb s ^{(2)}}^{r ^{(2)},S ^{(2)},l ^{(2)}}\,C_{ \pmb s ^{(1)}\cup\pmb s ^{(2)}}^{r ^{(3)},S ^{(3)},l ^{(3)}}\nn
&
=\sum_{\pmb s ^{(1)},\,\pmb s ^{(2)}}\, 
\langle r ^{(1)},S ^{(1)},l ^{(1)}|\pmb s ^{(1)}\rangle \,
\langle r ^{(2)},S ^{(2)},l ^{(2)}|\pmb s ^{(2)}\rangle \,
 \langle r ^{(3)},S ^{(3)},l ^{(3)}|\pmb s ^{(1)}\cup\pmb s ^{(2)} \rangle \nn
&
 =\sum_{\pmb s ^{(1)},\,\pmb s ^{(2)}}\, 
\left(
\langle r ^{(1)},S ^{(1)},l ^{(1)}|\otimes 
\langle r ^{(2)},S ^{(2)},l ^{(2)}|
\right)\,
\left(
|\pmb s ^{(1)}\rangle\otimes |\pmb s ^{(2)}\rangle
\right)\,
 \langle r ^{(3)},S ^{(3)},l ^{(3)}|\pmb s ^{(1)}\cup\pmb s ^{(2)} \rangle \nn
 &=
\langle r ^{(1)},S ^{(1)},l ^{(1)}|\otimes 
\langle r ^{(2)},S ^{(2)},l ^{(2)}|
\,
\left(
\sum_{\pmb s ^{(1)},\,\pmb s ^{(2)}}\,
|\pmb s ^{(1)}\rangle\otimes |\pmb s ^{(2)}\rangle
\,
 \langle\pmb s ^{(1)}|\otimes\langle \pmb s ^{(2)}|\right) |r ^{(3)},S ^{(3)},l ^{(3)} \rangle\nn
 &=
\left(\langle r ^{(1)},S ^{(1)},l ^{(1)}|\otimes 
\langle r ^{(2)},S ^{(2)},l ^{(2)}|\right)
|r ^{(3)},S ^{(3)},l ^{(3)} \rangle\nn
&=
\langle \{r ^{(1)},r ^{(2)}\},\,
		\{S ^{(1)},S ^{(2)}\},\,
		\{l ^{(1)},l ^{(2)}\}
		|r ^{(3)},S ^{(3)},l ^{(3)} \rangle
\end{align}
Since the generic state \(|r,S,l\rangle\in V_r^{S_n}\otimes V_r^{U(F)}\) is by definition the tensor product \(|r,S,l\rangle=|r,S\rangle \otimes|r,l\rangle\), we may separately decompose the two states $|r ^{(3)},S ^{(3)},l ^{(3)} \rangle$ and $| \{r ^{(1)},r ^{(2)}\},\,\allowbreak\{S ^{(1)},S ^{(2)}\}, \,\{l ^{(1)},l ^{(2)}\}\rangle$ as follows.
We factorise the former according to the decomposition \eqref{dec branch}, which in this case reads
\begin{align}\label{symm branch app}
V^{S_{n ^{(3)}}}_{r ^{(3)}}=
\bigoplus_{u ^{(1)}\vdash n ^{(1)}}\,
\bigoplus_{u ^{(2)}\vdash n ^{(2)}}
\left(V^{S_{n ^{(1)}}}_{u ^{(1)}}
\otimes
V^{S_{n ^{(2)}}}_{u ^{(2)}}\right)
\otimes
V^{u ^{(1)},u ^{(2)}}_{r ^{(3)}}
\end{align}
We then write
\begin{align}\label{S dec app}
&\left|
r ^{(3)},S ^{(3)},l ^{(3)}
\right\rangle
=
\left|r ^{(3)},S ^{(3)}\right\rangle\otimes \left|r ^{(3)},l ^{(3)}\right\rangle\nn
&\qquad\quad=
\sum_{u ^{(1)},\,u ^{(2
)}}\,
\sum_{p ^{(1)},\,p ^{(2
)}}\,
\sum_{\nu }\,
B_{l ^{(3)}\rightarrow p ^{(1)},p ^{(2)}}^{r ^{(3)}\rightarrow u ^{(1)},u ^{(2)};\nu }
\left|r ^{(3)},S ^{(3)}\right\rangle\otimes
\left|\{u ^{(1)},u ^{(2)}\},\{p ^{(1)},p ^{(2)}\};\nu \right\rangle\nn
&\qquad\quad=
\sum_{u ^{(1)},\,u ^{(2
)}}\,
\sum_{p ^{(1)},\,p ^{(2
)}}\,
\sum_{\nu }\,
B_{l ^{(3)}\rightarrow p ^{(1)},p ^{(2)}}^{r ^{(3)}\rightarrow u ^{(1)},u ^{(2)};\nu }
\left|\{u ^{(1)},u ^{(2)},r ^{(3)}\},\,\{p ^{(1)},p ^{(2)}\},\,S ^{(3)}\,;\nu \right\rangle
\end{align}
For the latter we use instead the the unitary group decomposition \eqref{U(N) dec}, which in this case takes the explicit form
\begin{align}\label{U(F) dec app}
V_{r ^{(1)}}^{U(F )}
\otimes
V_{r ^{(2)}}^{U(F )}
=
\bigoplus_{u ^{(3)}\vdash n ^{(3)}}\,
V_{u ^{(3)}}^{U(F )}
\otimes
V_{u ^{(3)}}^{r ^{(1)},r ^{(2)}}
\,,\qquad\qquad
n ^{(3)}=n ^{(1)}+n ^{(2)}
\end{align}
We therefore have
\begin{align}\label{C dec app}
&\left|\{r ^{(1)},r ^{(2)}\},\,
\{S ^{(1)},S ^{(2)}\},\,
\{l ^{(1)},l ^{(2)}\}\right\rangle
=
\left|\{r ^{(1)},r ^{(2)}\},\,
\{S ^{(1)},S ^{(2)}\}\right\rangle
\otimes
\left|\{r ^{(1)},r ^{(2)}\},\,\{l ^{(1)},l ^{(2)}\}\right\rangle
\nn
&\qquad\qquad=
\sum_{u ^{(3)}}\,\sum_{P ^{(3)}}\,\sum_{\tilde\nu }
C^{u ^{(3)};\tilde\nu \rightarrow r ^{(1)},r ^{(2)}}_{
		P ^{(3)} \rightarrow S ^{(1)},S ^{(2)}}
\left|
u ^{(3)},\,P ^{(3)}\,;\tilde\nu 
\right\rangle\otimes
\left|\{r ^{(1)},r ^{(2)}\},\,\{l ^{(1)},l ^{(2)}\}\right\rangle\nn
&\qquad\qquad=
\sum_{u ^{(3)}}\,\sum_{P ^{(3)}}\,\sum_{\tilde\nu }
C^{u ^{(3)};\tilde\nu \rightarrow r ^{(1)},r ^{(2)}}_{
P ^{(3)} \rightarrow S ^{(1)},S ^{(2)}}
\left|
\{r ^{(1)},r ^{(2)},u ^{(3)}\},\,\{l ^{(1)},l ^{(2)}\}
,\,P ^{(3)}\,;\tilde\nu 
\right\rangle
\end{align}
The vector spaces $V^{u ^{(1)},u ^{(2)}}_{r ^{(3)}}$ in \eqref{symm branch app} and $V_{u ^{(3)}}^{r ^{(1)},r ^{(2)}}$ in \eqref{U(F) dec app} are both multiplicity vector spaces. We recall that $\text{dim}(V_{r^{(3)}}^{r^{(1)},r^{(2)}})=g(r^{(1)},r^{(2)};r^{(3)})$, where $g$ is the Littlewood-Richardson coefficient. Notice that both the states on the far RHSs of \eqref{S dec app} and \eqref{C dec app} live in the tensor space $\mathcal W $, where
\begin{align}
\mathcal W =
V^{S_{n ^{(1)}}}_{r ^{(1)}}
\otimes
V^{S_{n ^{(2)}}}_{r ^{(2)}}
\otimes
V_{r ^{(3)}}^{U(F )}
\otimes
V_{r ^{(3)}}^{r ^{(1)},r ^{(2)}}
\end{align}

\begin{figure}[H]
\begin{center}\includegraphics[scale=1.3]{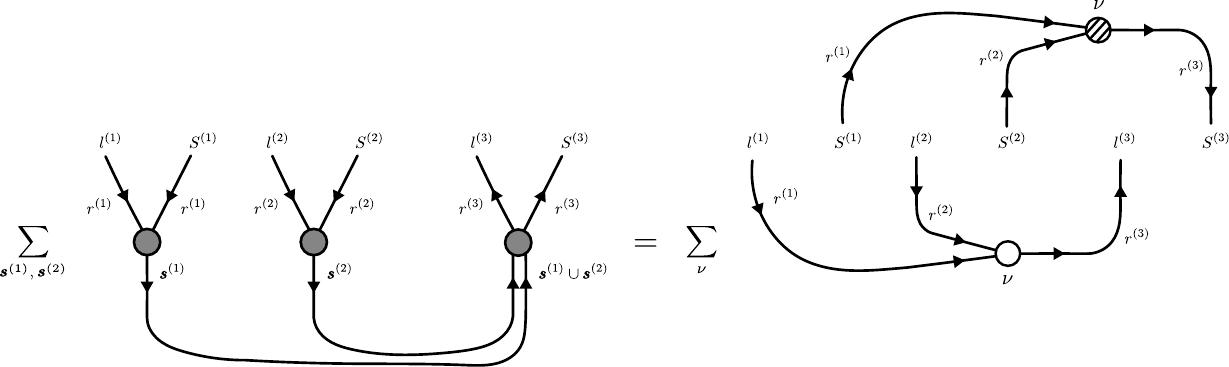}
\end{center}
\caption{Diagrammatic interpretation of eq. \eqref{app chiral ring clebsches rec}.}\label{fig: App_Chiral_Rings_Clebsches}
\end{figure}

Taking the scalar product of \eqref{S dec app} and \eqref{U(F) dec app} then gives
\begin{align}
\langle \{r ^{(1)},r ^{(2)}\},&\,
\{S ^{(1)},S ^{(2)}\},\,
\{l ^{(1)},l ^{(2)}\}
|r ^{(3)},S ^{(3)},l ^{(3)} \rangle\nn
&=
\left(\prod_{k=1}^3\,\sum_{u ^{(k)}}\right)\,
\left(\prod_{q=1}^2\,\sum_{p ^{(q)}}\right)\,
\sum_{P ^{(3)}}\,
\sum_{\nu ,\,\tilde\nu }\,\,
B_{l ^{(3)}\rightarrow p ^{(1)},p ^{(2)}}^{r ^{(3)}\rightarrow u ^{(1)},u ^{(2)};\nu }\,
C^{u ^{(3)};\tilde\nu \rightarrow r ^{(1)},r ^{(2)}}_{
		P ^{(3)} \rightarrow S ^{(1)},S ^{(2)}}\nn
&\qquad\qquad\times
\left(\prod_{k=1}^3\,\delta_{r ^{(k)},\,u ^{(k)}}\right)
\left(\prod_{q=1}^2\,\delta_{l ^{(q)},\,p ^{(q)}}\right)
\delta_{S ^{(3)},\,P ^{(3)}}\,
\delta_{\nu ,\,\tilde\nu }\nn
&=
\sum_{\nu }\,\,
B_{l ^{(3)}\rightarrow l ^{(1)},l ^{(2)}}^{r ^{(3)}\rightarrow r ^{(1)},r ^{(2)};\nu }\,
C^{r ^{(3)};\nu \rightarrow r ^{(1)},r ^{(2)}}_{
		S ^{(3)} \rightarrow S ^{(1)},S ^{(2)}}
\end{align}
We conclude that
\begin{align}\label{app chiral ring clebsches rec}
\sum_{\pmb s ^{(1)},\,\pmb s ^{(2)}}&\, C_{ \pmb s ^{(1)}}^{r ^{(1)},S ^{(1)},l ^{(1)}}\,C_{ \pmb s ^{(2)}}^{r ^{(2)},S ^{(2)},l ^{(2)}}\,C_{ \pmb s ^{(1)}\cup\pmb s ^{(2)}}^{r ^{(3)},S ^{(3)},l ^{(3)}}
=
\sum_{\nu }\,\,
B_{l ^{(3)}\rightarrow l ^{(1)},l ^{(2)}}^{r ^{(3)}\rightarrow r ^{(1)},r ^{(2)};\nu }\,
C^{r ^{(3)};\nu \rightarrow r ^{(1)},r ^{(2)}}_{
		S ^{(3)} \rightarrow S ^{(1)},S ^{(2)}}
\end{align}
The diagrammatic interpretation of eq. \eqref{app chiral ring clebsches rec} is drawn in Fig. \ref{fig: App_Chiral_Rings_Clebsches}.

Reintroducing the $a,\beta$ notation, we then obtain
\begin{align}\label{cg stcs 2}
\sum_{\pmb s_{a,\beta}^{(1)},\,\pmb s_{a,\beta}^{(2)}}\, C_{ \pmb s_{a,\beta}^{(1)}}^{r_{a,\beta}^{(1)},S_{a,\beta}^{(1)},l_{a,\beta}^{(1)}}\,C_{ \pmb s_{a,\beta}^{(2)}}^{r_{a,\beta}^{(2)},S_{a,\beta}^{(2)},l_{a,\beta}^{(2)}}\,C_{ \pmb s_{a,\beta}^{(1)}\cup\pmb s_{a,\beta}^{(2)}}^{r_{a,\beta}^{(3)},S_{a,\beta}^{(3)},l_{a,\beta}^{(3)}}=
\sum_{\nu_{a,\beta}}\,\,
B_{l_{a,\beta}^{(3)}\rightarrow l_{a,\beta}^{(1)},l_{a,\beta}^{(2)}}^{r_{a,\beta}^{(3)}\rightarrow r_{a,\beta}^{(1)},r_{a,\beta}^{(2)};\nu_{a,\beta}}\,
C^{r_{a,\beta}^{(3)};\nu_{a,\beta}\rightarrow r_{a,\beta}^{(1)},r_{a,\beta}^{(2)}}_{
		S_{a,\beta}^{(3)} \rightarrow S_{a,\beta}^{(1)},S_{a,\beta}^{(2)}}
\end{align}
Similarly, we can show that for \eqref{cg g stcs}
\begin{align}\label{cg g stcs 2}
\sum_{\pmb{\bar s}_{a,\gamma}^{(1)},\,\pmb{\bar s}_{a,\gamma}^{(2)}}\,C^{\pmb{\bar s}_{a,\gamma}^{(1)}}_{\bar r_{a,\gamma}^{(1)},\bar S_{a,\gamma}^{(1)},\bar l_{a,\gamma}^{(1)}}\,
C^{\pmb{\bar s}_{a,\gamma}^{(2)}}_{\bar r_{a,\gamma}^{(2)},\bar S_{a,\gamma}^{(2)},\bar l_{a,\gamma}^{(2)}}\,
C^{\pmb{\bar s}_{a,\gamma}^{(1)}\cup\pmb{\bar s}_{a,\gamma}^{(2)}}_{\bar r_{a,\gamma}^{(3)},\bar S_{a,\gamma}^{(3)},\bar l_{a,\gamma}^{(3)}}=
\sum_{\bar \nu_{a,\gamma}}\,\,
B_{\bar l_{a,\gamma}^{(3)}\rightarrow \bar l_{a,\gamma}^{(1)},\bar l_{a,\gamma}^{(2)}}^{\bar r_{a,\gamma}^{(3)}\rightarrow \bar r_{a,\gamma}^{(1)},\bar r_{a,\gamma}^{(2)};\bar \nu_{a,\gamma}}\,
C^{\bar r_{a,\gamma}^{(3)};\bar \nu_{a,\gamma}\rightarrow \bar r_{a,\gamma}^{(1)},\bar r_{a,\gamma}^{(2)}}_{\bar S_{a,\gamma}^{(3)}\rightarrow\bar S_{a,\gamma}^{(1)},\bar S_{a,\gamma}^{(2)}}
\end{align}
From eq. \eqref{cg stcs 2} and \eqref{cg g stcs 2} (or equivalently by considering Fig. \ref{fig: App_Chiral_Rings_Clebsches}) one can see the manifestation of another selection rule for the holomorphic GIO ring structure constants. In particular, the coefficients $B_{l_{a,\beta}^{(3)}\rightarrow l_{a,\beta}^{(1)},l_{a,\beta}^{(2)}}^{r_{a,\beta}^{(3)}\rightarrow r_{a,\beta}^{(1)},r_{a,\beta}^{(2)};\nu_{a,\beta}}$ are identically zero if the restriction of the $S_{n_{a,\beta}^{(1)}+n_{a,\beta}^{(2)}}$ representation $r_{a,\beta}^{(3)}$ to $S_{n_{a,\beta}^{(1)}}\times S_{n_{a,\beta}^{(2)}}$ does not contain the representation $r_{a,\beta}^{(1)}\otimes r_{a,\beta}^{(2)}$. A similar condition holds for the coefficients $B_{\bar l_{a,\gamma}^{(3)}\rightarrow \bar l_{a,\gamma}^{(1)},\bar l_{a,\gamma}^{(2)}}^{\bar r_{a,\gamma}^{(3)}\rightarrow \bar r_{a,\gamma}^{(1)},\bar r_{a,\gamma}^{(2)};\bar \nu_{a,\gamma}}$.

Inserting eqs. \eqref{ddd sigma}, \eqref{cg stcs 2} and \eqref{cg g stcs 2} into \eqref{first stcs} we then get
\begin{align}\label{second stcs}
G_{\pmb L^{(1)},\,\pmb L^{(2)},\,\pmb L^{(3)}}&=
c_{\pmb L^{(1)}}\,c_{\pmb L^{(2)}}\,c_{\pmb L^{(3)}}\,\prod_a\,
\frac{n_a^{(1)}!\,n_{a}^{(2)}!}{d(R_a^{(1)})\,d(R_a^{(2)})}\nn
&\qquad\qquad\times
\sum_{\mu_a}
\left(D^{R_a^{(3)}}_{i_a^{(3)},h_a^{(3)}}(\lambda_{a+}^{-1})\,B^{R_a^{(3)}\rightarrow R_a^{(1)},R_a^{(2)};\mu_a}_{h_a^{(3)}\rightarrow i_a^{(1)},i_{a}^{(2)}}\right)
\left(D^{R_a^{(3)}}_{j_a^{(3)},g_a^{(3)}}(\lambda_{a-})B^{R_a^{(3)}\rightarrow R_a^{(1)},R_a^{(2)};\mu_a}_{g_a^{(3)}\rightarrow j_a^{(1)},j_{a}^{(2)}}\right)\displaybreak[0]\nn
&\qquad\qquad\times\left(\prod_{p=1}^3 B
^{R_a^{(p)}\rightarrow\cup_{b,\alpha}r_{ab,\alpha}^{(p)}\cup_\beta r_{a,\beta}^{(p)};\nu_a^{-(p)}}
_{j_a^{(p)}\rightarrow \cup_{b,\alpha}l_{ab,\alpha}^{(p)}\cup_\beta l_{a,\beta}^{(p)}}
B
^{R_a^{(p)}\rightarrow\cup_{b,\alpha}r_{ba,\alpha}^{(p)}\cup_\gamma\bar r_{a,\gamma}^{(p)};\nu_a^{+(p)}}
_{i_a^{(p)}\rightarrow \cup_{b,\alpha}l_{ba,\alpha}^{(p)}\cup_\gamma\bar l_{a,\gamma}^{(p)}}
\right)\displaybreak[0]\nn
&\qquad\qquad\times\left(\prod_\beta \sum_{\nu_{a,\beta}}\,\,
B_{l_{a,\beta}^{(3)}\rightarrow l_{a,\beta}^{(1)},l_{a,\beta}^{(2)}}^{r_{a,\beta}^{(3)}\rightarrow r_{a,\beta}^{(1)},r_{a,\beta}^{(2)};\nu_{a,\beta}}\,
C^{r_{a,\beta}^{(3)};\nu_{a,\beta}\rightarrow r_{a,\beta}^{(1)},r_{a,\beta}^{(2)}}_{
		S_{a,\beta}^{(3)} \rightarrow S_{a,\beta}^{(1)},S_{a,\beta}^{(2)}} \right)\displaybreak[0]\nn
&\qquad\qquad\times\left(\prod_\gamma \sum_{\bar \nu_{a,\gamma}}\,\,
B_{\bar l_{a,\gamma}^{(3)}\rightarrow \bar l_{a,\gamma}^{(1)},\bar l_{a,\gamma}^{(2)}}^{\bar r_{a,\gamma}^{(3)}\rightarrow \bar r_{a,\gamma}^{(1)},\bar r_{a,\gamma}^{(2)};\bar \nu_{a,\gamma}}\,
C^{\bar r_{a,\gamma}^{(3)};\bar \nu_{a,\gamma}\rightarrow \bar r_{a,\gamma}^{(1)},\bar r_{a,\gamma}^{(2)}}_{\bar S_{a,\gamma}^{(3)}\rightarrow\bar S_{a,\gamma}^{(1)},\bar S_{a,\gamma}^{(2)}} \right)
\end{align}

\subsubsection*{5) Fusing the bi-fundamental edges and factorising the $\pm$ nodes }
The two tasks of this last step are to fuse the edges corresponding to the bi-fundamental fields and to factorise the positive and negative node of the split-node quiver. 
We start by considering the product 
\begin{align}\label{prod b}
D^{R_a^{(3)}}_{j_a^{(3)},g_a^{(3)}}(\lambda_{a-})\,B^{R_a^{(3)}\rightarrow R_a^{(1)},R_a^{(2)};\mu_a}_{g_a^{(3)}\rightarrow j_a^{(1)},j_{a}^{(2)}}
\left(\prod_{p=1}^3 B
^{R_a^{(p)}\rightarrow\cup_{b,\alpha}r_{ab,\alpha}^{(p)}\cup_\beta r_{a,\beta}^{(p)};\nu_a^{-(p)}}
_{j_a^{(p)}\rightarrow \cup_{b,\alpha}l_{ab,\alpha}^{(p)}\cup_\beta l_{a,\beta}^{(p)}}\right)
\end{align}
which appears in eq. \eqref{second stcs}. We want to decompose this term into a product of branching coefficients of the form \(B^{r_{ab,\alpha}^{(3)}\rightarrow r_{ab,\alpha}^{(1)},r_{ab,\alpha}^{(2)} }_{l_{ab,\alpha}^{(3)}\rightarrow l_{ab,\alpha}^{(1)},l_{ab,\alpha}^{(2)}}\).\\
First we notice that the equivariance property of the branching coefficients% (eq. \eqref{B a D})
\begin{align}\label{B a D 2}
D^R_{k,j}(\times_a\gamma_a)\,B_{j\rightarrow\cup_a l_a}^{R\rightarrow \cup_a r_a;\nu_a}
=\left(\prod_a\,D_{l_a',l_a}^{r_a}(\gamma_a)\right)\,B_{k\rightarrow \cup_{a} l_a'}^{R\rightarrow\cup_a r_a;\nu_a}
\end{align}
also implies
\begin{align}
B_{i\rightarrow\cup_a l_a}^{R\rightarrow \cup_a r_a;\nu_a}
=D^R_{i,k}(\times_a\gamma_a)\left(\prod_a\,D_{l_a,l_a'}^{r_a}(\gamma_a)\right)\,\,B_{k\rightarrow \cup_{a} l_a'}^{R\rightarrow\cup_a r_a;\nu_a}
\end{align}
for a collection of permutations \(\cup_a\{\gamma_a\in S_{n_a}\}\), where each $r_a$ is a partition of the integer \(n_a\). We can use this identity to write \eqref{prod b} as
\begin{align}\label{B a D 3}
D^{R_a^{(3)}}_{j_a^{(3)},g_a^{(3)}}&(\lambda_{a-})\,B^{R_a^{(3)}\rightarrow R_a^{(1)},R_a^{(2)};\mu_a}_{g_a^{(3)}\rightarrow j_a^{(1)},j_{a}^{(2)}}
\left(\prod_{p=1}^3 B
^{R_a^{(p)}\rightarrow\cup_{b,\alpha}r_{ab,\alpha}^{(p)}\cup_\beta r_{a,\beta}^{(p)};\nu_a^{-(p)}}
_{j_a^{(p)}\rightarrow \cup_{b,\alpha}l_{ab,\alpha}^{(p)}\cup_\beta l_{a,\beta}^{(p)}}\right)\nn
&= \left(
D^{R_a^{(1)}}_{j_a^{(1)},k_a^{(1)}}(\times_{b,\alpha}\eta_{ab,\alpha}^{(1)}\times {1})\,
D^{R_a^{(2)}}_{j_a^{(2)},k_a^{(2)}}(\times_{b,\alpha}\eta_{ab,\alpha}^{(2)}\times {1})\,
\right.\nn
&\qquad\times\left. D^{R_a^{(3)}}_{j_a^{(3)},k_a^{(3)}}\left(\lambda_{a-}\left(\times_{b,\alpha}\eta_{ab,\alpha}^{(1)}\times{1}\times_{b,\alpha}\eta_{ab,\alpha}^{(2)}\times {1}\right)\right)\,\,
B_{k_a^{(3)}\rightarrow k_a^{(1)},k_a^{(2)}}^{R_a^{(3)}\rightarrow R_a^{(1)},R_a^{(2)};\mu_a}
\right)\nn
&\quad\quad\qquad\times \left(\prod_{p=1}^3 B
^{R_a^{(p)}\rightarrow\cup_{b,\alpha}r_{ab,\alpha}^{(p)}\cup_\beta r_{a,\beta}^{(p)};\nu_a^{-(p)}}
_{j_a^{(p)}\rightarrow \cup_{b,\alpha}l_{ab,\alpha}^{(p)}\cup_\beta l_{a,\beta}^{(p)}}\right)
\end{align}
where \((\times_{b,\alpha}\eta_{ab,\alpha}^{(p)}\times {1})\in S_{n_{a}^{(p)}}\) and \(\eta_{ab,\alpha}^{(p)}\in S_{n_{ab,\alpha}^{(p)}}\), for \(p=1,2\).

Let us now go back to the equation defining the \(\lambda_{a-}\) permutations, \eqref{lamb_{a-}}.
It is easy to see that
\begin{align}
\lambda_{a-}\left(\times_{b,\alpha}\eta_{ab,\alpha}^{(1)}\times_\beta\rho_{a,\beta}^{(1)}\times_{b,\alpha}\eta_{ab,\alpha}^{(2)}\times_\beta\rho_{a,\beta}^{(2)}\right)=
\left[\times_{b,\alpha}\left(\eta^{(1)}_{ab,\alpha}\times \eta^{(2)}_{ab,\alpha}\right)
\times_{\beta}\left(\rho^{(1)}_{a,\beta}\times \rho^{(2)}_{a,\beta}\right)\right]
\lambda_{a-}
\end{align}
We can use this identity in \eqref{B a D 3} to get
\begin{align}\label{B a D 4}
D^{R_a^{(3)}}_{j_a^{(3)},g_a^{(3)}}&(\lambda_{a-})\,B^{R_a^{(3)}\rightarrow R_a^{(1)},R_a^{(2)};\mu_a}_{g_a^{(3)}\rightarrow j_a^{(1)},j_{a}^{(2)}}
\left(\prod_{p=1}^3 B
^{R_a^{(p)}\rightarrow\cup_{b,\alpha}r_{ab,\alpha}^{(p)}\cup_\beta r_{a,\beta}^{(p)};\nu_a^{-(p)}}
_{j_a^{(p)}\rightarrow \cup_{b,\alpha}l_{ab,\alpha}^{(p)}\cup_\beta l_{a,\beta}^{(p)}}\right)\nn
&= \left(
D^{R_a^{(1)}}_{j_a^{(1)},k_a^{(1)}}(\times_{b,\alpha}\eta_{ab,\alpha}^{(1)}\times {1})\,
D^{R_a^{(2)}}_{j_a^{(2)},k_a^{(2)}}(\times_{b,\alpha}\eta_{ab,\alpha}^{(2)}\times {1})\,
\right.\nn
&\qquad\times\left. D^{R_a^{(3)}}_{j_a^{(3)},k_a^{(3)}}\left(\left(\times_{b,\alpha}\left(\eta^{(1)}_{ab,\alpha}\times \eta^{(2)}_{ab,\alpha}\right)
\times{1}\right)\,
\lambda_{a-}\right)\,\,
B_{k_a^{(3)}\rightarrow k_a^{(1)},k_a^{(2)}}^{R_a^{(3)}\rightarrow R_a^{(1)},R_a^{(2)};\mu_a}
\right)\nn
&\quad\quad\qquad\times \left(\prod_{p=1}^3 B
^{R_a^{(p)}\rightarrow\cup_{b,\alpha}r_{ab,\alpha}^{(p)}\cup_\beta r_{a,\beta}^{(p)};\nu_a^{-(p)}}
_{j_a^{(p)}\rightarrow \cup_{b,\alpha}l_{ab,\alpha}^{(p)}\cup_\beta l_{a,\beta}^{(p)}}\right)
\end{align}
Next we use the identity \eqref{B a D 2} in eq. \eqref{B a D 4} as follows, for $p=1,2$:
\begin{align}
D^{R_a^{(p)}}_{j_a^{(p)},k_a^{(p)}}&(\times_{b,\alpha}\eta_{ab,\alpha}^{(p)}\times {1})\,
B
^{R_a^{(p)}\rightarrow\cup_{b,\alpha}r_{ab,\alpha}^{(p)}\cup_\beta r_{a,\beta}^{(p)};\nu_a^{-(p)}}
_{j_a^{(p)}\rightarrow \cup_{b,\alpha}l_{ab,\alpha}^{(p)}\cup_\beta l_{a,\beta}^{(p)}}\nn
&=\left(
\prod_{b,\alpha}
D^{r_{ab,\alpha}^{(p)}}_{l_{ab,\alpha}^{(p)},q_{ab,\alpha}^{(p)}}
\left(\eta_{ab,\alpha}^{(p)}\right)
\right)
\,
\left(
\prod_\beta \delta_{l_{a,\beta}^{(p)},q_{a,\beta}^{(p)}}
\right)
B_{k_a^{(p)}\rightarrow \cup_{b,\alpha}q_{ab,\alpha}^{(p)}\cup_{\beta}q_{a,\beta}^{(p)}}^{R_a^{(p)}\rightarrow\cup_{b,\alpha}r_{ab,\alpha}^{(p)}\cup_\beta r_{a,\beta}^{(p)};\nu_a^{-(p)}}\nn
&=
\left(
\prod_{b,\alpha}
D^{r_{ab,\alpha}^{(p)}}_{l_{ab,\alpha}^{(p)},q_{ab,\alpha}^{(p)}}
\left(\eta_{ab,\alpha}^{(p)}\right)
\right)
\,
B_{k_a^{(p)}\rightarrow \cup_{b,\alpha}q_{ab,\alpha}^{(p)}\cup_{\beta}l_{a,\beta}^{(p)}}^{R_a^{(p)}\rightarrow\cup_{b,\alpha}r_{ab,\alpha}^{(p)}\cup_\beta r_{a,\beta}^{(p)};\nu_a^{-(p)}}
\end{align}
Similarly, we use \eqref{B a D 2} also for the term
\begin{align}
&D^{R_a^{(3)}}_{j_a^{(3)},k_a^{(3)}}\left(\left(\times_{b,\alpha}\left(\eta^{(1)}_{ab,\alpha}\times \eta^{(2)}_{ab,\alpha}\right)
\times{1}\right)\,
\lambda_{a-}\right)\,B
^{R_a^{(3)}\rightarrow\cup_{b,\alpha}r_{ab,\alpha}^{(3)}\cup_\beta r_{a,\beta}^{(3)};\nu_a^{-(3)}}
_{j_a^{(3)}\rightarrow \cup_{b,\alpha}l_{ab,\alpha}^{(3)}\cup_\beta l_{a,\beta}^{(3)}}\nn
&=
D^{R_a^{(3)}}_{g_a^{(3)},k_a^{(3)}}\left(\lambda_{a-}\right)
D^{R_a^{(3)}}_{j_a^{(3)},g_a^{(3)}}\left(\times_{b,\alpha}\left(\eta^{(1)}_{ab,\alpha}\times \eta^{(2)}_{ab,\alpha}\right)
\times{1}\right)\,
B
^{R_a^{(3)}\rightarrow\cup_{b,\alpha}r_{ab,\alpha}^{(3)}\cup_\beta r_{a,\beta}^{(3)};\nu_a^{-(3)}}
_{j_a^{(3)}\rightarrow \cup_{b,\alpha}l_{ab,\alpha}^{(3)}\cup_\beta l_{a,\beta}^{(3)}}\nn
&=
D^{R_a^{(3)}}_{g_a^{(3)},k_a^{(3)}}\left(\lambda_{a-}\right)
\left(
\prod_{b,\alpha}
D^{r_{ab,\alpha}^{(3)}}_{l_{ab,\alpha}^{(3)},q_{ab,\alpha}^{(3)}}
\left(\eta^{(1)}_{ab,\alpha}\times \eta^{(2)}_{ab,\alpha}\right)
\right)
\,
B_{g_a^{(3)}\rightarrow \cup_{b,\alpha}q_{ab,\alpha}^{(3)}\cup_{\beta}l_{a,\beta}^{(3)}}^{R_a^{(3)}\rightarrow\cup_{b,\alpha}r_{ab,\alpha}^{(3)}\cup_\beta r_{a,\beta}^{(3)};\nu_a^{-(3)}}
\end{align}
Putting these last equations together, we get to
\begin{align}
&D^{R_a^{(3)}}_{j_a^{(3)},g_a^{(3)}}(\lambda_{a-})\,B^{R_a^{(3)}\rightarrow R_a^{(1)},R_a^{(2)};\mu_a}_{g_a^{(3)}\rightarrow j_a^{(1)},j_{a}^{(2)}}
\left(\prod_{p=1}^3 B
^{R_a^{(p)}\rightarrow\cup_{b,\alpha}r_{ab,\alpha}^{(p)}\cup_\beta r_{a,\beta}^{(p)};\nu_a^{-(p)}}
_{j_a^{(p)}\rightarrow \cup_{b,\alpha}l_{ab,\alpha}^{(p)}\cup_\beta l_{a,\beta}^{(p)}}\right)\nn
&=
\left(
\prod_{b,\alpha}
D^{r_{ab,\alpha}^{(1)}}_{l_{ab,\alpha}^{(1)},q_{ab,\alpha}^{(1)}}
\left(\eta_{ab,\alpha}^{(1)}\right)
D^{r_{ab,\alpha}^{(2)}}_{l_{ab,\alpha}^{(2)},q_{ab,\alpha}^{(2)}}
\left(\eta_{ab,\alpha}^{(2)}\right)
D^{r_{ab,\alpha}^{(3)}}_{l_{ab,\alpha}^{(3)},q_{ab,\alpha}^{(3)}}
\left(\eta_{ab,\alpha}^{(1)}\times\eta_{ab,\alpha}^{(2)}\right)
\right)
B_{k_a^{(3)}\rightarrow k_a^{(1)},k_a^{(2)}}^{R_a^{(3)}\rightarrow R_a^{(1)},R_a^{(2)};\mu_a}\nn
&\qquad\times
\left[D^{R_a^{(3)}}_{g_a^{(3)},k_a^{(3)}}\left(\lambda_{a-}\right)
B_{g_a^{(3)}\rightarrow \cup_{b,\alpha}q_{ab,\alpha}^{(3)}\cup_{\beta}l_{a,\beta}^{(3)}}^{R_a^{(3)}\rightarrow\cup_{b,\alpha}r_{ab,\alpha}^{(3)}\cup_\beta r_{a,\beta}^{(3)};\nu_a^{-(3)}}\right]
\prod_{p=1}^2 B_{k_a^{(p)}\rightarrow \cup_{b,\alpha}q_{ab,\alpha}^{(p)}\cup_{\beta}l_{a,\beta}^{(p)}}^{R_a^{(p)}\rightarrow\cup_{b,\alpha}r_{ab,\alpha}^{(p)}\cup_\beta r_{a,\beta}^{(p)};\nu_a^{-(p)}}
\end{align}
Notice that the quantity on the LHS above is independent of the permutations \(\eta\). We can then sum over all possible permutations \(\eta\) on the RHS, provided we divide by the number of permutations themselves: we thus obtain
\begin{align}\label{cr bsim id}
&D^{R_a^{(3)}}_{j_a^{(3)},g_a^{(3)}}(\lambda_{a-})\,B^{R_a^{(3)}\rightarrow R_a^{(1)},R_a^{(2)};\mu_a}_{g_a^{(3)}\rightarrow j_a^{(1)},j_{a}^{(2)}}
\left(\prod_{p=1}^3 B
^{R_a^{(p)}\rightarrow\cup_{b,\alpha}r_{ab,\alpha}^{(p)}\cup_\beta r_{a,\beta}^{(p)};\nu_a^{-(p)}}
_{j_a^{(p)}\rightarrow \cup_{b,\alpha}l_{ab,\alpha}^{(p)}\cup_\beta l_{a,\beta}^{(p)}}\right)\nn
&=
\frac{1}{\prod_{b,\alpha }\,n_{ab,\alpha}^{(1)}!\,n_{ab,\alpha}^{(2)}!}
\left\{
\sum_{\cup_{b,\alpha}\{\eta_{b,\alpha}^{(1)},\,\eta_{ab,\alpha}^{(2)}\}}
\prod_{b,\alpha}
D^{r_{ab,\alpha}^{(1)}}_{l_{ab,\alpha}^{(1)},q_{ab,\alpha}^{(1)}}
\left(\eta_{ab,\alpha}^{(1)}\right)
D^{r_{ab,\alpha}^{(2)}}_{l_{ab,\alpha}^{(2)},q_{ab,\alpha}^{(2)}}
\left(\eta_{ab,\alpha}^{(2)}\right)\right.\nn
&\qquad\quad \times\left. D^{r_{ab,\alpha}^{(3)}}_{l_{ab,\alpha}^{(3)},q_{ab,\alpha}^{(3)}}
\left(\eta_{ab,\alpha}^{(1)}\times \eta_{ab,\alpha}^{(2)}\right)
\vphantom{\sum_{\cup_b\{\eta_b\}}}\right\}\,
B_{k_a^{(3)}\rightarrow k_a^{(1)},k_a^{(2)}}^{R_a^{(3)}\rightarrow R_a^{(1)},R_a^{(2)};\mu_a}\nn
&\times
\left[D^{R_a^{(3)}}_{g_a^{(3)},k_a^{(3)}}\left(\lambda_{a-}\right)
B_{g_a^{(3)}\rightarrow \cup_{b,\alpha}q_{ab,\alpha}^{(3)}\cup_{\beta}l_{a,\beta}^{(3)}}^{R_a^{(3)}\rightarrow\cup_{b,\alpha}r_{ab,\alpha}^{(3)}\cup_\beta r_{a,\beta}^{(3)};\nu_a^{-(3)}}\right]
\prod_{p=1}^2 B_{k_a^{(p)}\rightarrow \cup_{b,\alpha}q_{ab,\alpha}^{(p)}\cup_{\beta}l_{a,\beta}^{(p)}}^{R_a^{(p)}\rightarrow\cup_{b,\alpha}r_{ab,\alpha}^{(p)}\cup_\beta r_{a,\beta}^{(p)};\nu_a^{-(p)}}
\end{align}
The quantity inside the curvy brackets above has the same structure of the far LHS of eq. \eqref{ddd sigma}. Performing similar steps to the ones presented in that equation we obtain, dropping the $a,b,\alpha$ notation for improved clarity
\begin{align}
\sum_{\eta^{(1)},\,\eta ^{(2)}}
&D^{r ^{(1)}}_{l ^{(1)},q ^{(1)}}
\left(\eta ^{(1)}\right)
D^{r ^{(2)}}_{l ^{(2)},q ^{(2)}}
\left(\eta ^{(2)}\right)
D^{r ^{(3)}}_{l ^{(3)},q ^{(3)}}
\left(\eta ^{(1)}\times \eta ^{(2)}\right)
\vphantom{\sum_{\cup_b\{\eta_b\}}}\nn
&\qquad=
\frac{n ^{(1)}!\,n ^{(2)}!}{d(r ^{(1)})\,d(r ^{(2)})}\,\sum_{\nu }\,
B_{l ^{(3)}\rightarrow l ^{(1)},l ^{(2)}}^{r ^{(3)}\rightarrow r ^{(1)},r ^{(2)};\nu }\,\,
B_{q ^{(3)}\rightarrow q ^{(1)},q ^{(2)}}^{r ^{(3)}\rightarrow r ^{(1)},r ^{(2)};\nu }
\end{align}
Inserting this identity in \eqref{cr bsim id} we get
\begin{align}\label{cr -term pre}
&D^{R_a^{(3)}}_{j_a^{(3)},g_a^{(3)}}(\lambda_{a-})\,B^{R_a^{(3)}\rightarrow R_a^{(1)},R_a^{(2)};\mu_a}_{g_a^{(3)}\rightarrow j_a^{(1)},j_{a}^{(2)}}
\left(\prod_{p=1}^3 B
^{R_a^{(p)}\rightarrow\cup_{b,\alpha}r_{ab,\alpha}^{(p)}\cup_\beta r_{a,\beta}^{(p)};\nu_a^{-(p)}}
_{j_a^{(p)}\rightarrow \cup_{b,\alpha}l_{ab,\alpha}^{(p)}\cup_\beta l_{a,\beta}^{(p)}}\right)\nn
%
%&=
%\sum_{\cup_{b,\alpha}\{s_{ab,\alpha}^{(1)}\,s_{ab,\alpha}^{(2)}\}\atop \cup_{b,\alpha}\{\nu_{ab,\alpha}\}}
%\left\{\prod_{b,\alpha }\frac{1}{d(r_{ab,\alpha}^{(1)})\,d(r_{ab,\alpha}^{(2)})}\,\,
%
%B_{q_{ab,\alpha}^{(3)}\rightarrow h_{ab,\alpha}^{(1)},h_{ab,\alpha}^{(2)}}^{r_{ab,\alpha}^{(3)}\rightarrow s_{ab,\alpha}^{(1)},s_{ab,\alpha}^{(2)};\nu_{ab,\alpha}}\,\,
%
%B_{l_{ab,\alpha}^{(3)}\rightarrow t_{ab,\alpha}^{(1)},t_{ab,\alpha}^{(2)}}^{r_{ab,\alpha}^{(3)}\rightarrow s_{ab,\alpha}^{(1)},s_{ab,\alpha}^{(2)};\nu_{ab,\alpha}}\right.
%\nn
%
%&\times\left.\left(
%\prod_{w=1}^{2}
%\delta_{r_{ab,\alpha}^{(w)},s_{ab,\alpha}^{(w)}}\,
%\delta_{l_{ab,\alpha}^{(w)},t_{ab,\alpha}^{(w)}}\,
%\delta_{q_{ab,\alpha}^{(w)},h_{ab,\alpha}^{(w)}}
%\right)\right\}
%
%B_{k_a^{(3)}\rightarrow k_a^{(1)},k_a^{(2)}}^{R_a^{(3)}\rightarrow R_a^{(1)},R_a^{(2)};\mu_a}\,\nn
%
%
%&\times
%\left[D^{R_a^{(3)}}_{g_a^{(3)},k_a^{(3)}}\left(\lambda_{a-}^{-1}\right)
%B_{g_a^{(3)}\rightarrow \cup_{b,\alpha}q_{ab,\alpha}^{(3)}\cup_{\beta}l_{a,\beta}^{(3)}}^{R_a^{(3)}\rightarrow\cup_{b,\alpha}r_{ab,\alpha}^{(3)}\cup_\beta r_{a,\beta}^{(3)};\nu_a^{-(3)}}\right]
%
%\prod_{p=1}^2 B_{k_a^{(p)}\rightarrow \cup_{b,\alpha}q_{ab,\alpha}^{(p)}\cup_{\beta}l_{a,\beta}^{(p)}}^{R_a^{(p)}\rightarrow\cup_{b,\alpha}r_{ab,\alpha}^{(p)}\cup_\beta r_{a,\beta}^{(p)};\nu_a^{-(p)}}\nn
%
%
&=
\sum_{ \cup_{b,\alpha}\{\nu_{ab,\alpha}\}}
\left\{\prod_{b,\alpha }\frac{1}{d(r_{ab,\alpha}^{(1)})\,d(r_{ab,\alpha}^{(2)})}\,\,
B_{l_{ab,\alpha}^{(3)}\rightarrow l_{ab,\alpha}^{(1)},l_{ab,\alpha}^{(2)}}^{r_{ab,\alpha}^{(3)}\rightarrow r_{ab,\alpha}^{(1)},r_{ab,\alpha}^{(2)};\nu_{ab,\alpha}}\,\,
B_{q_{ab,\alpha}^{(3)}\rightarrow q_{ab,\alpha}^{(1)},q_{ab,\alpha}^{(2)}}^{r_{ab,\alpha}^{(3)}\rightarrow r_{ab,\alpha}^{(1)},r_{ab,\alpha}^{(2)};\nu_{ab,\alpha}}\right\}\,\,
B_{k_a^{(3)}\rightarrow k_a^{(1)},k_a^{(2)}}^{R_a^{(3)}\rightarrow R_a^{(1)},R_a^{(2)};\mu_a}
\nn
&\qquad\qquad\qquad\times
\left[D^{R_a^{(3)}}_{g_a^{(3)},k_a^{(3)}}\left(\lambda_{a-}\right)
B_{g_a^{(3)}\rightarrow \cup_{b,\alpha}q_{ab,\alpha}^{(3)}\cup_{\beta}l_{a,\beta}^{(3)}}^{R_a^{(3)}\rightarrow\cup_{b,\alpha}r_{ab,\alpha}^{(3)}\cup_\beta r_{a,\beta}^{(3)};\nu_a^{-(3)}}\right]\,\,
\prod_{p=1}^2 B_{k_a^{(p)}\rightarrow \cup_{b,\alpha}q_{ab,\alpha}^{(p)}\cup_{\beta}l_{a,\beta}^{(p)}}^{R_a^{(p)}\rightarrow\cup_{b,\alpha}r_{ab,\alpha}^{(p)}\cup_\beta r_{a,\beta}^{(p)};\nu_a^{-(p)}}
\end{align}
Using the substitutions $k_a^{(3)}\rightarrow t_a^{(3)}$ and $g_{a}^{(3)}\rightarrow k_a^{(3)}$ we can then write
\begin{align}\label{cr -term}
&D^{R_a^{(3)}}_{j_a^{(3)},g_a^{(3)}}(\lambda_{a-})\,B^{R_a^{(3)}\rightarrow R_a^{(1)},R_a^{(2)};\mu_a}_{g_a^{(3)}\rightarrow j_a^{(1)},j_{a}^{(2)}}
\left(\prod_{p=1}^3 B
^{R_a^{(p)}\rightarrow\cup_{b,\alpha}r_{ab,\alpha}^{(p)}\cup_\beta r_{a,\beta}^{(p)};\nu_a^{-(p)}}
_{j_a^{(p)}\rightarrow \cup_{b,\alpha}l_{ab,\alpha}^{(p)}\cup_\beta l_{a,\beta}^{(p)}}\right)\nn
&=
\sum_{ \cup_{b,\alpha}\{\nu_{ab,\alpha}\}}
\left\{\prod_{b,\alpha }\frac{1}{d(r_{ab,\alpha}^{(1)})\,d(r_{ab,\alpha}^{(2)})}\,\,
B_{l_{ab,\alpha}^{(3)}\rightarrow l_{ab,\alpha}^{(1)},l_{ab,\alpha}^{(2)}}^{r_{ab,\alpha}^{(3)}\rightarrow r_{ab,\alpha}^{(1)},r_{ab,\alpha}^{(2)};\nu_{ab,\alpha}}\,\,
B_{q_{ab,\alpha}^{(3)}\rightarrow q_{ab,\alpha}^{(1)},q_{ab,\alpha}^{(2)}}^{r_{ab,\alpha}^{(3)}\rightarrow r_{ab,\alpha}^{(1)},r_{ab,\alpha}^{(2)};\nu_{ab,\alpha}}\right\}
\nn
&\qquad\qquad\qquad\times
B_{t_a^{(3)}\rightarrow k_a^{(1)},k_a^{(2)}}^{R_a^{(3)}\rightarrow R_a^{(1)},R_a^{(2)};\mu_a}
\left[D^{R_a^{(3)}}_{k_a^{(3)},t_a^{(3)}}\left(\lambda_{a-}\right)
B_{k_a^{(3)}\rightarrow \cup_{b,\alpha}q_{ab,\alpha}^{(3)}\cup_{\beta}l_{a,\beta}^{(3)}}^{R_a^{(3)}\rightarrow\cup_{b,\alpha}r_{ab,\alpha}^{(3)}\cup_\beta r_{a,\beta}^{(3)};\nu_a^{-(3)}}\right]\nn
&\qquad\qquad\qquad\qquad\qquad\times
\prod_{p=1}^2 B_{k_a^{(p)}\rightarrow \cup_{b,\alpha}q_{ab,\alpha}^{(p)}\cup_{\beta}l_{a,\beta}^{(p)}}^{R_a^{(p)}\rightarrow\cup_{b,\alpha}r_{ab,\alpha}^{(p)}\cup_\beta r_{a,\beta}^{(p)};\nu_a^{-(p)}}\nn
&=
\sum_{ \cup_{b,\alpha}\{\nu_{ab,\alpha}\}}
\left\{\prod_{b,\alpha }\frac{1}{d(r_{ab,\alpha}^{(1)})\,d(r_{ab,\alpha}^{(2)})}\,\,
B_{l_{ab,\alpha}^{(3)}\rightarrow l_{ab,\alpha}^{(1)},l_{ab,\alpha}^{(2)}}^{r_{ab,\alpha}^{(3)}\rightarrow r_{ab,\alpha}^{(1)},r_{ab,\alpha}^{(2)};\nu_{ab,\alpha}}\,\,
B_{q_{ab,\alpha}^{(3)}\rightarrow q_{ab,\alpha}^{(1)},q_{ab,\alpha}^{(2)}}^{r_{ab,\alpha}^{(3)}\rightarrow r_{ab,\alpha}^{(1)},r_{ab,\alpha}^{(2)};\nu_{ab,\alpha}}\right\}
\nn
&\qquad\qquad\qquad\times
D^{R_a^{(3)}}_{k_a^{(3)},t_a^{(3)}}\left(\lambda_{a-}\right)\,\,
B_{t_a^{(3)}\rightarrow k_a^{(1)},k_a^{(2)}}^{R_a^{(3)}\rightarrow R_a^{(1)},R_a^{(2)};\mu_a}\,
\prod_{p=1}^3 B_{k_a^{(p)}\rightarrow \cup_{b,\alpha}q_{ab,\alpha}^{(p)}\cup_{\beta}l_{a,\beta}^{(p)}}^{R_a^{(p)}\rightarrow\cup_{b,\alpha}r_{ab,\alpha}^{(p)}\cup_\beta r_{a,\beta}^{(p)};\nu_a^{-(p)}}
\end{align}
We see here the manifestation of the last selection rule, enforced by the branching coefficients $B_{l_{ab,\alpha}^{(3)}\rightarrow l_{ab,\alpha}^{(1)},l_{ab,\alpha}^{(2)}}^{r_{ab,\alpha}^{(3)}\rightarrow r_{ab,\alpha}^{(1)},r_{ab,\alpha}^{(2)};\nu_{ab,\alpha}}$. These quantities are non zero only if the restriction of the $S_{n_{ab,\alpha}^{(1)}+n_{ab,\alpha}^{(2)}}$ representation $r_{ab,\alpha}^{(3)}$ to $S_{n_{ab,\alpha}^{(1)}}\times S_{n_{ab,\alpha}^{(2)}}$ contains the representation $r_{ab,\alpha}^{(1)}\otimes r_{ab,\alpha}^{(2)}$.

With the identity \eqref{cr -term} we have achieved a factorisation of the branching coefficients over all the nodes of the quiver. Moreover, the positive and negative node of every split-node $a$ are now disentangled. There are no symmetric group states $q_{ab,\alpha}^{(i)}$ ($i=1,2,3$), associated with the negative node of the split-node $a$, that mix with symmetric group states $l_{ab,\alpha}^{(i)}$ ($i=1,2,3$), associated with its positive node.

Plugging eq. \eqref{cr -term} into \eqref{second stcs}, we get
\begin{align}\label{third stcs}
G_{\pmb L^{(1)},\,\pmb L^{(2)},\,\pmb L^{(3)}}&=
c_{\pmb L^{(1)}}\,c_{\pmb L^{(2)}}\,c_{\pmb L^{(3)}}\,\prod_a\,
\frac{n_a^{(1)}!\,n_{a}^{(2)}!}{d(R_a^{(1)})\,d(R_a^{(2)})}\,\frac{1}{\prod_{b,\alpha }d(r_{ab,\alpha}^{(1)})\,d(r_{ab,\alpha}^{(2)})}\,\sum_{\mu_a}\nn
&\qquad\qquad\times
\left(
D^{R_a^{(3)}}_{k_a^{(3)},t_a^{(3)}}\left(\lambda_{a-}\right)\,\,
B_{t_a^{(3)}\rightarrow k_a^{(1)},k_a^{(2)}}^{R_a^{(3)}\rightarrow R_a^{(1)},R_a^{(2)};\mu_a}\,
\prod_{p=1}^3 B_{k_a^{(p)}\rightarrow \cup_{b,\alpha}q_{ab,\alpha}^{(p)}\cup_{\beta}l_{a,\beta}^{(p)}}^{R_a^{(p)}\rightarrow\cup_{b,\alpha}r_{ab,\alpha}^{(p)}\cup_\beta r_{a,\beta}^{(p)};\nu_a^{-(p)}}
\right)\nn
&\qquad\qquad\times
\left(
D^{R_a^{(3)}}_{i_a^{(3)},h_a^{(3)}}\left(\lambda_{a+}^{-1}\right)\,\,
B^{R_a^{(3)}\rightarrow R_a^{(1)},R_a^{(2)};\mu_a}_{h_a^{(3)}\rightarrow i_a^{(1)},i_{a}^{(2)}}\,
\prod_{p=1}^3 B
^{R_a^{(p)}\rightarrow\cup_{b,\alpha}r_{ba,\alpha}^{(p)}\cup_\gamma\bar r_{a,\gamma}^{(p)};\nu_a^{+(p)}}
_{i_a^{(p)}\rightarrow \cup_{b,\alpha}l_{ba,\alpha}^{(p)}\cup_\gamma\bar l_{a,\gamma}^{(p)}}
\right)\nn
&\qquad\qquad\times\left(
\prod_{b,\alpha}\,\sum_{\nu_{ab,\alpha}} B_{l_{ab,\alpha}^{(3)}\rightarrow l_{ab,\alpha}^{(1)},l_{ab,\alpha}^{(2)}}^{r_{ab,\alpha}^{(3)}\rightarrow r_{ab,\alpha}^{(1)},r_{ab,\alpha}^{(2)};\nu_{ab,\alpha}}\,\,
B_{q_{ab,\alpha}^{(3)}\rightarrow q_{ab,\alpha}^{(1)},q_{ab,\alpha}^{(2)}}^{r_{ab,\alpha}^{(3)}\rightarrow r_{ab,\alpha}^{(1)},r_{ab,\alpha}^{(2)};\nu_{ab,\alpha}} 
\right)\nn
&\qquad\qquad\times\left(\prod_\beta \sum_{\nu_{a,\beta}}\,\,
B_{l_{a,\beta}^{(3)}\rightarrow l_{a,\beta}^{(1)},l_{a,\beta}^{(2)}}^{r_{a,\beta}^{(3)}\rightarrow r_{a,\beta}^{(1)},r_{a,\beta}^{(2)};\nu_{a,\beta}}\,
C^{r_{a,\beta}^{(3)};\nu_{a,\beta}\rightarrow r_{a,\beta}^{(1)},r_{a,\beta}^{(2)}}_{
		S_{a,\beta}^{(3)} \rightarrow S_{a,\beta}^{(1)},S_{a,\beta}^{(2)}} \right)\nn
&\qquad\qquad\times\left(\prod_\gamma \sum_{\bar \nu_{a,\gamma}}\,\,
B_{\bar l_{a,\gamma}^{(3)}\rightarrow \bar l_{a,\gamma}^{(1)},\bar l_{a,\gamma}^{(2)}}^{\bar r_{a,\gamma}^{(3)}\rightarrow \bar r_{a,\gamma}^{(1)},\bar r_{a,\gamma}^{(2)};\bar \nu_{a,\gamma}}
C^{\bar r_{a,\gamma}^{(3)};\bar \nu_{a,\gamma}\rightarrow \bar r_{a,\gamma}^{(1)},\bar r_{a,\gamma}^{(2)}}_{\bar S_{a,\gamma}^{(3)}\rightarrow\bar S_{a,\gamma}^{(1)},\bar S_{a,\gamma}^{(2)}} \right)
\end{align}
The latter equation can be finally rewritten as
\begin{align}\label{eq for G}
&G_{\pmb L^{(1)},\,\pmb L^{(2)},\,\pmb L^{(3)}}=
c_{\pmb L^{(1)}}\,c_{\pmb L^{(2)}}\,c_{\pmb L^{(3)}}\,\prod_a\,
\frac{n_a^{(1)}!\,n_{a}^{(2)}!}{d(R_a^{(1)})\,d(R_a^{(2)})}\,\frac{1}{\prod_{b,\alpha }d(r_{ab,\alpha}^{(1)})\,d(r_{ab,\alpha}^{(2)})}\,\sum_{\mu_a}\left(\prod_{b,\alpha}\sum_{\nu_{ab,\alpha}}\right)\nn
&\qquad\times
\left[
D^{R_a^{(3)}}_{k_a^{(3)},t_a^{(3)}}\left(\lambda_{a-}\right)\,
B^{R_a^{(3)}\rightarrow R_a^{(1)},R_a^{(2)};\mu_a}_{t_a^{(3)}\rightarrow k_a^{(1)},k_{a}^{(2)}}\,
\left(\prod_{p=1}^3 B
^{R_a^{(p)}\rightarrow\cup_{b,\alpha}r_{ab,\alpha}^{(p)}\cup_\beta r_{a,\beta}^{(p)};\nu_a^{-(p)}}
_{k_a^{(p)}\rightarrow \cup_{b,\alpha}q_{ab,\alpha}^{(p)}\cup_\beta l_{a,\beta}^{(p)}}\right)
\right.\nn
&\qquad\qquad\times \left.
\left(
\prod_{b,\alpha} B_{q_{ab,\alpha}^{(3)}\rightarrow q_{ab,\alpha}^{(1)},q_{ab,\alpha}^{(2)}}^{r_{ab,\alpha}^{(3)}\rightarrow r_{ab,\alpha}^{(1)},r_{ab,\alpha}^{(2)};\nu_{ab,\alpha}} 
\right)\,\,
\left(\prod_\beta \sum_{\nu_{a,\beta}}\,\,
B_{l_{a,\beta}^{(3)}\rightarrow l_{a,\beta}^{(1)},l_{a,\beta}^{(2)}}^{r_{a,\beta}^{(3)}\rightarrow r_{a,\beta}^{(1)},r_{a,\beta}^{(2)};\nu_{a,\beta}}\,
C^{r_{a,\beta}^{(3)};\nu_{a,\beta}\rightarrow r_{a,\beta}^{(1)},r_{a,\beta}^{(2)}}_{
		S_{a,\beta}^{(3)} \rightarrow S_{a,\beta}^{(1)},S_{a,\beta}^{(2)}} \right)\right]
\nn
%
%
% END OF FIRST PIECE
%
%
&\qquad\times
\left[
D^{R_a^{(3)}}_{i_a^{(3)},h_a^{(3)}}\left(\lambda_{a+}^{-1}\right)\,
B^{R_a^{(3)}\rightarrow R_a^{(1)},R_a^{(2)};\mu_a}_{h_a^{(3)}\rightarrow i_a^{(1)},i_{a}^{(2)}}\,
\left(\prod_{p=1}^3 B^{R_a^{(p)}\rightarrow\cup_{b,\alpha}r_{ba,\alpha}^{(p)}\cup_\gamma\bar r_{a,\gamma}^{(p)};\nu_a^{+(p)}}
_{i_a^{(p)}\rightarrow \cup_{b,\alpha}l_{ba,\alpha}^{(p)}\cup_\gamma\bar l_{a,\gamma}^{(p)}}\right)
\right.\nn
&\qquad\qquad\times \left.
\left(
\prod_{b,\alpha} B_{l_{ba,\alpha}^{(3)}\rightarrow l_{ba,\alpha}^{(1)},l_{ba,\alpha}^{(2)}}^{r_{ba,\alpha}^{(3)}\rightarrow r_{ba,\alpha}^{(1)},r_{ba,\alpha}^{(2)};\nu_{ba,\alpha}}\right)\,\,
\left(\prod_\gamma \sum_{\bar \nu_{a,\gamma}}\,\,
B_{\bar l_{a,\gamma}^{(3)}\rightarrow \bar l_{a,\gamma}^{(1)},\bar l_{a,\gamma}^{(2)}}^{\bar r_{a,\gamma}^{(3)}\rightarrow \bar r_{a,\gamma}^{(1)},\bar r_{a,\gamma}^{(2)};\bar \nu_{a,\gamma}}
C^{\bar r_{a,\gamma}^{(3)};\bar \nu_{a,\gamma}\rightarrow \bar r_{a,\gamma}^{(1)},\bar r_{a,\gamma}^{(2)}}_{\bar S_{a,\gamma}^{(3)}\rightarrow\bar S_{a,\gamma}^{(1)},\bar S_{a,\gamma}^{(2)}} \right)
\right]
\end{align}
The last equation shows that, at each node $a$ in the quiver, the holomorphic GIO ring structure constant factorises into two components, one associated with the positive node and one associated with the negative node of the corresponding split node $a$. Figure \ref{fig: CRSC picture} shows a pictorial interpretation of this formula.

\subsection{Diagrammatic derivation for an $\N=2$ SQCD}\label{App_sub: CRSC derivation for N=2}

\begin{figure}[H]
\begin{center}\includegraphics[scale=1.45]{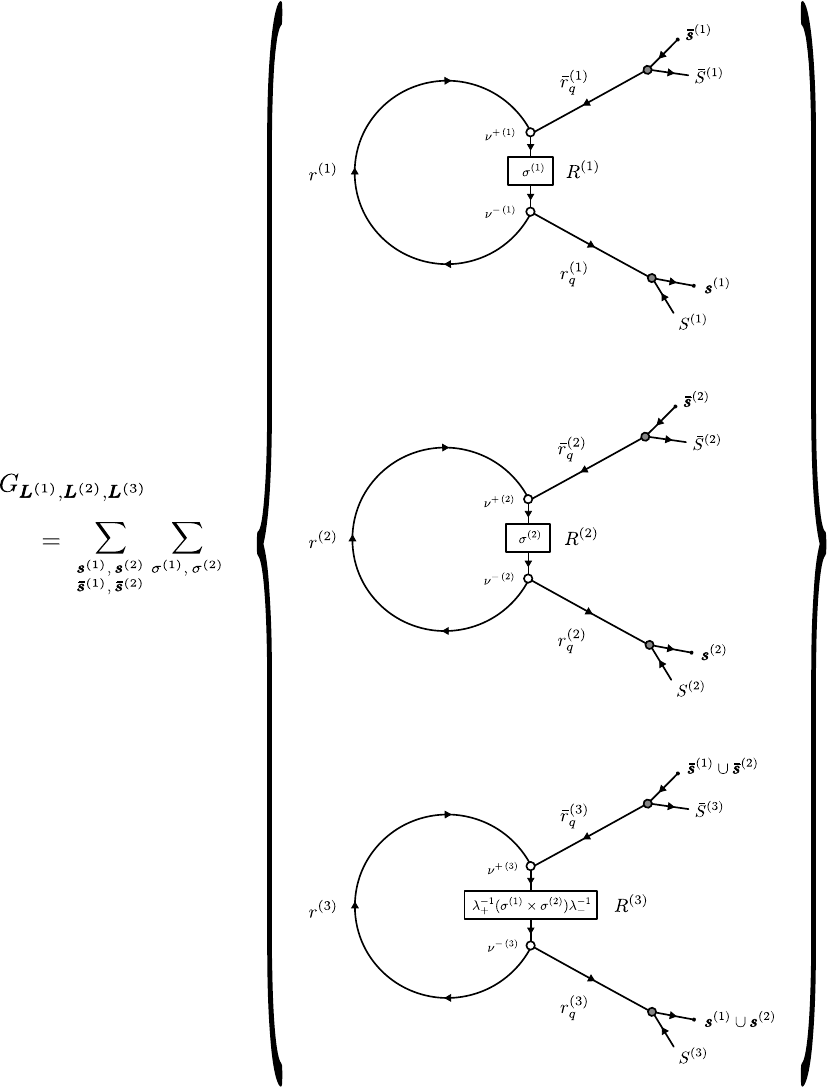}
\end{center}
\caption{Diagrammatic representation of the chiral ring structure constants for an $\N=2$ SQCD, corresponding to eq. \eqref{first stcs}.}\label{fig: CRSC Appendix example Step1}
\end{figure}

\begin{figure}[H]
\begin{center}\includegraphics[scale=1.6]{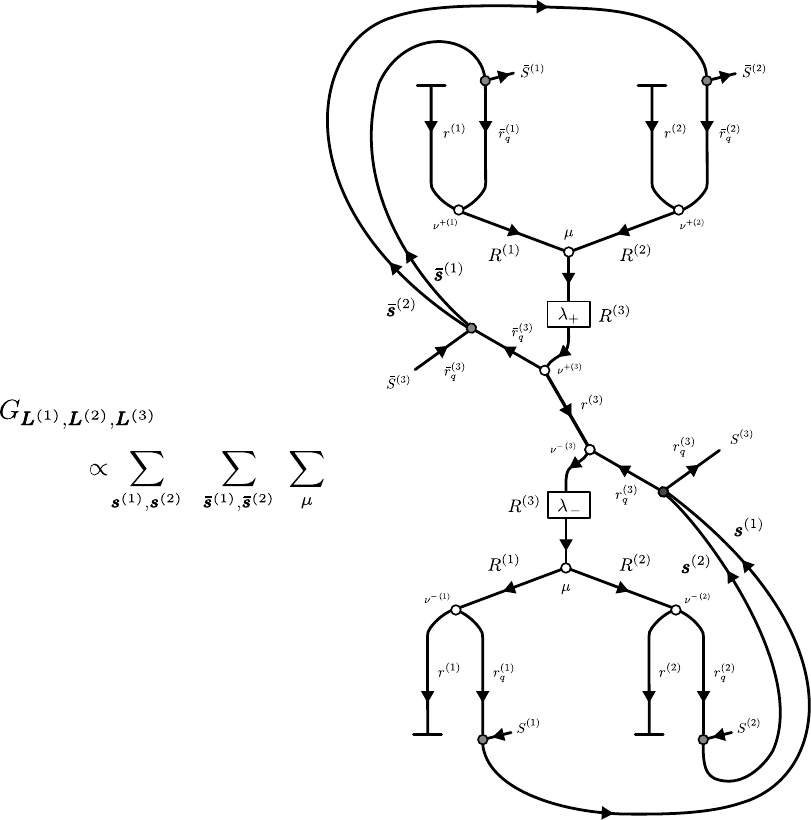}
\end{center}
\caption{The diagram for the chiral ring structure constants after using the identity \eqref{ddd sigma}. The horizontal bars are to be identified.}\label{fig: CRSC Appendix example Step2}
\end{figure}

We are now going to present a diagrammatic recap of this derivation, for the example of an $\N=2$ SQCD already discussed in section \ref{sec: Holomorphic Gauge Invariant Operator Ring Structure Constants}. Our starting point is \eqref{first stcs}, where each $\pmb L ^{(i)}$ has been simplified as in eq. \eqref{N=2 L set}. We can depict this quantity as in Fig. \ref{fig: CRSC Appendix example Step1}.
After using identity \eqref{ddd sigma}, which is represented in Fig. \ref{fig: App_Chiral_Rings_recombining}, the diagram is
transformed to the one in Fig. \ref{fig: CRSC Appendix example Step2}. We see that now the three disjoint diagrams of the previous Fig. \ref{fig: CRSC Appendix example Step1} are now joined into a single connected component.
Here we can see the relevance of the permutations $\lambda_-$ and $\lambda_+$, which were previously obtained 
in the explicit derivation. 
They allow the fusing of all the state indices of the three disjoint pieces of Fig. \ref{fig: CRSC Appendix example Step1}.
This can be understood by looking at Fig. \ref{fig: CRSC Appendix example Step2}. Let us follow the flow at the top of the diagram from $  r^{(1)} \otimes\bar r_q^{(1)} \otimes r^{(2)} \otimes \bar r_q^{(2)} $ to $R^{(3)}$. This corresponds to the embeddings 
\begin{align}\label{lambda appEx before}
S_{n^{(1)} } \times S_{n^{(1)}_q }  \times  S_{n^{(2)}} \times S_{ n_q^{(2)}}\rightarrow 
S_{n^{(1)}+n^{(1)}_q }\times S_{n^{(2)}+n_q^{(2)}}\rightarrow S_{n^{(1)}+n^{(2)}+n_q^{(1)}+n_q^{(2)}}
\end{align}
and 
\begin{align} \label{lambda appEx embedding before}
[n^{(1)}] \sqcup  [n^{(1)}_q ]  \sqcup [n^{(2 )}] \sqcup [n^{(2)}_q ] 
 \rightarrow  [n^{(1)}+n_q^{(1)}]\sqcup [n^{(2)}+n_q^{(2)}] \rightarrow [n^{(1)}+n_q^{(1)}+n^{(2)}+n_q^{(2)}]
\end{align} 
The second embedding corresponds to the branching coefficient labelled by $\mu$. 
In the branching \emph{after} the $\lambda_+$ permutation, \(R^{(3)}\) splits into \(r^{(3)}\) and \(r_q^{(3)}\). The relevant 
embedding is now 
\begin{align}\label{lambda appEx embedding after}
[n^{(1)}+n^{(2)}]\sqcup [n_q^{(1)}+n_q^{(2)}] \rightarrow [n^{(1)}+n_q^{(1)}+n^{(2)}+n_q^{(2)}]
\end{align}
which comes naturally from the construction of $ \cO ( \pmb L_3 ) $. 
The purpose of $\lambda_+$ is to allow the transition from \eqref{lambda appEx embedding before} to \eqref{lambda appEx embedding after}. A similar (but reversed) role is played by the permutation \(\lambda_-\).

\begin{figure}[H]
\begin{center}\includegraphics[scale=1.6]{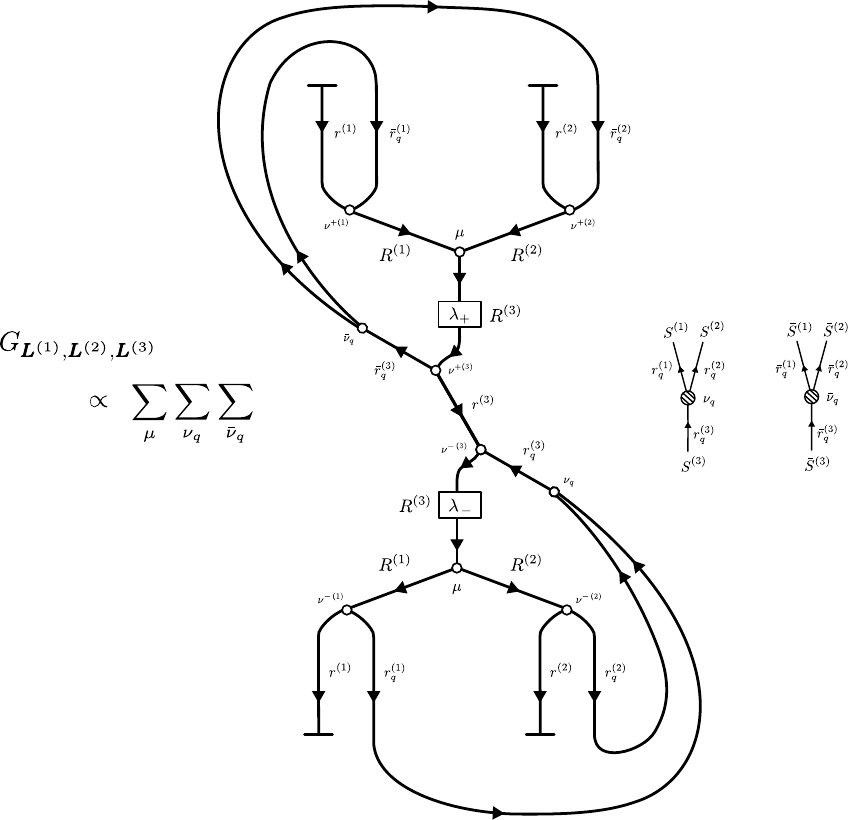}
\end{center}
\caption{The outcome of inserting the identity described by Fig. \ref{fig: App_Chiral_Rings_Clebsches} into Fig. \ref{fig: CRSC Appendix example Step2}. The horizontal bars are to be identified.}\label{fig: CRSC Appendix example Step3}
\end{figure}

\begin{figure}[H]
\begin{center}\includegraphics[scale=1.54]{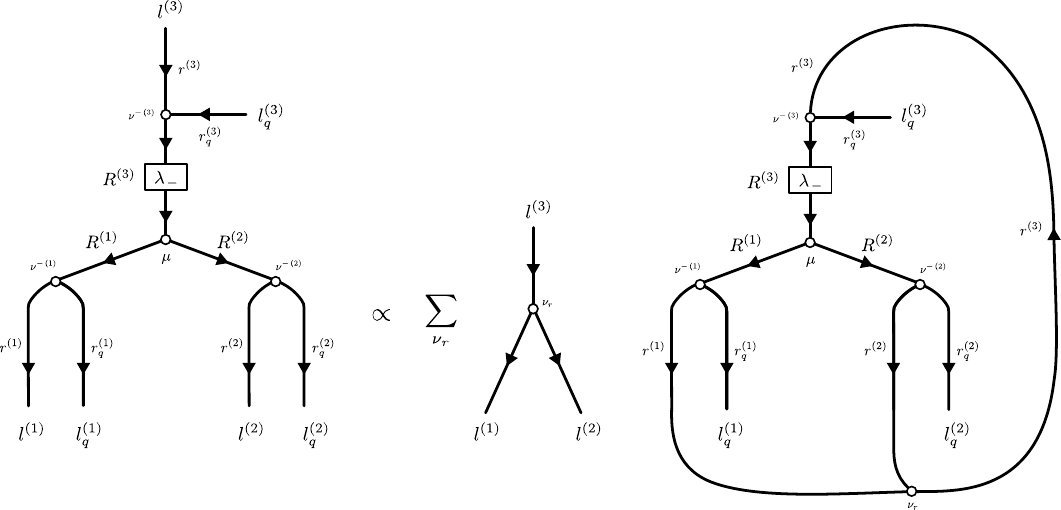}
\end{center}
\caption{Diagrammatic description of eq. \eqref{cr -term} for the $\N=2$ SQCD example.}\label{fig: CRSC Appendix example Step4}
\end{figure}

Now we use the relation in Fig. \ref{fig: App_Chiral_Rings_Clebsches} to separate the edges corresponding to the quark (and antiquark) fields from the rest of the diagram. We thus obtain Fig. \ref{fig: CRSC Appendix example Step3}.
The last step is to separate all the edges connected to the negative node of the split-node from all the edges connected to its positive node. As explained in the derivation above, this operation is achieved through the identity \eqref{cr -term}, which in this example takes the form depicted in Fig. \ref{fig: CRSC Appendix example Step4}.
Once this diagrammatic relation has been inserted into Fig. \ref{fig: CRSC Appendix example Step3}, we straightforwardly obtain the final diagram for the chiral ring structure constants for an $\N=2$ SQCD, depicted in Fig. \ref{fig: CRSC split node step4}.

\section{Quiver Restricted Schur Polynomials for an $\mathcal N=2$ SQCD: $\vec n=(2,2,2)$ Field Content}\label{app: SQCD example 222}

In this appendix we will summarise the main steps which led to the expression of the operators in \eqref{SQCD ex: 222 ops}. In particular we will derive all the fourteen different quiver characters, corresponding to the set of labels $\pmb L_i$ described in \eqref{sqcd ex 222 14 labels}, $i=1,2,...,14$. The operators \eqref{SQCD ex: 222 ops} are then readily obtained by using the definition \eqref{fourier2}. 

We start from $\cO(\pmb L_1)$ and $\cO(\pmb L_2)$. Their quiver characters can be immediately computed to be respectively
\begin{align}
&\chi(\pmb L_1,\vec s,\sigma)=\frac{1}{\sqrt{4!}}\,
C_{s_1,s_2}^{\scalebox{.6}{ \young(ij)}}\,\,C_{\bar s_1,\bar s_2}^{\overline{\scalebox{.6}{\young(pq)}}}\,,\qquad\qquad
\chi(\pmb L_2,\vec s,\sigma)=\frac{1}{\sqrt{4!}}\,\text{sign}(\sigma)\,
C_{s_1,s_2}^{\scalebox{.6}{ \young(i,j)}}\,\,C_{\bar s_1,\bar s_2}^{\overline{\scalebox{.6}{\young(p,q)}}}\end{align}
Here we used the Clebsch-Gordan coefficients already derived in \eqref{sqcd ex cg1}. We will keep using this notation for the rest of this appendix.

Let us now turn to the three dimensional representation $\tiny\yng(3,1)$ of $S_4$. We choose a basis $\{e_1,e_2,e_3\}$ in which the three Jucys-Murphy elements $(12)$, $(13)+(23)$, $(14)+(24)+(34)$ of $S_4$ have the eigenvalues in table \ref{table: App: 3,1 JM eigen}.
\begin{center}
\begin{table}[H]
\centering
\begin{tabular}{l|c|c|c|} 
 & \footnotesize{$(12)$} & \footnotesize $(13)+(23)$ & \footnotesize $(14)+(24)+(34)$ \\ \hline
$e_1$ & 1 & -1 & 2 \\ \hline
$e_2$ & -1 & 1 & 2 \\ \hline
$ e_3$ & 1 & 2 & -1 \\\hline 
\end{tabular}
\caption{Eigenvalues of the Jucys-Murphy elements $(12)$, $(13)+(23)$, $(14)+(24)+(34)$ on our chosen basis $\{e_1,e_2,e_3\}$ for the standard representation of $S_4$.}
\label{table: App: 3,1 JM eigen}
\end{table}
\end{center}
Alternatively, we can specify our basis choice with the standard Young tableaux
\begin{equation}\label{rep stand S4 def}
e_1\sim \small\young(124,3)\,\,,\qquad
e_2\sim \small\young(134,2)\,\,,\qquad
e_3\sim \small\young(123,4)\,
\end{equation}
%
%	-> this below is not really necessary, it just says that with our basis choice the decomposition below is already manifest
%
%With this basis choice, by restricting $S_4|_{S_3\times S_1}$, the $\tiny\yng(3,1)$ %decomposes as
%\begin{align}
%\left.\yng(3,1)\,\right|_{S_3\times S_1}=\yng(2,1)\otimes \yng(1)\,\,\,\oplus \,\,\,\yng(3)\otimes\yng(1)\,,
%\end{align}
%and the $\tiny\yng(2,1)$ representation of $S_3$ has the explicit form given in \eqref{S3 standard}. 
We now consider the group restriction $\left.S_4\right|_{ S_2\times S_2}=\{(1),(12),(34),(12)(34)\}$. Under this restriction, the $\tiny\yng(3,1)$ decomposes as
\begin{align}\label{App: 3,1 dec}
\left.\yng(3,1)\,\right|_{S_2\times S_2}\,\,= \yng(2)\otimes \yng(2)\,\,\,\,\oplus\,\,\,\, \yng(1,1)\otimes\yng(2)\,\,\,\,\oplus\,\,\,\,\yng(2)\otimes \yng(1,1)
\end{align}
The branching coefficients for this group reduction will then be the matrix elements of the orthogonal operator $B$ such that
\begin{equation}
\begin{array}{lll}
&B^{-1}D^{\scalebox{.2}{\yng(3,1)}}(\,(1)\,)B=\left(
\begin{array}{ccc}
1&0&0\\0&1&0\\0&0&1
\end{array}
\right)\,,\qquad\qquad
&B^{-1}D^{\scalebox{.2}{\yng(3,1)}}(\,(12)\,)B=\left(
\begin{array}{ccc}
1&0&0\\0&-1&0\\0&0&1
\end{array}
\right)\,,\\[14mm]
&B^{-1}D^{\scalebox{.2}{\yng(3,1)}}(\,(34)\,)B=\left(
\begin{array}{ccc}
1&0&0\\0&1&0\\0&0&-1
\end{array}
\right)\,,\qquad\qquad
&B^{-1}D^{\scalebox{.2}{\yng(3,1)}}(\,(12)(34)\,)B=\left(
\begin{array}{ccc}
1&0&0\\0&-1&0\\0&0&-1
\end{array}
\right)
\end{array}
\end{equation}
In our basis choice \eqref{rep stand S4 def} the matrix $B$ reads
\begin{equation}
B=\frac{1}{\sqrt 3}\left(
\begin{array}{ccc}
\sqrt 2&0&-1\\
0&\sqrt 3&0\\
1&0&\sqrt 2
\end{array}
\right)
\end{equation}
The branching coefficient for \eqref{App: 3,1 dec} are then
\begin{equation}
\begin{array}{llll}
&B_{1\rightarrow 1,1}^{\scalebox{.3}{\yng(3,1)}\,\rightarrow\,\scalebox{.3}{\yng(2)}\,,\,\scalebox{.3}{\yng(2)}}=\sqrt\frac{2}{3}\,,\qquad\qquad\qquad
&B_{1\rightarrow 1,1}^{\scalebox{.3}{\yng(3,1)}\,\rightarrow\,\scalebox{.3}{\yng(1,1)}\,,\,\scalebox{.3}{\yng(2)}}=0\,,\qquad\qquad\qquad
%\qquad
&
B_{1\rightarrow 1,1}^{\scalebox{.3}{\yng(3,1)}\,\rightarrow\,\scalebox{.3}{\yng(2)}\,,\,\scalebox{.3}{\yng(1,1)}}=-\frac{1}{\sqrt 3}\,,\\[5mm]
&B_{2\rightarrow 1,1}^{\scalebox{.3}{\yng(3,1)}\,\rightarrow\,\scalebox{.3}{\yng(2)}\,,\,\scalebox{.3}{\yng(2)}}=0\,,\qquad\qquad\qquad
&B_{2\rightarrow 1,1}^{\scalebox{.3}{\yng(3,1)}\,\rightarrow\,\scalebox{.3}{\yng(1,1)}\,,\,\scalebox{.3}{\yng(2)}}=1\,,\qquad\qquad\qquad
%\qquad
&
B_{2\rightarrow 1,1}^{\scalebox{.3}{\yng(3,1)}\,\rightarrow\,\scalebox{.3}{\yng(2)}\,,\,\scalebox{.3}{\yng(1,1)}}=0\,,\\[5mm]
&B_{3\rightarrow 1,1}^{\scalebox{.3}{\yng(3,1)}\,\rightarrow\,\scalebox{.3}{\yng(2)}\,,\,\scalebox{.3}{\yng(2)}}=1\,,\qquad\qquad\qquad
&B_{3\rightarrow 1,1}^{\scalebox{.3}{\yng(3,1)}\,\rightarrow\,\scalebox{.3}{\yng(1,1)}\,,\,\scalebox{.3}{\yng(2)}}=0\,,\qquad\qquad\qquad
%\qquad
&
B_{3\rightarrow 1,1}^{\scalebox{.3}{\yng(3,1)}\,\rightarrow\,\scalebox{.3}{\yng(2)}\,,\,\scalebox{.3}{\yng(1,1)}}=\sqrt\frac{2}{3}\\[3mm]
\end{array}
\end{equation}
We now define the orthogonal projectors 
\begin{align}
&P_{i,j}^{\scalebox{.3}{\yng(3,1)}\,\rightarrow\,\scalebox{.3}{\yng(2)}\,,\,\scalebox{.3}{\yng(2)}}=B_{i\rightarrow 1,1}^{\scalebox{.3}{\yng(3,1)}\,\rightarrow\,\scalebox{.3}{\yng(2)}\,,\,\scalebox{.3}{\yng(2)}}\,B_{j\rightarrow 1,1}^{\scalebox{.3}{\yng(3,1)}\,\rightarrow\,\scalebox{.3}{\yng(2)}\,,\,\scalebox{.3}{\yng(2)}}\,,\qquad\qquad
P_{i,j}^{\scalebox{.3}{\yng(3,1)}\,\rightarrow\,\scalebox{.3}{\yng(1,1)}\,,\,\scalebox{.3}{\yng(2)}}=B_{i\rightarrow 1,1}^{\scalebox{.3}{\yng(3,1)}\,\rightarrow\,\scalebox{.3}{\yng(1,1)}\,,\,\scalebox{.3}{\yng(2)}}\,B_{j\rightarrow 1,1}^{\scalebox{.3}{\yng(3,1)}\,\rightarrow\,\scalebox{.3}{\yng(1,1)}\,,\,\scalebox{.3}{\yng(2)}}\,,\nn
&\qquad\qquad\qquad\qquad\qquad\quad
P_{i,j}^{\scalebox{.3}{\yng(3,1)}\,\rightarrow\,\scalebox{.3}{\yng(2)}\,,\,\scalebox{.3}{\yng(1,1)}}=B_{i\rightarrow 1,1}^{\scalebox{.3}{\yng(3,1)}\,\rightarrow\,\scalebox{.3}{\yng(2)}\,,\,\scalebox{.3}{\yng(1,1)}}\,B_{j\rightarrow 1,1}^{\scalebox{.3}{\yng(3,1)}\,\rightarrow\,\scalebox{.3}{\yng(2)}\,,\,\scalebox{.3}{\yng(1,1)}}
\end{align}
which project the $\tiny \yng(3,1)$ of $S_4$ on the $\tiny \yng(2)\otimes\tiny\yng(2)$, on the $\tiny \yng(1,1)\otimes\tiny\yng(2)$ and on the $\tiny \yng(2)\otimes\tiny\yng(1,1)$ of $S_2\times S_2$ respectively. We also define a fourth operator, that we label $T$, as
\begin{align}\label{App: 3,1 dec matrix T}
T_{i,j}=
B_{i\rightarrow 1,1}^{\scalebox{.3}{\yng(3,1)}\,\rightarrow\,\scalebox{.3}{\yng(2)}\,,\,\scalebox{.3}{\yng(2)}}\,B_{j\rightarrow 1,1}^{\scalebox{.3}{\yng(3,1)}\,\rightarrow\,\scalebox{.3}{\yng(2)}\,,\,\scalebox{.3}{\yng(1,1)}}
\end{align}
These matrices explicitly read
\begin{equation}\label{App 222 projector for 31}
\begin{array}{lll}
&P^{\scalebox{.3}{\yng(3,1)}\,\rightarrow\,\scalebox{.3}{\yng(2)}\,,\,\scalebox{.3}{\yng(2)}}=\frac{1}{3}\left(
\begin{array}{ccc}
2&0&\sqrt 2\\
0&0&0\\
\sqrt 2&0&1
\end{array}
\right)\,,\qquad\quad\qquad\quad
&P^{\scalebox{.3}{\yng(3,1)}\,\rightarrow\,\scalebox{.3}{\yng(1,1)}\,,\,\scalebox{.3}{\yng(2)}}=\left(
\begin{array}{ccc}
0&0&0\\
0&1&0\\
0&0&0\\
\end{array}
\right)\,,\\[13mm]
&P^{\scalebox{.3}{\yng(3,1)}\,\rightarrow\,\scalebox{.3}{\yng(2)}\,,\,\scalebox{.3}{\yng(1,1)}}=\frac{1}{3}\left(
\begin{array}{ccc}
1&0&-\sqrt 2\\
0&0&0\\
-\sqrt 2&0&2
\end{array}
\right)\,,\qquad\quad
&T=\frac{1}{3}\left(
\begin{array}{ccc}
-\sqrt 2&0& 2\\
0&0&0\\
-1&0&\sqrt 2
\end{array}
\right)
\end{array}
\end{equation}

The quiver character for $\cO(\pmb L_{3})$, $\cO(\pmb L_{4})$, $\cO(\pmb L_5)$, $\cO(\pmb L_{6})$, $\cO(\pmb L_{7})$ are then
\begin{align}
&\chi(\pmb L_3, \vec s, \sigma)=\frac{1}{2\sqrt 2}\,\,\Tr\left[D^{\scalebox{.2}{\yng(3,1)}}(\sigma)\,P^{\scalebox{.2}{\yng(3,1)}\,\rightarrow\,\scalebox{.2}{\yng(2)}\,,\,\scalebox{.2}{\yng(2)}}\right]\,\,C_{s_1,s_2}^{\scalebox{.6}{ \young(ij)}}\,\,C_{\bar s_1,\bar s_2}^{\overline{\scalebox{.6}{\young(pq)}}}\,,\nn
&\chi(\pmb L_4, \vec s, \sigma)=\frac{1}{2\sqrt 2}\,\,\Tr\left[D^{\scalebox{.2}{\yng(3,1)}}(\sigma)\,P^{\scalebox{.2}{\yng(3,1)}\,\rightarrow\,\scalebox{.2}{\yng(1,1)}\,,\,\scalebox{.2}{\yng(2)}}\right]\,\,C_{s_1,s_2}^{\scalebox{.6}{ \young(ij)}}\,\,C_{\bar s_1,\bar s_2}^{\overline{\scalebox{.6}{\young(pq)}}}\,,\nn
&\chi(\pmb L_5, \vec s, \sigma)=\frac{1}{2\sqrt 2}\,\,\Tr\left[D^{\scalebox{.2}{\yng(3,1)}}(\sigma)\,P^{\scalebox{.2}{\yng(3,1)}\,\rightarrow\,\scalebox{.2}{\yng(2)}\,,\,\scalebox{.2}{\yng(1,1)}}\right]\,\,C_{s_1,s_2}^{\scalebox{.6}{ \young(i,j)}}\,\,C_{\bar s_1,\bar s_2}^{\overline{\scalebox{.6}{\young(p,q)}}}\,,\displaybreak[0]\\[3mm]
&\chi(\pmb L_6, \vec s, \sigma)=\frac{1}{2\sqrt 2}\,\,\Tr\left[D^{\scalebox{.2}{\yng(3,1)}}(\sigma)\,T\right]\,\,C_{s_1,s_2}^{\scalebox{.6}{ \young(ij)}}\,\,C_{\bar s_1,\bar s_2}^{\overline{\scalebox{.6}{\young(p,q)}}}\,,\nn
&\chi(\pmb L_7, \vec s, \sigma)=\frac{1}{2\sqrt 2}\,\,\Tr\left[D^{\scalebox{.2}{\yng(3,1)}}(\sigma)\,T^t\right]\,\,C_{s_1,s_2}^{\scalebox{.6}{ \young(i,j)}}\,\,C_{\bar s_1,\bar s_2}^{\overline{\scalebox{.6}{\young(pq)}}}\nonumber
\end{align}
Here $T^t$ is the transpose of the matrix $T$ in \eqref{App: 3,1 dec matrix T}.

We now focus on the $\tiny\yng(2,1,1)$ representation of $S_4$. This representation can be obtained by tensoring together the standard and the sign representation of $S_4$:
\begin{align}\label{rep stand x alt S4 def}
\yng(2,1,1)=\yng(3,1)\otimes \yng(1,1,1,1)
\end{align}
In the following, we will continue to use \eqref{rep stand S4 def} as our basis choice for the standard representation $\tiny\yng(3,1)$.
Under the group restriction $\left.S_4\right|_{ S_2\times S_2}=\{(1),(12),(34),(12)(34)\}$, the $\tiny\yng(2,1,1)$ decomposes as
\begin{align}\label{App: 2,1,1 dec}
\left.\yng(2,1,1)\,\right|_{S_2\times S_2}\,\,= \yng(2)\otimes \yng(1,1)\,\,\,\,
\oplus\,\,\,\,
\yng(1,1)\otimes \yng(2)\,\,\,\,\oplus\,\,\,\, \yng(1,1)\otimes\yng(1,1)
\end{align}
As in the previous instance, the branching coefficients for this group reduction are the matrix elements of the orthogonal operator $B$, such that
\begin{equation}
\begin{array}{lll}
&B^{-1}D^{\scalebox{.2}{\yng(2,1,1)}}(\,(1)\,)B=\left(
\begin{array}{ccc}
1&0&0\\0&1&0\\0&0&1
\end{array}
\right)\,,\qquad\qquad
&B^{-1}D^{\scalebox{.2}{\yng(2,1,1)}}(\,(12)\,)B=\left(
\begin{array}{ccc}
1&0&0\\0&-1&0\\0&0&-1
\end{array}
\right)\,,\\[14mm]
&B^{-1}D^{\scalebox{.2}{\yng(2,1,1)}}(\,(34)\,)B=\left(
\begin{array}{ccc}
-1&0&0\\0&1&0\\0&0&-1
\end{array}
\right)\,,\qquad\qquad
&B^{-1}D^{\scalebox{.2}{\yng(2,1,1)}}(\,(12)(34)\,)B=\left(
\begin{array}{ccc}
-1&0&0\\0&-1&0\\0&0&1
\end{array}
\right)
\end{array}
\end{equation}
In our basis choice, the matrix $B$ reads
\begin{equation}
B=\frac{1}{\sqrt 3}\left(
\begin{array}{ccc}
0&-1&\sqrt 2\\
\sqrt 3&0&0\\
0&\sqrt 2&1
\end{array}
\right)
\end{equation}
The branching coefficient for \eqref{App: 2,1,1 dec} are thus
\begin{equation}
\begin{array}{llll}
&B_{1\rightarrow 1,1}^{\scalebox{.3}{\yng(2,1,1)}\,\rightarrow\,\scalebox{.3}{\yng(2)}\,,\,\scalebox{.3}{\yng(1,1)}}=0\,,\qquad\qquad\qquad
&B_{1\rightarrow 1,1}^{\scalebox{.3}{\yng(2,1,1)}\,\rightarrow\,\scalebox{.3}{\yng(1,1)}\,,\,\scalebox{.3}{\yng(2)}}=-\frac{1}{\sqrt 3}\,,\qquad\qquad\qquad
%\qquad
&
B_{1\rightarrow 1,1}^{\scalebox{.3}{\yng(2,1,1)}\,\rightarrow\,\scalebox{.3}{\yng(1,1)}\,,\,\scalebox{.3}{\yng(1,1)}}=\sqrt\frac{2}{3}\,,\\[5mm]
&B_{2\rightarrow 1,1}^{\scalebox{.3}{\yng(2,1,1)}\,\rightarrow\,\scalebox{.3}{\yng(2)}\,,\,\scalebox{.3}{\yng(1,1)}}=1\,,\qquad\qquad\qquad
&B_{2\rightarrow 1,1}^{\scalebox{.3}{\yng(2,1,1)}\,\rightarrow\,\scalebox{.3}{\yng(1,1)}\,,\,\scalebox{.3}{\yng(2)}}=0\,,\qquad\qquad\qquad
%\qquad
&
B_{2\rightarrow 1,1}^{\scalebox{.3}{\yng(2,1,1)}\,\rightarrow\,\scalebox{.3}{\yng(1,1)}\,,\,\scalebox{.3}{\yng(1,1)}}=0\,,\\[5mm]
&B_{3\rightarrow 1,1}^{\scalebox{.3}{\yng(2,1,1)}\,\rightarrow\,\scalebox{.3}{\yng(2)}\,,\,\scalebox{.3}{\yng(1,1)}}=0\,,\qquad\qquad\qquad
&B_{3\rightarrow 1,1}^{\scalebox{.3}{\yng(2,1,1)}\,\rightarrow\,\scalebox{.3}{\yng(1,1)}\,,\,\scalebox{.3}{\yng(2)}}=\sqrt\frac{2}{3}\,,\qquad\qquad\qquad
%\qquad
&
B_{3\rightarrow 1,1}^{\scalebox{.3}{\yng(2,1,1)}\,\rightarrow\,\scalebox{.3}{\yng(1,1)}\,,\,\scalebox{.3}{\yng(1,1)}}=\frac{1}{\sqrt 3}\\[3mm]
\end{array}
\end{equation}
Closely following the procedure of the previous paragraph, we define the orthogonal projectors 
\begin{align}
&P_{i,j}^{\scalebox{.3}{\yng(2,1,1)}\,\rightarrow\,\scalebox{.3}{\yng(2)}\,,\,\scalebox{.3}{\yng(1,1)}}=B_{i\rightarrow 1,1}^{\scalebox{.3}{\yng(2,1,1)}\,\rightarrow\,\scalebox{.3}{\yng(2)}\,,\,\scalebox{.3}{\yng(1,1)}}\,B_{j\rightarrow 1,1}^{\scalebox{.3}{\yng(2,1,1)}\,\rightarrow\,\scalebox{.3}{\yng(2)}\,,\,\scalebox{.3}{\yng(1,1)}}\,,\qquad\qquad
P_{i,j}^{\scalebox{.3}{\yng(2,1,1)}\,\rightarrow\,\scalebox{.3}{\yng(1,1)}\,,\,\scalebox{.3}{\yng(2)}}=B_{i\rightarrow 1,1}^{\scalebox{.3}{\yng(2,1,1)}\,\rightarrow\,\scalebox{.3}{\yng(1,1)}\,,\,\scalebox{.3}{\yng(2)}}\,B_{j\rightarrow 1,1}^{\scalebox{.3}{\yng(2,1,1)}\,\rightarrow\,\scalebox{.3}{\yng(1,1)}\,,\,\scalebox{.3}{\yng(2)}}\,,\nn
&\qquad\qquad\qquad\qquad\qquad\quad
P_{i,j}^{\scalebox{.3}{\yng(2,1,1)}\,\rightarrow\,\scalebox{.3}{\yng(1,1)}\,,\,\scalebox{.3}{\yng(1,1)}}=B_{i\rightarrow 1,1}^{\scalebox{.3}{\yng(2,1,1)}\,\rightarrow\,\scalebox{.3}{\yng(1,1)}\,,\,\scalebox{.3}{\yng(1,1)}}\,B_{j\rightarrow 1,1}^{\scalebox{.3}{\yng(2,1,1)}\,\rightarrow\,\scalebox{.3}{\yng(1,1)}\,,\,\scalebox{.3}{\yng(1,1)}}
\end{align}
These operators project the $\tiny \yng(2,1,1)$ of $S_4$ on the $\tiny \yng(2)\otimes\tiny\yng(1,1)$, on the $\tiny \yng(1,1)\otimes\tiny\yng(2)$ and on the $\tiny \yng(1,1)\otimes\tiny\yng(1,1)$ of $S_2\times S_2$ respectively. We also introduce the operator $V$:
\begin{align}\label{App: 3,1 dec matrix V}
V_{i,j}=
B_{i\rightarrow 1,1}^{\scalebox{.3}{\yng(2,1,1)}\,\rightarrow\,\scalebox{.3}{\yng(1,1)}\,,\,\scalebox{.3}{\yng(2)}}\,B_{j\rightarrow 1,1}^{\scalebox{.3}{\yng(2,1,1)}\,\rightarrow\,\scalebox{.3}{\yng(1,1)}\,,\,\scalebox{.3}{\yng(1,1)}}
\end{align}
These matrices explicitly read
\begin{equation}
\begin{array}{lll}
&P^{\scalebox{.3}{\yng(2,1,1)}\,\rightarrow\,\scalebox{.3}{\yng(2)}\,,\,\scalebox{.3}{\yng(1,1)}}=\left(
\begin{array}{ccc}
0&0&0\\
0&1&0\\
0&0&0\\
\end{array}
\right)\,,\qquad\quad\qquad\quad
&P^{\scalebox{.3}{\yng(2,1,1)}\,\rightarrow\,\scalebox{.3}{\yng(1,1)}\,,\,\scalebox{.3}{\yng(2)}}=\frac{1}{3}\left(
\begin{array}{ccc}
1&0&-\sqrt 2\\
0&0&0\\
-\sqrt 2&0&2
\end{array}
\right)\,,\\[13mm]
&P^{\scalebox{.3}{\yng(2,1,1)}\,\rightarrow\,\scalebox{.3}{\yng(1,1)}\,,\,\scalebox{.3}{\yng(1,1)}}=\frac{1}{3}\left(
\begin{array}{ccc}
 2&0&\sqrt 2\\
0&0&0\\
-\sqrt 2&0& 1
\end{array}
\right)\,,\qquad\quad
&V=\frac{1}{3}\left(
\begin{array}{ccc}
-\sqrt 2&0&-1\\
0&0&0\\
2&0&\sqrt 2
\end{array}
\right)
\end{array}
\end{equation}
Notice that $V=T^t$, where $T$ is the matrix defined in \eqref{App 222 projector for 31}. The quiver character for $\cO(\pmb L_{8})$, $\cO(\pmb L_{9})$, $\cO(\pmb L_{10})$, $\cO(\pmb L_{11})$, $\cO(\pmb L_{12})$ are therefore
\begin{align}
&\chi(\pmb L_8, \vec s, \sigma)=\frac{1}{2\sqrt 2}\,\,\Tr\left[D^{\scalebox{.2}{\yng(2,1,1)}}(\sigma)\,P^{\scalebox{.2}{\yng(2,1,1)}\,\rightarrow\,\scalebox{.2}{\yng(2)}\,,\,\scalebox{.2}{\yng(1,1)}}\right]\,\,C_{s_1,s_2}^{\scalebox{.6}{ \young(i,j)}}\,\,C_{\bar s_1,\bar s_2}^{\overline{\scalebox{.6}{\young(p,q)}}}\,,\nn
&\chi(\pmb L_9, \vec s, \sigma)=\frac{1}{2\sqrt 2}\,\,\Tr\left[D^{\scalebox{.2}{\yng(2,1,1)}}(\sigma)\,P^{\scalebox{.2}{\yng(2,1,1)}\,\rightarrow\,\scalebox{.2}{\yng(1,1)}\,,\,\scalebox{.2}{\yng(2)}}\right]\,\,C_{s_1,s_2}^{\scalebox{.6}{ \young(ij)}}\,\,C_{\bar s_1,\bar s_2}^{\overline{\scalebox{.6}{\young(pq)}}}\,,\displaybreak[0]\\[3mm]
&\chi(\pmb L_{10}, \vec s, \sigma)=\frac{1}{2\sqrt 2}\,\,\Tr\left[D^{\scalebox{.2}{\yng(2,1,1)}}(\sigma)\,P^{\scalebox{.2}{\yng(2,1,1)}\,\rightarrow\,\scalebox{.2}{\yng(1,1)}\,,\,\scalebox{.2}{\yng(1,1)}}\right]\,\,C_{s_1,s_2}^{\scalebox{.6}{ \young(i,j)}}\,\,C_{\bar s_1,\bar s_2}^{\overline{\scalebox{.6}{\young(p,q)}}}\,,\nn
&\chi(\pmb L_{11}, \vec s, \sigma)=\frac{1}{2\sqrt 2}\,\,\Tr\left[D^{\scalebox{.2}{\yng(2,1,1)}}(\sigma)\,V\right]\,\,C_{s_1,s_2}^{\scalebox{.6}{ \young(ij)}}\,\,C_{\bar s_1,\bar s_2}^{\overline{\scalebox{.6}{\young(p,q)}}}\,,\nn
&\chi(\pmb L_{12}, \vec s, \sigma)=\frac{1}{2\sqrt 2}\,\,\Tr\left[D^{\scalebox{.2}{\yng(2,1,1)}}(\sigma)\,V^t\right]\,\,C_{s_1,s_2}^{\scalebox{.6}{ \young(i,j)}}\,\,C_{\bar s_1,\bar s_2}^{\overline{\scalebox{.6}{\young(pq)}}}\nonumber
\end{align}

Two operators still remain. They can be obtained by considering the $S_4$ $\tiny\yng(2,2)$ representation branching
\begin{align}\label{App: 2,2 dec}
\left.\yng(2,2)\,\right|_{S_2\times S_2}\,\,= \yng(2)\otimes \yng(2)\,\,\,\,\oplus\,\,\,\, \yng(1,1)\otimes\yng(1,1)
\end{align}
The $\tiny\yng(2,2)$ representation of $S_4$ is really a representation of the quotient group $S_4/\{(1),\break (12)(34),(13)(24),(14)(23)\}$, which in turn is isomorphic to $S_3$. This representation is thus just the standard representation of $S_3$ pulled back to $S_4$ via this quotient \cite{FulHar}. We choose a basis $\{e_1,e_2\}$ in which the Jucys-Murphy elements $(12)$, $(13)+(23)$, $(14)+(24)+(34)$ of $S_4$ have the eigenvalues in table \ref{table: App: 2,2 JM eigen}. 
\begin{center}
\begin{table}[H]
\centering
\begin{tabular}{l|c|c|c|} 
 & \footnotesize{$(12)$} & \footnotesize $(13)+(23)$ & \footnotesize $(14)+(24)+(34)$ \\ \hline
$e_1$ & 1 & -1 & 0 \\ \hline
$e_2$ & -1 & 1 & 0 \\ \hline
\end{tabular}
\caption{Eigenvalues of the Jucys-Murphy elements $(12)$, $(13)+(23)$, $(14)+(24)+(34)$ on our chosen basis $\{e_1,e_2\}$ for the two-dimensional representation of $S_4$.}
\label{table: App: 2,2 JM eigen}
\end{table}
\end{center}
The standard Young tableaux labelling of this basis is
\begin{equation}\label{rep 2,2 S4 def}
e_1\sim \small\young(12,34)\,\,,\qquad\qquad
e_2\sim \small\young(13,24)\,
\end{equation}
An explicit representation of $\tiny\yng(2,2)$ is therefore obtained by considering the set of matrices
\begin{equation}
\begin{array}{ll}
&D^{\scalebox{.3}{\yng(2,2)}}(\,(1)\,)=
D^{\scalebox{.3}{\yng(2,2)}}(\,(12)(34)\,)=
D^{\scalebox{.3}{\yng(2,2)}}(\,(13)(24)\,)=
D^{\scalebox{.3}{\yng(2,2)}}(\,(14)(23)\,)=
\left(
\begin{array}{cc}
1&0\\0&1
\end{array}
\right)\,,\\[6mm]
&D^{\scalebox{.3}{\yng(2,2)}}(\,(12)\,)=
D^{\scalebox{.3}{\yng(2,2)}}(\,(34)\,)=
D^{\scalebox{.3}{\yng(2,2)}}(\,(1324)\,)=
D^{\scalebox{.3}{\yng(2,2)}}(\,(1423)\,)=
\left(
\begin{array}{cc}
1&0\\0&-1
\end{array}
\right)\,,\\[6mm]
&D^{\scalebox{.3}{\yng(2,2)}}(\,(13)\,)=
D^{\scalebox{.3}{\yng(2,2)}}(\,(24)\,)=
D^{\scalebox{.3}{\yng(2,2)}}(\,(1234)\,)=
D^{\scalebox{.3}{\yng(2,2)}}(\,(1432)\,)=
\left(
\begin{array}{cc}
 -\frac{1}{2} & -\frac{\sqrt{3}}{2} \\
 -\frac{\sqrt{3}}{2} & \frac{1}{2} \\
\end{array}
\right)\,,\\[6mm]
&D^{\scalebox{.3}{\yng(2,2)}}(\,(23)\,)=
D^{\scalebox{.3}{\yng(2,2)}}(\,(14)\,)=
D^{\scalebox{.3}{\yng(2,2)}}(\,(1342)\,)=
D^{\scalebox{.3}{\yng(2,2)}}(\,(1243)\,)=
\left(
\begin{array}{cc}
 -\frac{1}{2} & \frac{\sqrt{3}}{2} \\
 \frac{\sqrt{3}}{2} & \frac{1}{2} \\
\end{array}
\right)\,,\\[6mm]
&D^{\scalebox{.3}{\yng(2,2)}}(\,(123)\,)=
D^{\scalebox{.3}{\yng(2,2)}}(\,(243)\,)=
D^{\scalebox{.3}{\yng(2,2)}}(\,(142)\,)=
D^{\scalebox{.3}{\yng(2,2)}}(\,(134)\,)=
\left(
\begin{array}{cc}
 -\frac{1}{2} & -\frac{\sqrt{3}}{2} \\
 \frac{\sqrt{3}}{2} & -\frac{1}{2} \\
\end{array}
\right)\,,\\[6mm]
&D^{\scalebox{.3}{\yng(2,2)}}(\,(132)\,)=
D^{\scalebox{.3}{\yng(2,2)}}(\,(143)\,)=
D^{\scalebox{.3}{\yng(2,2)}}(\,(234)\,)=
D^{\scalebox{.3}{\yng(2,2)}}(\,(124)\,)=
\left(
\begin{array}{cc}
 -\frac{1}{2} & \frac{\sqrt{3}}{2} \\
 -\frac{\sqrt{3}}{2} & -\frac{1}{2} \\
\end{array}
\right)
\end{array}
\end{equation}
With this basis choice, under the group restriction $\left.S_4\right|_{ S_2\times S_2}=\{(1),(12),(34),(12)(34)\}$, we have
\begin{equation}
\begin{array}{lll}
&D^{\scalebox{.2}{\yng(2,2)}}(\,(1)\,)=\left(
\begin{array}{cc}
1&0\\0&1
\end{array}
\right)\,,\qquad\qquad
&D^{\scalebox{.2}{\yng(2,2)}}(\,(12)\,)=\left(
\begin{array}{cc}
1&0\\0&-1
\end{array}
\right)\,,\\[14mm]
&D^{\scalebox{.2}{\yng(2,2)}}(\,(34)\,)=\left(
\begin{array}{cc}
1&0\\0&-1
\end{array}
\right)\,,\qquad\qquad
&D^{\scalebox{.2}{\yng(2,2)}}(\,(12)(34)\,)=\left(
\begin{array}{cc}
1&0\\0&1
\end{array}
\right)
\end{array}
\end{equation}
The decomposition \eqref{App: 2,2 dec} is already manifest. The branching coefficients for this reduction are then
\begin{align}
B_{j\rightarrow 1,1}^{\scalebox{.3}{\yng(2,2)}\,\rightarrow\,\scalebox{.3}{\yng(2)}\,,\,\scalebox{.3}{\yng(2)}}=\delta_{j,1}\,,\qquad\qquad\quad
B_{j\rightarrow 1,1}^{\scalebox{.3}{\yng(2,2)}\,\rightarrow\,\scalebox{.3}{\yng(1,1)}\,,\,\scalebox{.3}{\yng(1,1)}}=\delta_{j,2}\,,\qquad\quad j=1,2
\end{align}
We can now write the orthogonal projectors
\begin{equation}
\begin{array}{llll}
&P_{i,j}^{\scalebox{.3}{\yng(2,2)}\,\rightarrow\,\scalebox{.3}{\yng(2)}\,,\,\scalebox{.3}{\yng(2)}}=B_{i\rightarrow 1,1}^{\scalebox{.3}{\yng(2,2)}\,\rightarrow\,\scalebox{.3}{\yng(2)}\,,\,\scalebox{.3}{\yng(2)}}\,B_{j\rightarrow 1,1}^{\scalebox{.3}{\yng(2,2)}\,\rightarrow\,\scalebox{.3}{\yng(2)}\,,\,\scalebox{.3}{\yng(2)}}\qquad&\longrightarrow\qquad P^{\scalebox{.3}{\yng(2,2)}\,\rightarrow\,\scalebox{.3}{\yng(2)}\,,\,\scalebox{.3}{\yng(2)}}&=
\left(
\begin{array}{cc}
1&0\\0&0
\end{array}
\right)\,,
\\[9mm]
&P_{i,j}^{\scalebox{.3}{\yng(2,2)}\,\rightarrow\,\scalebox{.3}{\yng(1,1)}\,,\,\scalebox{.3}{\yng(1,1)}}=B_{i\rightarrow 1,1}^{\scalebox{.3}{\yng(2,2)}\,\rightarrow\,\scalebox{.3}{\yng(1,1)}\,,\,\scalebox{.3}{\yng(1,1)}}\,B_{j\rightarrow 1,1}^{\scalebox{.3}{\yng(2,2)}\,\rightarrow\,\scalebox{.3}{\yng(1,1)}\,,\,\scalebox{.3}{\yng(1,1)}}\qquad&\longrightarrow\qquad P^{\scalebox{.3}{\yng(2,2)}\,\rightarrow\,\scalebox{.3}{\yng(1,1)}\,,\,\scalebox{.3}{\yng(1,1)}}&=
\left(
\begin{array}{cc}
0&0\\0&1
\end{array}
\right)
\end{array}
\end{equation}
projecting the $\tiny \yng(2,2)$ of $S_4$ on the $\tiny \yng(2)\otimes\tiny\yng(2)$ and on the $\tiny \yng(1,1)\otimes\tiny\yng(1,1)$ of $S_2\times S_2$ respectively.
The quiver characters for the remaining two operators, $\cO(\pmb L_{13})$ and $\cO(\pmb L_{14})$, are then
\begin{align}
&\chi(\pmb L_{13}, \vec s, \sigma)=\frac{1}{2\sqrt 3}\,\,\Tr\left[D^{\scalebox{.2}{\yng(2,2)}}(\sigma)\,P^{\scalebox{.2}{\yng(2,2)}\,\rightarrow\,\scalebox{.2}{\yng(2)}\,,\,\scalebox{.2}{\yng(2)}}\right]\,\,C_{s_1,s_2}^{\scalebox{.6}{ \young(ij)}}\,\,C_{\bar s_1,\bar s_2}^{\overline{\scalebox{.6}{\young(pq)}}}\,,\nn
&\chi(\pmb L_{14}, \vec s, \sigma)=\frac{1}{2\sqrt 3}\,\,\Tr\left[D^{\scalebox{.2}{\yng(2,2)}}(\sigma)\,P^{\scalebox{.2}{\yng(2,2)}\,\rightarrow\,\scalebox{.2}{\yng(1,1)}\,,\,\scalebox{.2}{\yng(1,1)}}\right]\,\,C_{s_1,s_2}^{\scalebox{.6}{ \young(i,j)}}\,\,C_{\bar s_1,\bar s_2}^{\overline{\scalebox{.6}{\young(p,q)}}}
\end{align}

\bibliographystyle{utphys}
\bibliography{biblio_3pf.bib}

\end{document}